\newcommand{\printstyle}{reprint}

\newcommand{\thirdheight}{4.14cm}
\newcommand{\figsep}{\qquad}
\newcommand{\figsepsm}{\figsep}

\documentclass[aps,pra,superscriptaddress,floatfix,\printstyle ,nofootinbib]{revtex4-2}

\usepackage{color}

\usepackage[caption=false]{subfig}
\usepackage{graphicx,booktabs,array}
\usepackage{multirow}
\usepackage{tabularx} 
\DeclareGraphicsExtensions{.pdf,.png,.jpg,.eps,.ps}
\usepackage{epsfig}
\usepackage{epstopdf}
\usepackage{amssymb}
\usepackage{amsmath}
\usepackage{amsfonts}
\usepackage{mathrsfs}
\usepackage{amsthm}
\usepackage{float}
\usepackage{amsbsy}
\usepackage{bm}
\usepackage{grffile}
\usepackage{hyperref}
\hypersetup{
    colorlinks = true,
    citecolor = blue,
    urlcolor = blue,
    linkcolor = blue,
}
\usepackage{courier}
\usepackage{upgreek}
\usepackage{xspace}
\usepackage{xcolor}
\usepackage{calc} 
\usepackage{soul} 

\newcommand{\eg}{\textit{e}.\textit{g}. }
\newcommand{\etc}{\textit{etc.}\xspace} 

\newcommand{\Alfven}{Alfv\'{e}n\xspace}
\newcommand{\Alfvenic}{Alfv\'{e}nic\xspace}

\newcommand{\vs}{v_S}

\newcommand{\fci}{f_{ci}}

\newcommand{\betae}{\beta_e}

\newcommand{\vtherme}{v_{th,e}}

\newcommand{\Te}{T_e}
\newcommand{\Ti}{T_i}

\newcommand{\betap}{\beta_p}
\newcommand{\exb}{E \cross B}
\newcommand{\nustar}{\nu_*}
\newcommand{\ptsolver}{PT\_SOLVER\xspace}

\newcommand{\betan}{\beta_N}
\newcommand{\LTe}{L_{T_e}}
\newcommand{\taue}{\tau_E}

\newcommand{\teti}{\Te/\Ti}

\newcommand{\Qie}{Q_{ie}}

\newcommand{\ten}[1]{\cdot 10^{#1}}
\newcommand{\abs}[1]{\left|#1\right|}
\let\overdot\dot
\renewcommand{\dot}{\cdot}
\newcommand{\cross}{\times}

\newcommand{\grad}{\nabla}

\newcommand{\like}{\sim}

\newcommand{\approptoinn}[2]{\mathrel{\vcenter{
  \offinterlineskip\halign{\hfil$##$\cr
    #1\propto\cr\noalign{\kern2pt}#1\sim\cr\noalign{\kern-2pt}}}}}

\newcommand{\tto}{\text{ to }}

\newcommand{\pderiv}[2]{\frac{\partial #1}{\partial #2}}

\newcommand{\figref}[1]{Fig.\xspace\ref{#1}}
\renewcommand{\eqref}[1]{Eq.\xspace\ref{#1}}
\newcommand{\secref}[1]{Sec.\xspace\ref{#1}}
\newcommand{\citeref}[1]{Ref.\xspace\onlinecite{#1}}

\newcommand{\tabref}[1]{Table\xspace\ref{#1}}

\newcommand{\myname}{J.B. Lestz}

\newcommand{\Stan}{S.M. Kaye}
\newcommand{\Kathreen}{K.E. Thome}
\newcommand{\Galina}{G. Avdeeva}
\newcommand{\Joey}{J. McClenaghan}
\newcommand{\Federico}{F.D. Halpern}
\newcommand{\Alexei}{A.Y. Pankin}
\newcommand{\Marina}{M.V. Gorelenkova}

\newcommand{\PPPL}{Princeton Plasma Physics Lab, Princeton, NJ 08543, USA}

\newcommand{\GA}{General Atomics, San Diego, CA, 92121, USA}

\newcommand{\Podesta}{Podest{\`a}}

\definecolor{darkgreen}{rgb}{0,0.5,0}

\newcommand{\gadisclaimer}{This report was prepared as an account of work sponsored by an agency of the United States Government. Neither the United States Government nor any agency thereof, nor any of their employees, makes any warranty, express or implied, or assumes any legal liability or responsibility for the accuracy, completeness, or usefulness of any information, apparatus, product, or process disclosed, or represents that its use would not infringe privately owned rights. Reference herein to any specific commercial product, process, or service by trade name, trademark, manufacturer, or otherwise does not necessarily constitute or imply its endorsement, recommendation, or favoring by the United States Government or any agency thereof. The views and opinions of authors expressed herein do not necessarily state or reflect those of the United States Government or any agency thereof.}
\renewcommand{\vs}{{\xspace}versus\xspace}

\newcommand{\ktext}[1]{\textcolor{black}{#1}}
\newcommand{\rev}[1]{\ktext{#1}}

\newcommand{\twotab}[2]{\begin{tabular}{c} #1 \\ #2 \end{tabular}}

\begin{document}

\title{Sensitivities of time-dependent temperature profile predictions for NSTX with the Multi-Mode Model}

\author{\myname}
\email{lestzj@fusion.gat.com}
\affiliation{\GA}
\author{\Galina}
\affiliation{\GA}
\author{\Stan}
\affiliation{\PPPL}
\author{\Marina}
\affiliation{\PPPL}
\author{\Federico}
\affiliation{\GA}
\author{\Joey}
\affiliation{\GA}
\author{\Alexei}
\affiliation{\PPPL}
\author{\Kathreen}
\affiliation{\GA}
\date{\today}
\begin{abstract}

The Multi-Mode Model (MMM) for turbulent transport was applied to a large set of well-analyzed discharges from the National Spherical Torus Experiment (NSTX) in order to evaluate its sensitivities to a wide range of plasma conditions. MMM calculations were performed for hundreds of milliseconds in each discharge by performing time-dependent predictive simulations with the 1.5D tokamak integrated modeling code TRANSP. A closely related study \cite{Lestz2025pre1} concluded that MMM predicted electron and ion temperature profiles that were in reasonable agreement with NSTX observations, generally outperforming a different reduced transport model, TGLF. This finding motivates the more thorough investigation of the characteristics of the MMM predictions conducted in this work. The simulations with MMM have electron energy transport dominated by electron temperature gradient modes in the examined discharges with relatively low plasma $\beta$ (ratio of kinetic plasma pressure to magnetic field pressure) and high collisionality, transitioning to a mixture of different modes for higher $\beta$ and lower collisionality. The thermal ion diffusivity predicted by MMM is much smaller than the neoclassical contribution, in line with previous experimental analysis of NSTX. Nonetheless, the electron and ion temperature profiles are coupled via collisional energy exchange and thus sensitive to which transport channels are predicted. The time-dependent simulations with MMM are robust to the simulation start time, converging to remarkably similar temperature profiles later during the discharge. MMM typically overpredicts confinement relative to NSTX observations, leading to the prediction of overly steep temperature profiles. Plasmas with spatially broader temperature profiles, higher plasma $\beta$, and longer energy confinement times tend to be predicted by MMM with better agreement with the experiment. These findings provide useful context for understanding the regime-dependent tendencies of MMM in anticipation of self-consistent, time-dependent predictive simulations of NSTX-U discharges with these same modeling tools. 

\end{abstract}
\maketitle

\section{Introduction}
\label{sec:intro}

The spherical tokamak concept is an appealing design for a future fusion reactor due to its characteristic high power density, favorable confinement, and reduced construction cost relative to a large aspect ratio tokamak \cite{Menard2011NF,Menard2016NF,Menard2019RSA,Menard2022NF}. The National Spherical Torus Experiment (NSTX) \cite{Ono2000NF,Sabbagh2013NF} and its subsequent upgrade (NSTX-U) \cite{Menard2012NF,Menard2017NF,Berkery2024NF} were designed to improve the understanding of key spherical tokamak physics issues on the path to economical commercial fusion energy production \cite{Buxton2019PPCF,Wilson2020book,Waldon2024PTRSA,Meyer2024PTRA,Kingham2024POP}. Spherical tokamaks operate at much higher $\beta$ (ratio of kinetic plasma pressure to magnetic field pressure) than modern conventional tokamaks \cite{Ono2015POP} and 
\rev{have observed enhanced confinement scaling at low $\nustar$ (electron collisionality normalized to the bounce frequency) in multiple devices}  \cite{Kaye2007NF,Valovic2009NF,Valovic2011NF,Kaye2013NF,Kurskiev2019NF,Kurskiev2022NF,Kaye2021PPCF}. Hence, we are motivated to understand the sensitivity of transport models to these parameters to better interpret physics experiments and forecast to future devices. In particular, microtearing modes (MTMs) and kinetic ballooning modes (KBMs) are expected to drive stronger transport in the spherical tokamak parameter regime than in conventional tokamaks 
\cite{Guttenfelder2012POP,Guttenfelder2013NF,Kaye2014POP,Clauser2022POP,Patel2022NF,McClenaghan2023POP,Kennedy2023NF,Kennedy2024NF,Giacomin2024PPCF,Dominski2024POP,McClenaghan2025PPCF,Singh2025NF},  while ion temperature gradient modes (ITGs) and trapped electron modes (TEMs) are strongly suppressed by larger $\exb$ flow shear in spherical tokamaks \cite{Roach2009PPCF,Kaye2021PPCF}. 
\rev{Recent work has demonstrated that the electron thermal transport can exhibit a strong sensitivity to $\exb$ shear in spherical tokamaks \cite{Patel2025NF,Avdeeva2025pre}.}

Full discharge simulations are enabled by computationally inexpensive, reduced, physics-based or surrogate models which can provide many orders of magnitude speedup over first principles simulations. Thus, it is important to validate such models against a wide range of experimental conditions to understand their capabilities and limitations, such that they can be relied upon when extrapolating to new parameter regimes, plasma scenarios, or even new devices. Time-dependent validation of reduced transport models, as opposed to time slice analysis, is especially important for NSTX since the plasma profiles often evolved for much of the discharge. Long pulse reactors also require reliable time-dependent integrated transport modeling in order to optimize the trajectory to the high performing steady state and reliably model the controlled ramp down phase. 

This work entails a detailed examination of the time-dependent accuracy of the Multi-Mode Model (MMM) turbulent transport code \cite{Rafiq2013POP,Luo2013CPC} in reproducing temperature profiles observed in a large set of well-analyzed discharges from NSTX, covering multiple scenario regimes. NSTX operated with major radius of 0.85 m and inverse aspect ratio of around 1.4. The large set of NSTX discharges investigated in this work had on-axis toroidal fields of $0.3 - 0.6$ T, plasma current of $0.5 - 1.3$ MA, on-axis densities of $3\ten{19} - 10^{20} \text{ m}^{-3}$, on-axis temperatures of $0.5 - 1.5$ keV, pulse lengths of around 1 s, volume averaged thermal toroidal $\beta = 2\mu_0 P_\text{th} / B_T^2$ of $3 - 15$\%, and effective collisionality evaluated at $\rho = 0.7$ of $\nustar = \nu_e/\omega_b \propto q R n_e/\Te^2\epsilon^{3/2} = 0.06 - 2.1$. Here, $\rho$ is a normalized flux surface label, $P_\text{th}$ is the thermal plasma pressure, $B_T$ is the toroidal field, $\nu_e$ and $\omega_b$ are the electron collision and bounce frequencies, $q$ is the safety factor, $R$ is the major radius, $n_e$ is the electron density, $\Te$ is the electron temperature, and $\epsilon$ is the inverse aspect ratio. The upgrade to NSTX-U  involved approximately doubling the toroidal field, plasma current, and available neutral beam power, with expected pulse lengths of up to 5 s. 

A closely related study to this work compared the ability of MMM to reproduce experimentally measured NSTX temperature profiles against a different reduced turbulent transport model, TGLF \cite{Staebler2007POP,Kinsey2008POP}. There it was found that that MMM's predictions were both more robust and significantly less computationally expensive than those from TGLF \cite{Lestz2025pre1}. Typical MMM temperature profiles differed from the observed profiles by approximately $15 - 40\%$, with a strong tendency for overpredicting confinement. While electromagnetic TGLF had comparable performance when predicting ion temperatures, its electron temperature ($\Te$) predictions less accurately reproduced the experimental profiles (characteristic disagreement of $15 - 75\%$) and simultaneous underprediction of ion temperature ($\Ti$) profiles resulted in a nearly factor of two higher $\teti$. Variation of the TGLF settings yielded significant variation in the predicted profiles, but no set of TGLF settings was found that consistently outperformed MMM across the range of well-analyzed NSTX discharges that were modeled. Based on those findings, a more detailed investigation into the properties and sensitivities of the MMM predictions is motivated, to fully understand the characteristics of the model in preparation for applying it for self-consistent, time-dependent scenario development for NSTX-U. 

MMM is a multi-mode model for turbulent transport capable of predicting electron and ion temperature profiles, electron and impurity density profiles, and rotation profiles. For each transport channel, diffusivities are predicted separately for each of the four individual models and then summed to return the total predicted diffusivity profile. The four distinct models in MMM are: 1) an electromagnetic model for electron temperature gradient modes (ETGs) \cite{Rafiq2022POP}, 2) a model for microtearing modes \cite{Rafiq2016POP}, 3) the Weiland model for trapped electron modes, ion temperature gradient modes, kinetic ballooning modes, and high-$n$ MHD modes \cite{Weiland2012text}, and 4) a model for drift resistive inertial ballooning modes (DRIBM) \cite{Rafiq2010POP}. The DRIBM model is excluded for NSTX simulations based on previous analysis indicating that these modes should be stable in NSTX \cite{Rafiq2024NF}. 
\rev{The ETG model includes a calibration against NSTX data that was implemented during its development and is used to offset simplifying assumptions used in its derivation \cite{Rafiq2022POP,Rafiq2024NF}. This calibration was not modified for this study, as it is not intended to be changed when using MMM to model different NSTX discharges or different tokamaks.}
MMM does not currently incorporate transport due to low-$n$ MHD modes or anomalous fast ion transport, though the latter may be included in a future MMM version \cite{Weiland2023POP,Rafiq2023APS}. MMM has been used extensively for NSTX analysis, for instance finding improved agreement with observed $\Te$ profiles when accounting for MTM transport \cite{Rafiq2021POP}, reproducing trends observed in gyrokinetic simulations for ETG turbulence \cite{Rafiq2022POP}, and assessing the influence of flow shear stabilization at low and high collisionality \cite{Rafiq2024NF}.   

Time-dependent simulations with MMM are performed with the TRANSP 1.5D tokamak power balance and transport code \cite{Hawryluk1980transp,Goldston1981JCP,transp2018,Grierson2018FST,Pankin2025CPC}. In the simulations presented here, TRANSP assumes a fully prescribed 2D axisymmetric equilibrium and performs transport calculations on a 1D radial grid. When run predictively (referred to as predictive TRANSP, or PTRANSP), an iterative, implicit transport solver \ptsolver can be employed to time evolve the plasma's temperature, density, and/or rotation profiles by interfacing with a range of physics models \cite{Jardin2008JCP,Yuan2011APS,Yuan2012APS,Yuan2013APS}. Additional description of predictive TRANSP can be found in Sec. 9 of \citeref{Pankin2025CPC} and Sec. II of \citeref{Lestz2025pre1}. In spherical tokamaks, predictive TRANSP simulations with MMM have been previously used for scenario development, optimization, and scoping studies for NSTX \cite{Poli2015NF,Lopez2018PPCF} and SMART \cite{Podesta2024PPCF,CruzZabala2024NF}. Similar workflows have also been used to validate MMM against discharges from several different conventional tokamaks \cite{Rafiq2023plasma} and to study advanced scenarios in DIII-D \cite{Pankin2018POP}. 

Building on the assessment of time-averaged agreement with experiment performed in \citeref{Lestz2025pre1}, this work demonstrates MMM's ability to closely track the time evolution of a discharge. Moreover, the agreement of the predicted temperature gradients with experimental fits is examined, finding that steeper local gradients are predicted with better agreement, even while NSTX discharges with broader profiles tend to be better reproduced by MMM. Special attention is paid to MMM's profile predictions with respect to $\beta$ and $\nustar$, finding that the level of agreement with experiments is more sensitive to $\beta$, with the few poorly predicted plasmas being characterized by relatively low $\beta$. Fixing the $\Ti$ profile to its experimentally measured value reduces the $\Te$ prediction disagreement by about half relative to MMM simulations that predict both $\Te$ and $\Ti$ simultaneously. However, with $\Ti$ specified this way, $\Te$ can be ``predicted'' with the same level of experimental agreement as in MMM by instead setting $\Te = \Ti$ without performing any simulation. 

The rest of the paper is organized as follows. Additional details about the large set of well-analyzed NSTX discharges that are simulated with MMM and the focus of most of the paper are discussed in \secref{sec:character}, along with a discussion of the character of the electron and ion transport in those discharges. \secref{sec:time} considers the degree to which predictive TRANSP simulations with MMM are able to reproduce the observed temperature evolution over time, including sensitivity to initial conditions. The MMM predictions of the measured normalized temperature gradient scale lengths and profile peaking  in relation to NSTX observations is analyzed in \secref{sec:gradpeak}. \secref{sec:betanu} examines trends in the profile predictions with respect to $\beta$ and $\nustar$ from two different approaches. \secref{sec:betanuJBL} discusses trends in the main set of MMM simulations presented throughout this paper. A second set of simulations of NSTX discharges chosen specifically to decouple the usual operational dependency of $\beta$ and $\nustar$ is presented in \secref{sec:skaye}. \secref{sec:teonly} compares the consequences of using MMM within TRANSP to predict both $\Te$ and $\Ti$ profiles simultaneously as opposed to using the experimental $\Ti$ profile as an additional input and predicting $\Te$ alone. Lastly, a summary and conclusions are provided in \secref{sec:conclusion}. 

\section{Character of Electron and Ion Energy Transport}
\label{sec:character}

\begin{figure}[tb]
\includegraphics[width = \columnwidth]{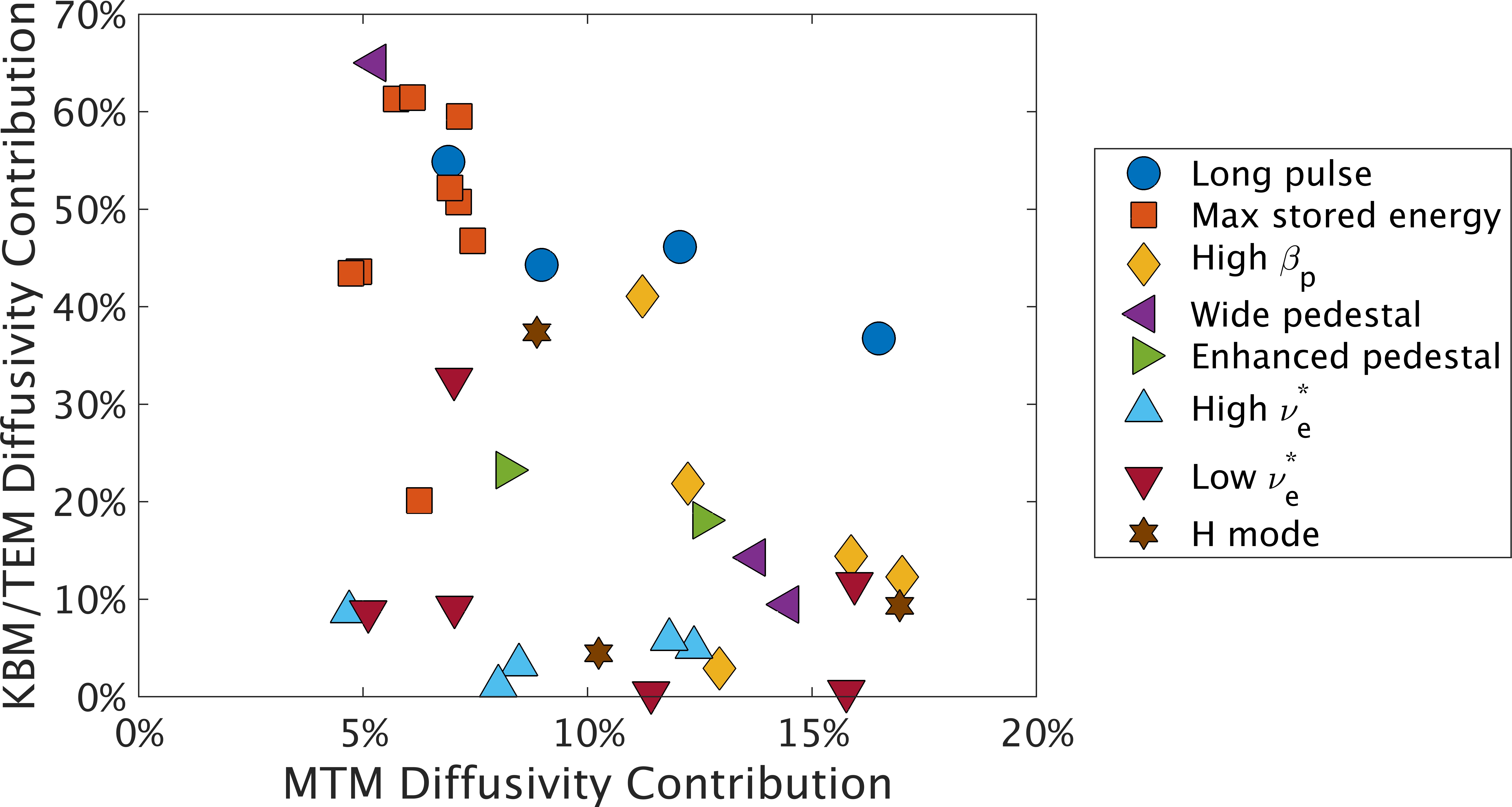}
\caption{Collection of well-analyzed NSTX discharges focused on in this study, classified by the property which made them noteworthy in previous NSTX transport studies. The abscissa and ordinate show the MMM calculation of the fraction of the electron heat diffusivity that is attributed to MTMs and the Weiland model (KBMs, TEMs, ITGs), respectively, averaged over the prediction region $\rho = 0 - 0.7$. The full list of discharges can be found in \tabref{tab:runids}.}
\label{fig:shots_mtm_w19}
\end{figure}

In this work, predictive TRANSP simulations with MMM are used to time evolve the electron and ion temperature profiles, with an assessment of the prediction of density and rotation profiles left to future work. The same input settings for \ptsolver and MMM were used for all simulations, which also use the NCLASS neoclassical transport solver \cite{Houlberg1997POP}. Moreover, the profile predictions are restricted to $\rho < 0.7$ (or $\rho < 0.8$ for the simulations discussed in \secref{sec:skaye}) so that a reliable boundary condition can be imposed with experimental measurements. 
\rev{At larger $\rho$ in NSTX, the experimental data has larger uncertainties, which would introduce an additional source of error if the profile prediction boundary were extended further towards the plasma edge \cite{Avdeeva2023NF,Avdeeva2024PPCF}}
\rev{\ptsolver iteratively solves the following equation in order to evolve $\Te$ in time:}

\begin{multline}
\pderiv{}{t}\left[\frac{3}{2}V^\prime n_e \Te\right] + \pderiv{}{\rho}\left[V^\prime\langle\abs{\grad\rho}^2\rangle n_e \left(\chi_e \pderiv{\Te}{\rho} - \Te v_e\right)\right] \\ 
- \overdot{\xi}\pderiv{}{\rho}\left[\rho V^\prime \frac{3}{2} n_e \Te\right] = S_e V^\prime.
\label{eq:pt_te}
\end{multline}

\rev{The toroidal magnetic flux is given by $\Phi = \pi\rho^2 B_0$, with $\rho$ being a radial coordinate with dimensions. The normalized flux surface label is defined as $\xi = \rho/\rho_\text{sep} = \sqrt{\Phi/\Phi_\text{sep}}$. The transformation to flux coordinates results in the appearance of geometric factors $V^\prime$ and $\langle\abs{\grad\rho}^2\rangle$. Electron thermal conductivity and convective velocities are given by $\chi_e$ and $v_e$, respectively. All electron sources and sinks are combined into $S_e$, which includes auxiliary heating, Ohmic heating, collisional electron-ion coupling, radiated power loss, and neutral ionization, as detailed in Sec. 4 of \citeref{Pankin2025CPC}. The term proportional to $\overdot{\xi}$ accounts for the time evolution of the flux surfaces. Note that \eqref{eq:pt_te} and the following two equations adopt the same symbols used in the \ptsolver references \cite{Yuan2011APS,Yuan2012APS,Yuan2013APS,Pankin2025CPC} to avoid confusion with those documents, in contrast to the rest of this paper which uses the variable $\rho$ to represent the normalized flux surface label when discussing radial profiles.}

\begin{figure*}[tb]
\subfloat[\label{fig:xke_blow}]{\includegraphics[height = \thirdheight]{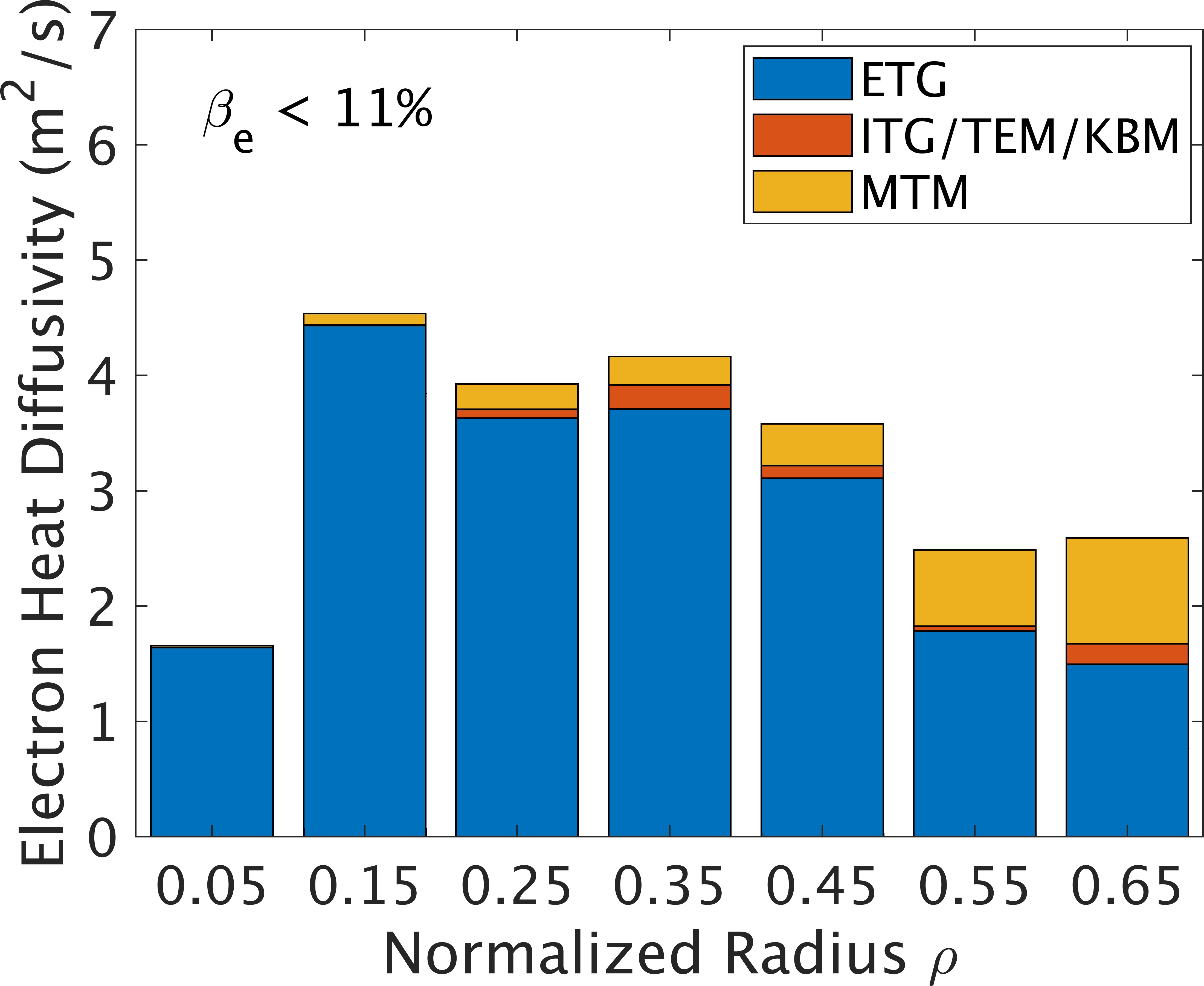}} \figsep
\subfloat[\label{fig:xke_bhigh}]{\includegraphics[height = \thirdheight]{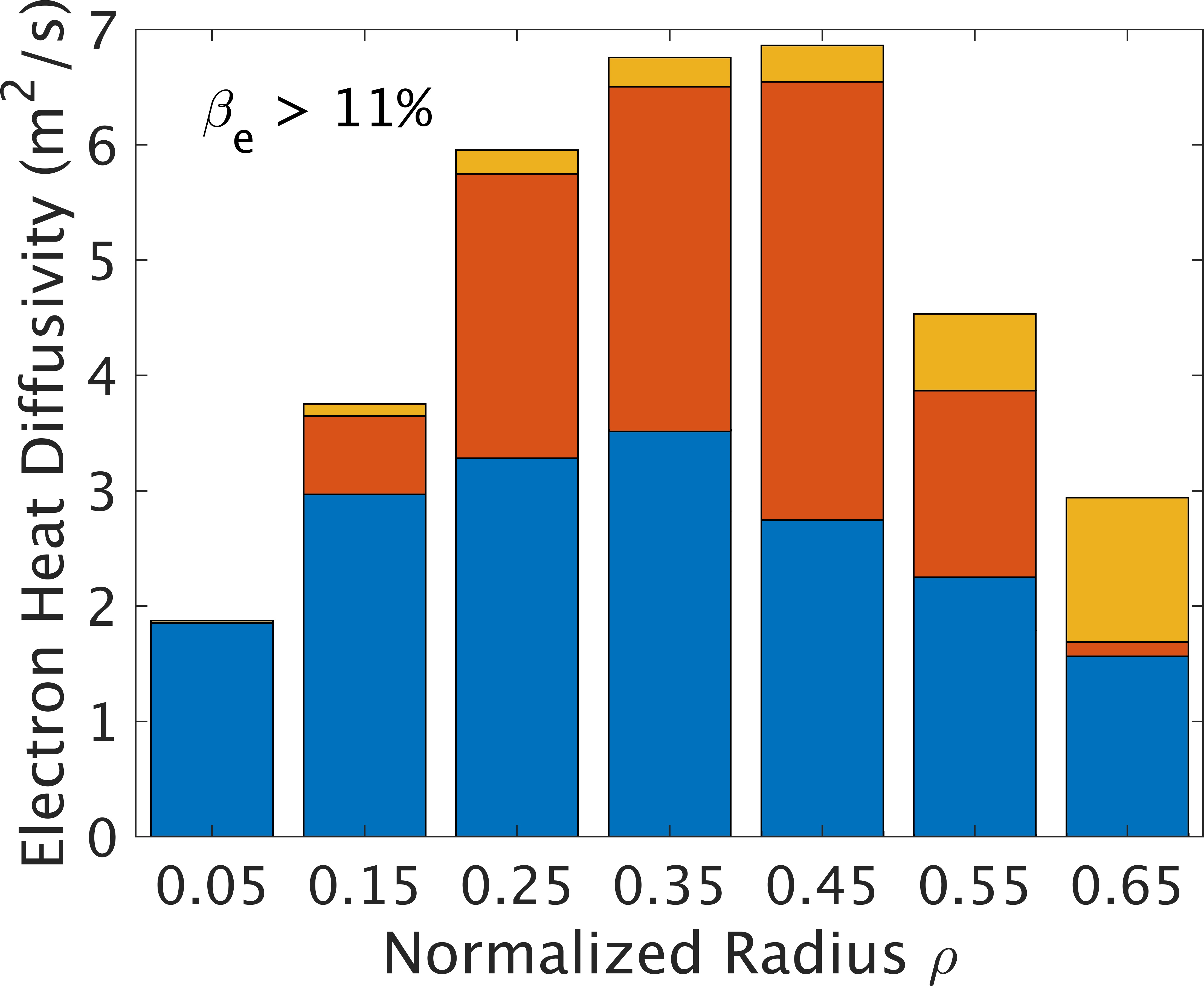}} \\
\subfloat[\label{fig:xke_nlow}]{\includegraphics[height = \thirdheight]{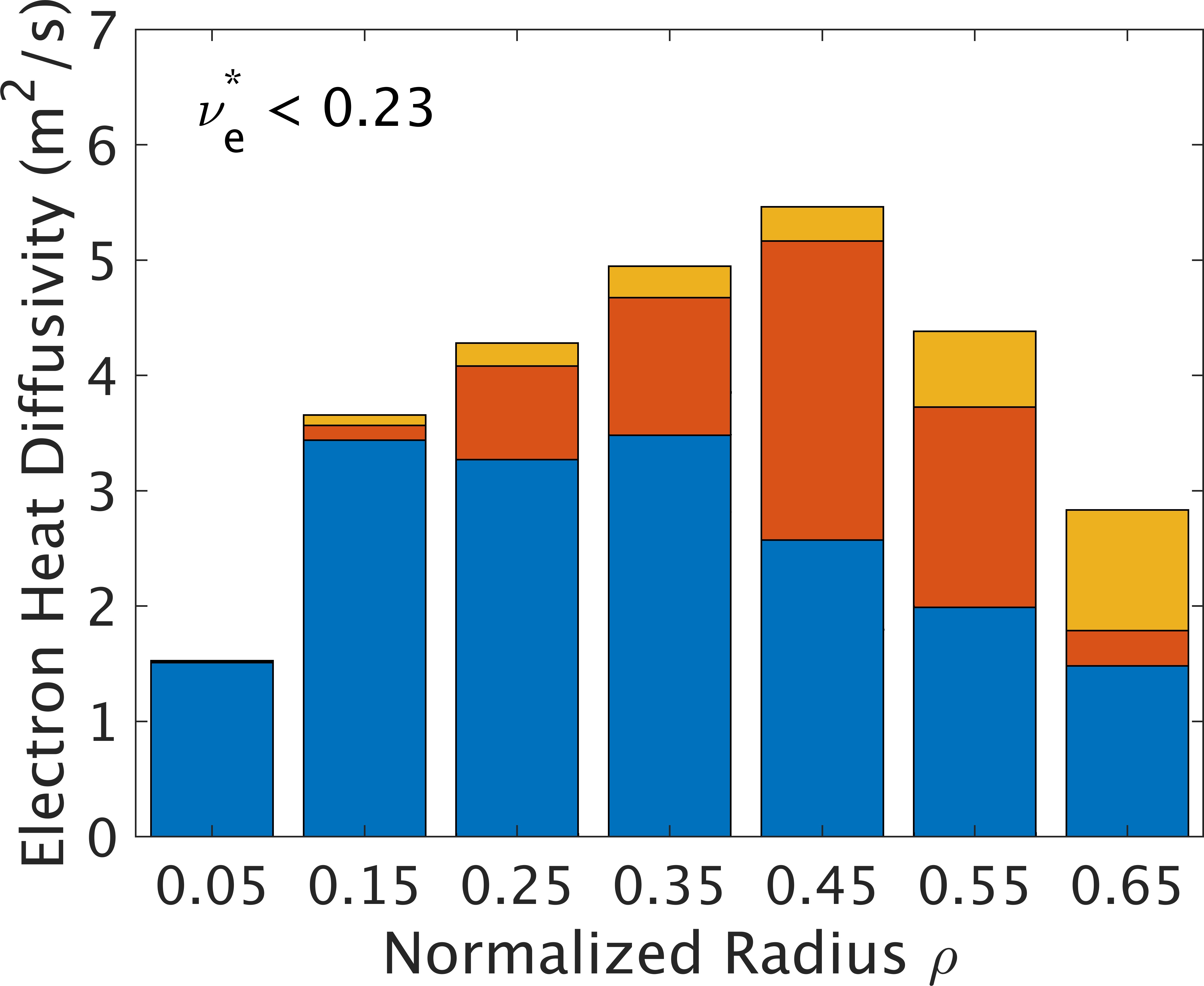}} \figsep
\subfloat[\label{fig:xke_nhigh}]{\includegraphics[height = \thirdheight]{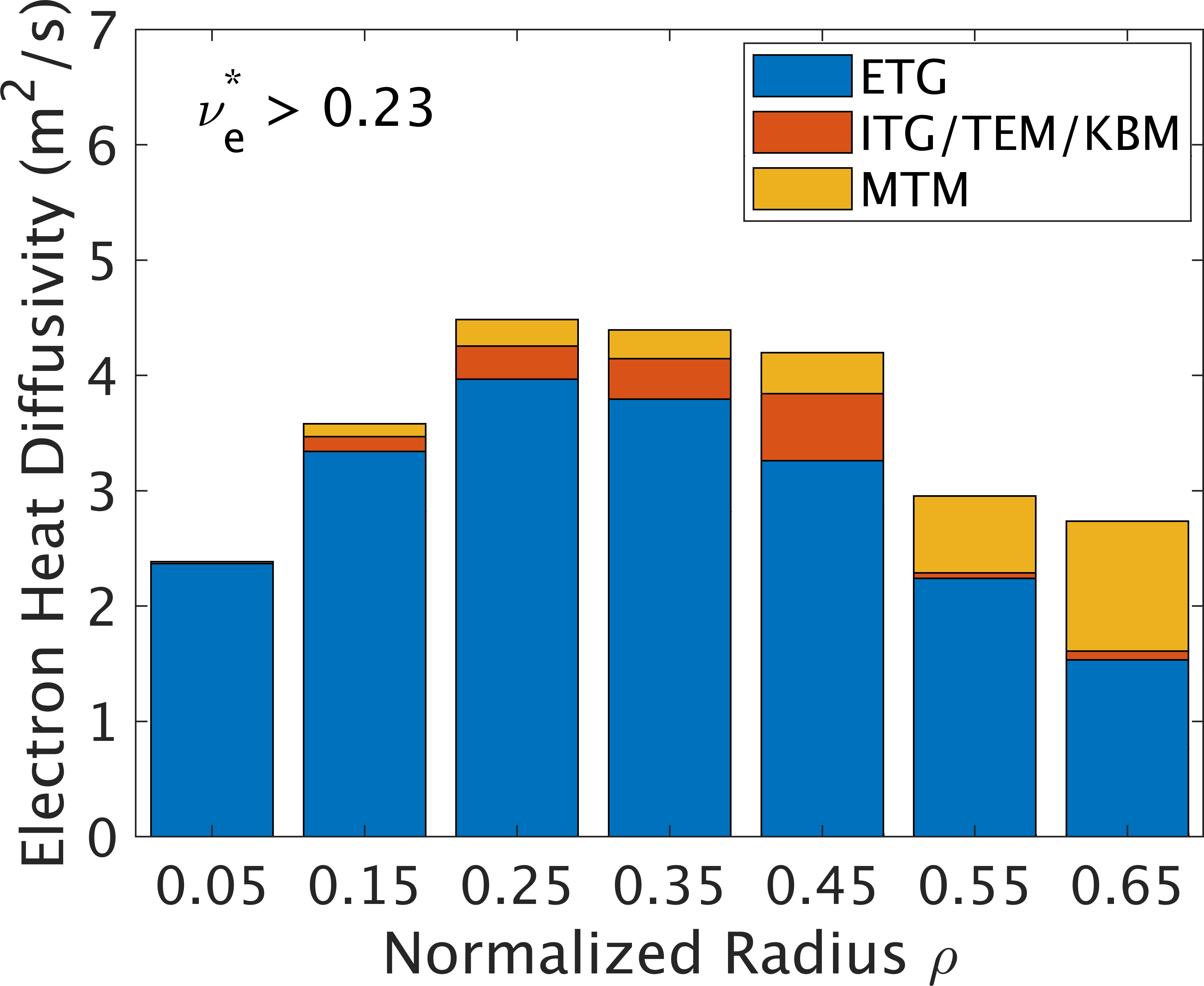}} 
\caption{Median electron energy diffusivity as predicted by MMM in predictive TRANSP simulations for discharges with (a) relatively low $\betae < 11\%$ on axis, (b) high $\beta > 11\%$ on axis, (c) relatively low $\nustar < 0.23$ at $\rho = 0.7$, and (d) high $\nustar > 0.23$ at $\rho = 0.7$, averaged over all examined discharges and decomposed into contributions from different submodels within MMM.}
\label{fig:xke_beta_nu}
\end{figure*}

Most of the analysis in this paper focuses on a large set of beam-heated NSTX discharges that are well-analyzed in existing work
\cite{Gerhardt2011NFat,Gerhardt2011NFcur,Ren2012POP,Guttenfelder2013NF,Ren2013NF,Gerhardt2014NF,Kaye2014POP,RuizRuiz2019PPCF,
 Ren2020NF,Battaglia2020POP,Clauser2022POP,Avdeeva2023NF,Dominski2024POP,Rafiq2024NF,McClenaghan2023POP,Clauser2025POP,McClenaghan2025PPCF}, spanning a range of operational regimes such as long pulse length, large stored energy, sustained high $\betap$ (kinetic plasma pressure normalized to poloidal field pressure), wide pedestal, enhanced pedestal, high $\nustar$, low $\nustar$, and some well-analyzed H modes. The range of plasma parameters included in this set of discharges was listed in \secref{sec:intro}, and further details can be found in \citeref{Lestz2025pre1}, which made use of the same collection of 37 discharges. The main exception is \secref{sec:skaye}, which models a different set of NSTX discharges with the intention of decoupling operational dependencies on $\beta$ and $\nustar$. 
\rev{In the examined discharges, $2 - 6$ MW of beam power was injected at $65 - 95$ keV}. 
 
\rev{Although the plasma density tends to evolve throughout NSTX discharges, analysis windows were chosen in each discharge to avoid low $n$ MHD instabilities and rapid changes in the stored energy or neutron rate, which could indicate substantial energy transport that would not be captured by the  employed models.}
\tabref{tab:runids} lists the time windows used for each discharge, with a median duration of 200 ms. When time-averaging over the analysis windows and spatially averaging over the profiles, standard figures of merit are used for the relative offset and root mean square error (RMSE) between the temperature profiles predicted by MMM and experimental observations \cite{ITER1999NF,Abbate2024POP}. Hence, all instances of language such as accuracy, error, \etc when making these comparisons are intended as convenient shorthand for the quantitative agreement between the simulations and measurements. Measurement errors in the raw NSTX data and systematic errors in profile fitting methods increase the uncertainty in these comparisons, though they are not quantified here. Evaluating the agreement between the MMM predictions within TRANSP and first principles simulations is outside the scope of this work. 

To quantify the strength of trends, correlation coefficients between plasma parameters and figures of merit will be considered at times. In addition to the usual linear (Pearson) correlation coefficient, in some cases, the Spearman rank correlation coefficient \cite{Spearman1904AJP} will also be considered. The Spearman correlation coefficient is equivalent to the linear correlation between the ordered rank of the two variables. The primary benefits of considering this alternative metric are reducing the influence of outliers, which can have an outsized effect on linear correlation coefficients, and also better representing monotonic trends that are not linear. 

\subsection{Electron Energy Transport}
\label{sec:char_e}

To give a sense of the conditions in the discharges that are being examined by MMM, it is instructive to first consider the character of the electron and ion energy transport present in these discharges. For each examined discharge, \figref{fig:shots_mtm_w19} plots the fractional contribution to the electron heat diffusivity from different MMM models, averaged over the prediction region $\rho = 0 - 0.7$. Not shown is the contribution from the ETG model, which is responsible for all remaining transport. Since the DRIBM model is not used for NSTX simulations, the total electron heat diffusivity predicted in MMM is equal to the sum of transport from MTMs (abscissa), the Weiland model (TEM and KBMs, ordinate), and ETGs. While there are no discharges included that are predicted to be completely dominated by MTMs, this is because MTMs are only relevant in a small region near the edge of the prediction region, whereas instabilities from the Weiland model are present in a broad mid-radius region and ETGs drive significantly transport across the entire plasma radius. As will be discussed in this section, MTMs can drive a substantial fraction or even majority of the electron heat transport near the pedestal. Thus, the set of discharge used in this study samples a variety of regimes in terms of the composition of the electron energy transport. 

When averaging over all of the examined NSTX discharges, trends emerge in the contributions to the electron heat diffusivity from the different types of instabilities present in the MMM simulations. \figref{fig:xke_beta_nu} shows the median diffusivity, averaged in radial bins of width $\Delta\rho = 0.1$, and grouped into discharges with relatively low on-axis $\betae < 11\%$ (\figref{fig:xke_blow}) and relatively high $\betae > 11\%$ (\figref{fig:xke_bhigh}). Here, $\betae = 2\mu_0 n_e T_e /B_T^2$ is the ratio of electron pressure to magnetic field pressure. An on-axis $\betae$ of 11\% is chosen as a threshold since it divides the examined discharges into roughly equal groups, but is otherwise not physically significant. Of the examined NSTX discharges, 19 have $\betae < 11\%$ and the remaining 18 have $\betae > 11\%$. As in some other statistical analysis in this work, the median is chosen instead of the mean in order to illustrate representative values and trends, whereas the mean can overweight outliers and is also susceptible to overweighting similar discharges that were studied, since the set of modeled discharges is neither large enough nor proportionally sampled in parameter space to serve as a balanced database of all high performing NSTX discharges. From these figures, it is clear that near the axis, transport is always dominated by ETGs. At sufficiently low $\betae$, ETGs continue to dominate until at least mid-radius, with MTMs becoming nontrivial closer to the pedestal. However, in the higher $\betae$ discharges, contributions from the Weiland model (ITG, TEM, and KBM) can become comparable to or exceed the ETG diffusivity in the mid radius region, whereas the Weiland diffusivity is negligible at lower $\betae$. Even at high $\beta$, the MTM transport predicted by MMM is essentially unchanged, aside from a modest tendency for greater transport near the pedestal. Overall, the increase in diffusivity from ITGs, TEMs, and KBMs significantly increases the total electron heat diffusivity predicted by MMM for NSTX discharges with relatively high $\beta$. 

\begin{figure}[tb]
\subfloat[\label{fig:xke_betap_34}]{\includegraphics[height = \thirdheight]{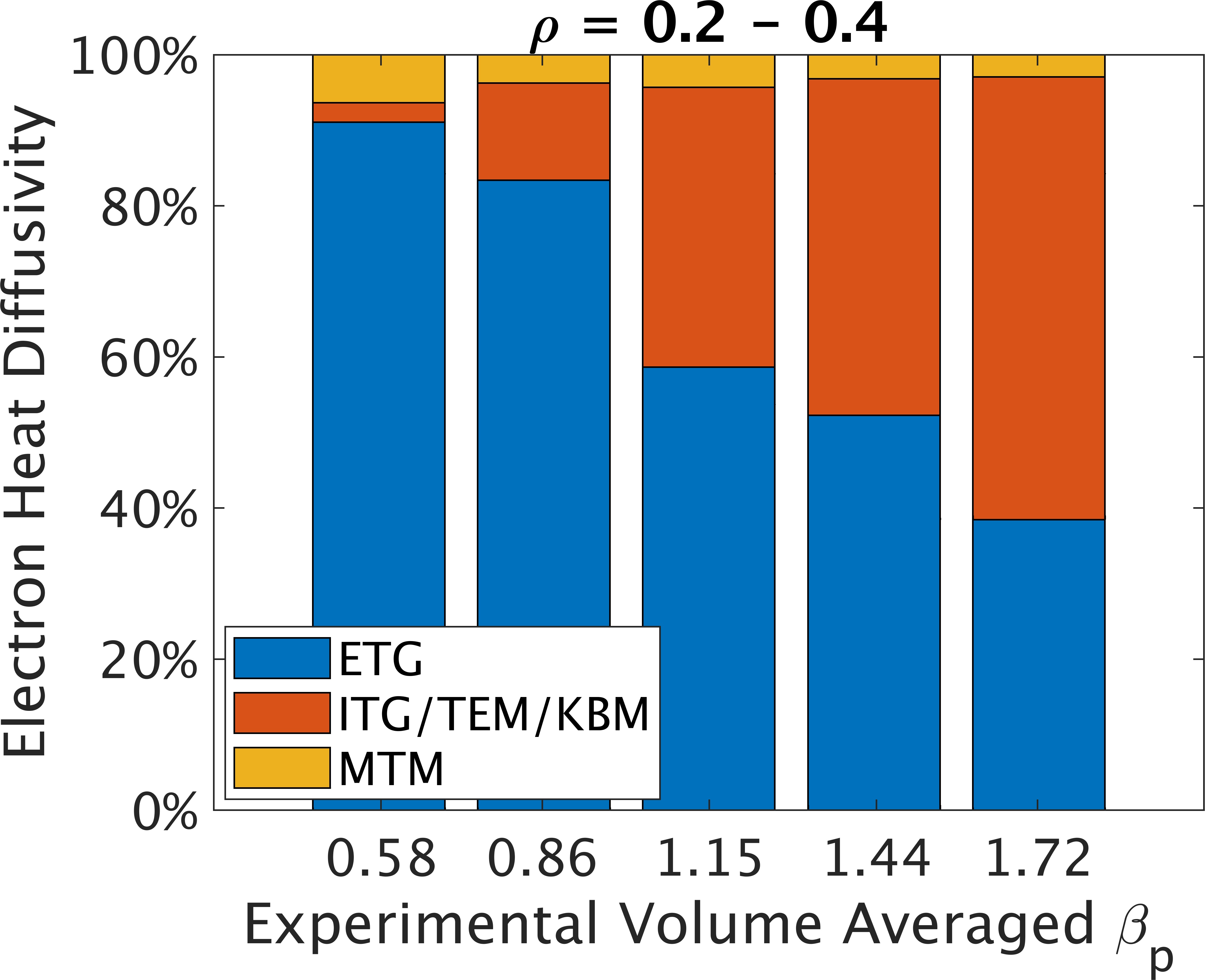}} \\
\subfloat[\label{fig:xke_betap_7}]{\includegraphics[height = \thirdheight]{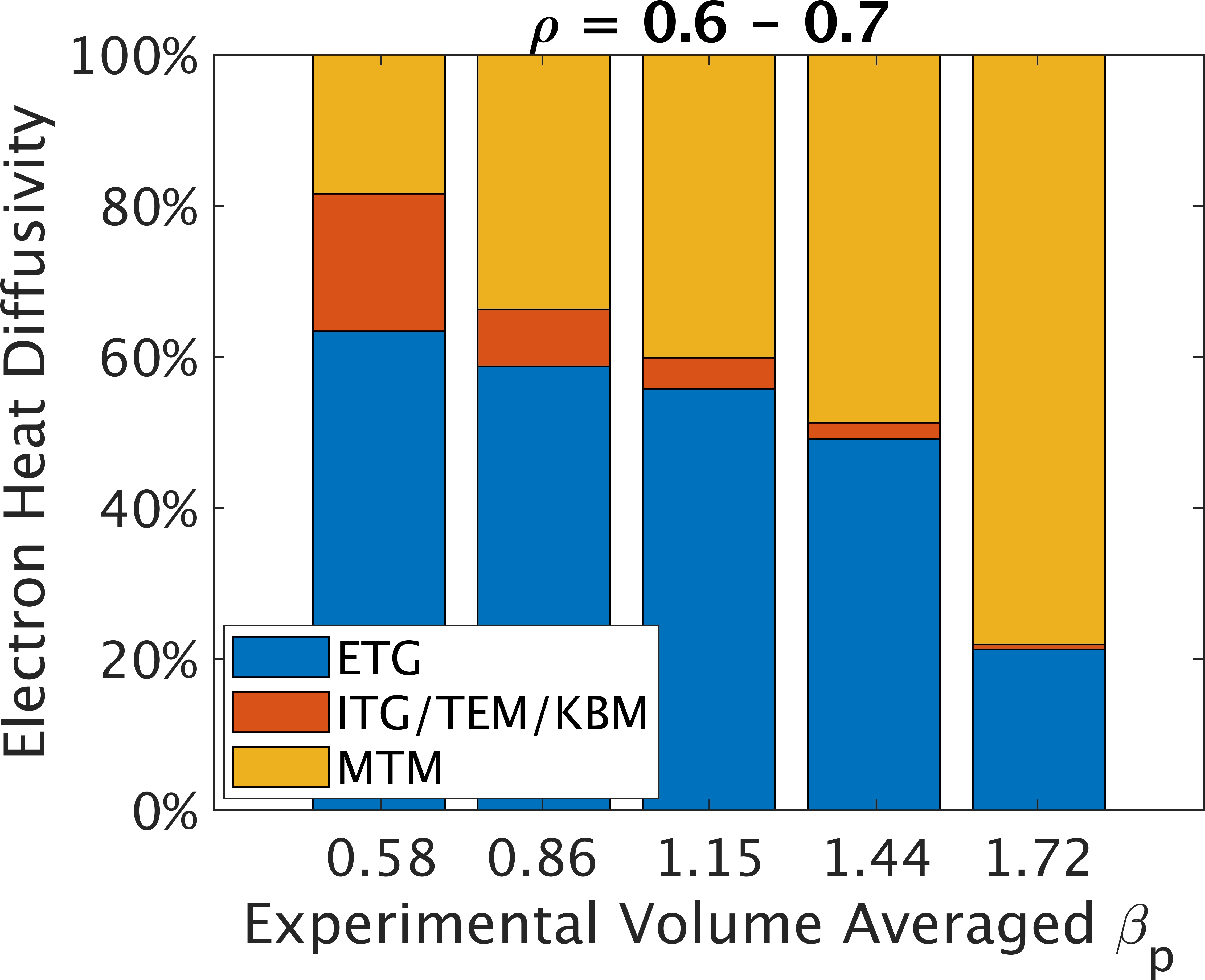}} 
\caption{Median electron energy diffusivity fraction from different submodels within MMM as a function of volume averaged $\betap$ for (a) $\rho = 0.2 - 0.4$ and (b) $\rho = 0.6 - 0.7$. Each bar is labeled with the central $\betap$ value of the equally spaced bins, which span the minimum to maximum values for the discharges studied in this work. $\betap$ is averaged over the entire plasma volume, not the range of $\rho$ that the diffusivities are restricted to in each panel.}
\label{fig:xke_betap}
\end{figure}

\begin{figure*}[tb]
\subfloat[\label{fig:xki_hist_all}]{\includegraphics[height = \thirdheight]{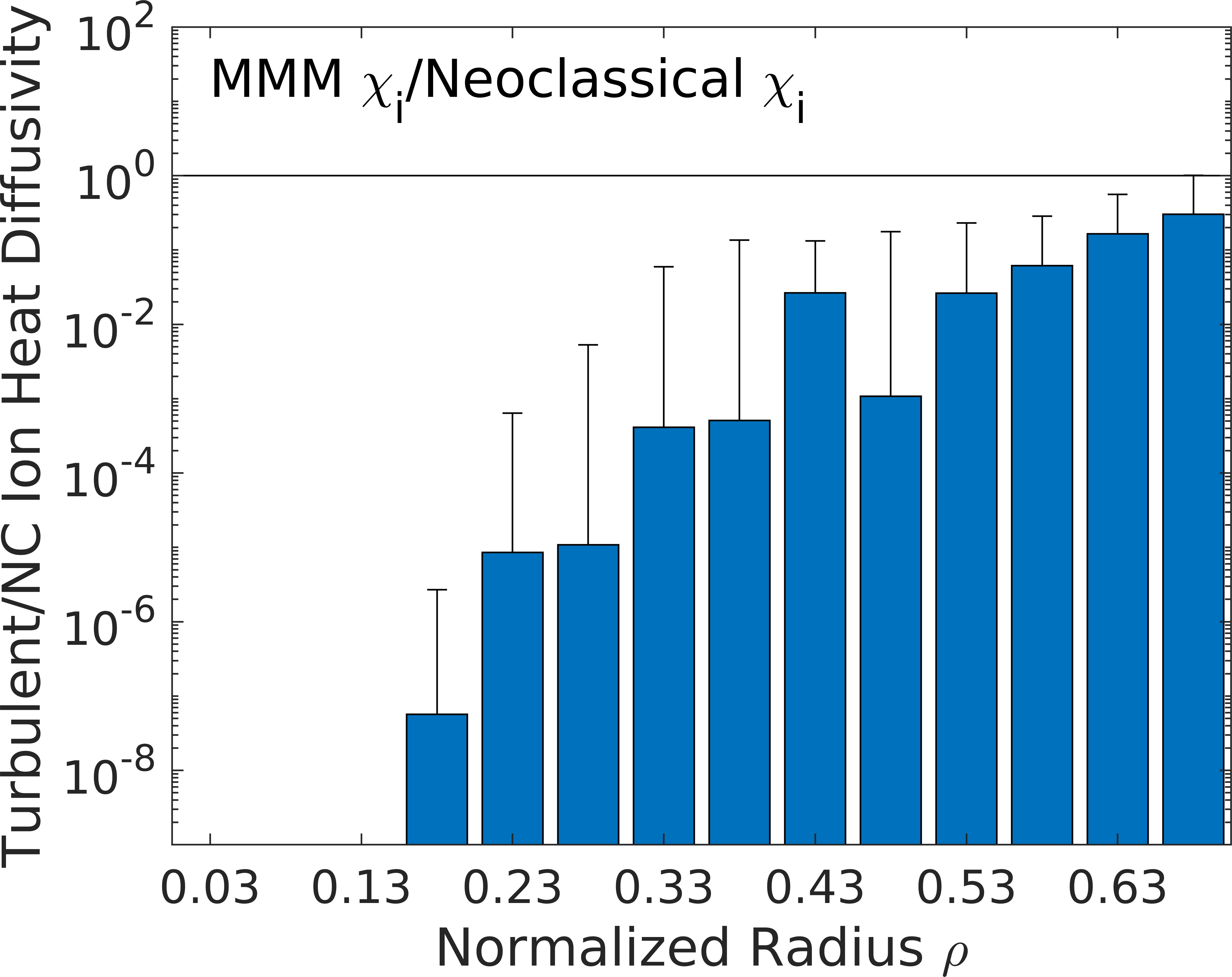}} \figsepsm
\subfloat[\label{fig:xki_rho}]{\includegraphics[height = \thirdheight]{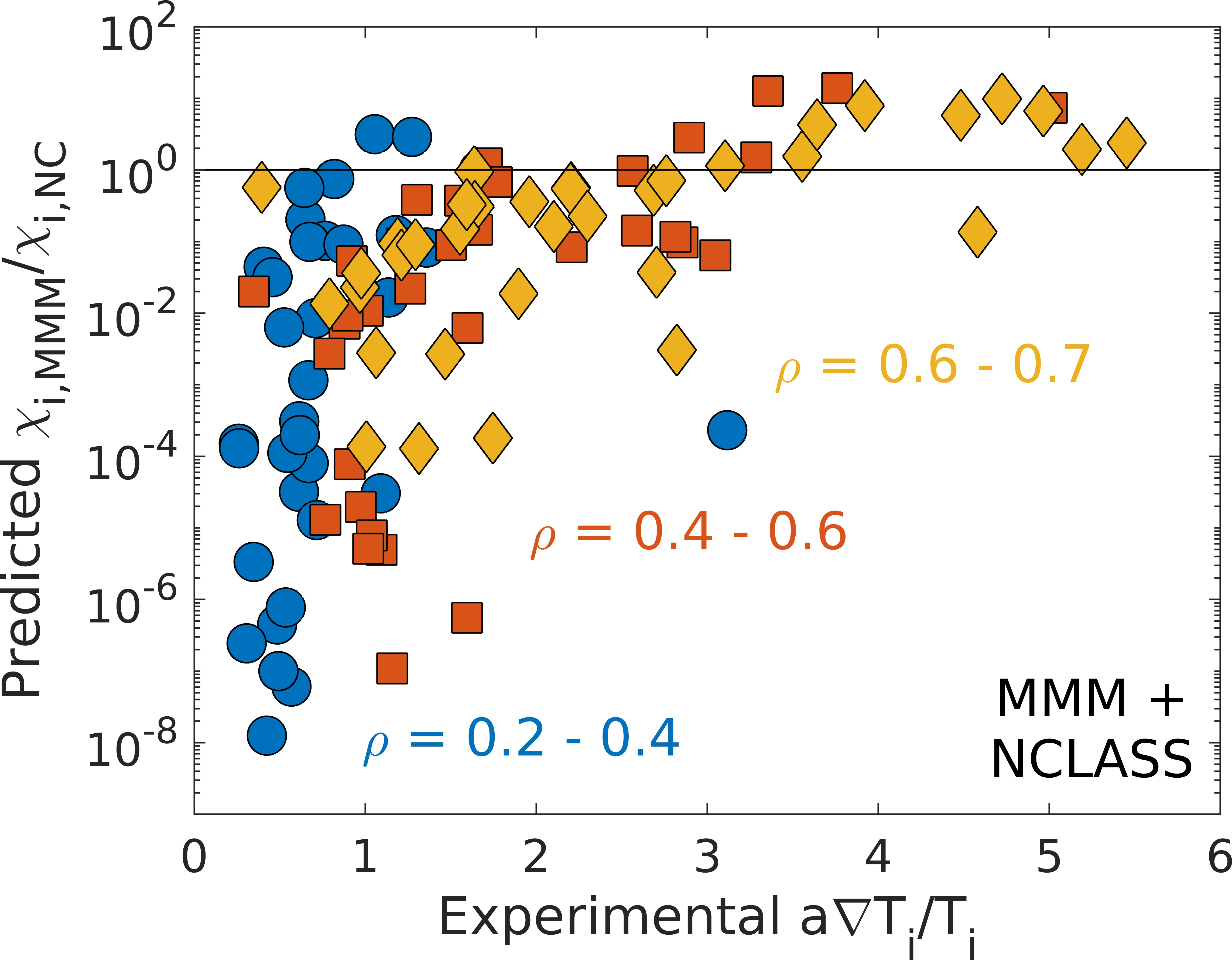}} \figsepsm
\subfloat[\label{fig:xki_beta}]{\includegraphics[height = \thirdheight]{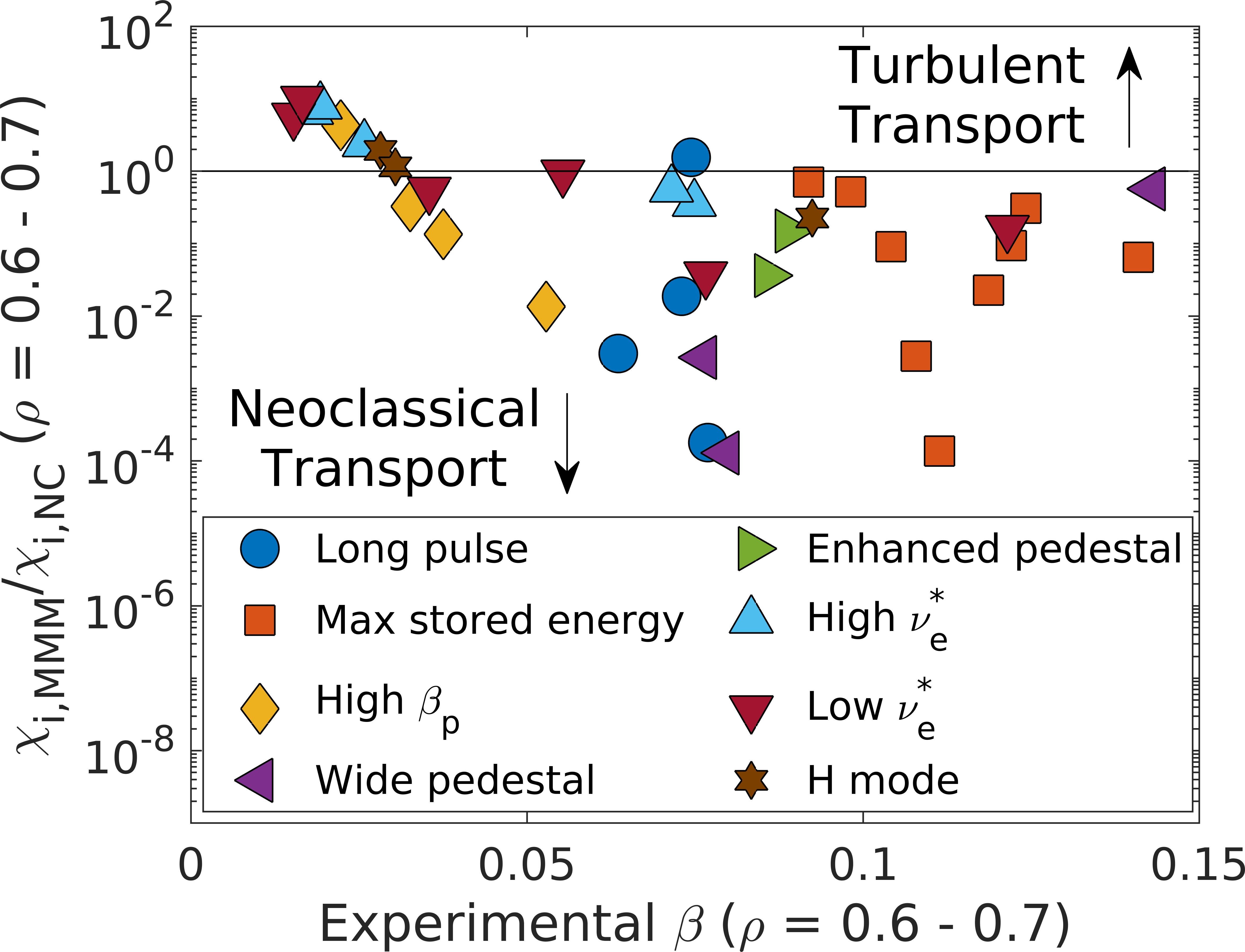}}
\caption{Ratio of turbulent ion diffusivity (predicted by MMM) to neoclassical ion diffusivity (calculated by NCLASS) as a function of three different variables. (a) Dependence on $\rho$, averaged over all examined discharges, with errorbars showing the half interquartile range across the discharges. (b) Dependence on normalized ion temperature gradient for each discharge, averaged over different radial locations. (c) Dependence on $\beta$ for all examined discharges, averaged over $\rho = 0.6 - 0.7$. The solid line is for reference only, indicating equal contributions from neoclassical and turbulent ion transport.}
\label{fig:xki_plots}
\end{figure*}

A similar comparison is shown in \figref{fig:xke_nlow} and \figref{fig:xke_nhigh}, distinguishing discharges with relatively low and high normalized collisionality evaluated at $\rho = 0.7$. A threshold of $\nustar = 0.23$ is chosen to divide the discharges into roughly equal groups, with 18 plasmas having smaller $\nustar$ and 19 plasmas having larger $\nustar$. These figures demonstrate the relative importance of electron transport from the Weiland model at low collisionality, whereas the MTM contribution is increased only slightly in the higher $\nustar$ discharges. The relative insensitivity of the MTM transport to collisionality may be due to two competing effects. Higher collisionality tends to increase the linear drive for MTMs as the electrons become more resistive, increasing the parallel electric field. However, high collisionality can also reduce MTM transport due to decorrelation effects, such that the competition between these two effects leads to a complex dependence of MTM-induced electron heat diffusivity. In contrast to the low \vs high $\betae$ comparison where the total diffusivity was significantly higher in the high $\betae$ discharges, the low \vs high $\nustar$ discharges have similar levels of overall transport, as evidently the increased TEM and KBM transport at low $\nustar$ is offset by a tendency for reduced ETG transport in the MMM calculations for these discharges. Trends for the experimental agreement of MMM predictions with respect to $\betae$ and $\nustar$ will be discussed in \secref{sec:betanu}. 

Independent of the $\betae$ and $\nustar$ trends shown in \figref{fig:xke_beta_nu}, the relative importance of each type of instability in MMM has a strong dependence on volume averaged $\betap$. \figref{fig:xke_betap} separates the discharges into equal width bins of $\betap$ and then plots the median fractional contribution to the total electron heat diffusivity from each submodel in MMM, averaging over all of the discharges in each bin. \figref{fig:xke_betap_34} shows this calculation for the diffusivity in $\rho = 0.2 - 0.4$, excluding the region very close to the axis that typically has very shallow gradients. \figref{fig:xke_betap_7} does the same for $\rho = 0.6 - 0.7$, examining a region near the prediction boundary and pedestal. In the core region of $\rho = 0.2 - 0.4$, contributions from the Weiland model tend to increase with increasing $\betap$, while the fractional contribution from ETG transport decreases by a similar amount. MTM transport is insignificant in this spatial region and its minor fractional contribution is largely insensitive to $\betap$. In contrast, the edge region of $\rho = 0.6 - 0.7$ is strongly influenced by MTM transport, which becomes responsible for an even larger fraction of the electron heat diffusivity at higher $\betap$. As $\betap$ increases, the fractional transport from ETGs tends to decrease (similar to the tradeoff between the Weiland model and ETG diffusivity fractions in the core), with Weiland model transport remaining quite small in this region. 

The fractional MTM electron heat diffusivity is more strongly correlated with $\betap$ than $\betae$, indicating sensitivity to the magnetic geometry. Specifically, the linear correlation coefficient between MTM diffusivity fraction for $\rho = 0.6 - 0.7$ and $\betap$ is $r = 0.68$, while it is $r = 0.02$ for the on-axis $\betae$, essentially uncorrelated. This can be understood by considering the dependence for the MTM diffusivity in the MMM model, $\chi_e^\text{MTM} \like \left(\delta B/B\right)^2 \vtherme L_c$, where $\delta B$ is the fluctuation amplitude, $\vtherme$ is the electron thermal velocity, and $L_c$ is the decorrelation length \cite{Rafiq2016POP,Rafiq2021POP}. In the limit of low collisionality (specifically, $\vtherme/\nu_e > q R$), the decorrelation length can be approximated as $L_c \approx q R$. Physically, longer $L_c$ corresponds to electrons streaming farther along the stochastic field lines, yielding stronger electron heat transport. Hence $\chi_e^\text{MTM} \propto q \propto 1 / B_P \propto \sqrt{\beta_p}$, such that MTM transport would be expected to increase specifically with $\betap$, as found in the presented simulations. In contrast, the Weiland model diffusivity fraction has similar correlations with both $\betap$ and $\betae$ in the examined region, consistent with the findings of \figref{fig:xke_beta_nu}. 

The main picture that emerges for electron heat transport, as modeled by MMM, in these well-analyzed NSTX plasmas can be summarized as follows. Very close to the axis ($\rho < 0.2$), ETGs dominate, albeit with fairly low levels of transport due to shallow profile gradients limiting the free energy available to drive instabilities. In the core and mid-radius regions, spanning $\rho = 0.2 - 0.6$, both ETGs and the Weiland model (ITG, TEM, and KBM), can drive significant transport, with ETGs continuing to dominate at relatively low $\betae$ and high $\nustar$, while transport from TEMs and KBMs becomes relatively more important in this region with lower collisionality, increased $\betae$, and to some degree increased $\betap$. Closer to the edge ($\rho = 0.6 - 0.7$), electron transport is driven by both ETGs and MTMs, with MTMs dominating at high $\betap$ with less sensitivity to $\betae$ or $\nustar$. Moreover, the level of agreement between the $\Te$ profiles predicted by MMM and those observed in experiments is uncorrelated with the fractional diffusivity driven by each instability, suggesting that none of the submodels within MMM is systematically less accurate than the others. 

\newcommand{\figsepb}{\hspace{1.1em}}
\begin{figure*}[tb]
\subfloat[\label{fig:133964_te}]{\includegraphics[height = \thirdheight]{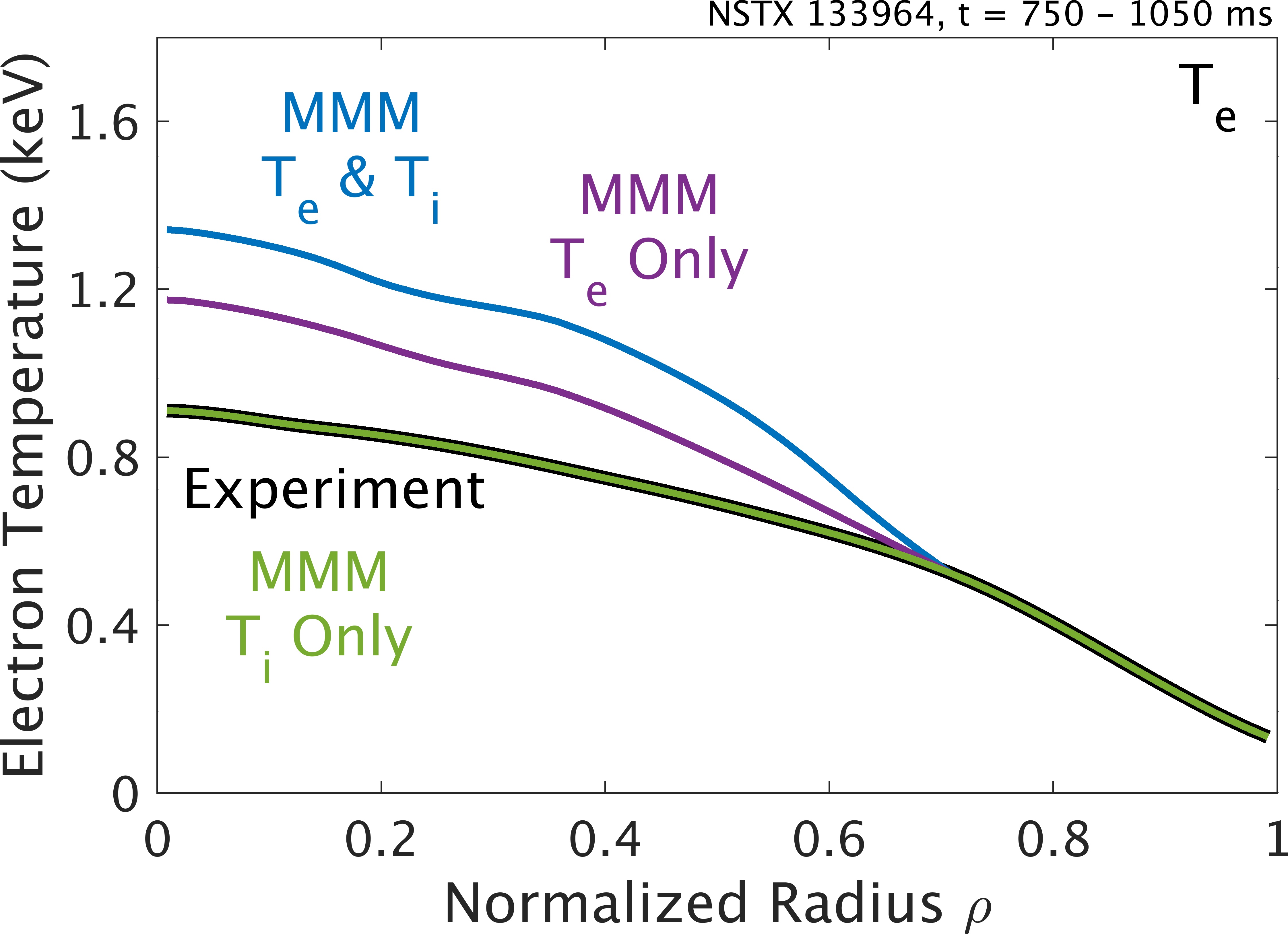}} \figsepb
\subfloat[\label{fig:133964_ti}]{\includegraphics[height = \thirdheight]{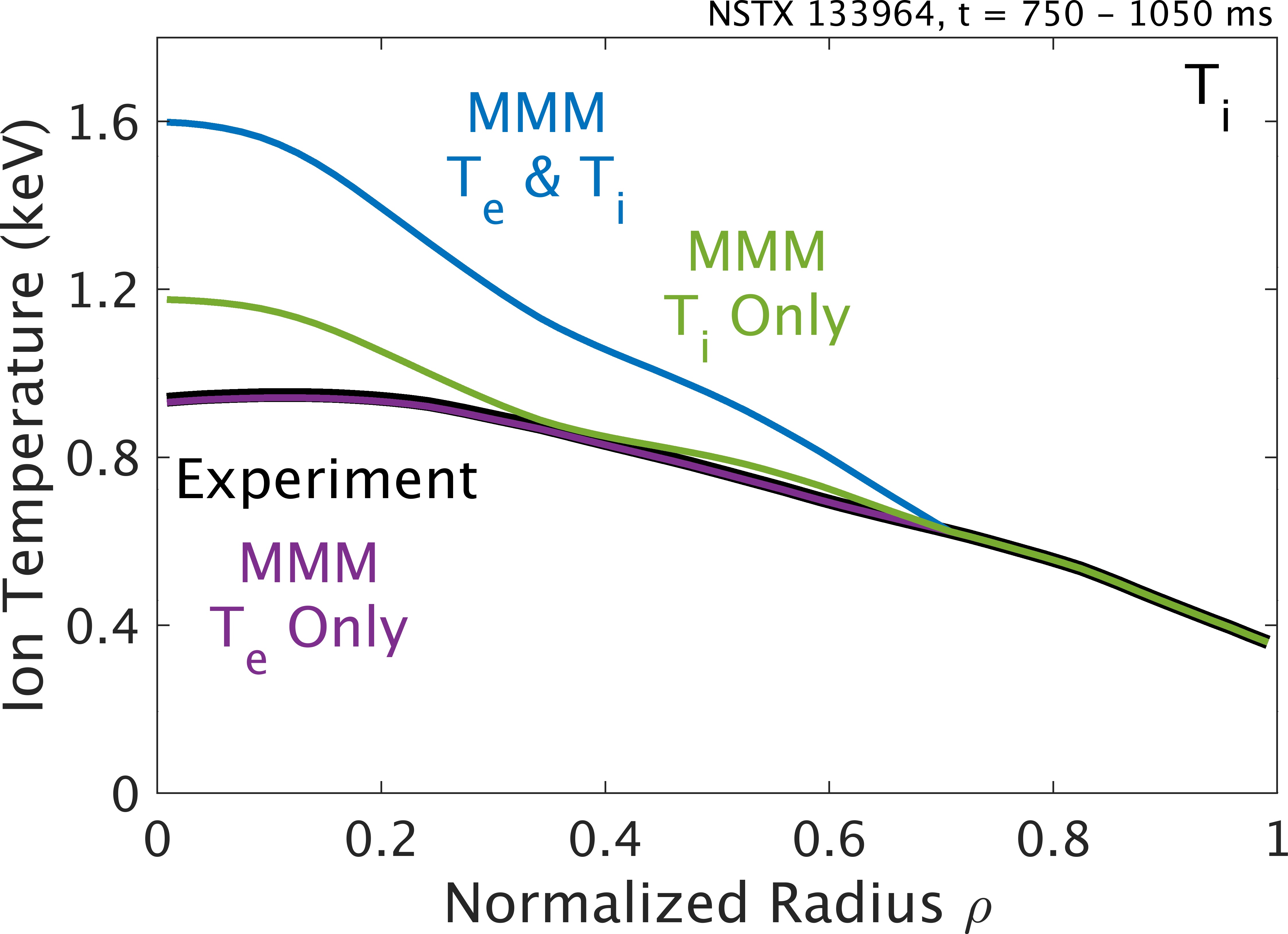}} \\
\subfloat[\label{fig:qbeam_comp}]{\includegraphics[height = \thirdheight]{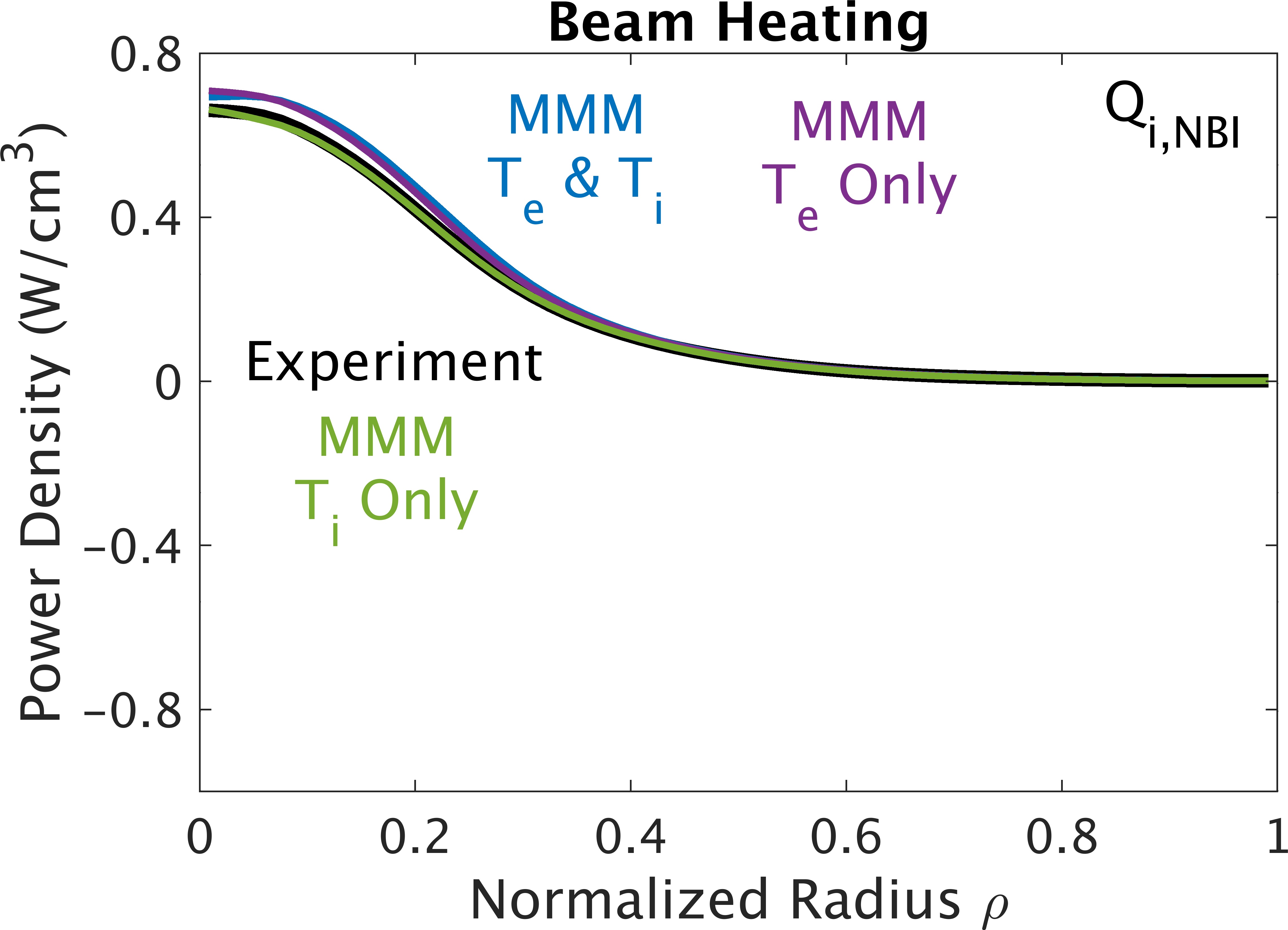}} \figsepb
\subfloat[\label{fig:qcond_comp}]{\includegraphics[height = \thirdheight]{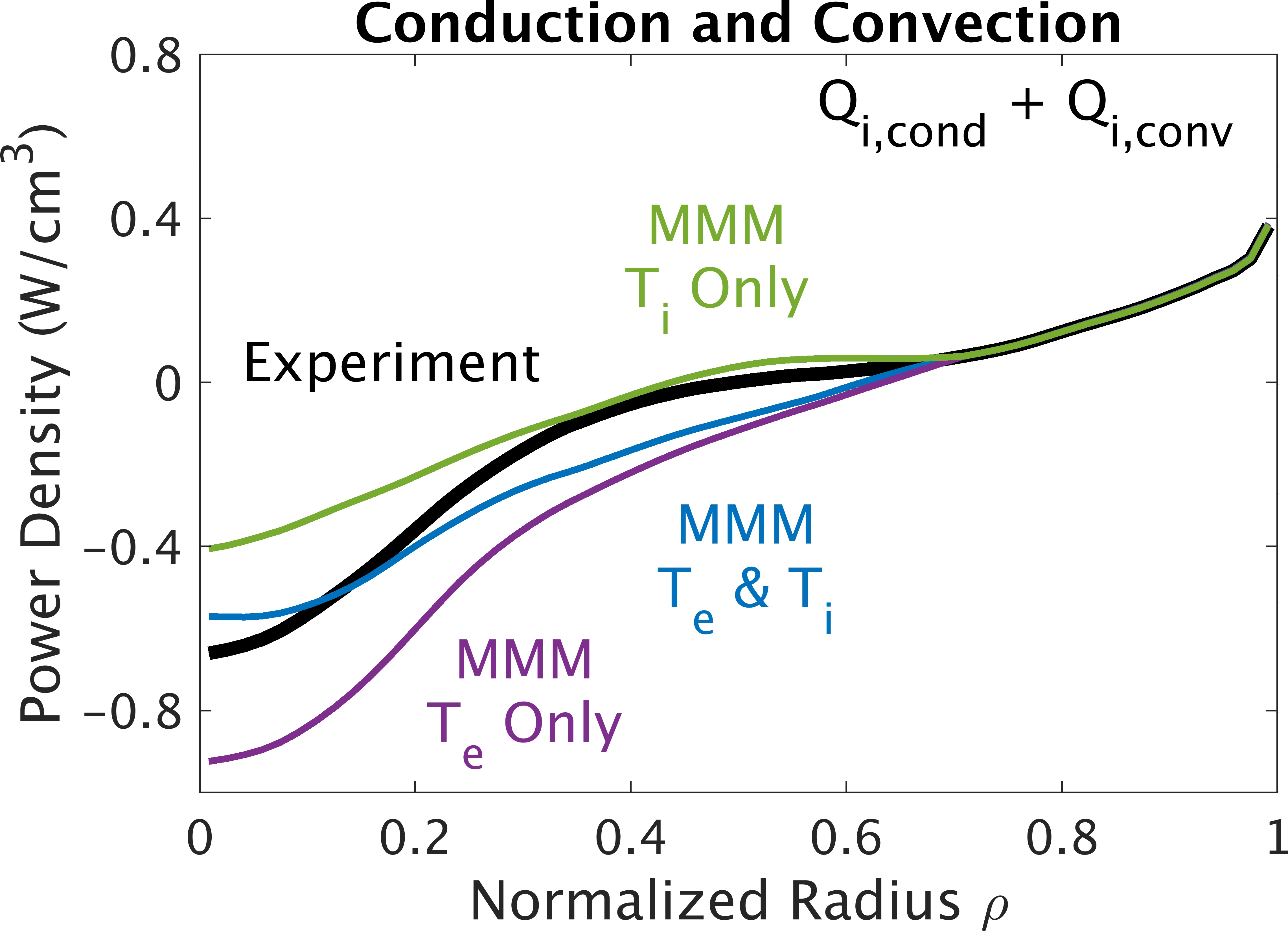}} \figsepb
\subfloat[\label{fig:qie_comp}]{\includegraphics[height = \thirdheight]{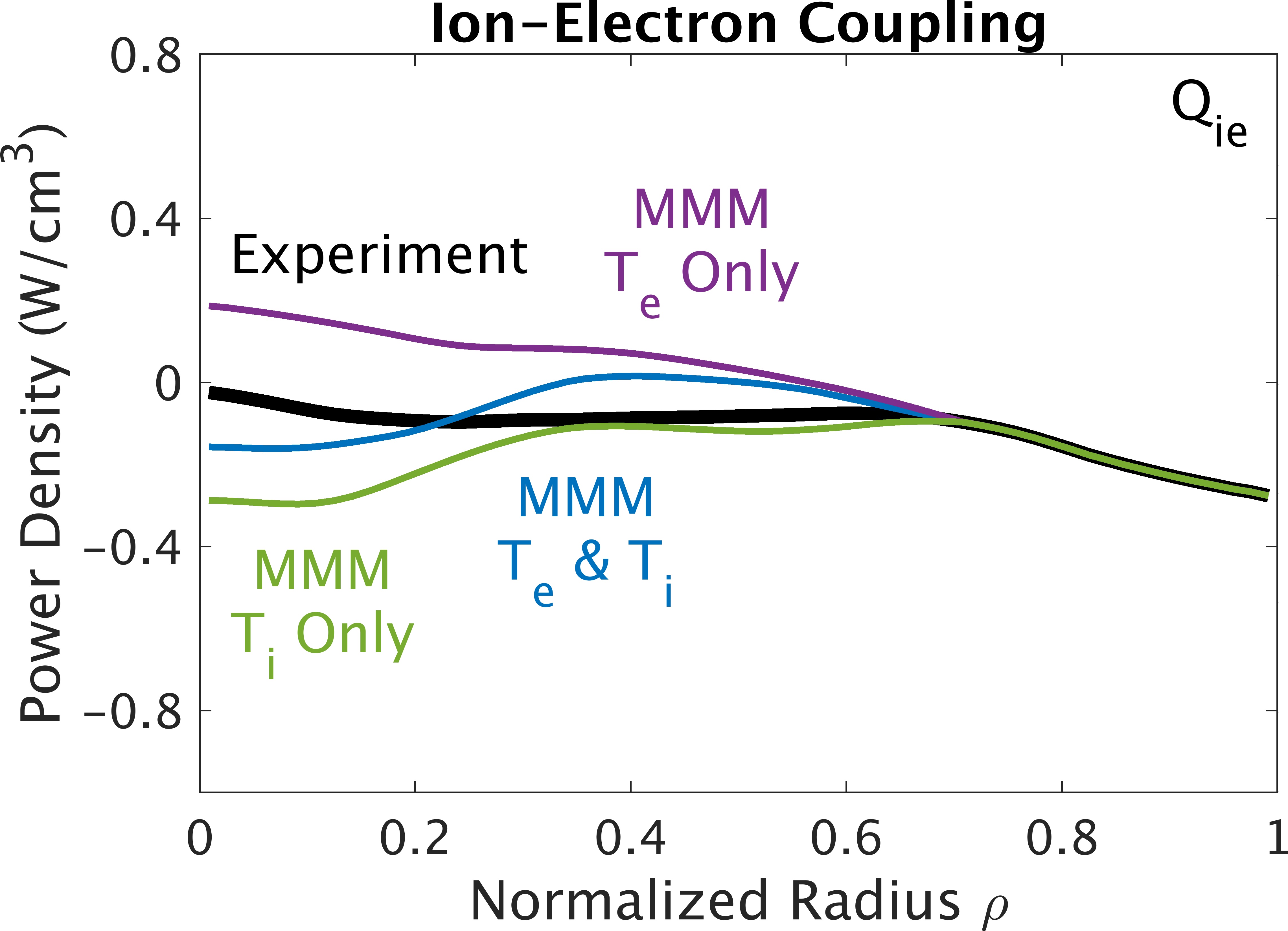}}
\caption{Temperature profiles and ion power balance profiles for NSTX discharge 133964. (a) electron temperature profile, (b) ion temperature profile, (c) neutral beam heating power to ions, (d) conduction and convection ion losses, and (e) ion-electron coupling. In (c - e), positive values indicate energy flows into the ions. In all plots, the thick black curves indicate the experimentally measured temperature profile or power balance profiles inferred from experimental inputs. The different curves correspond to different predictive TRANSP simulations that used MMM to predict different combinations of profiles: $\Te$ and $\Ti$ simultaneously (blue), $\Te$ only (purple), and $\Ti$ only (green).}
\label{fig:iebal}
\end{figure*}

\subsection{Ion Energy Transport}
\label{sec:char_i}

For the ion energy transport, it is expected that turbulent transport from ITGs and TEMs should be suppressed by large $\exb$ shear in NSTX \cite{Kinsey2007POP,Roach2009PPCF,Kaye2021PPCF}, leaving primarily the neoclassical component. \figref{fig:xki_hist_all} shows the ratio of the ion heat diffusivity from turbulent transport (as calculated by MMM) to the neoclassical component (from NCLASS), averaged into spatial bins of width $\Delta\rho = 0.05$. The height of the bars shows the median across all simulated discharges and the error bars show the half interquartile range, which only extend in the positive direction since the lower error bars would terminate at negative values, incompatible with the log scale. Hence, the $\Ti$ predictions in TRANSP in typical NSTX conditions are indeed dominated by neoclassical transport, with the turbulent ion contribution only becoming nontrivial closer to the edge. Note that some portion of the very steep drop in this ratio near the axis is unphysical, resulting from the neoclassical diffusion coefficient spiking near the axis due to a breakdown of the drift ordering as $\rho \rightarrow 0$. This is ultimately inconsequential due to the $\Ti$ profile gradients vanishing near the axis. As expected, the turbulent contribution to the ion heat diffusivity becomes stronger for steeper ion temperature gradients, illustrated in \figref{fig:xki_rho}, which tend to occur closer to the plasma edge. When examining the individual discharges in greater detail, it becomes apparent that turbulent ion transport becomes more relevant at lower values of $\beta$, as shown in \figref{fig:xki_beta}. In MMM, the turbulent ion transport can be attributed to the Weiland model, which notably includes ITGs, TEMs, and KBMs. In particular, TEMs are fundamentally electrostatic instabilities that do not require large $\beta$ to be unstable and drive significant transport. This suggests TEMs as a potential culprit for the relatively high fraction of turbulent ion transport in discharges with lower $\beta$. 
Aside from the low $\beta$ discharges, there are several high performing discharges shown on \figref{fig:xki_beta} which have nontrivial turbulent ion transport, including discharges from each of the considered categories, suggesting that turbulent ion transport can not be neglected completely when analyzing high performing NSTX discharges with MMM, even if it is atypical. 

\rev{The ratio of turbulent to neoclassical ion heat diffusivities shown in \figref{fig:xki_plots} is mostly consistent with an analogous ratio of ion heat fluxes calculated in \citeref{Lestz2025pre1} for profiles predicted for the same set of discharges with machine learning surrogate models of TGLF and the gyrokinetic code CGYRO \cite{Candy2016JCP}. In that work, a time slice flux matching solver within the FUSE framework \cite{Meneghini2024arxiv,Neiser2024APS} was used instead of TRANSP, employing the neoclassical transport code NEO \cite{Belli2008PPCF,Belli2011PPCF} instead of NCLASS. There, the median turbulent ion heat flux was between 10\% and 100\% of the neoclassical ion heat flux, becoming more comparable closer to the prediction boundary of $\rho = 0.7$ and also near the axis. At large $\rho$, the results with TRANSP/MMM/NCLASS shown in \figref{fig:xki_hist_all} are similar to those from \citeref{Lestz2025pre1} with FUSE/TGLF/NEO. At mid radius and closer to the axis, MMM predicts that the neoclassical transport completely dominates the turbulent contribution, conflicting with the TGLF-based prediction where the median ratio remains around 10\%. This would be consistent with the ITGs being very weakly unstable, such that the two different turbulent models predict the ITGs to be just below or just above the stability boundary, leading to vastly different ratios with respect to the predicted neoclassical ion transport.}

Although the ion energy transport is dominantly neoclassical in these NSTX discharges, the MMM predictions in TRANSP are nonetheless sensitive to which transport channels are included in the calculations due to nontrivial ion-electron energy coupling in the examined NSTX discharges. As a concrete example, consider \figref{fig:iebal}, which shows the temperature profiles and ion power balance for NSTX discharge 133964, averaged over a 350 ms time window late in the flat top, for several different TRANSP simulations. This was a high performing discharge that had the lowest flat-top average loop voltage of any NSTX plasma and achieved a sustained $\betan > 4.5$ \cite{Gerhardt2011NFcur}, where $\betan = \beta a B_T/I_p$ and $I_p$ is the plasma current. In \figref{fig:133964_te} and \figref{fig:133964_ti}, the $\Te$ and $\Ti$ profiles are shown for this shot, where the thick black curves indicate the experimental profiles and the colored curves show predictive TRANSP simulations that predict all the $\Te$ and/or $\Ti$ profiles. Blue curves use MMM to predict $\Te$ and $\Ti$ simultaneously, purple curves fix $\Ti$ to its experimental value and predict only $\Te$ with MMM, and green curves represent simulations that predict only $\Ti$. In all of the simulations, the same neoclassical model (NCLASS) is used to calculate the neoclassical transport. The predicted ion temperature profiles are substantially different when $\Te$ is set to its experimental value or also predicted simultaneously by MMM, where the former predicts an on-axis $\Ti$ that is 36\% higher than the latter simulation. As an aside, the overpredictions shown in \figref{fig:133964_te} and \figref{fig:133964_ti} for the simultaneous prediction of $\Te$ and $\Ti$ in this discharge are somewhat larger than was typically found across all of the examined discharges, which had median RMSE overpredictions of 28\% and 27\% for $\Te$ and $\Ti$, respectively \cite{Lestz2025pre1}. 

Superficially, one might expect that if the turbulent ion diffusivity is much smaller than the neoclassical diffusivity (true for this discharge and characteristic of most of the examined NSTX discharges), then the simultaneous prediction of $\Te$ should not influence the resulting $\Ti$ profile, conflicting with the example shown in \figref{fig:iebal}. This can be resolved by considering the ion power balance for this discharge. The three largest terms in the ion power balance for the prediction region of $\rho < 0.7$ are the auxiliary heating on thermal ions from neutral beam injection (NBI), power lost through ion conduction and convection, and collisional ion-electron coupling ($\Qie$). The radial profiles for these power sources and sinks in each of the different TRANSP simulations are shown in \figref{fig:qbeam_comp}, \figref{fig:qcond_comp}, and \figref{fig:qie_comp}, respectively, with positive values indicating power flowing into the ions. 

The beam heating is very similar in all four simulations, since this represents the classical beam power deposition as calculated by the Monte Carlo neutral beam injection code NUBEAM \cite{Pankin2004CPC}. Slight differences in the ion heating from NBI results from the different $\Te$ profiles, as the fraction of beam power transferred to thermal ions \vs electrons depends on the ratio of the beam voltage to electron temperature. With little difference in the auxiliary heating between the simulations, ion power balance then requires that any changes in $\Qie$ must be balanced by offsetting changes to the ion power losses through conduction and convection. Since the different simulations predict different temperature profiles, $\Qie \propto \Ti - \Te$ also across the simulations, resulting in different ion energy losses in the calculations, even when neoclassical. 

Insights can be gained by further examining the temperature profiles in each of the MMM simulations. For the simulations that predict both $\Te$ and $\Ti$ (blue curves in \figref{fig:iebal}), $\Te$ and $\Ti$ are untethered from their experimental values, and MMM finds solution with high $\Te$ and $\Ti$, generally overpredicting confinement. When instead one of $\Ti$ or $\Te$ are pinned to their experimental values (purple and green curves, respectively) while the other is predicted on its own, this tends to drag down the predicted profile relative to when both $\Te$ and $\Ti$ are predicted simultaneously. This is because the experimental profiles tend to be much lower than the predicted ones, creating a large $\abs{\Ti - \Te}$ difference that in turn creates a large $\Qie$, which gets progressively more difficult to consistently balance with neoclassical ion heat transport as the difference increases, constraining the solution. For instance, in the simulation that predicts only $\Te$ (purple curve), $\Te > \Ti$ over the entire predicted region, resulting in $\Qie > 0$ everywhere, effectively creating a new ion heat source. Consequently, the ion energy losses must increase through conduction and convection, even though MMM predicts the turbulent ion transport to be insignificant in all of the simulations shown for this discharge. Hence, the comparison of these TRANSP simulations with MMM illustrates how the electron and ion temperature profile predictions are strongly coupled and therefore sensitive to which profiles are predicted together. Note also that this property is not unique to the MMM model, as it is a generic feature that will be present whenever the ion-electron energy exchange is nontrivial, regardless of the specific turbulent transport model. A comprehensive comparison of the $\Te$ predictions both with and without simultaneous prediction of $\Ti$ will be discussed in \secref{sec:teonly}. 

\begin{figure}[tb]  
\subfloat[\label{fig:133964time_te}]{\includegraphics[height = \thirdheight]{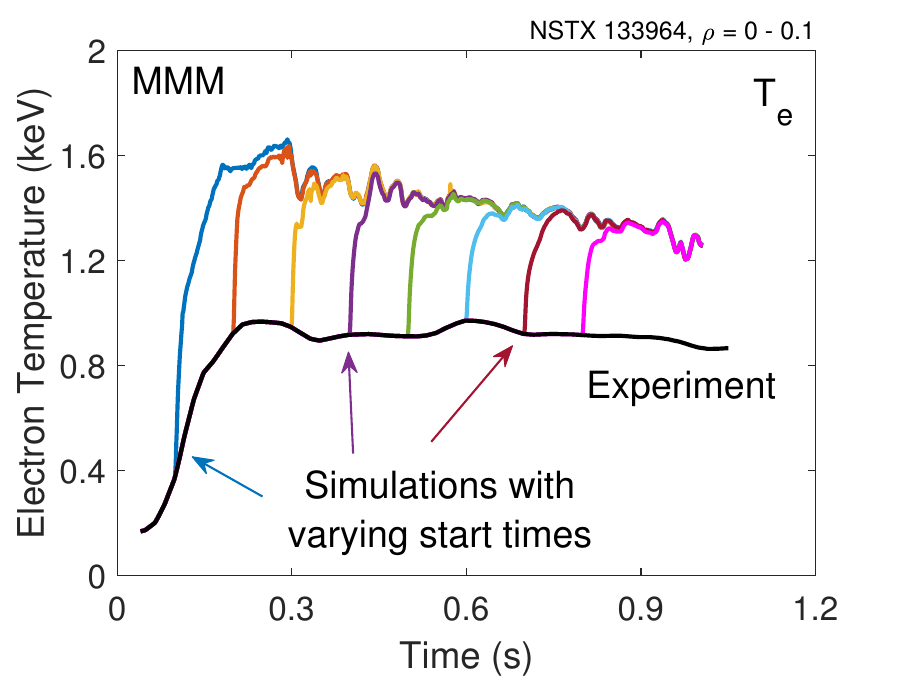}} \\ 
\subfloat[\label{fig:133964time_ti}]{\includegraphics[height = \thirdheight]{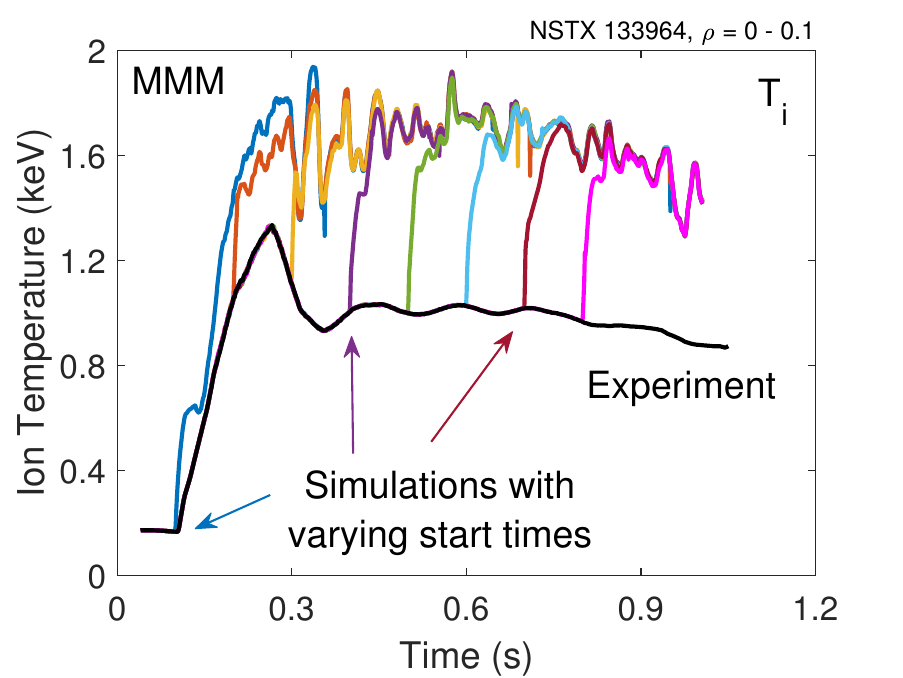}}
\caption{Time dependent predictions of (a) $\Te$ and (b) $\Ti$, averaged from $\rho = 0 - 0.1$ for many different TRANSP simulations of discharge 133964 where MMM was activated with different start times (colored curves). Black curve shows the experimental trace. Each simulation predicts $\Te$ and $\Ti$ simultaneously.}
\label{fig:133964time}
\end{figure}

\section{Time-Dependent Behavior}
\label{sec:time}

Most of the analysis in a closely related work that compared MMM to TGLF \cite{Lestz2025pre1} involved averaging over time windows of $\like 200$ ms and evaluating how well the TRANSP predictions reproduce the experimental profiles averaged over the same window. A complementary approach is to examine how well the predicted temperatures track the measurements in time within some fixed spatial region. For instance, given that the profiles predicted by \ptsolver at each time step inherently depend on the profiles predicted at the previous time step, one might be concerned that the errors could accumulate over time, leading to progressively worse agreement over time or a sensitivity to the initial condition. Fortunately, this does not seem to be the case, as the MMM temperature profile predictions appear to be extremely robust to the initial condition, converging to essentially the same solution independent of how long the simulation has run. 

This property is illustrated in \figref{fig:133964time} for NSTX discharge 133964, where the same TRANSP simulation was performed several times, changing only the time in the simulation when \ptsolver was activated with MMM. The traces show $\Te$ averaged over $\rho = 0 - 0.1$, though the trends are similar when considering any radial location. Each TRANSP simulation started in interpretive mode at 40 ms, in order to give the fast ion density from neutral beam injection some time to build up before the predictions begin. Then at some later time in each simulation, \ptsolver is activated with MMM predicting both $\Te$ and $\Ti$ simultaneously. For instance, the dark blue curves in \figref{fig:133964time} corresponds to a simulation where MMM is activated at 100 ms (during the current ramp), whereas in the simulation for the purple curves, MMM is not used until 400 ms (well into the flat top). The black curves show the observed near-axis $\Te$ (\figref{fig:133964time_te}) and $\Ti$ (\figref{fig:133964time_ti}) data that is input to TRANSP and is used up until the time that \ptsolver is enabled. In each simulation, there is an initial period of approximately $50 - 100$ ms where the predicted temperature rapidly changes, as the profile transitions from the experimental input to the transport model's preferred solution. After this initial startup period, the predicted temperature tracks the experimental temperature reasonably well, aside from offset due to the model overpredicting the profile in general. 

Moreover, \figref{fig:133964time_te} shows the striking feature that the predicted electron temperature at any given time is essentially independent of when the prediction started, as $\Te$ at 900 ms is almost identical for the simulations where \ptsolver began at 700 ms \vs 200 ms, for instance. This property can be thought of as a limitation on the system's memory since the predicted temperature seems to only depend on the most recent $50 - 100$ ms at most, which is similar to the calculated energy confinement time. Hence, the very robust $\Te$ predictions over time, without accumulated errors or sensitivity to the initial condition, can be understood as a consequence of finite time correlations present in the system, due to energy transport and replenishment by auxiliary heating sources (in this case, neutral beam injection). The corresponding $\Ti$ predictions in the same series of simulations are similarly robust to the simulation start time, even with larger variability in the predicted $\Ti$ over time, and exhibit a similar timescale for the transition from the experimental profile to the steady predicted one. While \figref{fig:133964time} compares the temperature predictions near the axis for simulations with different start times, the same result is found for the predicted temperature at all radial locations in the plasma. 

\begin{figure}[tb]
\includegraphics[height = \thirdheight]{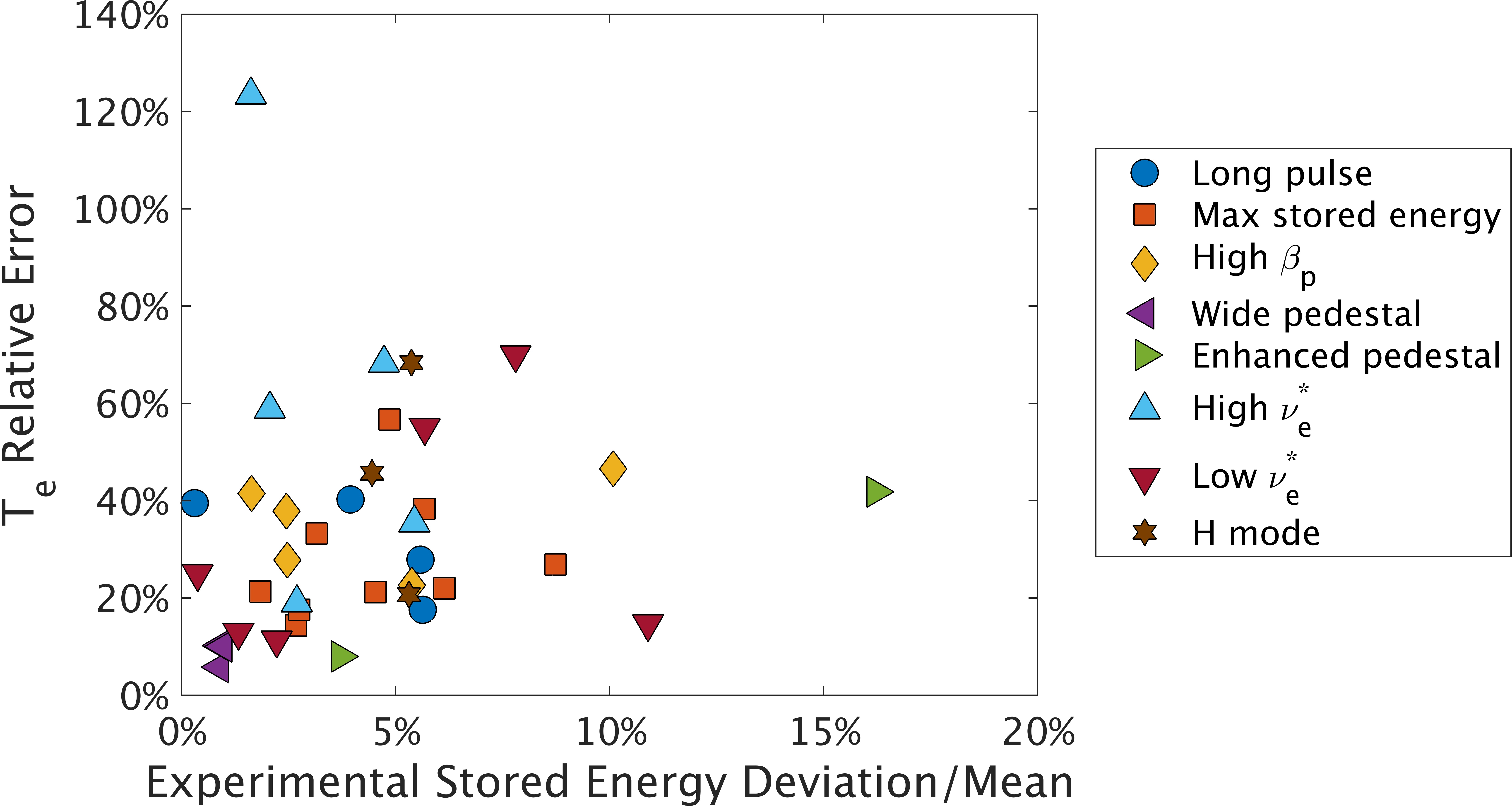}
\caption{Relative error in the predicted $\Te$ profile as a function of the standard deviation of the experimental stored energy over time divided by its mean value over the chosen analysis time window for each discharge. The stored energy is volume integrated from $\rho = 0 - 0.7$.}
\label{fig:utotl_var}
\end{figure}

One challenge of making time-dependent profile predictions over long time periods instead of individual time slices is that there can be transient dynamics in the real experiment that are not self-consistently captured in the TRANSP simulations. For instance, fast ion transport and losses due to \Alfven eigenmodes are not currently included in MMM, nor are low-$n$ MHD modes which can cause significant energy and particle transport. Since the experimental density profiles are used as inputs to the TRANSP simulations in this work, those profiles contain some information about these events, even if their self-consistent effects are not included. To test the sensitivity to this possible source of error, the coefficient of variation of the stored energy was calculated for each analyzed time window, defined as the standard deviation divided by the mean with respect to time, which had an average value of 5\% for the examined discharges. Here and elsewhere in this work, the stored energy is volume integrated from $\rho = 0 - 0.7$ in order to exclude the large contribution at larger radius which is not predicted in these simulations. A weak correlation is found between the time variation in the stored energy (as defined by the coefficient of variation) and the agreement of the $\Te$ profile predictions with experiment, as shown in \figref{fig:utotl_var}. This indicates that transient effects in the plasma are having some influence on the predictions, though not a strong one. There is likewise no correlation between how closely MMM's predictions agree with a given experimental profile and the duration of the chosen analysis windows, which spanned $20 - 750$ ms, depending on the longest quiescent window identified in each discharge. Moreover, a second set of simulations will be discussed in \secref{sec:teonly} which focused on analysis windows of only 20 ms, much shorter than the $\like 200$ ms analysis windows used in the rest of the work. Comparable levels of experimental agreement were found in the two sets of simulations. Hence, trends found in the profile predictions in comparison to experimental observations when averaging over large time windows in this work and \citeref{Lestz2025pre1} can be considered meaningful. 

\begin{figure}[tb]
\subfloat[\label{fig:linfit_133964}]{\includegraphics[height = \thirdheight]{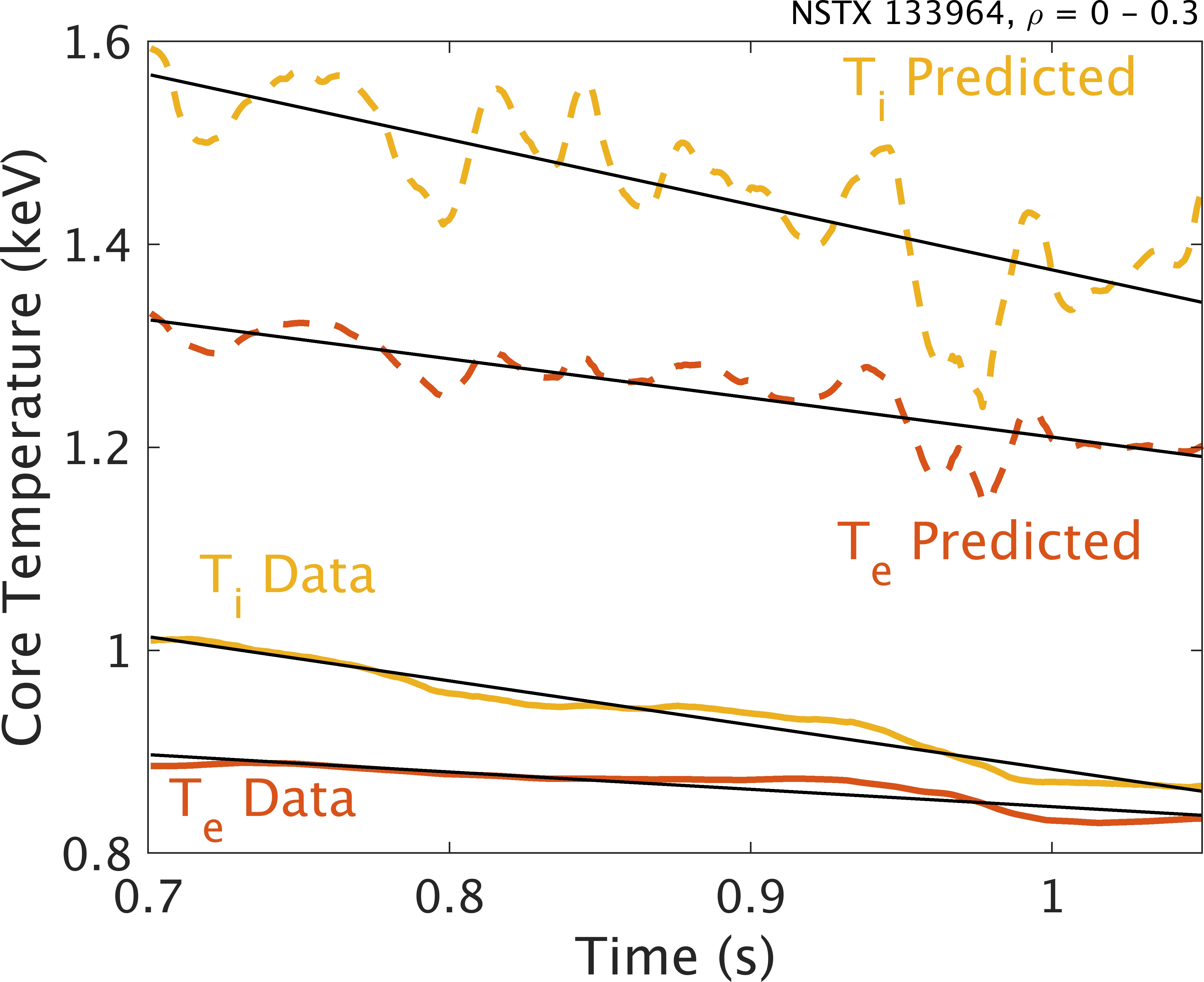}} \\
\subfloat[\label{fig:linfit_scatter}]{\includegraphics[height = \thirdheight]{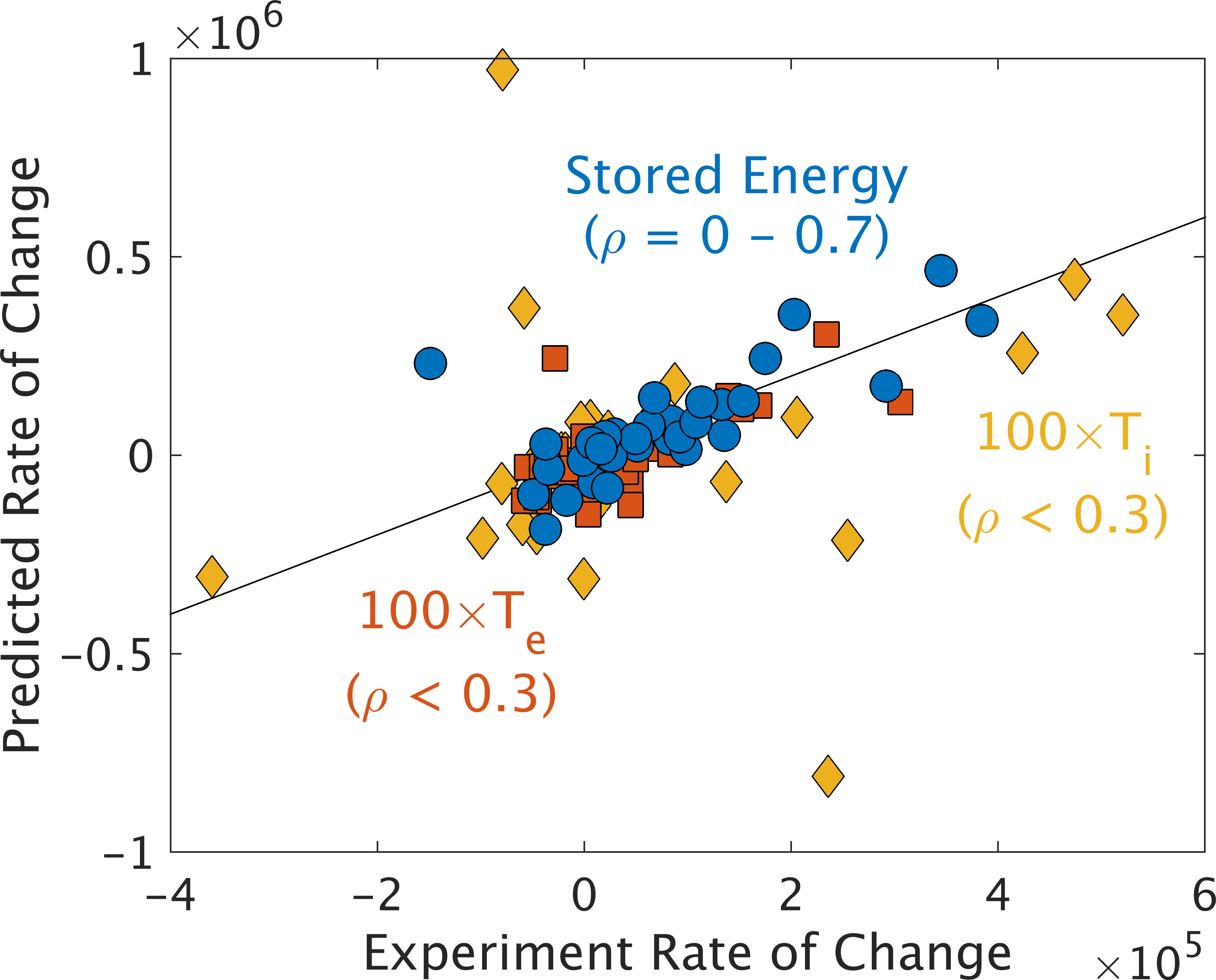}} 
\caption{Comparison of time-dependent rate of change for experimental data \vs MMM predictions. (a) An example of the linear fits for discharge 133964 for $\Te$ (orange) and $\Ti$ (gold), both averaged over $\rho < 0.3$. Solid curves are experimental data, dashed are the profile predictions, and thin black lines are the linear regressions. (b) Comparison of slopes of linear fits for stored energy (blue), integrated from $\rho = 0 - 0.7$, $T_e$ (orange), and $T_i$ (gold). Units are $J/s$ for stored energy and $eV/s$ for temperatures. The solid black line is for reference only, indicating zero offset between experimental and predicted values.}
\label{fig:linfit}
\end{figure}

\begin{figure*}[tb]
\subfloat[\label{fig:gradte_expred}]{\includegraphics[height = \thirdheight]{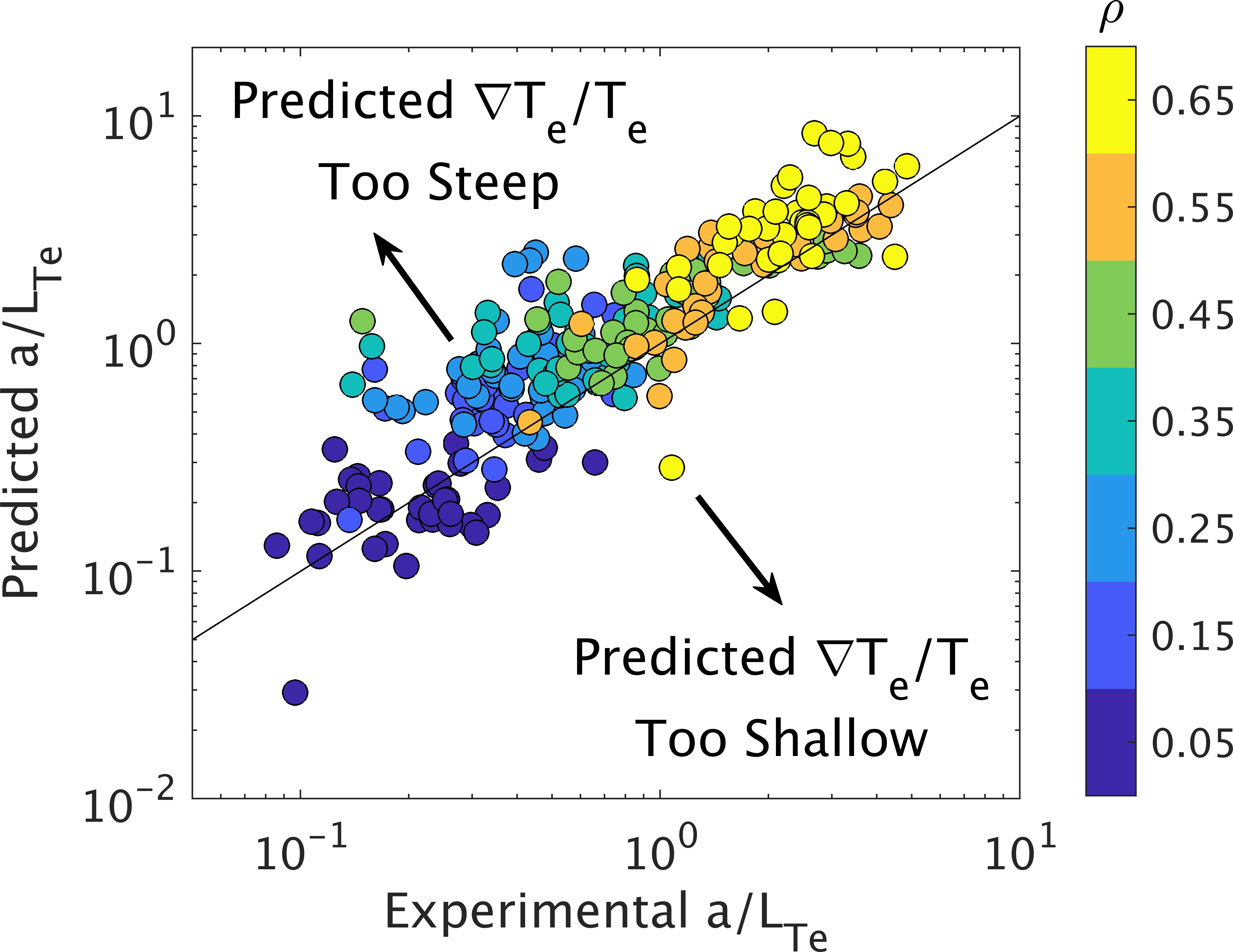}} \figsep
\subfloat[\label{fig:gradti_expred}]{\includegraphics[height = \thirdheight]{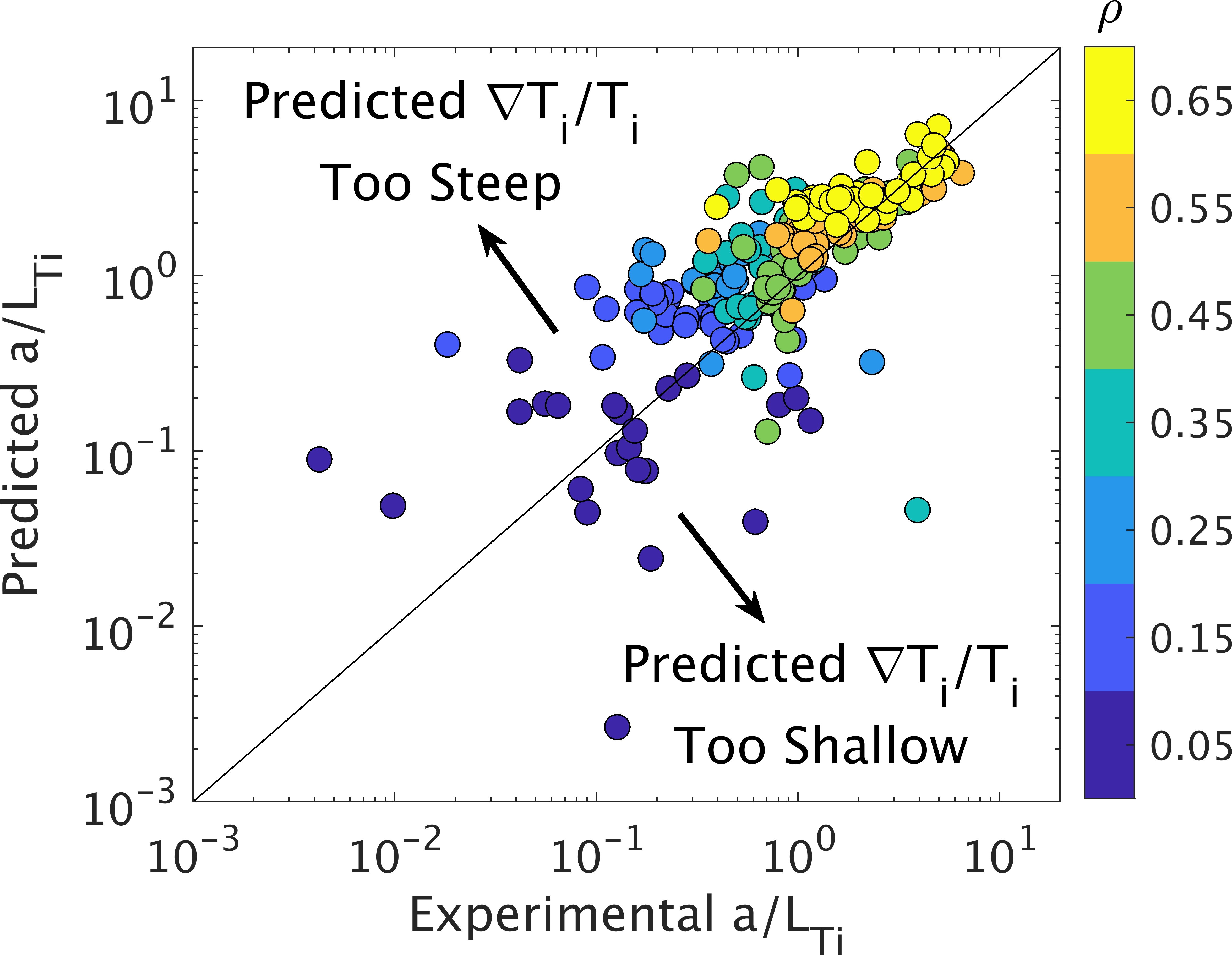}} \\ 
\subfloat[\label{fig:gradte_err}]{\includegraphics[height = \thirdheight]{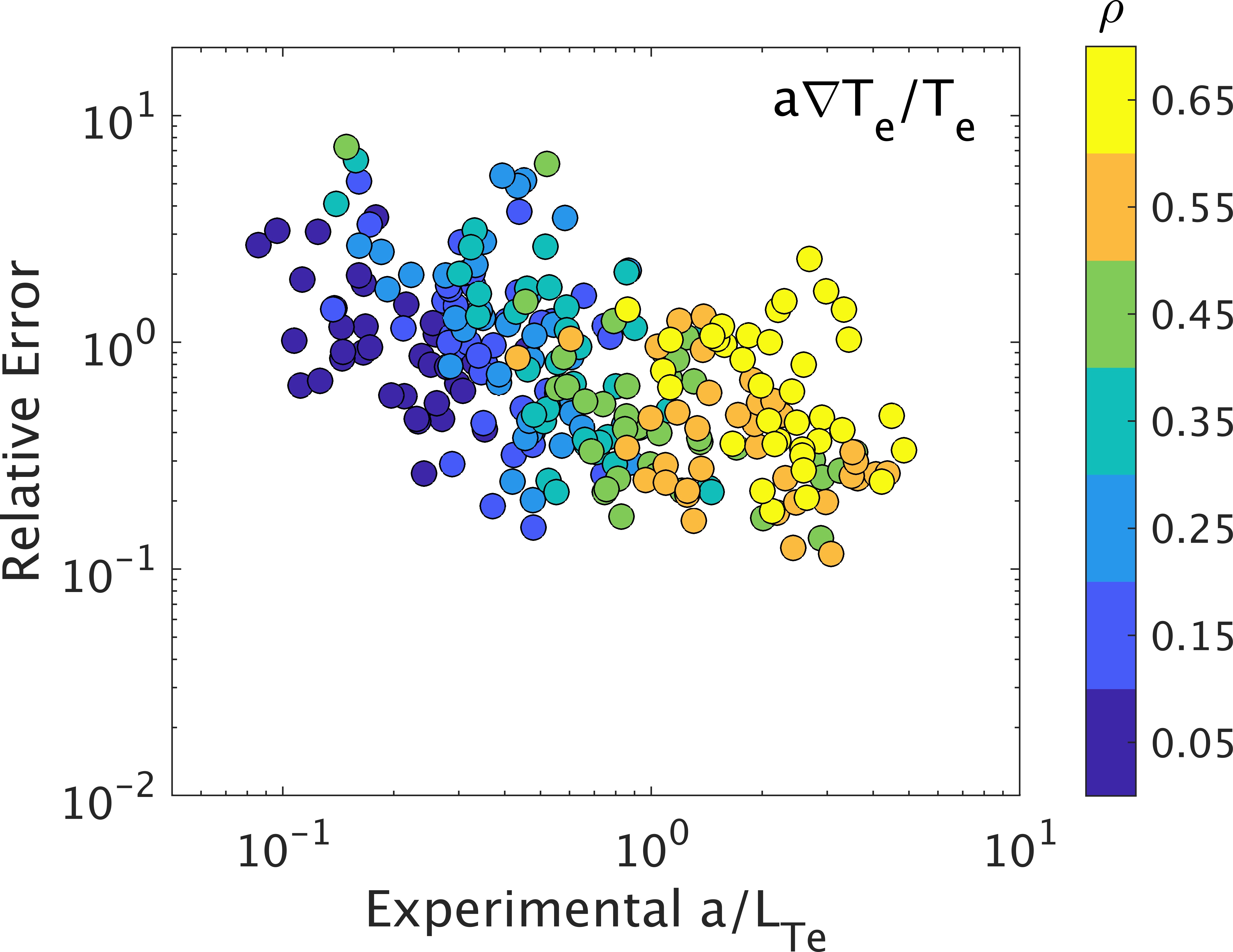}} \figsep
\subfloat[\label{fig:gradti_err}]{\includegraphics[height = \thirdheight]{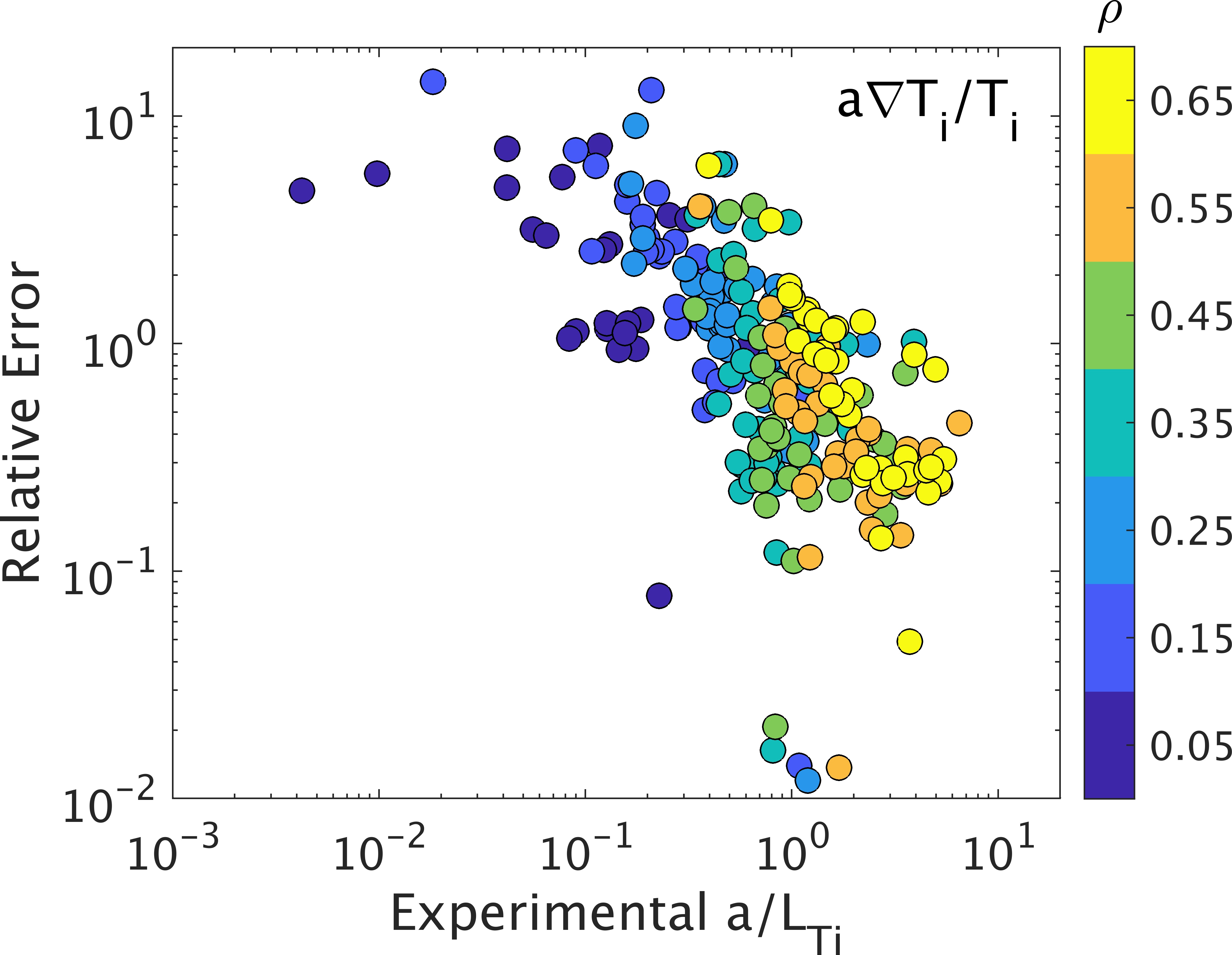}}
\caption{Top row: comparison of experimental \vs predicted gradient scale lengths for (a) $\Te$ and (b) $\Ti$. The solid lines are for reference only, indicating zero offset between experimental and predicted gradients. Bottom row: relative error of the predicted gradient scale lengths as a function of the experimental scale lengths for (c) $\Te$ and (d) $\Ti$. Each point represents a separate NSTX discharge in one radial region. The gradient scale length for each discharge is averaged over a region of width $\Delta\rho = 0.1$. }
\label{fig:graderr}
\end{figure*}

Beyond the average prediction agreement as a function of the variability of conditions during the analysis window, it is useful to understand how well the predicted temperatures are tracking the experimentally measured ones in time, independent of their offset. In essence, this is a test directly targeting the time-dependent nature of the predictions. Abstractly, this is a challenging time series analysis problem to determine if one time series is causally related to another one, where rigorous statistical tests have been developed with applications, for instance, in finance. Given the nature of the reduced transport models that are the subject of this work, the focus will be primarily on the broad qualitative trends over time, instead of quantifying their fine details. Hence for each TRANSP run, linear regressions were performed with respect to time in order to find the best fit lines representing the core $\Te$, core $\Ti$, and total stored energy evolution over the analyzed time window. These average slopes were subsequently compared for the predictions and experimental observations. For $\Te$ and $\Ti$, the temperature was averaged for $\rho < 0.3$, while the stored energy was integrated over the prediction region of $\rho = 0 - 0.7$. Since the analysis windows averaged a few hundred ms during the flat top, the slow evolution of the plasma is qualitatively captured reasonably well despite such a simple fit. For instance, the resulting fits for $\Te$ and $\Ti$ for discharge 133964 are shown in \figref{fig:linfit_133964}. 

In general, a very strong correlation was found between the predicted and measured time evolution for the rate of change plasma stored energy based on these linear fits, with a linear correlation coefficient of 0.95. A comparison of the slopes for each high performing discharge for the stored energy and core temperatures is shown in \figref{fig:linfit_scatter}. The correlations are somewhat lower for the predicted $\Te$ and $\Ti$ evolution compared to experiment, with linear correlation coefficients of 0.66 and 0.23, respectively. Especially for $\Ti$, it is evident that the lower correlation is due at least in part due to several outliers in the predicted rate of change. Whereas the predicted rates of change of the stored energy and core $\Ti$ are faster or slower than the experimental evolution of these quantities in about the same number of discharges, the predicted rate of change of the core $\Te$ is smaller than the experimental rate in 78\% of the discharges. Qualitatively, the reasonable correlation between the experimental and predicted rates of change demonstrates that the TRANSP simulations with MMM do a satisfactory job at reproducing the overall time evolution of the plasma temperature and stored energy, despite consistently overpredicting the temperature profiles at a given time. 

\section{Predicted Gradient Scale Lengths and Profile Steepness} 
\label{sec:gradpeak}

\subsection{Gradient Scale Lengths}
\label{sec:gradient}

\begin{figure}[tb]
\includegraphics[height = \thirdheight]{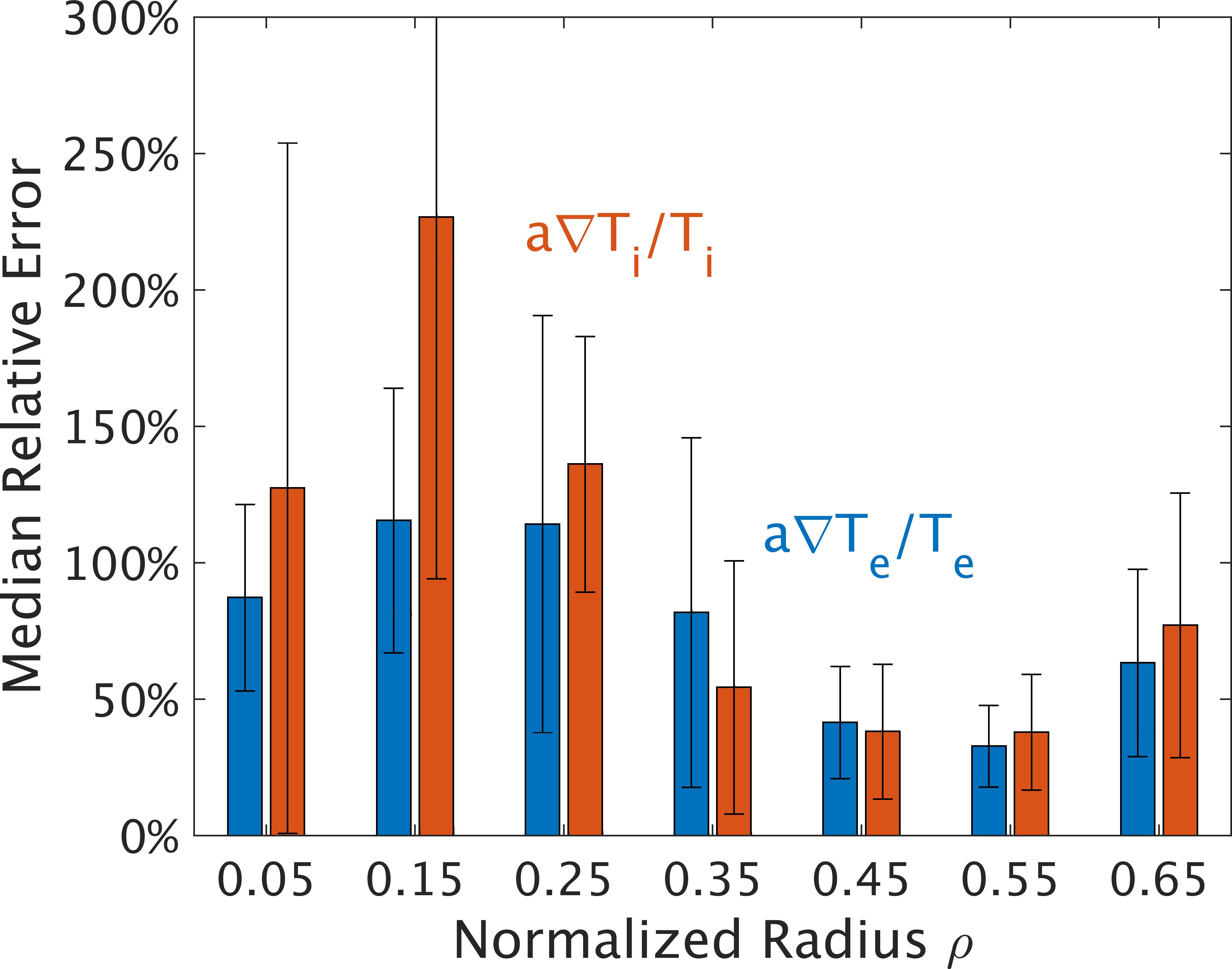}
\caption{Median relative error of the predicted $\Te$ (blue) and $\Ti$ (orange) gradient scale lengths as a function of $\rho$, where error bars show the interquartile range for all discharges.}
\label{fig:graderr_bar}
\end{figure}

\begin{figure*}[tb]
\subfloat[\label{fig:141133_te}]{\includegraphics[height = \thirdheight]{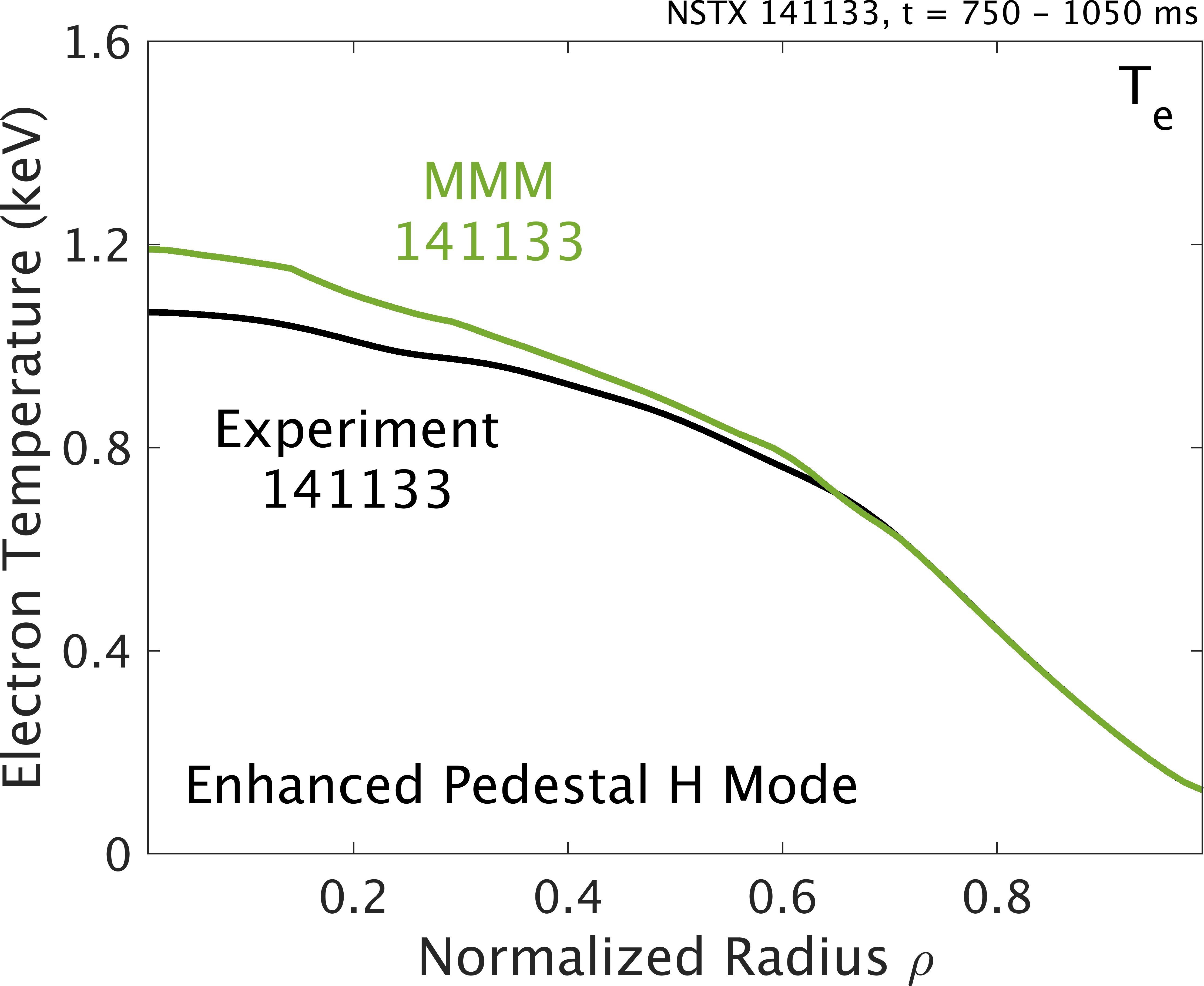}} \figsep
\subfloat[\label{fig:peak_teerr}]{\includegraphics[height = \thirdheight]{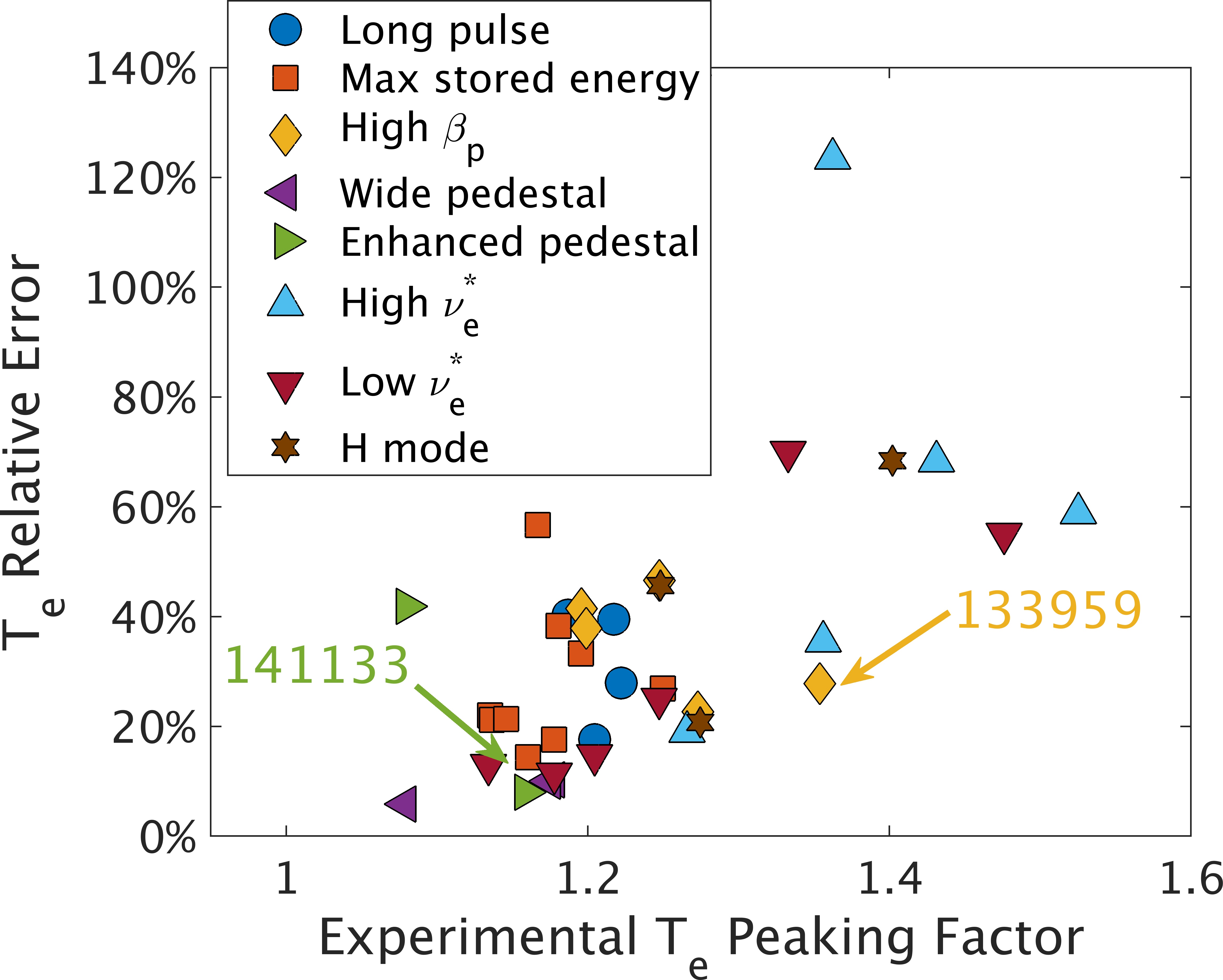}} \figsep
\subfloat[\label{fig:133959_te}]{\includegraphics[height = \thirdheight]{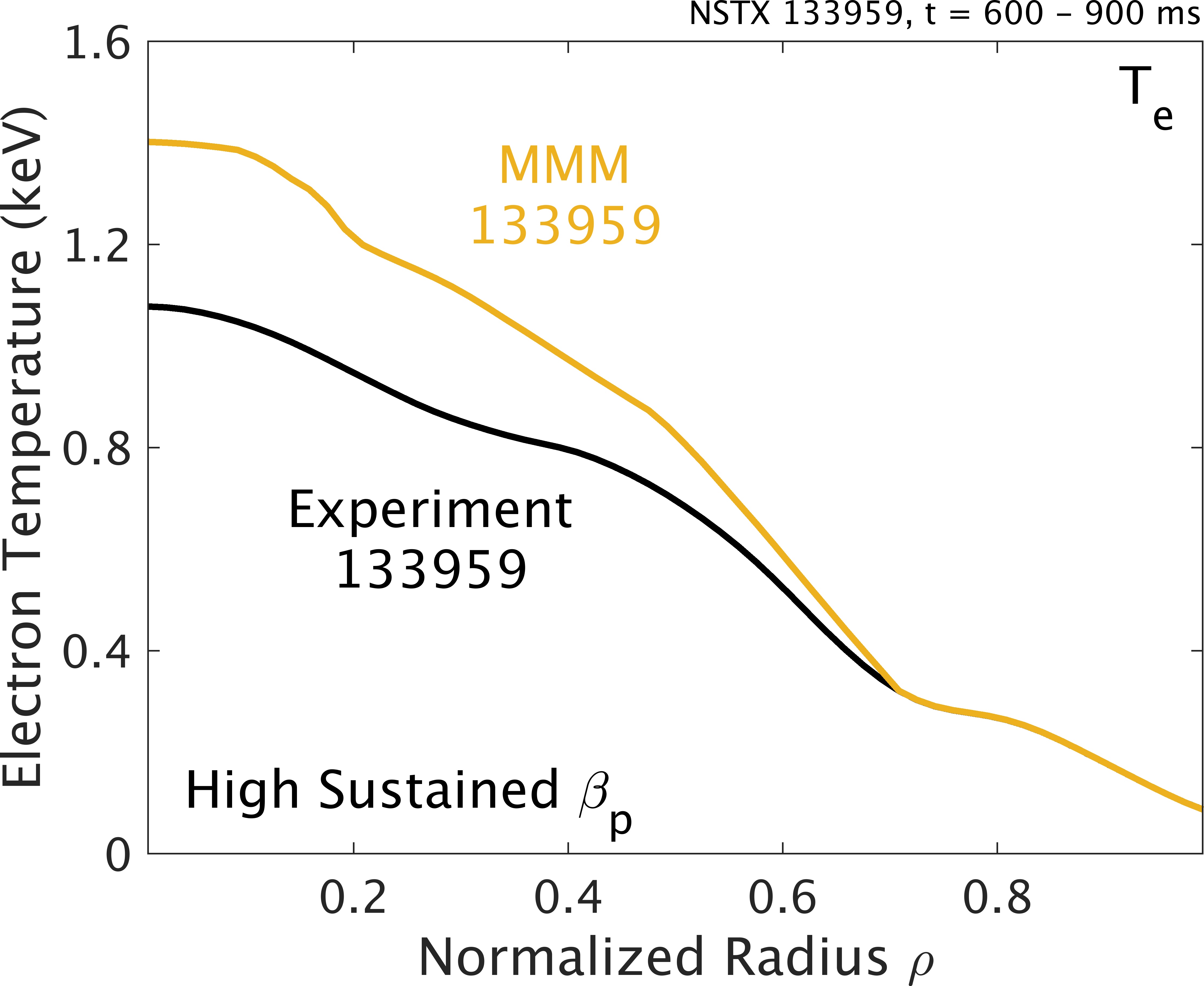}} 
\caption{(a) Comparison of experimental \vs predicted $\Te$ profile for enhanced pedestal H mode discharge 141133. (b) Relative error of the predicted $\Te$ profile as a function of the profile peaking factor. (c) Comparison of experimental \vs predicted $\Te$ profile for high $\betap$ discharge 133959.}
\label{fig:broadprofs}
\end{figure*}

The majority of the companion paper (\citeref{Lestz2025pre1}) focused on whether the temperature profiles predicted by reduced transport models such as MMM and TGLF can reliably reproduce those observed in high performing NSTX experiments. Since turbulent transport is fundamentally driven by kinetic profile gradients at each radial location, it is of interest to also directly assess MMM's predictions of local gradients. A comparison of the experimental temperature gradients to those predicted by MMM in each of the examined discharges is shown in \figref{fig:graderr}. Since the local gradient is a noisy quantity that is also sensitive to the grid resolution used for the transport predictions, the gradient is averaged over radial regions of width $\Delta\rho = 0.1$, with the color indicating the central $\rho$ value in the averaging region. Moreover, the normalized gradient scale length, $-a/\LTe = -(a/\Te)(\partial \Te/\partial r)$, is used instead of the raw $\grad\Te$. As illustrated in \figref{fig:gradte_expred}, the $\Te$ gradient scale lengths predicted by MMM in TRANSP generally track the experimental gradients. However, $\grad\Te$ is consistently predicted to be too steep at essentially all radii, in agreement with the finding that the $\Te$ profile itself is nearly always overpredicted by MMM \cite{Lestz2025pre1}. The main exception to this trend is very close to the axis, where the experimental gradients tend to be flatter and consequently local turbulent transport models may break down. For $\rho > 0.1$, $75 - 95\%$ of the investigated discharges have predicted gradients that are steeper than the experimental values, depending on the specific radial location, whereas for $\rho < 0.1$, only $46\%$ of simulations predicted overly steep gradients. For $\grad\Ti$, the story is qualitatively similar, as \figref{fig:gradti_expred} shows that the $\Ti$ gradients are dominantly overpredicted by MMM, though the overprediction is less pronounced at larger $\rho$, which is not the case for the offset of the predicted $\Te$ gradient. Physically, the overly steep predicted gradients are again reflecting the finding that the predictive TRANSP simulations with MMM are not generating enough thermal transport to match the experiment. 

Moreover, \figref{fig:gradte_err} and \figref{fig:gradti_err} demonstrate a clear trend that the MMM predicted gradients are closer to experiment in conditions where the experimental gradients are steeper. Since temperature gradients tend to become progressively steeper at larger radii, a direct consequence is that MMM tends to find better agreement with experimental gradients near the pedestal than near the axis. This is a strong trend for both $\Te$ and $\Ti$ gradient predictions. One notable difference between the electron and ion temperature gradient predictions is shown in \figref{fig:graderr_bar}, where the median gradient RMSE averaged across all discharges is shown as a function of radius for both $\grad\Te$ and $\grad\Ti$. In aggregate, the predicted normalized ion temperature gradient have relatively larger disagreement with experiment for $\rho < 0.3$ than the electron temperature gradients do. The error bars indicate the interquartile range for the gradient RMSE across all discharges, demonstrating substantial spread, especially in the core where relatively flat gradients amplify similar absolute differences into large relative differences. 

\subsection{Peaked Profiles}
\label{sec:peak}

Based on the findings of the previous section, that steeper experimental temperature gradients are predicted with better agreement by TRANSP simulations with MMM, one might intuitively infer that plasmas with more peaked temperature profiles are the ones that are more successfully reproduced. However, analysis of the database of discharges indicates that the opposite is true for $\Te$ -- the better predicted discharges tend to be the ones with very broad electron temperature profiles. To characterize how peaked a profile is, define a profile peaking factor as the ratio of the on-axis value to its radial average over the predicted region ($\rho = 0 - 0.7$ for these simulations). For monotonic profiles, a peaking factor of 1 corresponds to a completely flat profile, while the maximum temperature peaking factor is $T_{0}/T_{\text{edge}}$. \figref{fig:peak_teerr} shows that plasmas with more peaked experimental $\Te$ profiles tend to have a larger relative error in the predicted $\Te$ profile, with a linear correlation coefficient of $r = 0.59$. 

\begin{figure*}[tb]
\subfloat[\label{fig:teerr_beta}]{\includegraphics[height = \thirdheight]{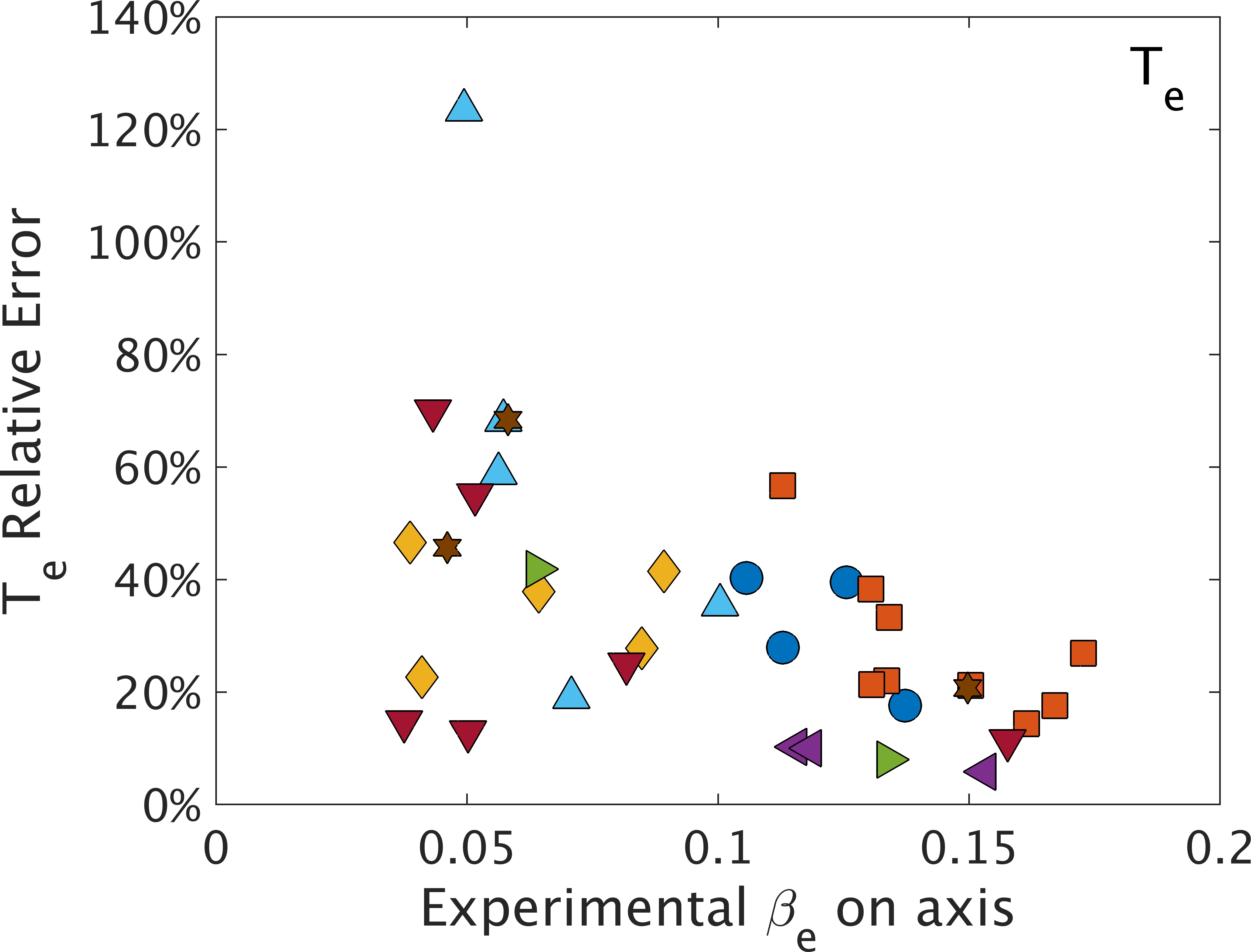}} \figsepsm
\subfloat[\label{fig:teerr_taue}]{\includegraphics[height = \thirdheight]{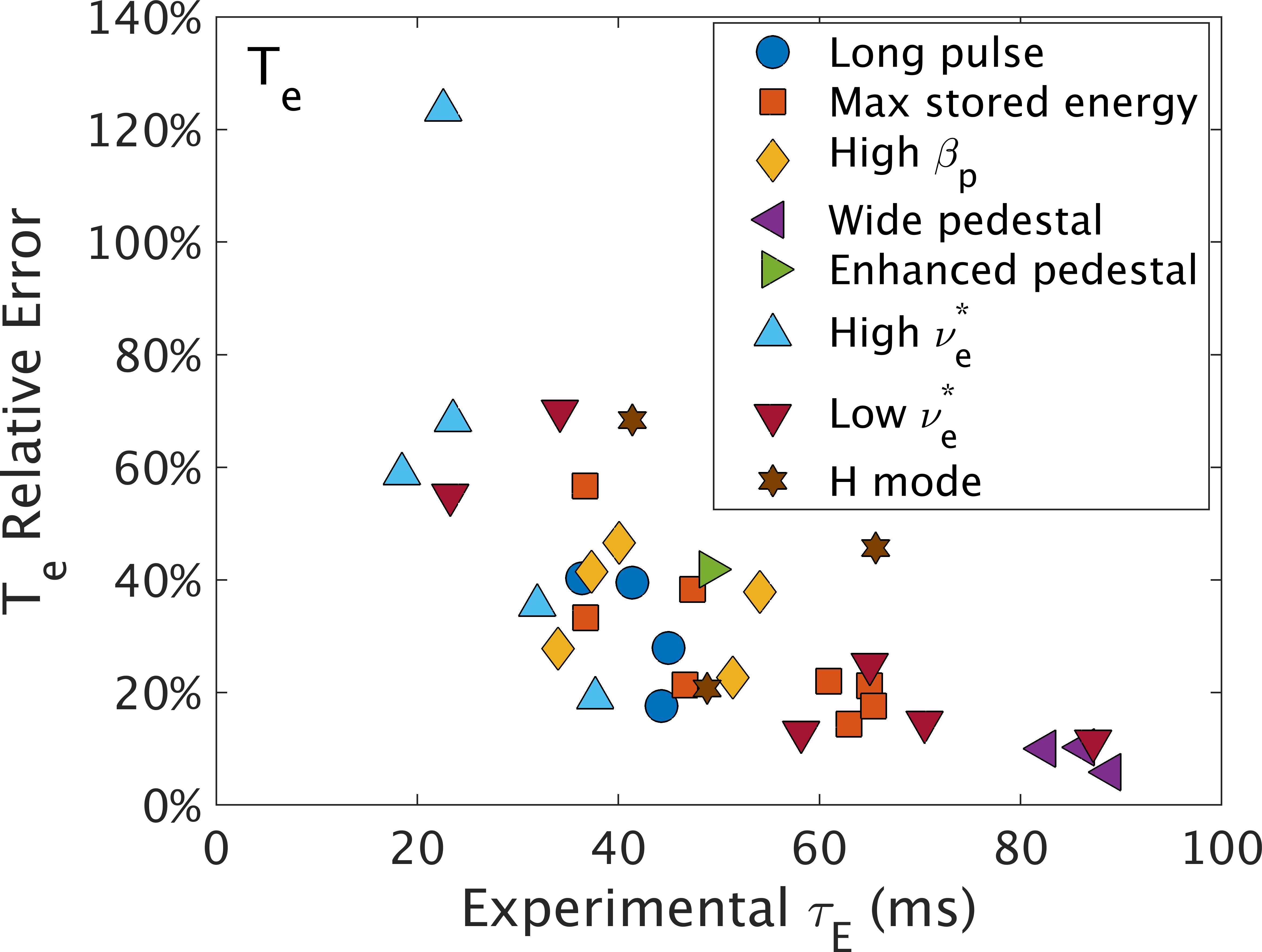}} \figsepsm
\subfloat[\label{fig:teerr_nustar}]{\includegraphics[height = \thirdheight]{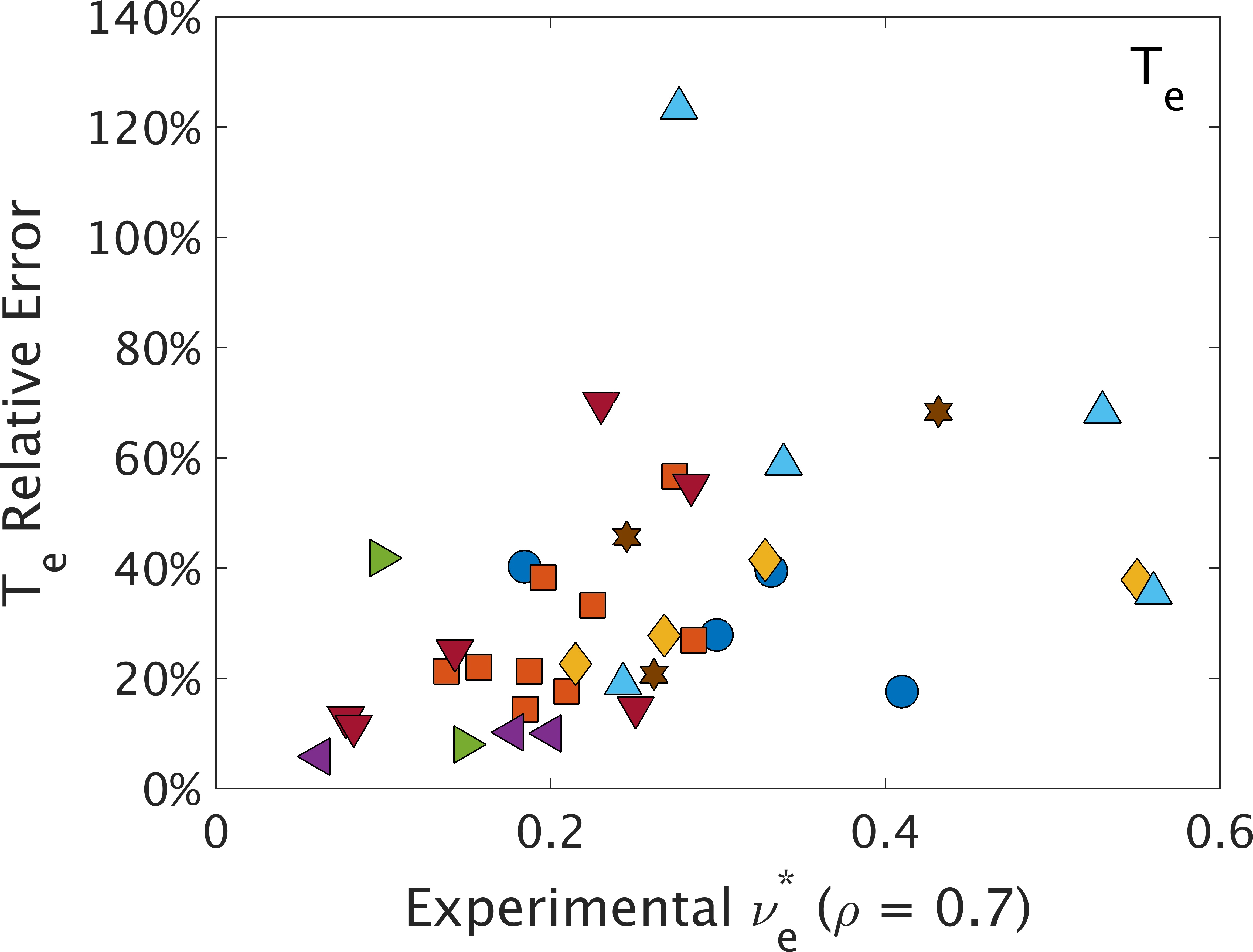}}
\caption{Relative error in $\Te$ profile predicted by MMM as a function of experimental values of (a) $\beta_e$ on axis, (b) global energy confinement time $\taue$, and (c) $\nustar$ evaluated at $\rho = 0.7$.}
\label{fig:beta_taue_nustar}
\end{figure*}

As two representative examples, consider the enhanced pedestal (EP) H mode discharge 141133 shown in \figref{fig:141133_te} in comparison to the high $\betap$ discharge 133959 in \figref{fig:133959_te}. The EP H mode scenario on NSTX features a very wide pedestal with several distinctive features \cite{Gerhardt2014NF,Battaglia2020POP}. For this EP H mode discharge, the MMM prediction for the $\Te$ profile is excellent, whereas the high $\betap$ discharge with a rather peaked profile has a somewhat larger disagreement in the predicted profile. Notably, the two discharges have very similar on-axis electron temperatures of approximately 1.1 keV, but very different temperatures at the prediction boundary of $\rho = 0.7$, which is responsible for the very different peaking factors. Although there are only a few discharges in each category, \figref{fig:peak_teerr} hints that this may be a general trend, as some of the discharges that are predicted most closely to experiment are those with wide or enhanced pedestals, whereas the high $\betap$ discharges are more peaked and tend to have larger disagreement. 
In particular, the discharges with $\Te$ profiles that are most poorly predicted by MMM have some of the the largest experimental peaking factors, consistent with the overall trend. No such trend exists for how peaked the experimental $\Ti$ profile is and how well it is predicted by MMM. 

So how can one resolve the apparent paradox between steeper local gradients being predicted with better experimental agreement while more peaked profiles are predicted with worse agreement? One interpretation is that broader profiles such as discharge 141133 are mathematically easier to predict when using the relative, \eg fractional, error as the figure of merit. To see this, consider experimental profiles similar to the two examples shown in \figref{fig:peak_teerr}, where one has a very shallow gradient at the edge and the other has a relatively steep gradient. Since the underlying transport model fundamentally predicts the local gradient, the transport model would be considered to perform equally well in the two situations if it overpredicted both gradients by  50\%. However, overpredicting the steep edge gradient case by 50\% will result in a much steeper gradient at the edge than overpredicting the shallow edge gradient case by the same fraction. Consequently, the steep edge gradient case will result in a much larger overshoot in the core temperature as a fraction of the experimental profile than the shallow edge gradient case will. Thus, the findings that MMM predicts steep gradients and also broader profiles more reliably can be simultaneously true since fractional errors in the local gradients and global profiles are not directly linked, and in fact depend on the shape of the profile being predicted. 

This interpretation is corroborated by examining correlations between the relative error of the predicted $\Te$ gradient in different regions with the overall agreement between experimentally measured and predicted profiles. 
\rev{The RMSE for the predicted edge $\Te$ gradient ($\rho = 0.6 - 0.7$) is well correlated with the relative error in the $\Te$ profile, with a linear correlation coefficient of $r = 0.65$, such that MMM overpredicting the edge gradient often results in overpredicting the entire $\Te$ profile relative to the experiment. Conversely, there is essentially no correlation of the relative error in the $\Te$ profile with the RMSE of $\Te$ gradient predictions in the mid-radius ($\rho = 0.4 - 0.6$) or core ($\rho = 0.2 - 0.4$) regions, which have $\abs{r} \leq 0.21$. The edge $\Te$ gradient from the experimental profile fit is likewise poorly correlated with the $\Te$ RMSE ($r = 0.24$).}
Hence, the relative error in the edge gradient is most influential in lifting the core $\Te$. Since MMM routinely predicts gradients that are too steep, and thus also core electron temperatures that are too hot, the model also tends to predict more peaked $\Te$ profiles than were observed in the experiment, with peaking factors that are about 10\% too large when compared to the experiment. 

\section{Influence of $\beta$, Energy Confinement Time, and Collisionality}
\label{sec:betanu}

An important parameter in transport models for spherical tokamaks is $\beta$, since high $\beta$ regimes can be qualitatively different from the low $\beta$ parameter space common for conventional tokamaks, due to the emergence of electromagnetic turbulence. As discussed in \secref{sec:character}, $\betae$ and $\betap$ have a strong influence on the types of instabilities that drive electron transport in MMM simulations, and the competition between neoclassical and turbulent ion transport is also sensitive to $\beta$. 
\rev{Beyond the unique high $\beta$ regime that spherical tokamaks operate in, more favorable confinement scaling at low collisionality has been observed in NSTX \cite{Kaye2007NF,Kaye2013NF}, MAST \cite{Valovic2009NF,Valovic2011NF}, and Globus-M(2) \cite{Kurskiev2019NF,Kurskiev2022NF} compared to conventional tokamaks.} 
Hence, a significant amount of experimental and modeling efforts have focused on comparing transport across a range of $\beta$ and collisionalities. In this section, results from two different sets of simulations will be presented to investigate sensitivities of MMM to $\beta$ and $\nustar$. \secref{sec:betanuJBL} will analyze the same set of simulations of well-analyzed NSTX discharges that have been presented up to this point. \secref{sec:skaye} will present simulations of a second set of NSTX discharges that are selected with the intention of performing independent scans of $\beta$ and $\nustar$, whereas these parameters are strongly interrelated in the initial set of discharges via the usual temperature dependence $\Te \propto \betae \propto 1/\sqrt{\nustar}$.  

\subsection{Trends in Large Set of Well-Analyzed Discharges}
\label{sec:betanuJBL}

For the predictive TRANSP simulations of the large set of well-analyzed NSTX discharges, the discharges where $\Te$ is most poorly predicted tend to be the ones with lowest $\beta$. A reasonable correlation is found between the on-axis electron $\beta$ and the $\Te$ profile prediction RMSE (linear coefficient of $r = -0.53$), as shown in \figref{fig:teerr_beta}. However, $\beta$ is correlated with other plasma properties in the examined NSTX discharges, such that it may not be the most salient trend. A stronger correlation is found between $\Te$ profile prediction RMSE and the global energy confinement time ($\taue$) than $\betae$, as shown in \figref{fig:teerr_taue} for the MMM simulations. In this work, $\taue$ is defined in the usual way, $\taue = U/(P_\text{aux} - dU/dt)$, where $U$ is the total stored energy and $P_\text{aux}$ is the auxiliary heating power. After removing two outlier points which TRANSP calculated unphysically large $\taue$ for even in the fully interpretive runs (not shown), the relative error in the $\Te$ profile prediction drops rapidly and consistently as $\taue$ is increased when comparing between high performing discharges in the database. The linear correlation is $r = -0.7$, and the rank correlation is even higher, $r = -0.75$, reflecting the nonlinear dependence between the $\Te$ error and $\taue$. 
The relative error in $\Ti$ profile prediction also tends to decrease with increasing $\taue$ in MMM simulations, though less prominently and with significantly more variation than is seen for the $\Te$ RMSE. The improvement of $\Te$ profile prediction in MMM simulations for larger energy confinement times is one of the most robust correlations found when investigating the database of well-analyzed NSTX discharges in this work. 

The interpretation of this result is complicated by parametric correlations in high performing discharges between the profile peaking factor, $\betae$, and $\taue$. In the database used for this work, plasmas with higher on-axis $\betae$ tend to have longer energy confinement times, and in turn, plasmas with longer energy confinement times tend to have less peaked profiles. Hence, it is not immediately obvious which of the trends discussed in this section and \secref{sec:gradpeak} with respect to the experimental agreement of the temperature profile predictions is the most fundamental sensitivity. However, there are some arguments to be made that the energy confinement time is the most relevant parameter. Quantitatively, the correlation between the experimental value of $\taue$ and the relative error in the $\Te$ profile is moderately stronger than the correlation with $\betae$ or the profile peaking factor. The relative error in $\Te$ has a rank correlation coefficient of $-0.75$ with $\taue$, compared to $-0.55$ with $\betae$ on axis and 0.59 with the profile peaking factor. The sign of the correlation coefficient indicates that the RMSE is reduced with increasing $\taue$ and $\betae$ and increases with more peaked profiles, as previously discussed. Physically, a strong dependence on the experimental energy confinement time is consistent with the main findings of this work. When used in predictive TRANSP simulations, MMM predicts electron temperatures that are consistently too high, with local gradients that are too steep, indicating insufficient transport to match the experiments. Hence, it could be the case that in NSTX discharges that have very good confinement, those experimental conditions could be closer to the range of outcomes that MMM tends to predict, reducing the overall disagreement. 

A modest correlation is found between the RMSE $\Te$ profile prediction error and normalized collisionality evaluated at $\rho = 0.7$ (linear correlation coefficient $r = 0.42$, rank correlation of $r = 0.58$), as shown in \figref{fig:teerr_nustar}. Based on this metric, the MMM model is found to more closely reproduce the observed $\Te$ profiles in NSTX discharges at lower normalized collisionality, suggesting that the model may remain reliable when extrapolating to high performing NSTX-U scenarios which will push to an even lower collisionality regime. However, similar to the discussion of trends with $\betae$ in the previous subsection, the underlying explanation for this collisionality trend merits further investigation in future work. In particular, there is an inverse correlation of experimentally measured $\nustar$ and $\betae$ for the discharges used in this database, and there is likewise a strong correlation between the $\Te$ profile peaking factor and $\nustar$. Hence, the observation of improved experimental agreement with the $\Te$ profiles predicted by MMM at low collisionality may be a consequence of other correlations. Additionally, this $\Te$ profile agreement appears to only be sensitive to the collisionality near the pedestal, as $\nustar$ averaged over regions near the core ($\rho = 0.2 - 0.4$) and mid-radius ($\rho = 0.4 - 0.6$) is essentially uncorrelated with the $\Te$ error, with linear correlation coefficients below $\abs{r} = 0.15$. Hence, MMM's prediction of $\Te$ profiles relative to their experimental fits appears to be less sensitive to collisionality than it is to $\betae$ or energy confinement time. 

\subsection{Dedicated $\beta$ and $\nustar$ Scans} 
\label{sec:skaye}

\begin{figure*}[tb] 
\subfloat[\label{fig:skaye_betatt_nustar50}]{\includegraphics[height = \thirdheight]{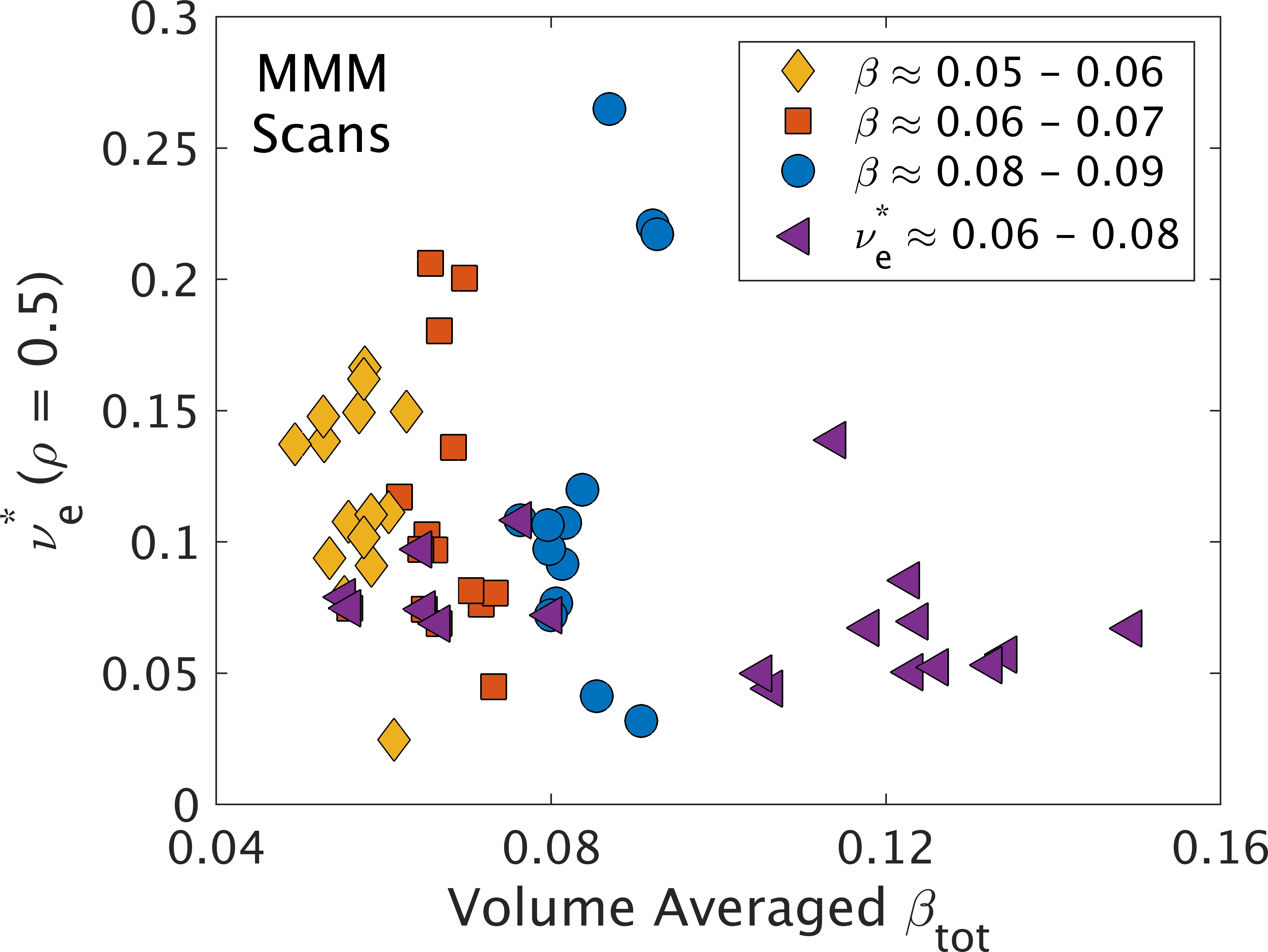}} \figsep
\subfloat[\label{fig:skaye_betatt_teoff20}]{\includegraphics[height = \thirdheight]{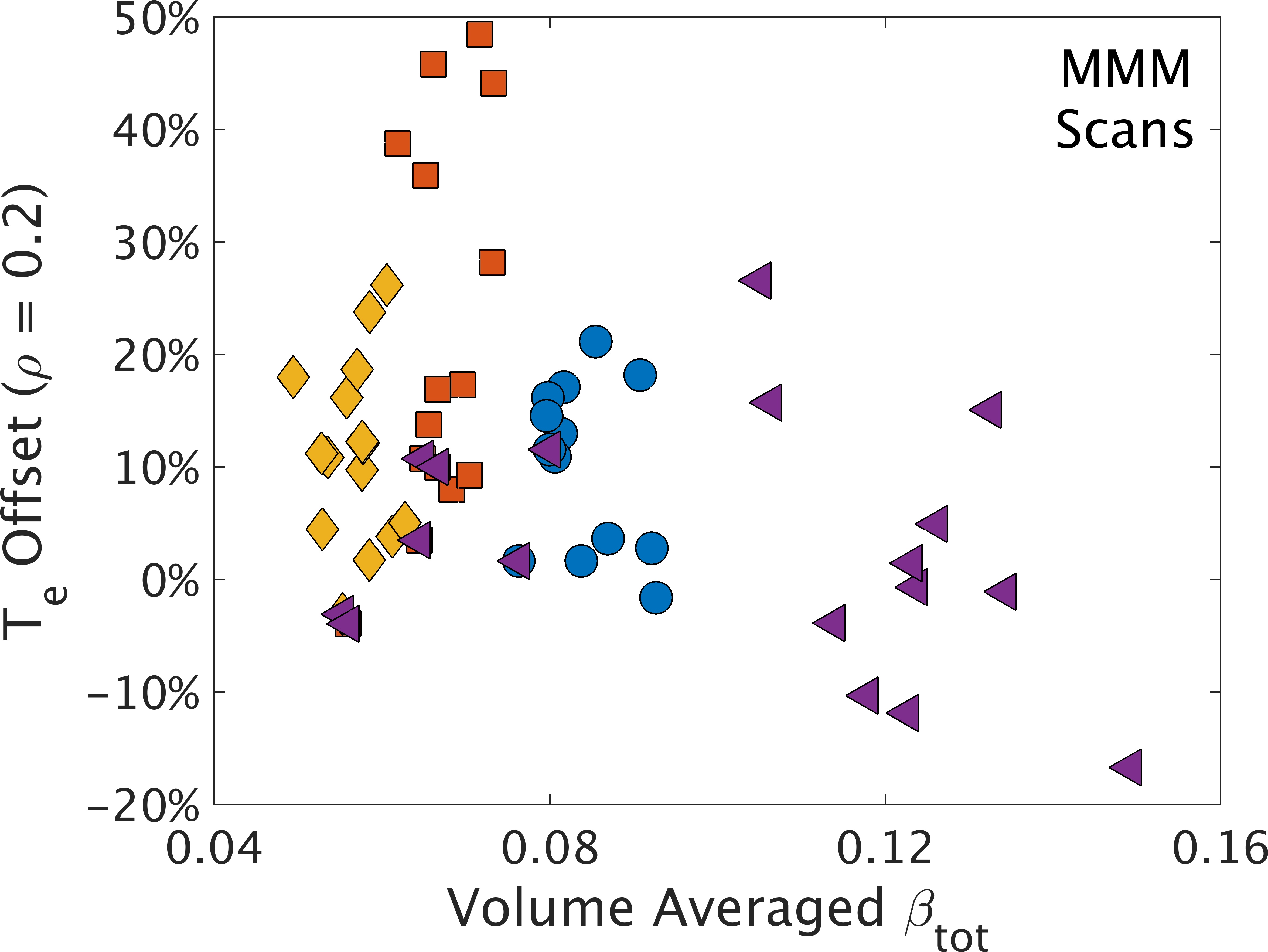}} \figsep
\subfloat[\label{fig:skaye_nustar50_teerr2080}]{\includegraphics[height = \thirdheight]{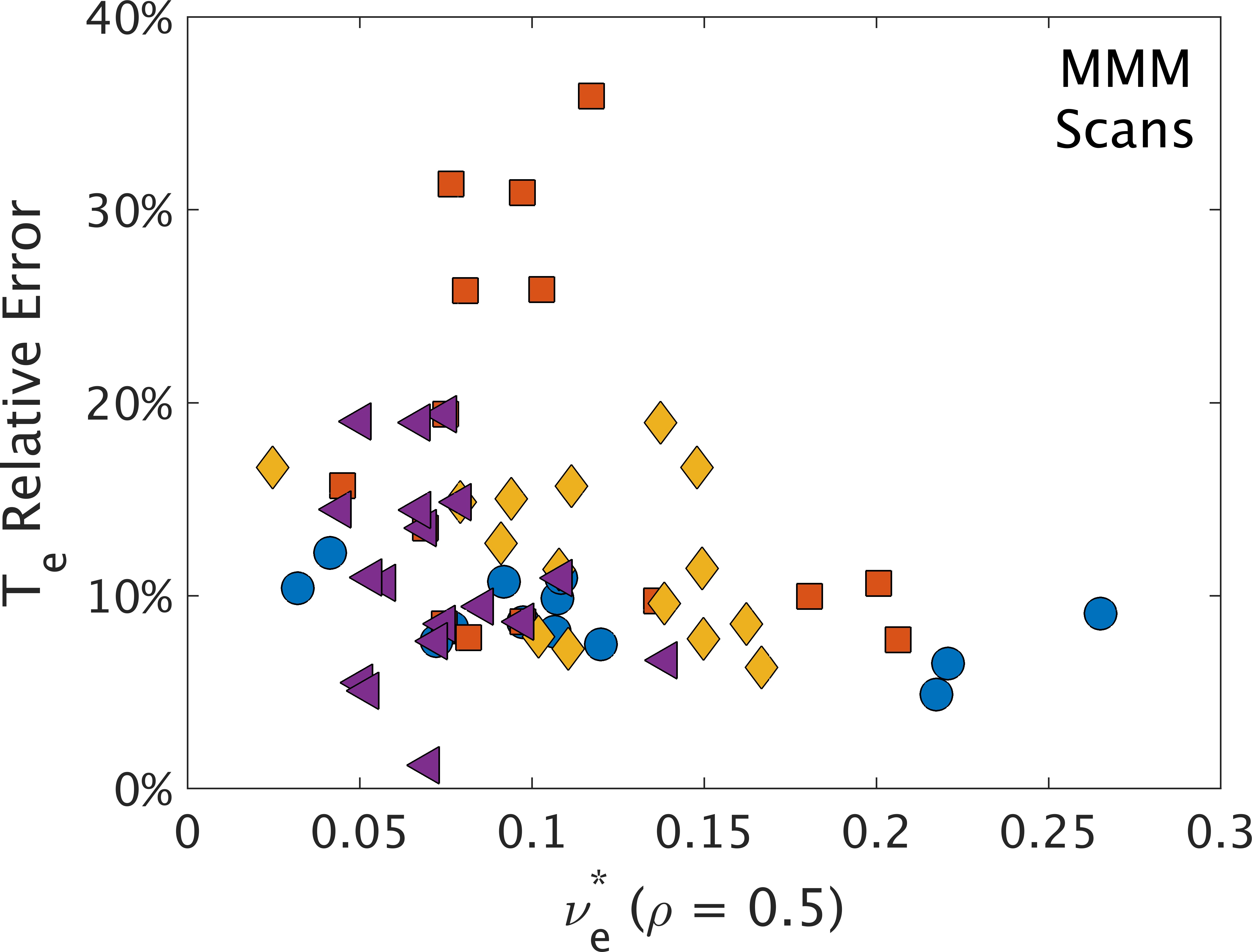}}  
\caption{TRANSP simulations using MMM to predict $\Te$ only, scanning $\beta$ and $\nustar$. (a) $\beta$ and $\nustar$ characterizing each discharge, with the legend labeling the median $\pm$ half interquartile range. (b) Relative offset of predicted $\Te$ at $\rho = 0.2$ as a function of $\beta$. (c) Relative error in predicted $\Te$ profile as a function of $\nustar$. In all subplots, $\beta$ is the thermal $\beta$, volume averaged over the entire plasma and $\nustar$ is the normalized electron collisionality evaluated at $\rho = 0.5$.}
\label{fig:skaye}
\end{figure*}

As a complementary approach to the investigation of a batch of well-analyzed NSTX discharges of interest, a second set of discharges was chosen in an attempt to decouple the effects of $\beta$ and collisionality when evaluating the tendencies of MMM in predicting the $\Te$ profile. As shown in \figref{fig:skaye_betatt_nustar50}, these discharges fall into four subgroups. Three of the groups scan a wide range of $\nustar$ while keeping $\beta$ relatively fixed and the fourth group scans $\beta$ for a narrow range of $\nustar$. In this section, $\beta$ usually refers to the thermal plasma toroidal $\beta$ averaged over the entire plasma volume, whereas in earlier sections $\beta$ was usually evaluated at a specific location in the plasma, such as on the axis or near the pedestal, as noted in those discussions. Also in this section, $\nustar$ denotes the normalized collisionality evaluated at $\rho = 0.5$ as a representative value. The predictive TRANSP simulations examined in this section use MMM to predict only the $\Te$ profile, with the prediction boundary set at $\rho = 0.8$, leaving $\Ti$ fixed to its experimental value. A comparison of simulations that predict both temperature profiles simultaneously \vs $\Te$ only will be made in \secref{sec:teonly}. 
The analysis in this section is restricted to specific 20 ms time slices chosen in each discharge instead of evaluating agreement over hundreds of ms as in the rest of this paper. Focusing on time slice analysis allows the selection of cases to analyze that more completely avoid forms of transport that are not accounted for in these simulations, such as sawteeth, \Alfven eigenmodes, and other macroscopic MHD instabilities. The discharges studied in this section are listed in \tabref{tab:skaye}. 

As in the previous set of well-analyzed NSTX discharges, a modest improvement in $\Te$ profile prediction agreement with experiment is found with increasing $\beta$ in this new set of discharges. Quantitatively, linear correlation coefficients of $r = -0.2 \tto -0.3$ are found between the $\Te$ RMSE and $\beta$ quantities, depending on calculating with $\beta$ on-axis or volume-averaged and total or electron $\beta$, with slightly higher correlations for electron $\beta$ than total $\beta$. Fewer than 10\% of the discharges have relative $\Te$ errors of greater than 20\%, and those outliers are characterized by relatively low $\beta$ and relatively low $q$ at mid-radius compared to the rest of the discharges (although there is greater variance in the discharges with low $q$, there is otherwise no consistent dependence of the RMSE on this quantity). Notably, the most poorly predicted discharges from the previously examined dataset were also characterized by low $\beta$ and low $q$, a similarity between the two complementary sets of simulations. The improved experimental agreement with increasing $\beta$ becomes more apparent when calculating the relative offset between the predicted and experimental value of $\Te$ at a fixed radial location like $\rho = 0.2$, as in \figref{fig:skaye_betatt_teoff20}, which strengthens the linear correlation coefficient to a range of $r = -0.4 \tto -0.5$. This trend is similar to that found in \citeref{Lestz2025pre1} for TGLF simulations, which found a very robust decrease in the $\Te$ offset with increasing $\beta$.  

A weak trend is found between the predicted $\Te$ error and collisionality, demonstrated in \figref{fig:skaye_nustar50_teerr2080}. The linear correlation coefficient is $r = -0.22$, improving slightly to $r = -0.3$ if the five poorly predicted outliers were removed. When examining scans within each of the distinct $\beta$ groups, the highest $\beta$ group (blue points) show a steady improvement in agreement with experiment with increasing $\nustar$, though the range of RMSE values in this group is small. The lowest $\beta$ group (gold) has some tendency for lower errors at higher $\nustar$, but with large variance, and the interpretation of the middle $\beta$ group (orange) is strongly influenced by the outliers, which all fall in this group. Moreover, a conflicting correlation was found in the set of discharges studied in \secref{sec:betanuJBL}, where a reasonable correlation ($r = 0.4 - 0.6$, depending on choice of linear or rank correlation) was found between $\nustar$ evaluated at the prediction boundary of $\rho = 0.7$ and the $\Te$ profile error. However, as noted in that section, this trend vanished when evaluating $\nustar$ instead at mid-radius or closer to the core, and there is likewise substantial variability when comparing the discharges in this section against $\nustar$ evaluated at other radial locations, reducing confidence in any of the trends with collisionality. Put together, comparison of the database of discharges constructed for this section with the one analyzed in the rest of this work bolsters the conclusion that the agreement between MMM $\Te$ profile predictions an experimental observations is more sensitive to $\beta$ than $\nustar$, across a wide range of discharges and conditions. 

\section{Comparison of Simulations that Predict Both $\Te$ and $\Ti$ Versus Only $\Te$ Profiles}
\label{sec:teonly}

\begin{figure*}[tb]
\subfloat[\label{fig:mean_te}]{\includegraphics[height = \thirdheight]{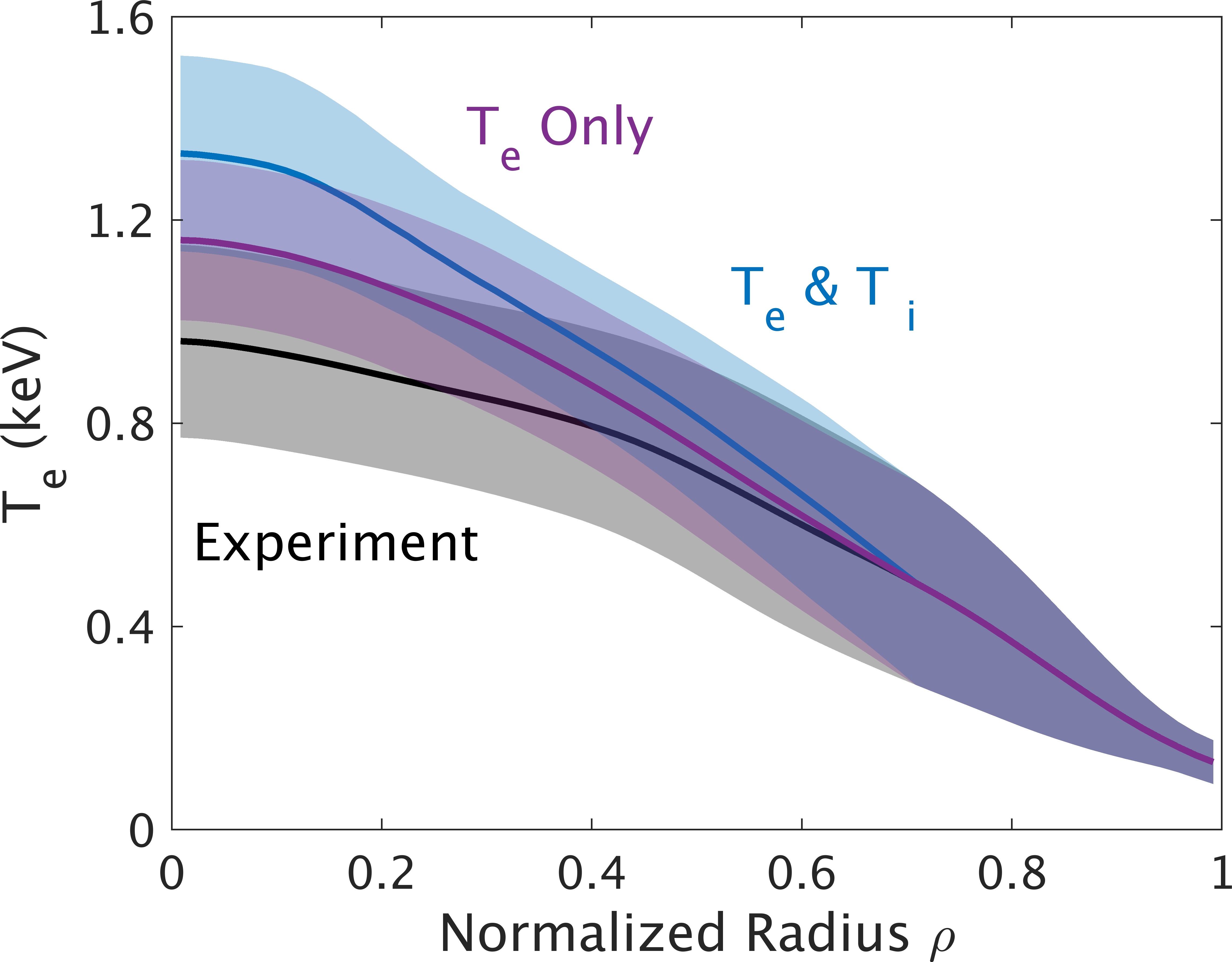}} \figsep
\subfloat[\label{fig:te_err_hist}]{\includegraphics[height = \thirdheight]{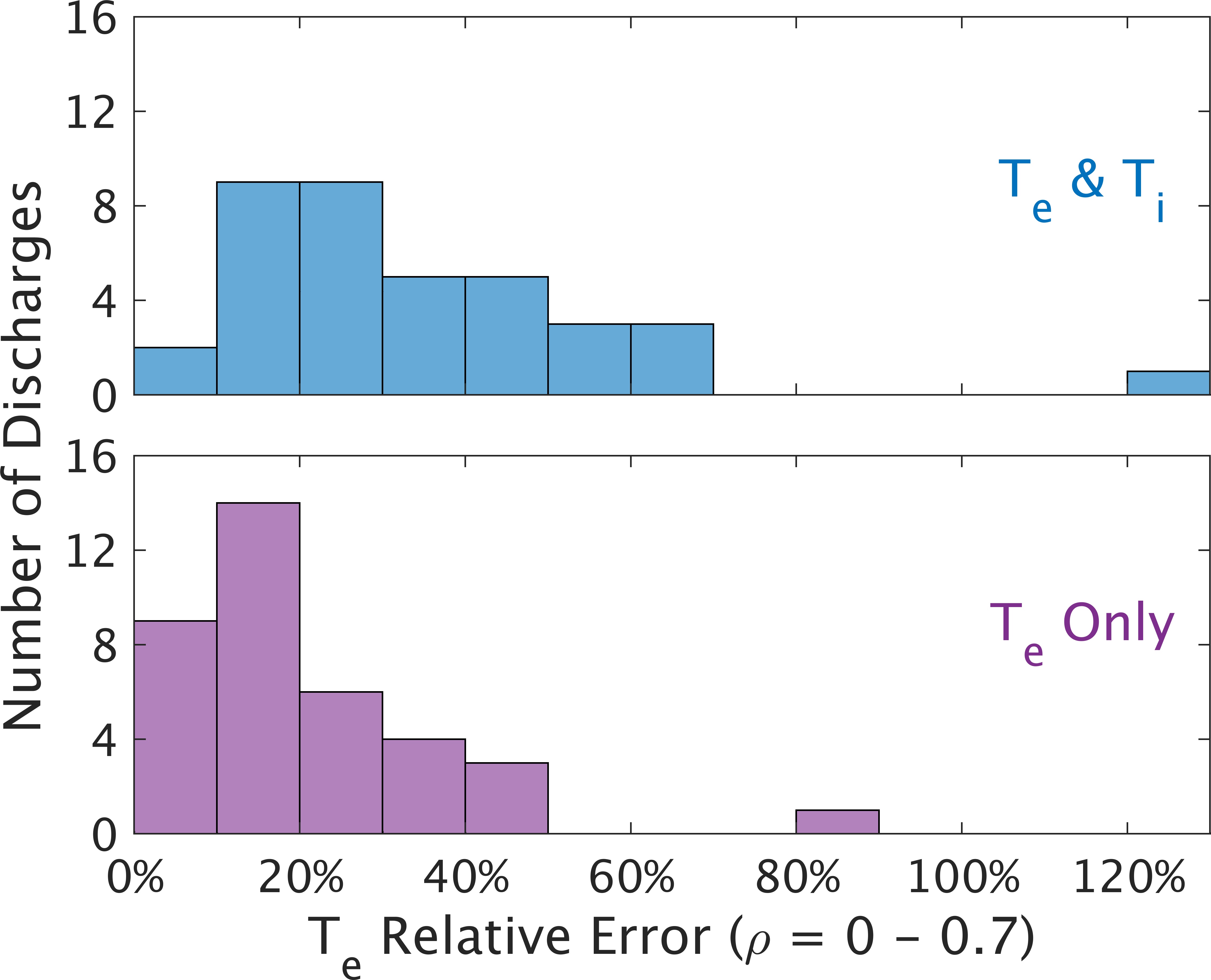}} \figsep
\subfloat[\label{fig:cputime}]{\includegraphics[height = \thirdheight]{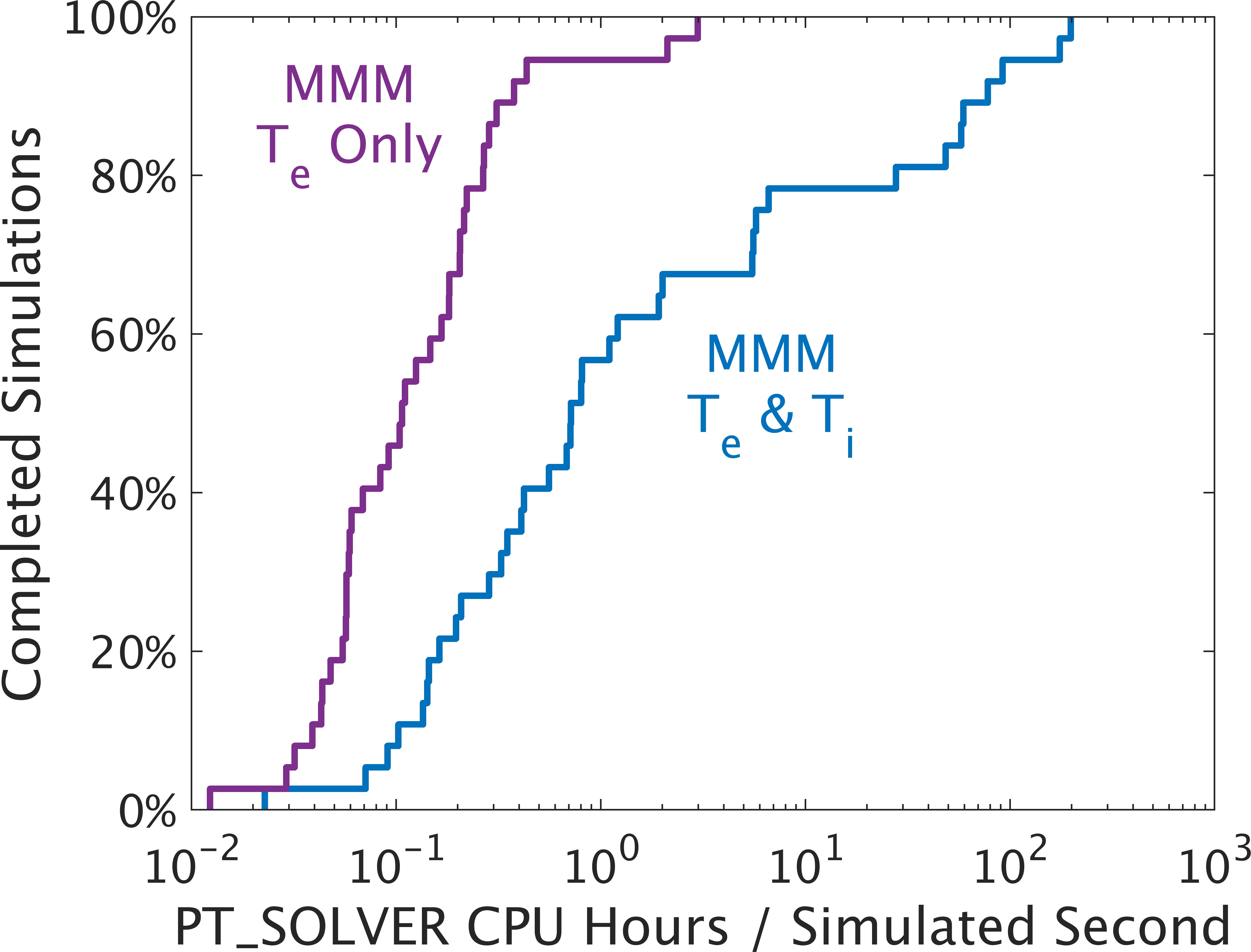}}
\caption{(a) Electron temperature profile predictions, averaged over all discharges. Thick curves show the mean value and shaded regions give the standard deviation. The average experimental $\Te$ profile is shown in black. In all plots, blue shows simulations that predicted both $\Te$ and $\Ti$ simultaneously and purple shows those that predicted $\Te$ only. (b) Histogram of relative error (RMSE over $\rho = 0 - 0.7$) between predicted \vs observed electron temperature profile for all discharges. (c) Comparison of computational cost, quantified as CPU hours spent on the \ptsolver portion of the TRANSP simulation, normalized to a full second of simulated plasma time.}
\label{fig:profs}
\end{figure*}

The bulk of this work and \citeref{Lestz2025pre1} focused on predictive TRANSP simulations that simultaneously predicted the $\Te$ and $\Ti$ profiles. This section evaluates the effect of evolving only $\Te$ by fixing $\Ti$ to its experimental value and then rerunning the predictive TRANSP simulations with MMM on all of the discharges listed in \tabref{tab:runids}. This investigation is motivated by the analysis of the ion power balance for discharge 133964 in \secref{sec:char_i}, which demonstrated a sensitivity to which combination of temperature profiles were predicted. \figref{fig:mean_te} shows the average $\Te$ profile across all examined discharges. The experimental profile average is given by the black curve, the blue curve shows the average of the $\Te$ profile from the simulations that evolve both $\Te$ and $\Ti$, and the purple curve gives the average $\Te$ profile when fixing $\Ti$ to its experimental value and predicting only $\Te$. The shaded regions show the standard deviation across all discharges. The $\Te$ profile prediction error is approximately cut in half when $\Ti$ is not predicted simultaneously, with the median $\pm$ half interquartile range reducing from $28 \pm 13\%$ to $14 \pm 8\%$. These and other statistics from the different types of simulations are summarized in \tabref{tab:allerrs}. A histogram of $\Te$ profile prediction errors is shown in \figref{fig:te_err_hist}, with the entire distribution of errors shifting to lower values when predicting only $\Te$. Hence, about half of the error in the MMM $\Te$ profile prediction is due to indirect effects from differences in the $\Ti$ profile. Otherwise, the predicted $\Te$ profiles are qualitatively similar, as both types of MMM simulations have a strong tendency to overestimate $\Te$ while being well-correlated with the experimental on-axis electron temperature. 

The MMM simulations that predict only $\Te$ while leaving $\Ti$ fixed were typically an order of magnitude faster than the MMM simulations that predicted both profiles. \figref{fig:cputime} shows the computation cost of the two types of simulations, based on the number of CPU hours spent in \ptsolver during the TRANSP simulation, per second of simulated discharge time. Each curve shows the fraction of simulations that have completed in less than a certain amount of computational time. The computation cost is further decomposed into the time spent on a single iteration of \ptsolver and the number of iterations needed per discharge second in \tabref{tab:allcpu}. Most of the speedup comes from each MMM calculation being significantly faster when excluding $\Ti$ prediction, as a similar number of iterations were required for \ptsolver to converge in both cases. 

\begin{figure*}[tb]
\subfloat[\label{fig:xke_teti}]{\includegraphics[height = \thirdheight]{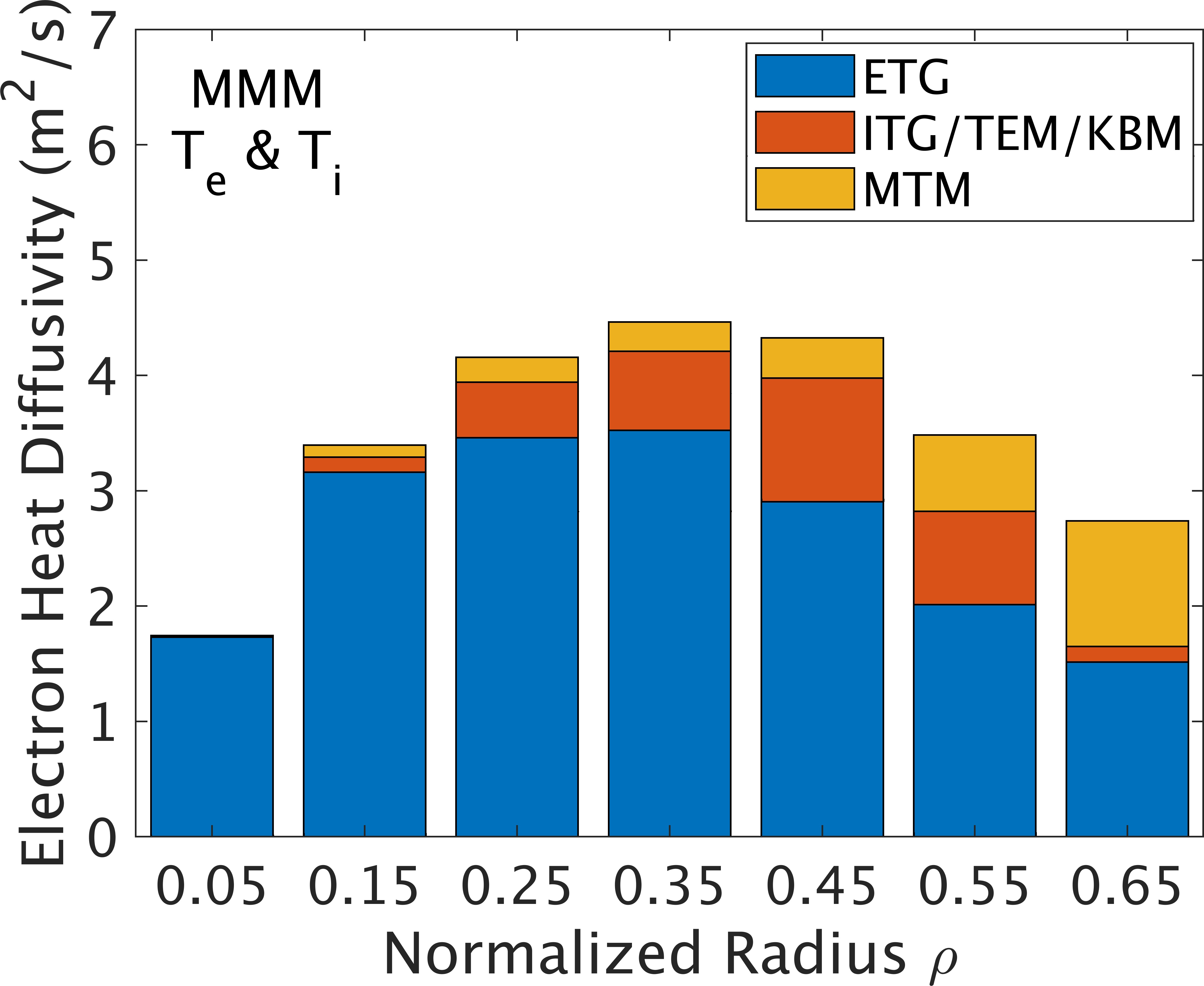}} \figsep
\subfloat[\label{fig:xke_teonly}]{\includegraphics[height = \thirdheight]{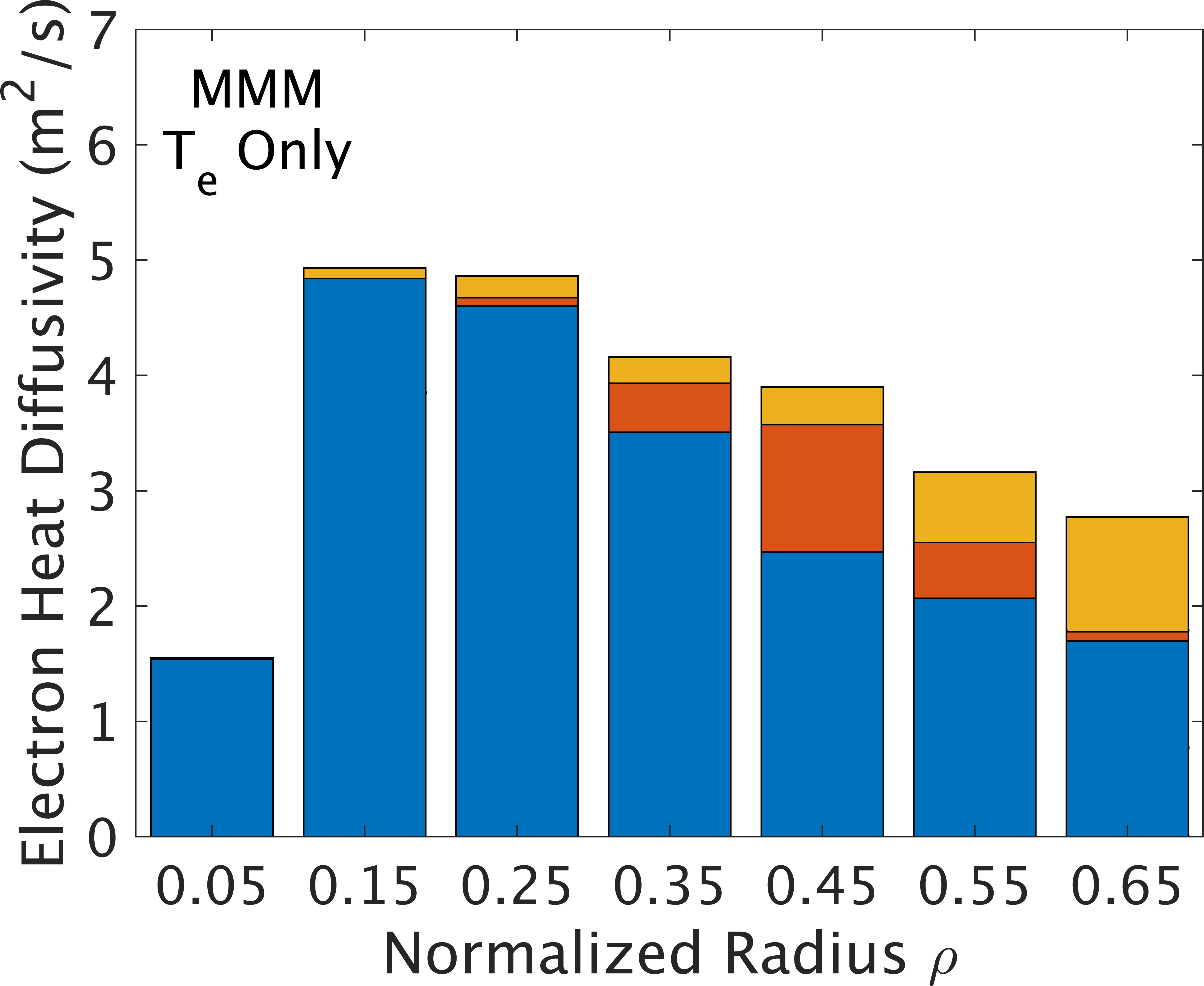}} \figsep
\subfloat[\label{fig:tegrad_comp}]{\includegraphics[height = \thirdheight]{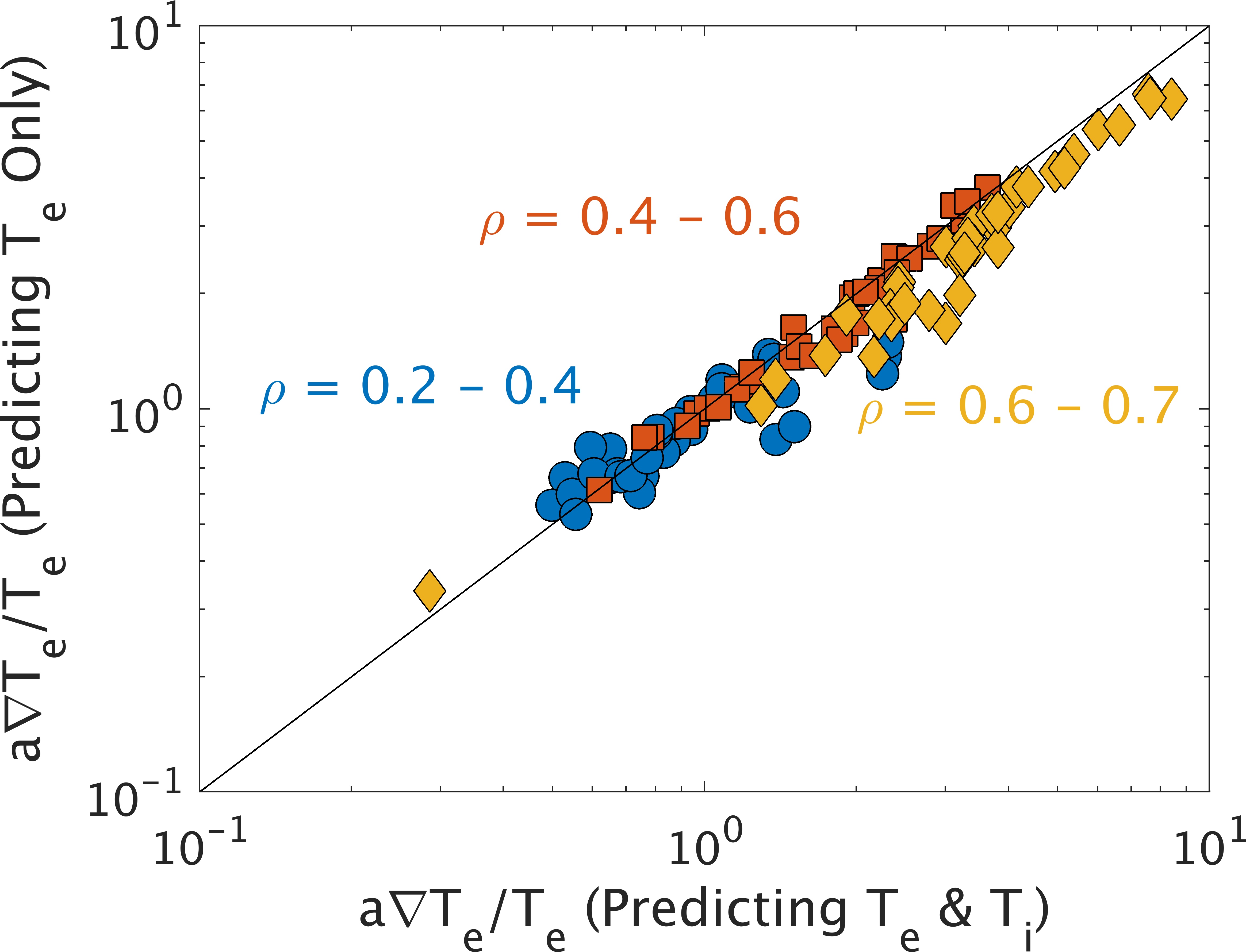}} 
\caption{(a) and (b): median electron energy diffusivity as calculated by MMM in TRANSP simulations which predicted (a) both $\Te$ and $\Ti$ simultaneously and (b) only $\Te$ while using the experimental $\Ti$ profile. The diffusivity profile is veraged over all examined discharges and decomposed into contributions from different submodels within MMM. (c) Comparison of electron temperature gradient scale lengths predicted in simulations that evolve both $\Te$ and $\Ti$ (abscissa) \vs only $\Te$ (ordinate), grouped into three radial regions. The solid line is for reference only, indicating zero offset between the gradients from the two sets of simulations.}
\label{fig:xke_comp}
\end{figure*}

The electron heat diffusivity profiles are also affected when using MMM to predict both $\Te$ and $\Ti$ or only the $\Te$ profile. A comparison similar to that shown in \figref{fig:xke_beta_nu} is shown in \figref{fig:xke_teti} and \figref{fig:xke_teonly}, where the median electron heat diffusivity from each group of instabilities is averaged into radial bins of width $\Delta\rho = 0.1$. While the peak diffusivity is similar for the two sets of simulations, the TRANSP runs that used the experimental $\Ti$ profile tend to have a diffusivity profile that peaks deeper in the core, whereas the simulations for $\Te$ and $\Ti$ have broader diffusivity profiles with a modest peak at mid-radius. When considering individual instabilities, ETGs drive stronger electron heat transport in the core in the ``$\Te$ only'' simulations, around 25\% higher for $\rho < 0.3$. Meanwhile, electron transport driven by the Weiland model (including KBMs and TEMs) is up to 40\% lower at mid-radius and MTM transport is about 10\% lower near the edge when predicting $\Te$ alone. 

The normalized gradient scale lengths from the two sets of simulations are shown in \figref{fig:tegrad_comp}, averaged over a radial region near the core ($\rho = 0.2 - 0.4$), mid-radius ($\rho = 0.4 - 0.6$), and closer to the edge ($\rho = 0.6 - 0.7$). Since the two sets of simulations broadly predict very similar $\Te$ profiles, it is unremarkable that the calculated gradients are very highly correlated. The noteworthy feature is that when MMM predicts both $\Te$ and $\Ti$ simultaneously, it tends to generate steeper profiles at larger radius than when only $\Te$ is predicted. Quantitatively, for $\rho = 0.2 - 0.4$, the gradients predicted in ``$\Te$ only'' simulations are shallower than those in the ``$\Te$ and $\Ti$'' simulations in 54\% of the examined discharges. For $\rho = 0.4 - 0.6$, this fraction increases to 73\%, and in the region closest to the edge of the simulation region of $\rho = 0.6 - 0.7$, this ratio is 97\%. Hence, the larger $\Te$ profile prediction error in TRANSP simulations using MMM to evolve both $\Te$ and $\Ti$ mostly results due to overpredicting the $\Te$ gradient close to the edge, which consequently lifts the core $\Te$. 

\begin{table}\centering
\begin{tabular}{cccc}
\hline\hline
Transport Model & $\Te$ Error \\ 
\hline
MMM $\Te$ \& $\Ti$   & $28 \pm 13\%$ \\
MMM $\Te$ Only       & $14 \pm 8\%$  \\
$\Te = \Ti$ Baseline & $11 \pm 5\%$  \\
\hline\hline
\end{tabular}
\caption{Statistics for temperature profile predictions for all simulated NSTX discharges. Reported as median error $\pm$ half interquartile range. The $\Te = \Ti$ ``model'' represents the baseline error from setting $\Te = \Ti$, if the $\Ti$ profile were known.}
\label{tab:allerrs}
\end{table}

\begin{table*}\centering
\begin{tabular}{cccc}
\hline\hline
Transport Model & \twotab{CPU Hours per}{Simulated Second} & \twotab{CPU Seconds per}{Newton Iteration} & \twotab{Newton Iterations per}{Simulated Second} \\ 
\hline
MMM $\Te$ \& $\Ti$ & $0.71 \pm 2.88$ & $1.51 \pm 5.44$  & $2,814  \pm 970$  \\ 
MMM $\Te$ Only     & $0.11 \pm 0.08$ & $0.13 \pm 0.22$  & $2,113  \pm 636$ \\
\hline\hline
\end{tabular}
\caption{Statistics for computational cost for all simulated NSTX discharges. Reported as median $\pm$ half interquartile range of each quantity.}
\label{tab:allcpu}
\end{table*}

Note that the simulations discussed in \secref{sec:skaye} predicted only the $\Te$ profile, as in the comparison made in this section, but with three key differences: 1) a different set of discharges (\tabref{tab:runids} \vs \tabref{tab:skaye}), 2) a different prediction boundary ($\rho = 0.7$ \vs 0.8) and radial resolution (typically 60 radial zones \vs 40), and 3) compared predicted \vs experimental $\Te$ profiles on much smaller time windows (duration $\like 200$ ms \vs 20 ms). Overall, MMM predicts the $\Te$ profiles evaluated in narrow time windows for the discharges from \secref{sec:skaye} with comparable RMSE as was found for the simulations discussed in this section which predicted only $\Te$ over much larger time windows. The discharges in that section have a median $\pm$ half interquartile range error of $11 \pm 4\%$, compared to $14 \pm 8\%$ for the discharges discussed here. Hence, there is essentially no penalty for averaging over 20 ms \vs 200 ms time windows when predicting the $\Te$ profile, despite the plasma evolving on these timescales. This further bolsters a conclusion in \secref{sec:time}, where it was demonstrated that the coefficient of variation of the stored energy over time in large analysis windows did not strongly influence the level of agreement with experiment. Furthermore, the similar RMSE ranges demonstrate a relative insensitivity to small changes in the prediction boundary or the spatial grid resolution. 

Although setting $\Ti$ to its experimental value clearly improves the agreement between the experimental and predicted $\Te$ profile, it is worth putting this into context. Namely, the computational cost of a physics model is generally justified by providing greater accuracy and/or physics insights than computationally cheaper alternatives. For instance, \citeref{Abbate2024POP} compared the temperature profile prediction RMSE of the trapped gyro-Landau fluid drift wave model TGLF for a large database of DIII-D discharges against an empirical two parameter linear regression based on a heuristic model motivated by ``profile consistency'' \cite{Coppi1988PLA}, with the latter having effectively zero computational cost. In that work, statistical hypothesis tests concluded that TGLF predictions had no statistical advantage over the regression method for the temperature profiles. For MMM simulations where the experimental $\Ti$ profile is treated as an input, one can compare the RMSE of MMM's $\Te$ predictions against a computation-free baseline alternative heuristic of setting $\Te = \Ti$. This is a fair comparison since both MMM and the baseline alternative have access to the same input data, namely the experimental $\Ti$ profile. The $\Te = \Ti$ baseline model has a median $\pm$ half interquartile $\Te$ prediction error of $11 \pm 6\%$, comparable to the $14 \pm 8\%$ for MMM when predicting only $\Te$. Hence, knowledge of the $\Ti$ profile allows $\Te$ to be modeled with similar experimental agreement with or without MMM, though MMM still provides important physics insights that a simple baseline model can not. In this sense, the improved $\Te$ profile predictions of MMM simulations that use the experimental $\Ti$ profile over those that predict both $\Te$ and $\Ti$ simultaneously has less significance than when the raw statistics are considered in isolation. 

\rev{Note that the choice of using $\Te = \Te$ as a baseline comparison is not intended to imply a general insight for NSTX or other low aspect ratio tokamaks, but rather was chosen as a simple relationship that reasonably well describes the beam-heated discharges examined in this work. In tokamaks with a different mixture of electron and ion heating (for instance, via radio frequency waves that preferentially damp on a specific species near a narrow resonance location), an appropriate baseline comparison could instead be constructed empirically, as done in \citeref{Abbate2024POP}.} 

\section{Summary and Conclusions} 
\label{sec:conclusion}

In this work, time-dependent TRANSP simulations were performed with the MMM turbulent transport model in order to identify trends in the model's level of agreement with experiment when predicting temperature profiles across a large set of well-analyzed NSTX discharges. Overall, MMM predicts the $\Te$ and $\Ti$ profiles with very reasonable agreement, especially for a computationally inexpensive reduced model. The set of analyzed discharges included a range of discharges where the electron energy transport was predicted to be dominated by each of the three different submodels of MMM: ETGs, the Weiland model (ITGs, TEMs, and KBMs), and MTMs. It was found that the Weiland model is most important at mid-radius, driving stronger transport at higher $\beta$ and lower normalized collisionality, $\nustar$, while transport driven by MTMs was restricted to the region closest to the prediction boundary ($\rho = 0.7$). In particular, MTM transport showed a much stronger sensitivity to $\betap$ than toroidal $\beta$. The neoclassical contribution to the ion energy transport dominated in all discharges, with nontrivial turbulent ion transport predicted to occur more often in discharges with relatively low $\beta$ and only close to the prediction boundary. Despite this, the $\Te$ and $\Ti$ profile predictions are strongly coupled through collisional energy exchange, resulting in a sensitivity of the resulting profiles to which transport channels are predicted. 

Regarding the time dependent behavior of the simulations, the profile predictions made by MMM were found to be very robust to the initial condition, converging to essentially the same profile at later times in the simulation regardless of whether the predictions began early in the plasma current ramp up or late in the flat top. It appears that the history of the system's evolution only influences predictions made over a timescale similar to the energy confinement time, beyond which it becomes irrelevant. The predicted temperature profiles also track the experimentally measured profile evolution well over time, as quantified by the rate of change over analysis windows of $\like 200$ ms. The closeness of the temperature profile predictions to experiment when averaged over the entire analysis windows was only weakly influenced by how much the plasma varied during that time. 

The steepness of the temperature profile and its local normalized gradients have a pronounced influence on how well the MMM predictions reproduce experimental measurements. MMM has a strong tendency to predict local gradients that are steeper than the experimental profile fits, corresponding to an overprediction of the energy confinement and consequently higher temperatures than were observed in the experiments. For gradients in both $\Te$ and $\Ti$, the steeper gradients near the pedestal are predicted with smaller relative error than shallower gradients that occur closer to the core. $\Ti$ gradients had relatively larger RMSE in the core than $\Te$ gradients. Despite steeper gradients being predicted in better agreement with experiment, broader experimental $\Te$ profiles tend to be better reproduced by MMM than those that are more peaked. This can be explained by smaller relative errors in steeper edge profiles having a larger absolute effect in lifting up the core $\Te$ than a larger relative error would in a shallower edge profile. Consequently, NSTX profiles with very wide pedestals, including enhanced pedestal (EP) and wide pedestal (WP) H modes are among the scenarios that are best reproduced by TRANSP simulations with MMM in this work. 

The sensitivity of the relative error in the $\Te$ prediction to $\beta$ and $\nu$ is evaluated in two ways. In the main set of well-analyzed NSTX discharges analyzed throughout this work, a reasonable improvement in RMSE is found with $\beta$, though an even stronger one is found with the energy confinement time, $\taue$. Given the correlation between $\beta$, $\taue$, and the $\Te$ profile peaking factor in the examined discharges, it is ambiguous which of these factors is the most relevant factor for the MMM simulations. Quantitatively, the strongest correlation is between the $\Te$ prediction error and the confinement time, though strong correlation coefficients are also found instead with $\beta$ or the profile peaking factor. A second set of discharges was constructed to intentionally separate parametric dependencies on $\beta$ and $\nu$. These simulations also tended to find better agreement with experiment at higher $\beta$, via driving stronger transport that reduced the characteristic overprediction of the $\Te$ profile. MMM's predictions were less sensitive to collisionality in both sets of simulations, finding a modest improvement in RMSE at lower $\nustar$ in the first set, conflicting with a modest improvement at higher $\nustar$ in the second set. Both of these trends were sensitive to the specific radial location where $\nustar$ was evaluated, underscoring their fragility in comparison to the much more robust trends with respect to $\beta$ and related confinement quantities. 

Lastly, a detailed comparison was made between MMM simulations that predicted both $\Te$ and $\Ti$ simultaneously \vs setting $\Ti$ to its experimental value and predicting only the $\Te$ profile. The simulations that use the experimental $\Ti$ profile as input are computationally faster and reduce the $\Te$ profile prediction error by about half compared to simulations that also predict $\Ti$. However, a baseline comparison of setting $\Te = \Ti$ predicts $\Te$ with comparable experimental agreement when $\Ti$ input data is provided, illustrating the high degree of certainty that is conferred from knowing $\Ti$, even absent a physics-based transport model. The ``$\Te$ only'' MMM simulations tend to predict shallower edge $\Te$ gradients and a larger fraction of the electron heat diffusivity due to ETGs than when also predicting $\Ti$. 

The results of this work serve as a useful foundation for interpreting future time-dependent predictive TRANSP modeling of NSTX-U with MMM and the corresponding degree of confidence one should have with regards to the quantitative temperature profile predictions. The robustness of the predictions in diverse plasma conditions and MMM's computational efficiency make it well suited to exploratory parameter scans and scenario optimization that can be further investigated with the judicious use of more expensive, higher fidelity, first-principles simulations. In particular, the improved agreement between predicted and observed temperature profiles in NSTX discharges with relatively high $\beta$ is promising for modeling high $\beta$ scenarios planned for NSTX-U. The demonstrated tendency for MMM to overpredict energy confinement in NSTX plasmas may serve as a guide for continued development of the MMM code, which could include exploring calibration strategies or more sophisticated geometric treatments. 

In the future, this study could be extended by evolving density and rotation profiles to evaluate how well predictive TRANSP simulations with MMM perform beyond the temperature profile predictions investigated here. Furthermore, comparisons could be made with semi-empirical baseline models, such as the ones described in \citeref{Abbate2024POP}, in order to more rigorously quantify the predictive power of MMM relative to simple alternatives. Given the sensitivity of the MMM $\Te$ profile predictions to the experimental profile's peaking factor, it would be interesting to apply MMM simulations to study high beam power NSTX discharges that featured anomalously flat $\Te$ profiles \cite{Stutman2009PRL,Ren2017NF}. These profiles could not be explained by gyrokinetic simulations, with the prevailing theories involving high frequency \Alfven eigenmodes ($f \lesssim \fci$, where $2\pi\fci = q_i B/m_i$ is the ion cyclotron frequency) destabilized by super-\Alfvenic neutral beam ions \cite{Stutman2009PRL,Gorelenkov2010NF,Kolesnichenko2010PRL,Kolesnichenko2010NF,Belova2015PRL,Belova2017POP,Lestz2021NF} or more recently low-$n$, pressure-driven MHD modes known as infernal modes \cite{Jardin2022PRL,Jardin2023POP}. Neither of these theories are mature enough to be ready to implement as a computationally efficient reduced model for integrated simulations. Thus, it would be interesting to determine how MMM tends to predict the $\Te$ profile for discharges in this regime, and especially whether a radially-dependent, ad hoc diffusivity enhancement factor could be used to approximately model the presently unexplained anomalous $\Te$ flattening in NSTX. 

\section{Acknowledgements}
\label{sec:acknowledgements}

The authors thank J. Abbate, X. Zhang, W. Choi, T. Rafiq, and X. Yuan for fruitful discussions. The simulations reported here were performed with computing resources at the Princeton Plasma Physics Lab. Part of the data analysis was performed using the OMFIT integrated modeling framework \cite{Meneghini2015NF}. The data required to generate the figures in this paper are archived in the NSTX-U Data Repository ARK at the following address: (placeholder).
This research was supported by the U.S. Department of Energy (contracts DE-SC0021113, DE-AC02-09CH11466).

\section{Disclaimer}
\label{sec:disclaimer}

\gadisclaimer

\appendix 

\section{TRANSP RunIDs}
\label{app:runids}

The NSTX discharges and corresponding TRANSP runIDs used for the majority of this study are listed in \tabref{tab:runids}, along with the category that each discharge was sorted into and previously published work that analyzed each discharge. The ``original'' interpretive ID is the TRANSP run where the equilibrium, profiles, input data, and most TRANSP settings were taken from, whereas the ``modernized'' interpretive ID is a recent rerun using the current version of TRANSP, with some settings tweaked for compatibility or consistency with other runs. The MMM predictive runs are listed in two separate columns for those that predicted $\Te$ and $\Ti$ simultaneously and $\Te$ only (with $\Ti$ prediction disabled). The TRANSP IDs for the second set of MMM simulations discussed in \secref{sec:skaye} for dedicated $\beta$ and $\nustar$ scans that only predicted $\Te$ are listed in \tabref{tab:skaye}. All runs are stored in MDSplus, access information available upon reasonable request.

\newcommand{\longpulse}{Long pulse\xspace}
\newcommand{\maxwmhd}{Max stored energy\xspace}
\newcommand{\highnustar}{High $\nu_e^*$\xspace}
\newcommand{\lownustar}{Low $\nu_e^*$\xspace}
\newcommand{\wphmode}{Wide pedestal\xspace}
\newcommand{\ephmode}{Enhanced pedestal\xspace}
\newcommand{\highbetap}{High $\beta_p$\xspace}
\newcommand{\lowbeta}{Low $\beta_e$\xspace}
\newcommand{\highbeta}{High $\beta_e$\xspace}
\newcommand{\hmode}{H mode\xspace}
\newcommand{\lmode}{L mode\xspace}
\newcommand{\stutman}{$\Te$ flattening\xspace}

\begin{table*}\centering
\begin{tabular}{ccccccc}
\hline\hline
\twotab{NSTX}{Discharge} & \twotab{Original}{Interpretive} & \twotab{Modernized}{Interpretive} & \twotab{MMM}{$\Te$ \& $\Ti$} & \twotab{MMM}{$\Te$ Only} & \twotab{Category}{and Reference} & \twotab{Analysis}{Time (ms)} \\ 
\hline
116313 & G12 & N10 & N03 & N05 & \longpulse \cite{Gerhardt2011NFat} & 500 - 1050 \\ 
117707 & A04 & J11 & J03 & J04 & \maxwmhd \cite{Gerhardt2011NFat} & 650 - 950 \\
120967 & A03 & J02 & J05 & J12 & \highnustar \cite{Guttenfelder2013NF,Kaye2014POP} & 300 - 600 \\
120968 & A02 & L18 & L05 & L09 & \highnustar \cite{Guttenfelder2013NF,Rafiq2021POP,Clauser2022POP,Rafiq2024NF,Clauser2025POP} & 260 - 400 \\
120982 & A09 & J11 & J03 & J05 & \lownustar \cite{Clauser2022POP,Rafiq2024NF,Clauser2025POP} & 590 - 650 \\
121123 & A02 & J08 & J03 & J05 & \maxwmhd \cite{Gerhardt2011NFat} & 500 - 1000 \\
129016 & A03 & J11 & J02 & J07 & \highnustar \cite{Guttenfelder2013NF,Rafiq2021POP,Rafiq2024NF,Clauser2025POP} & 300 - 500 \\
129017 & A04 & K36 & K31 & K33 & \hmode \cite{Guttenfelder2013NF,Avdeeva2023NF,Avdeeva2024PPCF} & 320 - 500 \\
129039 & A05 & J14 & J06 & J08 & \lownustar \cite{Rafiq2021POP} & 300 - 340 \\
129041 & A10 & K03 & J02 & J07 & \lownustar \cite{Guttenfelder2013NF,Kaye2014POP,Clauser2022POP,Clauser2025POP} & 250 - 370 \\
129125 & B08 & J08 & J03 & J05 & \longpulse \cite{Gerhardt2011NFat} & 500 - 1100 \\
132588 & B01 & J08 & J03 & J05 & \wphmode \cite{Dominski2024POP} & 530 - 790 \\
132911 & A01 & J02 & J03 & J06 & \maxwmhd \cite{Gerhardt2011NFat,Gerhardt2011NFcur} & 500 - 700 \\
132913 & B01 & J02 & J03 & J06 & \maxwmhd \cite{Gerhardt2011NFat} & 640 - 720 \\
133958 & G63 & K02 & K03 & K05 & \highbetap \cite{Gerhardt2011NFcur} & 345 - 365 \\
133959 & D45 & K11 & K03 & K05 & \highbetap \cite{Gerhardt2011NFcur} & 600 - 900 \\
133964 & D05 & I26 & I20 & I22 & \highbetap \cite{Gerhardt2011NFat,Gerhardt2011NFcur,McClenaghan2023POP,McClenaghan2025PPCF} & 700 - 1050 \\
134767 & A05 & K09 & K03 & K05 & \hmode \cite{Gerhardt2011NFat} & 500 - 1200 \\
134837 & A11 & K09 & K03 & K05 & \maxwmhd \cite{Gerhardt2011NFat} & 500 - 950 \\
135117 & A02 & J11 & J03 & J07 & \maxwmhd \cite{Gerhardt2011NFcur} & 600 - 900 \\
135129 & A02 & J09 & J03 & J05 & \maxwmhd \cite{Gerhardt2011NFat} & 600 - 1050 \\
135440 & S05 & J14 & J05 & J08 & \longpulse \cite{Gerhardt2011NFcur} & 600 - 880 \\
135445 & A04 & J08 & J03 & J05 & \longpulse \cite{Gerhardt2011NFat,Gerhardt2011NFcur} & 600 - 1350 \\
138536 & J01 & J28 & J07 & J15 & \lownustar \cite{Rafiq2022POP,Rafiq2024NF} & 550 - 660 \\
139517 & A04 & J08 & J03 & J05 & \maxwmhd \cite{Gerhardt2011NFcur} & 550 - 850 \\
140035 & A05 & J02 & J03 & J08 & \maxwmhd \cite{Gerhardt2011NFat} & 500 - 1180 \\
141007 & A03 & J08 & J03 & J05 & \lowbeta, \lownustar \cite{Ren2012POP} & 300 - 430 \\
141031 & S05 & J02 & J03 & J05 & \lowbeta, \highnustar \cite{Ren2012POP,Guttenfelder2013NF} & 260 - 300 \\
141032 & A04 & J02 & J03 & J05 & \lowbeta, \highnustar \cite{Ren2012POP} & 260 - 300 \\
141040 & A02 & J02 & J03 & J05 & \lowbeta, \lownustar \cite{Ren2012POP,Guttenfelder2013NF} & 330 - 390 \\
141125 & D04 & J10 & J03 & J06 & \wphmode \cite{Battaglia2020POP} & 700 - 900 \\
141131 & D04 & J02 & J03 & J05 & \wphmode \cite{Battaglia2020POP} & 600 - 900 \\
141133 & B11 & K09 & K03 & K06 & \ephmode \cite{Battaglia2020POP,Gerhardt2014NF} & 750 - 1050 \\
141340 & B04 & J02 & J03 & J05 & \ephmode \cite{Gerhardt2014NF} & 300 - 400 \\
141623 & A11 & J02 & J03 & J05 & \highbetap \cite{Gerhardt2011NFat} & 450 - 900 \\
141633 & A11 & J02 & J03 & J10 & \highbetap \cite{Gerhardt2011NFat} & 300 - 600 \\
141767 & B01 & J02 & J03 & J05 & \hmode \cite{RuizRuiz2019PPCF,Ren2020NF} & 300 - 560 \\
\hline\hline
\end{tabular}
\caption{TRANSP runs for the full set of NSTX discharges analyzed in most of this work.}
\label{tab:runids}
\end{table*}

\begin{table*}\centering
\begin{tabular}{ccccccc}
\hline\hline
\twotab{NSTX}{Discharge} & Interpretive & Predictive & $\quad\beta_\text{th}\quad$ & $\quad\nustar\quad$ & \twotab{Scan}{Group} & \twotab{Analysis}{Time (ms)} \\ 
\hline
120958 & A06 & P08 & 0.061 & 0.111 & $\nu_3$ & 570 \\ 
120959 & A07 & P07 & 0.073 & 0.045 & $\nu_2$ & 530 \\ 
120964 & A03 & P06 & 0.070 & 0.201 & $\nu_2$ & 510 \\ 
120967 & A03 & P35 & 0.092 & 0.221 & $\nu_1$ & 510 \\ 
120968 & A02 & P10 & 0.093 & 0.217 & $\nu_1$ & 550 \\ 
120982 & A09 & P08 & 0.085 & 0.041 & $\nu_1$ & 620 \\ 
120984 & A02 & P07 & 0.068 & 0.136 & $\nu_2$ & 590 \\ 
120985 & A02 & P07 & 0.053 & 0.138 & $\nu_3$ & 510 \\ 
120986 & A02 & P06 & 0.066 & 0.206 & $\nu_2$ & 510 \\ 
121003 & A03 & P13 & 0.091 & 0.032 & $\nu_1$ & 620 \\ 
121010 & A02 & P07 & 0.058 & 0.091 & $\nu_3$ & 610 \\ 
121013 & A02 & P07 & 0.067 & 0.181 & $\nu_2$ & 510 \\ 
121014 & A03 & P10 & 0.087 & 0.265 & $\nu_1$ & 510 \\ 
129015 & A03 & P03 & 0.053 & 0.094 & $\nu_3$ & 410 \\ 
129016 & A01 & P57 & 0.056 & 0.108 & $\nu_3$ & 460 \\ 
129017 & A02 & P03 & 0.058 & 0.110 & $\nu_3$ & 500 \\ 
129018 & A02 & P09 & 0.058 & 0.102 & $\nu_3$ & 450 \\ 
129019 & A06 & P08 & 0.053 & 0.148 & $\nu_3$ & 480 \\ 
129020 & A07 & P09 & 0.049 & 0.137 & $\nu_3$ & 450 \\ 
129021 & A02 & P11 & 0.065 & 0.103 & $\nu_2$ & 520 \\ 
129022 & A02 & P08 & 0.066 & 0.097 & $\nu_2$ & 570 \\ 
129023 & A02 & P08 & 0.072 & 0.076 & $\nu_2$ & 560 \\ 
129024 & A11 & P08 & 0.062 & 0.117 & $\nu_2$ & 580 \\ 
129027 & A08 & P08 & 0.065 & 0.074 & $\nu_2$, $\beta$ & 550 \\ 
129029 & A08 & P09 & 0.064 & 0.097 & $\nu_2$, $\beta$ & 590 \\ 
129030 & A01 & P07 & 0.073 & 0.081 & $\nu_2$ & 540 \\ 
129031 & A01 & P06 & 0.067 & 0.069 & $\nu_2$, $\beta$ & 460 \\ 
129038 & A16 & P09 & 0.055 & 0.079 & $\nu_3$, $\beta$ & 470 \\ 
129039 & A03 & P15 & 0.061 & 0.025 & $\nu_3$ & 310 \\ 
129041 & A10 & P07 & 0.056 & 0.075 & $\nu_2$, $\beta$ & 490 \\ 
138536 & K01 & P07 & 0.134 & 0.057 & $\beta$ & 630 \\ 
138542 & K01 & P08 & 0.124 & 0.070 & $\beta$ & 610 \\ 
138543 & K01 & P08 & 0.123 & 0.050 & $\beta$ & 600 \\ 
138544 & K01 & P09 & 0.126 & 0.052 & $\beta$ & 610 \\ 
138545 & K02 & P07 & 0.084 & 0.120 & $\nu_1$ & 550 \\ 
138547 & K01 & P08 & 0.070 & 0.082 & $\nu_2$ & 600 \\ 
138548 & K03 & P07 & 0.058 & 0.167 & $\nu_3$ & 550 \\ 
138555 & K02 & P11 & 0.082 & 0.107 & $\nu_1$ & 500 \\ 
138556 & K02 & P08 & 0.081 & 0.092 & $\nu_1$ & 500 \\ 
138558 & K03 & P07 & 0.063 & 0.150 & $\nu_3$ & 550 \\ 
138559 & K02 & P07 & 0.057 & 0.149 & $\nu_3$ & 550 \\ 
138560 & K03 & P07 & 0.058 & 0.162 & $\nu_3$ & 550 \\ 
138563 & K02 & P08 & 0.076 & 0.108 & $\nu_1$, $\beta$ & 500 \\ 
138565 & K01 & P08 & 0.118 & 0.067 & $\beta$ & 520 \\ 
138566 & K01 & P07 & 0.123 & 0.086 & $\beta$ & 630 \\ 
138567 & K01 & P08 & 0.114 & 0.139 & $\beta$ & 600 \\ 
138568 & K01 & P08 & 0.149 & 0.067 & $\beta$ & 520 \\ 
138843 & A03 & P07 & 0.081 & 0.077 & $\nu_1$ & 550 \\ 
138844 & A03 & P07 & 0.080 & 0.072 & $\nu_1$, $\beta$ & 550 \\ 
138846 & A03 & P07 & 0.080 & 0.097 & $\nu_1$ & 550 \\ 
138847 & A03 & P07 & 0.080 & 0.107 & $\nu_1$ & 550 \\ 
138848 & A01 & P09 & 0.106 & 0.044 & $\beta$ & 570 \\ 
138854 & A01 & P08 & 0.133 & 0.053 & $\beta$ & 580 \\ 
138855 & A01 & P08 & 0.105 & 0.050 & $\beta$ & 500 \\ 
\hline\hline
\end{tabular}
\caption{TRANSP runs for the NSTX discharges analyzed in the dedicated $\beta$ and $\nustar$ scans presented in \secref{sec:skaye}. $\beta_\text{th}$ is the volume averaged thermal $\beta$ and $\nustar$ is evaluated at $\rho = 0.5$.}
\label{tab:skaye}
\end{table*}

\bibliography{all_bib} 

\begin{thebibliography}{99}%
\makeatletter
\providecommand \@ifxundefined [1]{%
 \@ifx{#1\undefined}
}%
\providecommand \@ifnum [1]{%
 \ifnum #1\expandafter \@firstoftwo
 \else \expandafter \@secondoftwo
 \fi
}%
\providecommand \@ifx [1]{%
 \ifx #1\expandafter \@firstoftwo
 \else \expandafter \@secondoftwo
 \fi
}%
\providecommand \natexlab [1]{#1}%
\providecommand \enquote  [1]{``#1''}%
\providecommand \bibnamefont  [1]{#1}%
\providecommand \bibfnamefont [1]{#1}%
\providecommand \citenamefont [1]{#1}%
\providecommand \href@noop [0]{\@secondoftwo}%
\providecommand \href [0]{\begingroup \@sanitize@url \@href}%
\providecommand \@href[1]{\@@startlink{#1}\@@href}%
\providecommand \@@href[1]{\endgroup#1\@@endlink}%
\providecommand \@sanitize@url [0]{\catcode `\\12\catcode `\$12\catcode
  `\&12\catcode `\#12\catcode `\^12\catcode `\_12\catcode `\%12\relax}%
\providecommand \@@startlink[1]{}%
\providecommand \@@endlink[0]{}%
\providecommand \url  [0]{\begingroup\@sanitize@url \@url }%
\providecommand \@url [1]{\endgroup\@href {#1}{\urlprefix }}%
\providecommand \urlprefix  [0]{URL }%
\providecommand \Eprint [0]{\href }%
\providecommand \doibase [0]{https://doi.org/}%
\providecommand \selectlanguage [0]{\@gobble}%
\providecommand \bibinfo  [0]{\@secondoftwo}%
\providecommand \bibfield  [0]{\@secondoftwo}%
\providecommand \translation [1]{[#1]}%
\providecommand \BibitemOpen [0]{}%
\providecommand \bibitemStop [0]{}%
\providecommand \bibitemNoStop [0]{.\EOS\space}%
\providecommand \EOS [0]{\spacefactor3000\relax}%
\providecommand \BibitemShut  [1]{\csname bibitem#1\endcsname}%
\let\auto@bib@innerbib\@empty
\bibitem [{\citenamefont {Lestz}\ \emph {et~al.}(tted)\citenamefont {Lestz},
  \citenamefont {Avdeeva}, \citenamefont {Neiser}, \citenamefont {Gorelenkova},
  \citenamefont {Halpern}, \citenamefont {Kaye}, \citenamefont {McClenaghan},
  \citenamefont {Pankin},\ and\ \citenamefont {Thome}}]{Lestz2025pre1}%
  \BibitemOpen
  \bibfield  {author} {\bibinfo {author} {\bibfnamefont {J.~B.}\ \bibnamefont
  {Lestz}}, \bibinfo {author} {\bibfnamefont {G.}~\bibnamefont {Avdeeva}},
  \bibinfo {author} {\bibfnamefont {T.~F.}\ \bibnamefont {Neiser}}, \bibinfo
  {author} {\bibfnamefont {M.~V.}\ \bibnamefont {Gorelenkova}}, \bibinfo
  {author} {\bibfnamefont {F.~D.}\ \bibnamefont {Halpern}}, \bibinfo {author}
  {\bibfnamefont {S.~M.}\ \bibnamefont {Kaye}}, \bibinfo {author}
  {\bibfnamefont {J.}~\bibnamefont {McClenaghan}}, \bibinfo {author}
  {\bibfnamefont {A.~Y.}\ \bibnamefont {Pankin}},\ and\ \bibinfo {author}
  {\bibfnamefont {K.~E.}\ \bibnamefont {Thome}},\ }\bibfield  {title} {\bibinfo
  {title} {Assessing time-dependent temperature profile predictions using
  reduced transport models for high performing {NSTX} plasmas},\ }\href@noop {}
  {\bibfield  {journal} {\bibinfo  {journal} {Plasma Physics and Controlled
  Fusion}\ } (\bibinfo {year} {submitted})}\BibitemShut {NoStop}%
\bibitem [{\citenamefont {Menard}\ \emph {et~al.}(2011)\citenamefont {Menard},
  \citenamefont {Bromberg}, \citenamefont {Brown}, \citenamefont {Burgess},
  \citenamefont {Dix}, \citenamefont {El-Guebaly}, \citenamefont {Gerrity},
  \citenamefont {Goldston}, \citenamefont {Hawryluk}, \citenamefont {Kastner},
  \citenamefont {Kessel}, \citenamefont {Malang}, \citenamefont {Minervini},
  \citenamefont {Neilson}, \citenamefont {Neumeyer}, \citenamefont {Prager},
  \citenamefont {Sawan}, \citenamefont {Sheffield}, \citenamefont {Sternlieb},
  \citenamefont {Waganer}, \citenamefont {Whyte},\ and\ \citenamefont
  {Zarnstorff}}]{Menard2011NF}%
  \BibitemOpen
  \bibfield  {author} {\bibinfo {author} {\bibfnamefont {J.~E.}\ \bibnamefont
  {Menard}}, \bibinfo {author} {\bibfnamefont {L.}~\bibnamefont {Bromberg}},
  \bibinfo {author} {\bibfnamefont {T.}~\bibnamefont {Brown}}, \bibinfo
  {author} {\bibfnamefont {T.}~\bibnamefont {Burgess}}, \bibinfo {author}
  {\bibfnamefont {D.}~\bibnamefont {Dix}}, \bibinfo {author} {\bibfnamefont
  {L.}~\bibnamefont {El-Guebaly}}, \bibinfo {author} {\bibfnamefont
  {T.}~\bibnamefont {Gerrity}}, \bibinfo {author} {\bibfnamefont {R.~J.}\
  \bibnamefont {Goldston}}, \bibinfo {author} {\bibfnamefont {R.~J.}\
  \bibnamefont {Hawryluk}}, \bibinfo {author} {\bibfnamefont {R.}~\bibnamefont
  {Kastner}}, \bibinfo {author} {\bibfnamefont {C.}~\bibnamefont {Kessel}},
  \bibinfo {author} {\bibfnamefont {S.}~\bibnamefont {Malang}}, \bibinfo
  {author} {\bibfnamefont {J.}~\bibnamefont {Minervini}}, \bibinfo {author}
  {\bibfnamefont {G.~H.}\ \bibnamefont {Neilson}}, \bibinfo {author}
  {\bibfnamefont {C.~L.}\ \bibnamefont {Neumeyer}}, \bibinfo {author}
  {\bibfnamefont {S.}~\bibnamefont {Prager}}, \bibinfo {author} {\bibfnamefont
  {M.}~\bibnamefont {Sawan}}, \bibinfo {author} {\bibfnamefont
  {J.}~\bibnamefont {Sheffield}}, \bibinfo {author} {\bibfnamefont
  {A.}~\bibnamefont {Sternlieb}}, \bibinfo {author} {\bibfnamefont
  {L.}~\bibnamefont {Waganer}}, \bibinfo {author} {\bibfnamefont
  {D.}~\bibnamefont {Whyte}},\ and\ \bibinfo {author} {\bibfnamefont
  {M.}~\bibnamefont {Zarnstorff}},\ }\bibfield  {title} {\bibinfo {title}
  {Prospects for pilot plants based on the tokamak, spherical tokamak and
  stellarator},\ }\href {https://doi.org/10.1088/0029-5515/51/10/103014}
  {\bibfield  {journal} {\bibinfo  {journal} {Nuclear Fusion}\ }\textbf
  {\bibinfo {volume} {51}},\ \bibinfo {pages} {103014} (\bibinfo {year}
  {2011})}\BibitemShut {NoStop}%
\bibitem [{\citenamefont {Menard}\ \emph {et~al.}(2016)\citenamefont {Menard},
  \citenamefont {Brown}, \citenamefont {El-Guebaly}, \citenamefont {Boyer},
  \citenamefont {Canik}, \citenamefont {Colling}, \citenamefont {Raman},
  \citenamefont {Wang}, \citenamefont {Zhai}, \citenamefont {Buxton},
  \citenamefont {Covele}, \citenamefont {D'Angelo}, \citenamefont {Davis},
  \citenamefont {Gerhardt}, \citenamefont {Gryaznevich}, \citenamefont {Harb},
  \citenamefont {Hender}, \citenamefont {Kaye}, \citenamefont {Kingham},
  \citenamefont {Kotschenreuther}, \citenamefont {Mahajan}, \citenamefont
  {Maingi}, \citenamefont {Marriott}, \citenamefont {Meier}, \citenamefont
  {Mynsberge}, \citenamefont {Neumeyer}, \citenamefont {Ono}, \citenamefont
  {Park}, \citenamefont {Sabbagh}, \citenamefont {Soukhanovskii}, \citenamefont
  {Valanju},\ and\ \citenamefont {Woolley}}]{Menard2016NF}%
  \BibitemOpen
  \bibfield  {author} {\bibinfo {author} {\bibfnamefont {J.~E.}\ \bibnamefont
  {Menard}}, \bibinfo {author} {\bibfnamefont {T.}~\bibnamefont {Brown}},
  \bibinfo {author} {\bibfnamefont {L.}~\bibnamefont {El-Guebaly}}, \bibinfo
  {author} {\bibfnamefont {M.}~\bibnamefont {Boyer}}, \bibinfo {author}
  {\bibfnamefont {J.}~\bibnamefont {Canik}}, \bibinfo {author} {\bibfnamefont
  {B.}~\bibnamefont {Colling}}, \bibinfo {author} {\bibfnamefont
  {R.}~\bibnamefont {Raman}}, \bibinfo {author} {\bibfnamefont
  {Z.}~\bibnamefont {Wang}}, \bibinfo {author} {\bibfnamefont {Y.}~\bibnamefont
  {Zhai}}, \bibinfo {author} {\bibfnamefont {P.}~\bibnamefont {Buxton}},
  \bibinfo {author} {\bibfnamefont {B.}~\bibnamefont {Covele}}, \bibinfo
  {author} {\bibfnamefont {C.}~\bibnamefont {D'Angelo}}, \bibinfo {author}
  {\bibfnamefont {A.}~\bibnamefont {Davis}}, \bibinfo {author} {\bibfnamefont
  {S.}~\bibnamefont {Gerhardt}}, \bibinfo {author} {\bibfnamefont
  {M.}~\bibnamefont {Gryaznevich}}, \bibinfo {author} {\bibfnamefont
  {M.}~\bibnamefont {Harb}}, \bibinfo {author} {\bibfnamefont {T.~C.}\
  \bibnamefont {Hender}}, \bibinfo {author} {\bibfnamefont {S.}~\bibnamefont
  {Kaye}}, \bibinfo {author} {\bibfnamefont {D.}~\bibnamefont {Kingham}},
  \bibinfo {author} {\bibfnamefont {M.}~\bibnamefont {Kotschenreuther}},
  \bibinfo {author} {\bibfnamefont {S.}~\bibnamefont {Mahajan}}, \bibinfo
  {author} {\bibfnamefont {R.}~\bibnamefont {Maingi}}, \bibinfo {author}
  {\bibfnamefont {E.}~\bibnamefont {Marriott}}, \bibinfo {author}
  {\bibfnamefont {E.~T.}\ \bibnamefont {Meier}}, \bibinfo {author}
  {\bibfnamefont {L.}~\bibnamefont {Mynsberge}}, \bibinfo {author}
  {\bibfnamefont {C.}~\bibnamefont {Neumeyer}}, \bibinfo {author}
  {\bibfnamefont {M.}~\bibnamefont {Ono}}, \bibinfo {author} {\bibfnamefont
  {J.-K.}\ \bibnamefont {Park}}, \bibinfo {author} {\bibfnamefont {S.~A.}\
  \bibnamefont {Sabbagh}}, \bibinfo {author} {\bibfnamefont {V.}~\bibnamefont
  {Soukhanovskii}}, \bibinfo {author} {\bibfnamefont {P.}~\bibnamefont
  {Valanju}},\ and\ \bibinfo {author} {\bibfnamefont {R.}~\bibnamefont
  {Woolley}},\ }\bibfield  {title} {\bibinfo {title} {Fusion nuclear science
  facilities and pilot plants based on the spherical tokamak},\ }\href
  {https://doi.org/10.1088/0029-5515/56/10/106023} {\bibfield  {journal}
  {\bibinfo  {journal} {Nuclear Fusion}\ }\textbf {\bibinfo {volume} {56}},\
  \bibinfo {pages} {106023} (\bibinfo {year} {2016})}\BibitemShut {NoStop}%
\bibitem [{\citenamefont {Menard}(2019)}]{Menard2019RSA}%
  \BibitemOpen
  \bibfield  {author} {\bibinfo {author} {\bibfnamefont {J.~E.}\ \bibnamefont
  {Menard}},\ }\bibfield  {title} {\bibinfo {title} {Compact steady-state
  tokamak performance dependence on magnet and core physics limits},\ }\href
  {https://doi.org/10.1098/rsta.2017.0440} {\bibfield  {journal} {\bibinfo
  {journal} {Philosophical Transactions of the Royal Society A}\ }\textbf
  {\bibinfo {volume} {377}},\ \bibinfo {pages} {20170440} (\bibinfo {year}
  {2019})}\BibitemShut {NoStop}%
\bibitem [{\citenamefont {Menard}\ \emph {et~al.}(2022)\citenamefont {Menard},
  \citenamefont {Grierson}, \citenamefont {Brown}, \citenamefont {Rana},
  \citenamefont {Zhai}, \citenamefont {Poli}, \citenamefont {Maingi},
  \citenamefont {Guttenfelder},\ and\ \citenamefont {Snyder}}]{Menard2022NF}%
  \BibitemOpen
  \bibfield  {author} {\bibinfo {author} {\bibfnamefont {J.}~\bibnamefont
  {Menard}}, \bibinfo {author} {\bibfnamefont {B.}~\bibnamefont {Grierson}},
  \bibinfo {author} {\bibfnamefont {T.}~\bibnamefont {Brown}}, \bibinfo
  {author} {\bibfnamefont {C.}~\bibnamefont {Rana}}, \bibinfo {author}
  {\bibfnamefont {Y.}~\bibnamefont {Zhai}}, \bibinfo {author} {\bibfnamefont
  {F.}~\bibnamefont {Poli}}, \bibinfo {author} {\bibfnamefont {R.}~\bibnamefont
  {Maingi}}, \bibinfo {author} {\bibfnamefont {W.}~\bibnamefont
  {Guttenfelder}},\ and\ \bibinfo {author} {\bibfnamefont {P.}~\bibnamefont
  {Snyder}},\ }\bibfield  {title} {\bibinfo {title} {Fusion pilot plant
  performance and the role of a sustained high power density tokamak},\ }\href
  {https://doi.org/10.1088/1741-4326/ac49aa} {\bibfield  {journal} {\bibinfo
  {journal} {Nuclear Fusion}\ }\textbf {\bibinfo {volume} {62}},\ \bibinfo
  {pages} {036026} (\bibinfo {year} {2022})}\BibitemShut {NoStop}%
\bibitem [{\citenamefont {Ono}\ \emph {et~al.}(2000)\citenamefont {Ono},
  \citenamefont {Kaye}, \citenamefont {Peng}, \citenamefont {Barnes},
  \citenamefont {Blanchard}, \citenamefont {Carter}, \citenamefont
  {Chrzanowski}, \citenamefont {Dudek}, \citenamefont {Ewig}, \citenamefont
  {Gates}, \citenamefont {Hatcher}, \citenamefont {Jarboe}, \citenamefont
  {Jardin}, \citenamefont {Johnson}, \citenamefont {Kaita}, \citenamefont
  {Kalish}, \citenamefont {Kessel}, \citenamefont {Kugel}, \citenamefont
  {Maingi}, \citenamefont {Majeski}, \citenamefont {Manickam}, \citenamefont
  {McCormack}, \citenamefont {Menard}, \citenamefont {Mueller}, \citenamefont
  {Nelson}, \citenamefont {Nelson}, \citenamefont {Neumeyer}, \citenamefont
  {Oliaro}, \citenamefont {Paoletti}, \citenamefont {Parsells}, \citenamefont
  {Perry}, \citenamefont {Pomphrey}, \citenamefont {Ramakrishnan},
  \citenamefont {Raman}, \citenamefont {Rewoldt}, \citenamefont {Robinson},
  \citenamefont {Roquemore}, \citenamefont {Ryan}, \citenamefont {Sabbagh},
  \citenamefont {Swain}, \citenamefont {Synakowski}, \citenamefont {Viola},
  \citenamefont {Williams}, \citenamefont {Wilson},\ and\ \citenamefont {{NSTX
  Team}}}]{Ono2000NF}%
  \BibitemOpen
  \bibfield  {author} {\bibinfo {author} {\bibfnamefont {M.}~\bibnamefont
  {Ono}}, \bibinfo {author} {\bibfnamefont {S.}~\bibnamefont {Kaye}}, \bibinfo
  {author} {\bibfnamefont {Y.-K.}\ \bibnamefont {Peng}}, \bibinfo {author}
  {\bibfnamefont {G.}~\bibnamefont {Barnes}}, \bibinfo {author} {\bibfnamefont
  {W.}~\bibnamefont {Blanchard}}, \bibinfo {author} {\bibfnamefont
  {M.}~\bibnamefont {Carter}}, \bibinfo {author} {\bibfnamefont
  {J.}~\bibnamefont {Chrzanowski}}, \bibinfo {author} {\bibfnamefont
  {L.}~\bibnamefont {Dudek}}, \bibinfo {author} {\bibfnamefont
  {R.}~\bibnamefont {Ewig}}, \bibinfo {author} {\bibfnamefont {D.}~\bibnamefont
  {Gates}}, \bibinfo {author} {\bibfnamefont {R.}~\bibnamefont {Hatcher}},
  \bibinfo {author} {\bibfnamefont {T.}~\bibnamefont {Jarboe}}, \bibinfo
  {author} {\bibfnamefont {S.}~\bibnamefont {Jardin}}, \bibinfo {author}
  {\bibfnamefont {D.}~\bibnamefont {Johnson}}, \bibinfo {author} {\bibfnamefont
  {R.}~\bibnamefont {Kaita}}, \bibinfo {author} {\bibfnamefont
  {M.}~\bibnamefont {Kalish}}, \bibinfo {author} {\bibfnamefont
  {C.}~\bibnamefont {Kessel}}, \bibinfo {author} {\bibfnamefont
  {H.}~\bibnamefont {Kugel}}, \bibinfo {author} {\bibfnamefont
  {R.}~\bibnamefont {Maingi}}, \bibinfo {author} {\bibfnamefont
  {R.}~\bibnamefont {Majeski}}, \bibinfo {author} {\bibfnamefont
  {J.}~\bibnamefont {Manickam}}, \bibinfo {author} {\bibfnamefont
  {B.}~\bibnamefont {McCormack}}, \bibinfo {author} {\bibfnamefont
  {J.}~\bibnamefont {Menard}}, \bibinfo {author} {\bibfnamefont
  {D.}~\bibnamefont {Mueller}}, \bibinfo {author} {\bibfnamefont
  {B.}~\bibnamefont {Nelson}}, \bibinfo {author} {\bibfnamefont
  {B.}~\bibnamefont {Nelson}}, \bibinfo {author} {\bibfnamefont
  {C.}~\bibnamefont {Neumeyer}}, \bibinfo {author} {\bibfnamefont
  {G.}~\bibnamefont {Oliaro}}, \bibinfo {author} {\bibfnamefont
  {F.}~\bibnamefont {Paoletti}}, \bibinfo {author} {\bibfnamefont
  {R.}~\bibnamefont {Parsells}}, \bibinfo {author} {\bibfnamefont
  {E.}~\bibnamefont {Perry}}, \bibinfo {author} {\bibfnamefont
  {N.}~\bibnamefont {Pomphrey}}, \bibinfo {author} {\bibfnamefont
  {S.}~\bibnamefont {Ramakrishnan}}, \bibinfo {author} {\bibfnamefont
  {R.}~\bibnamefont {Raman}}, \bibinfo {author} {\bibfnamefont
  {G.}~\bibnamefont {Rewoldt}}, \bibinfo {author} {\bibfnamefont
  {J.}~\bibnamefont {Robinson}}, \bibinfo {author} {\bibfnamefont
  {A.}~\bibnamefont {Roquemore}}, \bibinfo {author} {\bibfnamefont
  {P.}~\bibnamefont {Ryan}}, \bibinfo {author} {\bibfnamefont {S.}~\bibnamefont
  {Sabbagh}}, \bibinfo {author} {\bibfnamefont {D.}~\bibnamefont {Swain}},
  \bibinfo {author} {\bibfnamefont {E.}~\bibnamefont {Synakowski}}, \bibinfo
  {author} {\bibfnamefont {M.}~\bibnamefont {Viola}}, \bibinfo {author}
  {\bibfnamefont {M.}~\bibnamefont {Williams}}, \bibinfo {author}
  {\bibfnamefont {J.}~\bibnamefont {Wilson}},\ and\ \bibinfo {author}
  {\bibnamefont {{NSTX Team}}},\ }\bibfield  {title} {\bibinfo {title}
  {Exploration of spherical torus physics in the {NSTX} device},\ }\href
  {https://doi.org/10.1088/0029-5515/40/3Y/316} {\bibfield  {journal} {\bibinfo
   {journal} {Nuclear Fusion}\ }\textbf {\bibinfo {volume} {40}},\ \bibinfo
  {pages} {557} (\bibinfo {year} {2000})}\BibitemShut {NoStop}%
\bibitem [{\citenamefont {Sabbagh}\ \emph {et~al.}(2013)\citenamefont
  {Sabbagh}, \citenamefont {Ahn}, \citenamefont {Allain}, \citenamefont
  {Andre}, \citenamefont {Balbaky}, \citenamefont {Bastasz}, \citenamefont
  {Battaglia}, \citenamefont {Bell}, \citenamefont {Bell}, \citenamefont
  {Beiersdorfer}, \citenamefont {Belova}, \citenamefont {Berkery},
  \citenamefont {Betti}, \citenamefont {Bialek}, \citenamefont {Bigelow},
  \citenamefont {Bitter}, \citenamefont {Boedo}, \citenamefont {Bonoli},
  \citenamefont {Boozer}, \citenamefont {Bortolon}, \citenamefont {Boyle},
  \citenamefont {Brennan}, \citenamefont {Breslau}, \citenamefont {Buttery},
  \citenamefont {Canik}, \citenamefont {Caravelli}, \citenamefont {Chang},
  \citenamefont {Crocker}, \citenamefont {Darrow}, \citenamefont {Davis},
  \citenamefont {Delgado-Aparicio}, \citenamefont {Diallo}, \citenamefont
  {Ding}, \citenamefont {D'Ippolito}, \citenamefont {Domier}, \citenamefont
  {Dorland}, \citenamefont {Ethier}, \citenamefont {Evans}, \citenamefont
  {Ferron}, \citenamefont {Finkenthal}, \citenamefont {Foley}, \citenamefont
  {Fonck}, \citenamefont {Frazin}, \citenamefont {Fredrickson}, \citenamefont
  {Fu}, \citenamefont {Gates}, \citenamefont {Gerhardt}, \citenamefont
  {Glasser}, \citenamefont {Gorelenkov}, \citenamefont {Gray}, \citenamefont
  {Guo}, \citenamefont {Guttenfelder}, \citenamefont {Hahm}, \citenamefont
  {Harvey}, \citenamefont {Hassanein}, \citenamefont {Heidbrink}, \citenamefont
  {Hill}, \citenamefont {Hirooka}, \citenamefont {Hooper}, \citenamefont
  {Hosea}, \citenamefont {Humphreys}, \citenamefont {Indireshkumar},
  \citenamefont {Jaeger}, \citenamefont {Jarboe}, \citenamefont {Jardin},
  \citenamefont {Jaworski}, \citenamefont {Kaita}, \citenamefont {Kallman},
  \citenamefont {Katsuro-Hopkins}, \citenamefont {Kaye}, \citenamefont
  {Kessel}, \citenamefont {Kim}, \citenamefont {Kolemen}, \citenamefont
  {Kramer}, \citenamefont {Krasheninnikov}, \citenamefont {Kubota},
  \citenamefont {Kugel}, \citenamefont {Haye}, \citenamefont {Lao},
  \citenamefont {LeBlanc}, \citenamefont {Lee}, \citenamefont {Lee},
  \citenamefont {Leuer}, \citenamefont {Levinton}, \citenamefont {Liang},
  \citenamefont {Liu}, \citenamefont {Lore}, \citenamefont {Jr}, \citenamefont
  {Maingi}, \citenamefont {Majeski}, \citenamefont {Manickam}, \citenamefont
  {Mansfield}, \citenamefont {Maqueda}, \citenamefont {Mazzucato},
  \citenamefont {McLean}, \citenamefont {McCune}, \citenamefont {McGeehan},
  \citenamefont {McKee}, \citenamefont {Medley}, \citenamefont {Meier},
  \citenamefont {Menard}, \citenamefont {Menon}, \citenamefont {Meyer},
  \citenamefont {Mikkelsen}, \citenamefont {Miloshevsky}, \citenamefont
  {Mueller}, \citenamefont {Munsat}, \citenamefont {Myra}, \citenamefont
  {Nelson}, \citenamefont {Nishino}, \citenamefont {Nygren}, \citenamefont
  {Ono}, \citenamefont {Osborne}, \citenamefont {Park}, \citenamefont {Park},
  \citenamefont {Park}, \citenamefont {Paul}, \citenamefont {Peebles},
  \citenamefont {Penaflor}, \citenamefont {Perkins}, \citenamefont {Phillips},
  \citenamefont {Pigarov}, \citenamefont {\Podesta}, \citenamefont
  {Preinhaelter}, \citenamefont {Raman}, \citenamefont {Ren}, \citenamefont
  {Rewoldt}, \citenamefont {Rognlien}, \citenamefont {Ross}, \citenamefont
  {Rowley}, \citenamefont {Ruskov}, \citenamefont {Russell}, \citenamefont
  {Ruzic}, \citenamefont {Ryan}, \citenamefont {Schaffer}, \citenamefont
  {Schuster}, \citenamefont {Scotti}, \citenamefont {Shaing}, \citenamefont
  {Shevchenko}, \citenamefont {Shinohara}, \citenamefont {Sizyuk},
  \citenamefont {Skinner}, \citenamefont {Smirnov}, \citenamefont {Smith},
  \citenamefont {Snyder}, \citenamefont {Solomon}, \citenamefont {Sontag},
  \citenamefont {Soukhanovskii}, \citenamefont {Stoltzfus-Dueck}, \citenamefont
  {Stotler}, \citenamefont {Stratton}, \citenamefont {Stutman}, \citenamefont
  {Takahashi}, \citenamefont {Takase}, \citenamefont {Tamura}, \citenamefont
  {Tang}, \citenamefont {Taylor}, \citenamefont {Taylor}, \citenamefont
  {Tritz}, \citenamefont {Tsarouhas}, \citenamefont {Umansky}, \citenamefont
  {Urban}, \citenamefont {Untergberg}, \citenamefont {Walker}, \citenamefont
  {Wampler}, \citenamefont {Wang}, \citenamefont {Whaley}, \citenamefont
  {White}, \citenamefont {Wilgen}, \citenamefont {Wilson}, \citenamefont
  {Wong}, \citenamefont {Wright}, \citenamefont {Xia}, \citenamefont
  {Youchison}, \citenamefont {Yu}, \citenamefont {Yuh}, \citenamefont
  {Zakharov}, \citenamefont {Zemlyanov}, \citenamefont {Zimmer},\ and\
  \citenamefont {Zweben}}]{Sabbagh2013NF}%
  \BibitemOpen
  \bibfield  {author} {\bibinfo {author} {\bibfnamefont {S.~A.}\ \bibnamefont
  {Sabbagh}}, \bibinfo {author} {\bibfnamefont {J.-W.}\ \bibnamefont {Ahn}},
  \bibinfo {author} {\bibfnamefont {J.}~\bibnamefont {Allain}}, \bibinfo
  {author} {\bibfnamefont {R.}~\bibnamefont {Andre}}, \bibinfo {author}
  {\bibfnamefont {A.}~\bibnamefont {Balbaky}}, \bibinfo {author} {\bibfnamefont
  {R.}~\bibnamefont {Bastasz}}, \bibinfo {author} {\bibfnamefont
  {D.}~\bibnamefont {Battaglia}}, \bibinfo {author} {\bibfnamefont
  {M.}~\bibnamefont {Bell}}, \bibinfo {author} {\bibfnamefont {R.}~\bibnamefont
  {Bell}}, \bibinfo {author} {\bibfnamefont {P.}~\bibnamefont {Beiersdorfer}},
  \bibinfo {author} {\bibfnamefont {E.}~\bibnamefont {Belova}}, \bibinfo
  {author} {\bibfnamefont {J.}~\bibnamefont {Berkery}}, \bibinfo {author}
  {\bibfnamefont {R.}~\bibnamefont {Betti}}, \bibinfo {author} {\bibfnamefont
  {J.}~\bibnamefont {Bialek}}, \bibinfo {author} {\bibfnamefont
  {T.}~\bibnamefont {Bigelow}}, \bibinfo {author} {\bibfnamefont
  {M.}~\bibnamefont {Bitter}}, \bibinfo {author} {\bibfnamefont
  {J.}~\bibnamefont {Boedo}}, \bibinfo {author} {\bibfnamefont
  {P.}~\bibnamefont {Bonoli}}, \bibinfo {author} {\bibfnamefont
  {A.}~\bibnamefont {Boozer}}, \bibinfo {author} {\bibfnamefont
  {A.}~\bibnamefont {Bortolon}}, \bibinfo {author} {\bibfnamefont
  {D.}~\bibnamefont {Boyle}}, \bibinfo {author} {\bibfnamefont
  {D.}~\bibnamefont {Brennan}}, \bibinfo {author} {\bibfnamefont
  {J.}~\bibnamefont {Breslau}}, \bibinfo {author} {\bibfnamefont
  {R.}~\bibnamefont {Buttery}}, \bibinfo {author} {\bibfnamefont
  {J.}~\bibnamefont {Canik}}, \bibinfo {author} {\bibfnamefont
  {G.}~\bibnamefont {Caravelli}}, \bibinfo {author} {\bibfnamefont
  {C.}~\bibnamefont {Chang}}, \bibinfo {author} {\bibfnamefont
  {N.}~\bibnamefont {Crocker}}, \bibinfo {author} {\bibfnamefont
  {D.}~\bibnamefont {Darrow}}, \bibinfo {author} {\bibfnamefont
  {B.}~\bibnamefont {Davis}}, \bibinfo {author} {\bibfnamefont
  {L.}~\bibnamefont {Delgado-Aparicio}}, \bibinfo {author} {\bibfnamefont
  {A.}~\bibnamefont {Diallo}}, \bibinfo {author} {\bibfnamefont
  {S.}~\bibnamefont {Ding}}, \bibinfo {author} {\bibfnamefont {D.}~\bibnamefont
  {D'Ippolito}}, \bibinfo {author} {\bibfnamefont {C.}~\bibnamefont {Domier}},
  \bibinfo {author} {\bibfnamefont {W.}~\bibnamefont {Dorland}}, \bibinfo
  {author} {\bibfnamefont {S.}~\bibnamefont {Ethier}}, \bibinfo {author}
  {\bibfnamefont {T.}~\bibnamefont {Evans}}, \bibinfo {author} {\bibfnamefont
  {J.}~\bibnamefont {Ferron}}, \bibinfo {author} {\bibfnamefont
  {M.}~\bibnamefont {Finkenthal}}, \bibinfo {author} {\bibfnamefont
  {J.}~\bibnamefont {Foley}}, \bibinfo {author} {\bibfnamefont
  {R.}~\bibnamefont {Fonck}}, \bibinfo {author} {\bibfnamefont
  {R.}~\bibnamefont {Frazin}}, \bibinfo {author} {\bibfnamefont
  {E.}~\bibnamefont {Fredrickson}}, \bibinfo {author} {\bibfnamefont
  {G.}~\bibnamefont {Fu}}, \bibinfo {author} {\bibfnamefont {D.}~\bibnamefont
  {Gates}}, \bibinfo {author} {\bibfnamefont {S.}~\bibnamefont {Gerhardt}},
  \bibinfo {author} {\bibfnamefont {A.}~\bibnamefont {Glasser}}, \bibinfo
  {author} {\bibfnamefont {N.}~\bibnamefont {Gorelenkov}}, \bibinfo {author}
  {\bibfnamefont {T.}~\bibnamefont {Gray}}, \bibinfo {author} {\bibfnamefont
  {Y.}~\bibnamefont {Guo}}, \bibinfo {author} {\bibfnamefont {W.}~\bibnamefont
  {Guttenfelder}}, \bibinfo {author} {\bibfnamefont {T.}~\bibnamefont {Hahm}},
  \bibinfo {author} {\bibfnamefont {R.}~\bibnamefont {Harvey}}, \bibinfo
  {author} {\bibfnamefont {A.}~\bibnamefont {Hassanein}}, \bibinfo {author}
  {\bibfnamefont {W.}~\bibnamefont {Heidbrink}}, \bibinfo {author}
  {\bibfnamefont {K.}~\bibnamefont {Hill}}, \bibinfo {author} {\bibfnamefont
  {Y.}~\bibnamefont {Hirooka}}, \bibinfo {author} {\bibfnamefont {E.~B.}\
  \bibnamefont {Hooper}}, \bibinfo {author} {\bibfnamefont {J.}~\bibnamefont
  {Hosea}}, \bibinfo {author} {\bibfnamefont {D.}~\bibnamefont {Humphreys}},
  \bibinfo {author} {\bibfnamefont {K.}~\bibnamefont {Indireshkumar}}, \bibinfo
  {author} {\bibfnamefont {F.}~\bibnamefont {Jaeger}}, \bibinfo {author}
  {\bibfnamefont {T.}~\bibnamefont {Jarboe}}, \bibinfo {author} {\bibfnamefont
  {S.}~\bibnamefont {Jardin}}, \bibinfo {author} {\bibfnamefont
  {M.}~\bibnamefont {Jaworski}}, \bibinfo {author} {\bibfnamefont
  {R.}~\bibnamefont {Kaita}}, \bibinfo {author} {\bibfnamefont
  {J.}~\bibnamefont {Kallman}}, \bibinfo {author} {\bibfnamefont
  {O.}~\bibnamefont {Katsuro-Hopkins}}, \bibinfo {author} {\bibfnamefont
  {S.}~\bibnamefont {Kaye}}, \bibinfo {author} {\bibfnamefont {C.}~\bibnamefont
  {Kessel}}, \bibinfo {author} {\bibfnamefont {J.}~\bibnamefont {Kim}},
  \bibinfo {author} {\bibfnamefont {E.}~\bibnamefont {Kolemen}}, \bibinfo
  {author} {\bibfnamefont {G.}~\bibnamefont {Kramer}}, \bibinfo {author}
  {\bibfnamefont {S.}~\bibnamefont {Krasheninnikov}}, \bibinfo {author}
  {\bibfnamefont {S.}~\bibnamefont {Kubota}}, \bibinfo {author} {\bibfnamefont
  {H.}~\bibnamefont {Kugel}}, \bibinfo {author} {\bibfnamefont {R.~J.~L.}\
  \bibnamefont {Haye}}, \bibinfo {author} {\bibfnamefont {L.}~\bibnamefont
  {Lao}}, \bibinfo {author} {\bibfnamefont {B.}~\bibnamefont {LeBlanc}},
  \bibinfo {author} {\bibfnamefont {W.}~\bibnamefont {Lee}}, \bibinfo {author}
  {\bibfnamefont {K.}~\bibnamefont {Lee}}, \bibinfo {author} {\bibfnamefont
  {J.}~\bibnamefont {Leuer}}, \bibinfo {author} {\bibfnamefont
  {F.}~\bibnamefont {Levinton}}, \bibinfo {author} {\bibfnamefont
  {Y.}~\bibnamefont {Liang}}, \bibinfo {author} {\bibfnamefont
  {D.}~\bibnamefont {Liu}}, \bibinfo {author} {\bibfnamefont {J.}~\bibnamefont
  {Lore}}, \bibinfo {author} {\bibfnamefont {N.~L.}\ \bibnamefont {Jr}},
  \bibinfo {author} {\bibfnamefont {R.}~\bibnamefont {Maingi}}, \bibinfo
  {author} {\bibfnamefont {R.}~\bibnamefont {Majeski}}, \bibinfo {author}
  {\bibfnamefont {J.}~\bibnamefont {Manickam}}, \bibinfo {author}
  {\bibfnamefont {D.}~\bibnamefont {Mansfield}}, \bibinfo {author}
  {\bibfnamefont {R.}~\bibnamefont {Maqueda}}, \bibinfo {author} {\bibfnamefont
  {E.}~\bibnamefont {Mazzucato}}, \bibinfo {author} {\bibfnamefont
  {A.}~\bibnamefont {McLean}}, \bibinfo {author} {\bibfnamefont
  {D.}~\bibnamefont {McCune}}, \bibinfo {author} {\bibfnamefont
  {B.}~\bibnamefont {McGeehan}}, \bibinfo {author} {\bibfnamefont
  {G.}~\bibnamefont {McKee}}, \bibinfo {author} {\bibfnamefont
  {S.}~\bibnamefont {Medley}}, \bibinfo {author} {\bibfnamefont
  {E.}~\bibnamefont {Meier}}, \bibinfo {author} {\bibfnamefont
  {J.}~\bibnamefont {Menard}}, \bibinfo {author} {\bibfnamefont
  {M.}~\bibnamefont {Menon}}, \bibinfo {author} {\bibfnamefont
  {H.}~\bibnamefont {Meyer}}, \bibinfo {author} {\bibfnamefont
  {D.}~\bibnamefont {Mikkelsen}}, \bibinfo {author} {\bibfnamefont
  {G.}~\bibnamefont {Miloshevsky}}, \bibinfo {author} {\bibfnamefont
  {D.}~\bibnamefont {Mueller}}, \bibinfo {author} {\bibfnamefont
  {T.}~\bibnamefont {Munsat}}, \bibinfo {author} {\bibfnamefont
  {J.}~\bibnamefont {Myra}}, \bibinfo {author} {\bibfnamefont {B.}~\bibnamefont
  {Nelson}}, \bibinfo {author} {\bibfnamefont {N.}~\bibnamefont {Nishino}},
  \bibinfo {author} {\bibfnamefont {R.}~\bibnamefont {Nygren}}, \bibinfo
  {author} {\bibfnamefont {M.}~\bibnamefont {Ono}}, \bibinfo {author}
  {\bibfnamefont {T.}~\bibnamefont {Osborne}}, \bibinfo {author} {\bibfnamefont
  {H.}~\bibnamefont {Park}}, \bibinfo {author} {\bibfnamefont {J.}~\bibnamefont
  {Park}}, \bibinfo {author} {\bibfnamefont {Y.~S.}\ \bibnamefont {Park}},
  \bibinfo {author} {\bibfnamefont {S.}~\bibnamefont {Paul}}, \bibinfo {author}
  {\bibfnamefont {W.}~\bibnamefont {Peebles}}, \bibinfo {author} {\bibfnamefont
  {B.}~\bibnamefont {Penaflor}}, \bibinfo {author} {\bibfnamefont {R.~J.}\
  \bibnamefont {Perkins}}, \bibinfo {author} {\bibfnamefont {C.}~\bibnamefont
  {Phillips}}, \bibinfo {author} {\bibfnamefont {A.}~\bibnamefont {Pigarov}},
  \bibinfo {author} {\bibfnamefont {M.}~\bibnamefont {\Podesta}}, \bibinfo
  {author} {\bibfnamefont {J.}~\bibnamefont {Preinhaelter}}, \bibinfo {author}
  {\bibfnamefont {R.}~\bibnamefont {Raman}}, \bibinfo {author} {\bibfnamefont
  {Y.}~\bibnamefont {Ren}}, \bibinfo {author} {\bibfnamefont {G.}~\bibnamefont
  {Rewoldt}}, \bibinfo {author} {\bibfnamefont {T.}~\bibnamefont {Rognlien}},
  \bibinfo {author} {\bibfnamefont {P.}~\bibnamefont {Ross}}, \bibinfo {author}
  {\bibfnamefont {C.}~\bibnamefont {Rowley}}, \bibinfo {author} {\bibfnamefont
  {E.}~\bibnamefont {Ruskov}}, \bibinfo {author} {\bibfnamefont
  {D.}~\bibnamefont {Russell}}, \bibinfo {author} {\bibfnamefont
  {D.}~\bibnamefont {Ruzic}}, \bibinfo {author} {\bibfnamefont
  {P.}~\bibnamefont {Ryan}}, \bibinfo {author} {\bibfnamefont {M.}~\bibnamefont
  {Schaffer}}, \bibinfo {author} {\bibfnamefont {E.}~\bibnamefont {Schuster}},
  \bibinfo {author} {\bibfnamefont {F.}~\bibnamefont {Scotti}}, \bibinfo
  {author} {\bibfnamefont {K.}~\bibnamefont {Shaing}}, \bibinfo {author}
  {\bibfnamefont {V.}~\bibnamefont {Shevchenko}}, \bibinfo {author}
  {\bibfnamefont {K.}~\bibnamefont {Shinohara}}, \bibinfo {author}
  {\bibfnamefont {V.}~\bibnamefont {Sizyuk}}, \bibinfo {author} {\bibfnamefont
  {C.~H.}\ \bibnamefont {Skinner}}, \bibinfo {author} {\bibfnamefont
  {A.}~\bibnamefont {Smirnov}}, \bibinfo {author} {\bibfnamefont
  {D.}~\bibnamefont {Smith}}, \bibinfo {author} {\bibfnamefont
  {P.}~\bibnamefont {Snyder}}, \bibinfo {author} {\bibfnamefont
  {W.}~\bibnamefont {Solomon}}, \bibinfo {author} {\bibfnamefont
  {A.}~\bibnamefont {Sontag}}, \bibinfo {author} {\bibfnamefont
  {V.}~\bibnamefont {Soukhanovskii}}, \bibinfo {author} {\bibfnamefont
  {T.}~\bibnamefont {Stoltzfus-Dueck}}, \bibinfo {author} {\bibfnamefont
  {D.}~\bibnamefont {Stotler}}, \bibinfo {author} {\bibfnamefont
  {B.}~\bibnamefont {Stratton}}, \bibinfo {author} {\bibfnamefont
  {D.}~\bibnamefont {Stutman}}, \bibinfo {author} {\bibfnamefont
  {H.}~\bibnamefont {Takahashi}}, \bibinfo {author} {\bibfnamefont
  {Y.}~\bibnamefont {Takase}}, \bibinfo {author} {\bibfnamefont
  {N.}~\bibnamefont {Tamura}}, \bibinfo {author} {\bibfnamefont
  {X.}~\bibnamefont {Tang}}, \bibinfo {author} {\bibfnamefont {G.}~\bibnamefont
  {Taylor}}, \bibinfo {author} {\bibfnamefont {C.}~\bibnamefont {Taylor}},
  \bibinfo {author} {\bibfnamefont {K.}~\bibnamefont {Tritz}}, \bibinfo
  {author} {\bibfnamefont {D.}~\bibnamefont {Tsarouhas}}, \bibinfo {author}
  {\bibfnamefont {M.}~\bibnamefont {Umansky}}, \bibinfo {author} {\bibfnamefont
  {J.}~\bibnamefont {Urban}}, \bibinfo {author} {\bibfnamefont
  {E.}~\bibnamefont {Untergberg}}, \bibinfo {author} {\bibfnamefont
  {M.}~\bibnamefont {Walker}}, \bibinfo {author} {\bibfnamefont
  {W.}~\bibnamefont {Wampler}}, \bibinfo {author} {\bibfnamefont
  {W.}~\bibnamefont {Wang}}, \bibinfo {author} {\bibfnamefont {J.}~\bibnamefont
  {Whaley}}, \bibinfo {author} {\bibfnamefont {R.}~\bibnamefont {White}},
  \bibinfo {author} {\bibfnamefont {J.}~\bibnamefont {Wilgen}}, \bibinfo
  {author} {\bibfnamefont {R.}~\bibnamefont {Wilson}}, \bibinfo {author}
  {\bibfnamefont {K.~L.}\ \bibnamefont {Wong}}, \bibinfo {author}
  {\bibfnamefont {J.}~\bibnamefont {Wright}}, \bibinfo {author} {\bibfnamefont
  {Z.}~\bibnamefont {Xia}}, \bibinfo {author} {\bibfnamefont {D.}~\bibnamefont
  {Youchison}}, \bibinfo {author} {\bibfnamefont {G.}~\bibnamefont {Yu}},
  \bibinfo {author} {\bibfnamefont {H.}~\bibnamefont {Yuh}}, \bibinfo {author}
  {\bibfnamefont {L.}~\bibnamefont {Zakharov}}, \bibinfo {author}
  {\bibfnamefont {D.}~\bibnamefont {Zemlyanov}}, \bibinfo {author}
  {\bibfnamefont {G.}~\bibnamefont {Zimmer}},\ and\ \bibinfo {author}
  {\bibfnamefont {S.~J.}\ \bibnamefont {Zweben}},\ }\bibfield  {title}
  {\bibinfo {title} {Overview of physics results from the conclusive operation
  of the {National Spherical Torus Experiment}},\ }\href
  {http://stacks.iop.org/0029-5515/53/i=10/a=104007} {\bibfield  {journal}
  {\bibinfo  {journal} {Nuclear Fusion}\ }\textbf {\bibinfo {volume} {53}},\
  \bibinfo {pages} {104007} (\bibinfo {year} {2013})}\BibitemShut {NoStop}%
\bibitem [{\citenamefont {Menard}\ \emph {et~al.}(2012)\citenamefont {Menard},
  \citenamefont {Gerhardt}, \citenamefont {Bell}, \citenamefont {Bialek},
  \citenamefont {Brooks}, \citenamefont {Canik}, \citenamefont {Chrzanowski},
  \citenamefont {Denault}, \citenamefont {Dudek}, \citenamefont {Gates},
  \citenamefont {Gorelenkov}, \citenamefont {Guttenfelder}, \citenamefont
  {Hatcher}, \citenamefont {Hosea}, \citenamefont {Kaita}, \citenamefont
  {Kaye}, \citenamefont {Kessel}, \citenamefont {Kolemen}, \citenamefont
  {Kugel}, \citenamefont {Maingi}, \citenamefont {Mardenfeld}, \citenamefont
  {Mueller}, \citenamefont {Nelson}, \citenamefont {Neumeyer}, \citenamefont
  {Ono}, \citenamefont {Perry}, \citenamefont {Ramakrishnan}, \citenamefont
  {Raman}, \citenamefont {Ren}, \citenamefont {Sabbagh}, \citenamefont {Smith},
  \citenamefont {Soukhanovskii}, \citenamefont {Stevenson}, \citenamefont
  {Strykowsky}, \citenamefont {Stutman}, \citenamefont {Taylor}, \citenamefont
  {Titus}, \citenamefont {Tresemer}, \citenamefont {Tritz}, \citenamefont
  {Viola}, \citenamefont {Williams}, \citenamefont {Woolley}, \citenamefont
  {Yuh}, \citenamefont {Zhang}, \citenamefont {Zhai}, \citenamefont
  {Zolfaghari},\ and\ \citenamefont {{the NSTX Team}}}]{Menard2012NF}%
  \BibitemOpen
  \bibfield  {author} {\bibinfo {author} {\bibfnamefont {J.~E.}\ \bibnamefont
  {Menard}}, \bibinfo {author} {\bibfnamefont {S.}~\bibnamefont {Gerhardt}},
  \bibinfo {author} {\bibfnamefont {M.}~\bibnamefont {Bell}}, \bibinfo {author}
  {\bibfnamefont {J.}~\bibnamefont {Bialek}}, \bibinfo {author} {\bibfnamefont
  {A.}~\bibnamefont {Brooks}}, \bibinfo {author} {\bibfnamefont
  {J.}~\bibnamefont {Canik}}, \bibinfo {author} {\bibfnamefont
  {J.}~\bibnamefont {Chrzanowski}}, \bibinfo {author} {\bibfnamefont
  {M.}~\bibnamefont {Denault}}, \bibinfo {author} {\bibfnamefont
  {L.}~\bibnamefont {Dudek}}, \bibinfo {author} {\bibfnamefont {D.~A.}\
  \bibnamefont {Gates}}, \bibinfo {author} {\bibfnamefont {N.}~\bibnamefont
  {Gorelenkov}}, \bibinfo {author} {\bibfnamefont {W.}~\bibnamefont
  {Guttenfelder}}, \bibinfo {author} {\bibfnamefont {R.}~\bibnamefont
  {Hatcher}}, \bibinfo {author} {\bibfnamefont {J.}~\bibnamefont {Hosea}},
  \bibinfo {author} {\bibfnamefont {R.}~\bibnamefont {Kaita}}, \bibinfo
  {author} {\bibfnamefont {S.}~\bibnamefont {Kaye}}, \bibinfo {author}
  {\bibfnamefont {C.}~\bibnamefont {Kessel}}, \bibinfo {author} {\bibfnamefont
  {E.}~\bibnamefont {Kolemen}}, \bibinfo {author} {\bibfnamefont
  {H.}~\bibnamefont {Kugel}}, \bibinfo {author} {\bibfnamefont
  {R.}~\bibnamefont {Maingi}}, \bibinfo {author} {\bibfnamefont
  {M.}~\bibnamefont {Mardenfeld}}, \bibinfo {author} {\bibfnamefont
  {D.}~\bibnamefont {Mueller}}, \bibinfo {author} {\bibfnamefont
  {B.}~\bibnamefont {Nelson}}, \bibinfo {author} {\bibfnamefont
  {C.}~\bibnamefont {Neumeyer}}, \bibinfo {author} {\bibfnamefont
  {M.}~\bibnamefont {Ono}}, \bibinfo {author} {\bibfnamefont {E.}~\bibnamefont
  {Perry}}, \bibinfo {author} {\bibfnamefont {R.}~\bibnamefont {Ramakrishnan}},
  \bibinfo {author} {\bibfnamefont {R.}~\bibnamefont {Raman}}, \bibinfo
  {author} {\bibfnamefont {Y.}~\bibnamefont {Ren}}, \bibinfo {author}
  {\bibfnamefont {S.}~\bibnamefont {Sabbagh}}, \bibinfo {author} {\bibfnamefont
  {M.}~\bibnamefont {Smith}}, \bibinfo {author} {\bibfnamefont
  {V.}~\bibnamefont {Soukhanovskii}}, \bibinfo {author} {\bibfnamefont
  {T.}~\bibnamefont {Stevenson}}, \bibinfo {author} {\bibfnamefont
  {R.}~\bibnamefont {Strykowsky}}, \bibinfo {author} {\bibfnamefont
  {D.}~\bibnamefont {Stutman}}, \bibinfo {author} {\bibfnamefont
  {G.}~\bibnamefont {Taylor}}, \bibinfo {author} {\bibfnamefont
  {P.}~\bibnamefont {Titus}}, \bibinfo {author} {\bibfnamefont
  {K.}~\bibnamefont {Tresemer}}, \bibinfo {author} {\bibfnamefont
  {K.}~\bibnamefont {Tritz}}, \bibinfo {author} {\bibfnamefont
  {M.}~\bibnamefont {Viola}}, \bibinfo {author} {\bibfnamefont
  {M.}~\bibnamefont {Williams}}, \bibinfo {author} {\bibfnamefont
  {R.}~\bibnamefont {Woolley}}, \bibinfo {author} {\bibfnamefont
  {H.}~\bibnamefont {Yuh}}, \bibinfo {author} {\bibfnamefont {H.}~\bibnamefont
  {Zhang}}, \bibinfo {author} {\bibfnamefont {Y.}~\bibnamefont {Zhai}},
  \bibinfo {author} {\bibfnamefont {A.}~\bibnamefont {Zolfaghari}},\ and\
  \bibinfo {author} {\bibnamefont {{the NSTX Team}}},\ }\bibfield  {title}
  {\bibinfo {title} {Overview of the physics and engineering design of {{NSTX}
  upgrade}},\ }\href {http://stacks.iop.org/0029-5515/52/i=8/a=083015}
  {\bibfield  {journal} {\bibinfo  {journal} {Nuclear Fusion}\ }\textbf
  {\bibinfo {volume} {52}},\ \bibinfo {pages} {083015} (\bibinfo {year}
  {2012})}\BibitemShut {NoStop}%
\bibitem [{\citenamefont {Menard}\ \emph {et~al.}(2017)\citenamefont {Menard},
  \citenamefont {Allain}, \citenamefont {Battaglia}, \citenamefont {Bedoya},
  \citenamefont {Bell}, \citenamefont {Belova}, \citenamefont {Berkery},
  \citenamefont {Boyer}, \citenamefont {Crocker}, \citenamefont {Diallo},
  \citenamefont {Ebrahimi}, \citenamefont {Ferraro}, \citenamefont
  {Fredrickson}, \citenamefont {Frerichs}, \citenamefont {Gerhardt},
  \citenamefont {Gorelenkov}, \citenamefont {Guttenfelder}, \citenamefont
  {Heidbrink}, \citenamefont {Kaita}, \citenamefont {Kaye}, \citenamefont
  {Kriete}, \citenamefont {Kubota}, \citenamefont {LeBlanc}, \citenamefont
  {Liu}, \citenamefont {Lunsford}, \citenamefont {Mueller}, \citenamefont
  {Myers}, \citenamefont {Ono}, \citenamefont {Park}, \citenamefont {\Podesta},
  \citenamefont {Raman}, \citenamefont {Reinke}, \citenamefont {Ren},
  \citenamefont {Sabbagh}, \citenamefont {Schmitz}, \citenamefont {Scotti},
  \citenamefont {Sechrest}, \citenamefont {Skinner}, \citenamefont {Smith},
  \citenamefont {Soukhanovskii}, \citenamefont {Stoltzfus-Dueck}, \citenamefont
  {Yuh}, \citenamefont {Wang}, \citenamefont {Waters}, \citenamefont {Ahn},
  \citenamefont {Andre}, \citenamefont {Barchfeld}, \citenamefont
  {Beiersdorfer}, \citenamefont {Bertelli}, \citenamefont {Bhattacharjee},
  \citenamefont {Brennan}, \citenamefont {Buttery}, \citenamefont {Capece},
  \citenamefont {Canal}, \citenamefont {Canik}, \citenamefont {Chang},
  \citenamefont {Darrow}, \citenamefont {Delgado-Aparicio}, \citenamefont
  {Domier}, \citenamefont {Ethier}, \citenamefont {Evans}, \citenamefont
  {Ferron}, \citenamefont {Finkenthal}, \citenamefont {Fonck}, \citenamefont
  {Gan}, \citenamefont {Gates}, \citenamefont {Goumiri}, \citenamefont {Gray},
  \citenamefont {Hosea}, \citenamefont {Humphreys}, \citenamefont {Jarboe},
  \citenamefont {Jardin}, \citenamefont {Jaworski}, \citenamefont {Koel},
  \citenamefont {Kolemen}, \citenamefont {Ku}, \citenamefont {Haye},
  \citenamefont {Levinton}, \citenamefont {Luhmann}, \citenamefont {Maingi},
  \citenamefont {Maqueda}, \citenamefont {McKee}, \citenamefont {Meier},
  \citenamefont {Myra}, \citenamefont {Perkins}, \citenamefont {Poli},
  \citenamefont {Rhodes}, \citenamefont {Riquezes}, \citenamefont {Rowley},
  \citenamefont {Russell}, \citenamefont {Schuster}, \citenamefont {Stratton},
  \citenamefont {Stutman}, \citenamefont {Taylor}, \citenamefont {Tritz},
  \citenamefont {Wang}, \citenamefont {Wirth}, \citenamefont {Zweben},\ and\
  \citenamefont {the {NSTX-U}~Team}}]{Menard2017NF}%
  \BibitemOpen
  \bibfield  {author} {\bibinfo {author} {\bibfnamefont {J.~E.}\ \bibnamefont
  {Menard}}, \bibinfo {author} {\bibfnamefont {J.~P.}\ \bibnamefont {Allain}},
  \bibinfo {author} {\bibfnamefont {D.~J.}\ \bibnamefont {Battaglia}}, \bibinfo
  {author} {\bibfnamefont {F.}~\bibnamefont {Bedoya}}, \bibinfo {author}
  {\bibfnamefont {R.~E.}\ \bibnamefont {Bell}}, \bibinfo {author}
  {\bibfnamefont {E.}~\bibnamefont {Belova}}, \bibinfo {author} {\bibfnamefont
  {J.~W.}\ \bibnamefont {Berkery}}, \bibinfo {author} {\bibfnamefont {M.~D.}\
  \bibnamefont {Boyer}}, \bibinfo {author} {\bibfnamefont {N.}~\bibnamefont
  {Crocker}}, \bibinfo {author} {\bibfnamefont {A.}~\bibnamefont {Diallo}},
  \bibinfo {author} {\bibfnamefont {F.}~\bibnamefont {Ebrahimi}}, \bibinfo
  {author} {\bibfnamefont {N.}~\bibnamefont {Ferraro}}, \bibinfo {author}
  {\bibfnamefont {E.}~\bibnamefont {Fredrickson}}, \bibinfo {author}
  {\bibfnamefont {H.}~\bibnamefont {Frerichs}}, \bibinfo {author}
  {\bibfnamefont {S.}~\bibnamefont {Gerhardt}}, \bibinfo {author}
  {\bibfnamefont {N.}~\bibnamefont {Gorelenkov}}, \bibinfo {author}
  {\bibfnamefont {W.}~\bibnamefont {Guttenfelder}}, \bibinfo {author}
  {\bibfnamefont {W.}~\bibnamefont {Heidbrink}}, \bibinfo {author}
  {\bibfnamefont {R.}~\bibnamefont {Kaita}}, \bibinfo {author} {\bibfnamefont
  {S.~M.}\ \bibnamefont {Kaye}}, \bibinfo {author} {\bibfnamefont {D.~M.}\
  \bibnamefont {Kriete}}, \bibinfo {author} {\bibfnamefont {S.}~\bibnamefont
  {Kubota}}, \bibinfo {author} {\bibfnamefont {B.~P.}\ \bibnamefont {LeBlanc}},
  \bibinfo {author} {\bibfnamefont {D.}~\bibnamefont {Liu}}, \bibinfo {author}
  {\bibfnamefont {R.}~\bibnamefont {Lunsford}}, \bibinfo {author}
  {\bibfnamefont {D.}~\bibnamefont {Mueller}}, \bibinfo {author} {\bibfnamefont
  {C.~E.}\ \bibnamefont {Myers}}, \bibinfo {author} {\bibfnamefont
  {M.}~\bibnamefont {Ono}}, \bibinfo {author} {\bibfnamefont {J.-K.}\
  \bibnamefont {Park}}, \bibinfo {author} {\bibfnamefont {M.}~\bibnamefont
  {\Podesta}}, \bibinfo {author} {\bibfnamefont {R.}~\bibnamefont {Raman}},
  \bibinfo {author} {\bibfnamefont {M.}~\bibnamefont {Reinke}}, \bibinfo
  {author} {\bibfnamefont {Y.}~\bibnamefont {Ren}}, \bibinfo {author}
  {\bibfnamefont {S.~A.}\ \bibnamefont {Sabbagh}}, \bibinfo {author}
  {\bibfnamefont {O.}~\bibnamefont {Schmitz}}, \bibinfo {author} {\bibfnamefont
  {F.}~\bibnamefont {Scotti}}, \bibinfo {author} {\bibfnamefont
  {Y.}~\bibnamefont {Sechrest}}, \bibinfo {author} {\bibfnamefont {C.~H.}\
  \bibnamefont {Skinner}}, \bibinfo {author} {\bibfnamefont {D.~R.}\
  \bibnamefont {Smith}}, \bibinfo {author} {\bibfnamefont {V.}~\bibnamefont
  {Soukhanovskii}}, \bibinfo {author} {\bibfnamefont {T.}~\bibnamefont
  {Stoltzfus-Dueck}}, \bibinfo {author} {\bibfnamefont {H.}~\bibnamefont
  {Yuh}}, \bibinfo {author} {\bibfnamefont {Z.}~\bibnamefont {Wang}}, \bibinfo
  {author} {\bibfnamefont {I.}~\bibnamefont {Waters}}, \bibinfo {author}
  {\bibfnamefont {J.-W.}\ \bibnamefont {Ahn}}, \bibinfo {author} {\bibfnamefont
  {R.}~\bibnamefont {Andre}}, \bibinfo {author} {\bibfnamefont
  {R.}~\bibnamefont {Barchfeld}}, \bibinfo {author} {\bibfnamefont
  {P.}~\bibnamefont {Beiersdorfer}}, \bibinfo {author} {\bibfnamefont
  {N.}~\bibnamefont {Bertelli}}, \bibinfo {author} {\bibfnamefont
  {A.}~\bibnamefont {Bhattacharjee}}, \bibinfo {author} {\bibfnamefont
  {D.}~\bibnamefont {Brennan}}, \bibinfo {author} {\bibfnamefont
  {R.}~\bibnamefont {Buttery}}, \bibinfo {author} {\bibfnamefont
  {A.}~\bibnamefont {Capece}}, \bibinfo {author} {\bibfnamefont
  {G.}~\bibnamefont {Canal}}, \bibinfo {author} {\bibfnamefont
  {J.}~\bibnamefont {Canik}}, \bibinfo {author} {\bibfnamefont {C.~S.}\
  \bibnamefont {Chang}}, \bibinfo {author} {\bibfnamefont {D.}~\bibnamefont
  {Darrow}}, \bibinfo {author} {\bibfnamefont {L.}~\bibnamefont
  {Delgado-Aparicio}}, \bibinfo {author} {\bibfnamefont {C.}~\bibnamefont
  {Domier}}, \bibinfo {author} {\bibfnamefont {S.}~\bibnamefont {Ethier}},
  \bibinfo {author} {\bibfnamefont {T.}~\bibnamefont {Evans}}, \bibinfo
  {author} {\bibfnamefont {J.}~\bibnamefont {Ferron}}, \bibinfo {author}
  {\bibfnamefont {M.}~\bibnamefont {Finkenthal}}, \bibinfo {author}
  {\bibfnamefont {R.}~\bibnamefont {Fonck}}, \bibinfo {author} {\bibfnamefont
  {K.}~\bibnamefont {Gan}}, \bibinfo {author} {\bibfnamefont {D.}~\bibnamefont
  {Gates}}, \bibinfo {author} {\bibfnamefont {I.}~\bibnamefont {Goumiri}},
  \bibinfo {author} {\bibfnamefont {T.}~\bibnamefont {Gray}}, \bibinfo {author}
  {\bibfnamefont {J.}~\bibnamefont {Hosea}}, \bibinfo {author} {\bibfnamefont
  {D.}~\bibnamefont {Humphreys}}, \bibinfo {author} {\bibfnamefont
  {T.}~\bibnamefont {Jarboe}}, \bibinfo {author} {\bibfnamefont
  {S.}~\bibnamefont {Jardin}}, \bibinfo {author} {\bibfnamefont {M.~A.}\
  \bibnamefont {Jaworski}}, \bibinfo {author} {\bibfnamefont {B.}~\bibnamefont
  {Koel}}, \bibinfo {author} {\bibfnamefont {E.}~\bibnamefont {Kolemen}},
  \bibinfo {author} {\bibfnamefont {S.}~\bibnamefont {Ku}}, \bibinfo {author}
  {\bibfnamefont {R.~J.~L.}\ \bibnamefont {Haye}}, \bibinfo {author}
  {\bibfnamefont {F.}~\bibnamefont {Levinton}}, \bibinfo {author}
  {\bibfnamefont {N.}~\bibnamefont {Luhmann}}, \bibinfo {author} {\bibfnamefont
  {R.}~\bibnamefont {Maingi}}, \bibinfo {author} {\bibfnamefont
  {R.}~\bibnamefont {Maqueda}}, \bibinfo {author} {\bibfnamefont
  {G.}~\bibnamefont {McKee}}, \bibinfo {author} {\bibfnamefont
  {E.}~\bibnamefont {Meier}}, \bibinfo {author} {\bibfnamefont
  {J.}~\bibnamefont {Myra}}, \bibinfo {author} {\bibfnamefont {R.}~\bibnamefont
  {Perkins}}, \bibinfo {author} {\bibfnamefont {F.}~\bibnamefont {Poli}},
  \bibinfo {author} {\bibfnamefont {T.}~\bibnamefont {Rhodes}}, \bibinfo
  {author} {\bibfnamefont {J.}~\bibnamefont {Riquezes}}, \bibinfo {author}
  {\bibfnamefont {C.}~\bibnamefont {Rowley}}, \bibinfo {author} {\bibfnamefont
  {D.}~\bibnamefont {Russell}}, \bibinfo {author} {\bibfnamefont
  {E.}~\bibnamefont {Schuster}}, \bibinfo {author} {\bibfnamefont
  {B.}~\bibnamefont {Stratton}}, \bibinfo {author} {\bibfnamefont
  {D.}~\bibnamefont {Stutman}}, \bibinfo {author} {\bibfnamefont
  {G.}~\bibnamefont {Taylor}}, \bibinfo {author} {\bibfnamefont
  {K.}~\bibnamefont {Tritz}}, \bibinfo {author} {\bibfnamefont
  {W.}~\bibnamefont {Wang}}, \bibinfo {author} {\bibfnamefont {B.}~\bibnamefont
  {Wirth}}, \bibinfo {author} {\bibfnamefont {S.~J.}\ \bibnamefont {Zweben}},\
  and\ \bibinfo {author} {\bibnamefont {the {NSTX-U}~Team}},\ }\bibfield
  {title} {\bibinfo {title} {Overview of {NSTX} {Upgrade} initial results and
  modelling highlights},\ }\href
  {http://stacks.iop.org/0029-5515/57/i=10/a=102006} {\bibfield  {journal}
  {\bibinfo  {journal} {Nuclear Fusion}\ }\textbf {\bibinfo {volume} {57}},\
  \bibinfo {pages} {102006} (\bibinfo {year} {2017})}\BibitemShut {NoStop}%
\bibitem [{\citenamefont {Berkery}\ \emph {et~al.}(2024)\citenamefont
  {Berkery}, \citenamefont {Adebayo-Ige}, \citenamefont {Khawaldeh},
  \citenamefont {Avdeeva}, \citenamefont {Baek}, \citenamefont {Banerjee},
  \citenamefont {Barada}, \citenamefont {Battaglia}, \citenamefont {Bell},
  \citenamefont {Belli}, \citenamefont {Belova}, \citenamefont {Bertelli},
  \citenamefont {Bisai}, \citenamefont {Bonoli}, \citenamefont {Boyer},
  \citenamefont {Butt}, \citenamefont {Candy}, \citenamefont {Chang},
  \citenamefont {Clauser}, \citenamefont {Rivera}, \citenamefont {Curie},
  \citenamefont {de~Vries}, \citenamefont {Diab}, \citenamefont {Diallo},
  \citenamefont {Dominski}, \citenamefont {Duarte}, \citenamefont {Emdee},
  \citenamefont {Ferraro}, \citenamefont {Fitzpatrick}, \citenamefont {Foley},
  \citenamefont {Fredrickson}, \citenamefont {Galante}, \citenamefont {Gan},
  \citenamefont {Gerhardt}, \citenamefont {Goldston}, \citenamefont
  {Guttenfelder}, \citenamefont {Hager}, \citenamefont {Hanson}, \citenamefont
  {Jardin}, \citenamefont {Jenkins}, \citenamefont {Kaye}, \citenamefont
  {Khodak}, \citenamefont {Kinsey}, \citenamefont {Kleiner}, \citenamefont
  {Kolemen}, \citenamefont {Ku}, \citenamefont {Lampert}, \citenamefont
  {Leard}, \citenamefont {LeBlanc}, \citenamefont {Lestz}, \citenamefont
  {Levinton}, \citenamefont {Liu}, \citenamefont {Looby}, \citenamefont
  {Lunsford}, \citenamefont {Macwan}, \citenamefont {Maingi}, \citenamefont
  {McClenaghan}, \citenamefont {Menard}, \citenamefont {Munaretto},
  \citenamefont {Ono}, \citenamefont {Pajares}, \citenamefont {Parisi},
  \citenamefont {Park}, \citenamefont {Parsons}, \citenamefont {Patel},
  \citenamefont {Petrov}, \citenamefont {Podest\'{a}}, \citenamefont {Poli},
  \citenamefont {Porcelli}, \citenamefont {Rafiq}, \citenamefont {Sabbagh},
  \citenamefont {Villar}, \citenamefont {Schuster}, \citenamefont {Schwartz},
  \citenamefont {Sharma}, \citenamefont {Shiraiwa}, \citenamefont {Sinha},
  \citenamefont {Smith}, \citenamefont {Smith}, \citenamefont {Soukhanovskii},
  \citenamefont {Staebler}, \citenamefont {Startsev}, \citenamefont {Stratton},
  \citenamefont {Thome}, \citenamefont {Tierens}, \citenamefont {Tobin},
  \citenamefont {Uzun-Kaymak}, \citenamefont {Compernolle}, \citenamefont
  {Wai}, \citenamefont {Wang}, \citenamefont {Wehner}, \citenamefont
  {Welander}, \citenamefont {Yang}, \citenamefont {Zamkovska}, \citenamefont
  {Zhang}, \citenamefont {Zhu},\ and\ \citenamefont {Zweben}}]{Berkery2024NF}%
  \BibitemOpen
  \bibfield  {author} {\bibinfo {author} {\bibfnamefont {J.}~\bibnamefont
  {Berkery}}, \bibinfo {author} {\bibfnamefont {P.}~\bibnamefont
  {Adebayo-Ige}}, \bibinfo {author} {\bibfnamefont {H.~A.}\ \bibnamefont
  {Khawaldeh}}, \bibinfo {author} {\bibfnamefont {G.}~\bibnamefont {Avdeeva}},
  \bibinfo {author} {\bibfnamefont {S.-G.}\ \bibnamefont {Baek}}, \bibinfo
  {author} {\bibfnamefont {S.}~\bibnamefont {Banerjee}}, \bibinfo {author}
  {\bibfnamefont {K.}~\bibnamefont {Barada}}, \bibinfo {author} {\bibfnamefont
  {D.}~\bibnamefont {Battaglia}}, \bibinfo {author} {\bibfnamefont
  {R.}~\bibnamefont {Bell}}, \bibinfo {author} {\bibfnamefont {E.}~\bibnamefont
  {Belli}}, \bibinfo {author} {\bibfnamefont {E.}~\bibnamefont {Belova}},
  \bibinfo {author} {\bibfnamefont {N.}~\bibnamefont {Bertelli}}, \bibinfo
  {author} {\bibfnamefont {N.}~\bibnamefont {Bisai}}, \bibinfo {author}
  {\bibfnamefont {P.}~\bibnamefont {Bonoli}}, \bibinfo {author} {\bibfnamefont
  {M.}~\bibnamefont {Boyer}}, \bibinfo {author} {\bibfnamefont
  {J.}~\bibnamefont {Butt}}, \bibinfo {author} {\bibfnamefont {J.}~\bibnamefont
  {Candy}}, \bibinfo {author} {\bibfnamefont {C.}~\bibnamefont {Chang}},
  \bibinfo {author} {\bibfnamefont {C.}~\bibnamefont {Clauser}}, \bibinfo
  {author} {\bibfnamefont {L.~C.}\ \bibnamefont {Rivera}}, \bibinfo {author}
  {\bibfnamefont {M.}~\bibnamefont {Curie}}, \bibinfo {author} {\bibfnamefont
  {P.}~\bibnamefont {de~Vries}}, \bibinfo {author} {\bibfnamefont
  {R.}~\bibnamefont {Diab}}, \bibinfo {author} {\bibfnamefont {A.}~\bibnamefont
  {Diallo}}, \bibinfo {author} {\bibfnamefont {J.}~\bibnamefont {Dominski}},
  \bibinfo {author} {\bibfnamefont {V.}~\bibnamefont {Duarte}}, \bibinfo
  {author} {\bibfnamefont {E.}~\bibnamefont {Emdee}}, \bibinfo {author}
  {\bibfnamefont {N.}~\bibnamefont {Ferraro}}, \bibinfo {author} {\bibfnamefont
  {R.}~\bibnamefont {Fitzpatrick}}, \bibinfo {author} {\bibfnamefont
  {E.}~\bibnamefont {Foley}}, \bibinfo {author} {\bibfnamefont
  {E.}~\bibnamefont {Fredrickson}}, \bibinfo {author} {\bibfnamefont
  {M.}~\bibnamefont {Galante}}, \bibinfo {author} {\bibfnamefont
  {K.}~\bibnamefont {Gan}}, \bibinfo {author} {\bibfnamefont {S.}~\bibnamefont
  {Gerhardt}}, \bibinfo {author} {\bibfnamefont {R.}~\bibnamefont {Goldston}},
  \bibinfo {author} {\bibfnamefont {W.}~\bibnamefont {Guttenfelder}}, \bibinfo
  {author} {\bibfnamefont {R.}~\bibnamefont {Hager}}, \bibinfo {author}
  {\bibfnamefont {M.}~\bibnamefont {Hanson}}, \bibinfo {author} {\bibfnamefont
  {S.}~\bibnamefont {Jardin}}, \bibinfo {author} {\bibfnamefont
  {T.}~\bibnamefont {Jenkins}}, \bibinfo {author} {\bibfnamefont
  {S.}~\bibnamefont {Kaye}}, \bibinfo {author} {\bibfnamefont {A.}~\bibnamefont
  {Khodak}}, \bibinfo {author} {\bibfnamefont {J.}~\bibnamefont {Kinsey}},
  \bibinfo {author} {\bibfnamefont {A.}~\bibnamefont {Kleiner}}, \bibinfo
  {author} {\bibfnamefont {E.}~\bibnamefont {Kolemen}}, \bibinfo {author}
  {\bibfnamefont {S.}~\bibnamefont {Ku}}, \bibinfo {author} {\bibfnamefont
  {M.}~\bibnamefont {Lampert}}, \bibinfo {author} {\bibfnamefont
  {B.}~\bibnamefont {Leard}}, \bibinfo {author} {\bibfnamefont
  {B.}~\bibnamefont {LeBlanc}}, \bibinfo {author} {\bibfnamefont
  {J.}~\bibnamefont {Lestz}}, \bibinfo {author} {\bibfnamefont
  {F.}~\bibnamefont {Levinton}}, \bibinfo {author} {\bibfnamefont
  {C.}~\bibnamefont {Liu}}, \bibinfo {author} {\bibfnamefont {T.}~\bibnamefont
  {Looby}}, \bibinfo {author} {\bibfnamefont {R.}~\bibnamefont {Lunsford}},
  \bibinfo {author} {\bibfnamefont {T.}~\bibnamefont {Macwan}}, \bibinfo
  {author} {\bibfnamefont {R.}~\bibnamefont {Maingi}}, \bibinfo {author}
  {\bibfnamefont {J.}~\bibnamefont {McClenaghan}}, \bibinfo {author}
  {\bibfnamefont {J.}~\bibnamefont {Menard}}, \bibinfo {author} {\bibfnamefont
  {S.}~\bibnamefont {Munaretto}}, \bibinfo {author} {\bibfnamefont
  {M.}~\bibnamefont {Ono}}, \bibinfo {author} {\bibfnamefont {A.}~\bibnamefont
  {Pajares}}, \bibinfo {author} {\bibfnamefont {J.}~\bibnamefont {Parisi}},
  \bibinfo {author} {\bibfnamefont {J.-K.}\ \bibnamefont {Park}}, \bibinfo
  {author} {\bibfnamefont {M.}~\bibnamefont {Parsons}}, \bibinfo {author}
  {\bibfnamefont {B.}~\bibnamefont {Patel}}, \bibinfo {author} {\bibfnamefont
  {Y.}~\bibnamefont {Petrov}}, \bibinfo {author} {\bibfnamefont
  {M.}~\bibnamefont {Podest\'{a}}}, \bibinfo {author} {\bibfnamefont
  {F.}~\bibnamefont {Poli}}, \bibinfo {author} {\bibfnamefont {M.}~\bibnamefont
  {Porcelli}}, \bibinfo {author} {\bibfnamefont {T.}~\bibnamefont {Rafiq}},
  \bibinfo {author} {\bibfnamefont {S.}~\bibnamefont {Sabbagh}}, \bibinfo
  {author} {\bibfnamefont {A.~S.}\ \bibnamefont {Villar}}, \bibinfo {author}
  {\bibfnamefont {E.}~\bibnamefont {Schuster}}, \bibinfo {author}
  {\bibfnamefont {J.}~\bibnamefont {Schwartz}}, \bibinfo {author}
  {\bibfnamefont {A.}~\bibnamefont {Sharma}}, \bibinfo {author} {\bibfnamefont
  {S.}~\bibnamefont {Shiraiwa}}, \bibinfo {author} {\bibfnamefont
  {P.}~\bibnamefont {Sinha}}, \bibinfo {author} {\bibfnamefont
  {D.}~\bibnamefont {Smith}}, \bibinfo {author} {\bibfnamefont
  {S.}~\bibnamefont {Smith}}, \bibinfo {author} {\bibfnamefont
  {V.}~\bibnamefont {Soukhanovskii}}, \bibinfo {author} {\bibfnamefont
  {G.}~\bibnamefont {Staebler}}, \bibinfo {author} {\bibfnamefont
  {E.}~\bibnamefont {Startsev}}, \bibinfo {author} {\bibfnamefont
  {B.}~\bibnamefont {Stratton}}, \bibinfo {author} {\bibfnamefont
  {K.}~\bibnamefont {Thome}}, \bibinfo {author} {\bibfnamefont
  {W.}~\bibnamefont {Tierens}}, \bibinfo {author} {\bibfnamefont
  {M.}~\bibnamefont {Tobin}}, \bibinfo {author} {\bibfnamefont
  {I.}~\bibnamefont {Uzun-Kaymak}}, \bibinfo {author} {\bibfnamefont {B.~V.}\
  \bibnamefont {Compernolle}}, \bibinfo {author} {\bibfnamefont
  {J.}~\bibnamefont {Wai}}, \bibinfo {author} {\bibfnamefont {W.}~\bibnamefont
  {Wang}}, \bibinfo {author} {\bibfnamefont {W.}~\bibnamefont {Wehner}},
  \bibinfo {author} {\bibfnamefont {A.}~\bibnamefont {Welander}}, \bibinfo
  {author} {\bibfnamefont {J.}~\bibnamefont {Yang}}, \bibinfo {author}
  {\bibfnamefont {V.}~\bibnamefont {Zamkovska}}, \bibinfo {author}
  {\bibfnamefont {X.}~\bibnamefont {Zhang}}, \bibinfo {author} {\bibfnamefont
  {X.}~\bibnamefont {Zhu}},\ and\ \bibinfo {author} {\bibfnamefont
  {S.}~\bibnamefont {Zweben}},\ }\bibfield  {title} {\bibinfo {title} {{NSTX-U}
  research advancing the physics of spherical tokamaks},\ }\href
  {https://doi.org/10.1088/1741-4326/ad3092} {\bibfield  {journal} {\bibinfo
  {journal} {Nuclear Fusion}\ }\textbf {\bibinfo {volume} {64}},\ \bibinfo
  {pages} {112004} (\bibinfo {year} {2024})}\BibitemShut {NoStop}%
\bibitem [{\citenamefont {Buxton}\ \emph {et~al.}(2019)\citenamefont {Buxton},
  \citenamefont {Connor}, \citenamefont {Costley}, \citenamefont
  {Gryaznevich},\ and\ \citenamefont {McNamara}}]{Buxton2019PPCF}%
  \BibitemOpen
  \bibfield  {author} {\bibinfo {author} {\bibfnamefont {P.~F.}\ \bibnamefont
  {Buxton}}, \bibinfo {author} {\bibfnamefont {J.~W.}\ \bibnamefont {Connor}},
  \bibinfo {author} {\bibfnamefont {A.~E.}\ \bibnamefont {Costley}}, \bibinfo
  {author} {\bibfnamefont {M.~P.}\ \bibnamefont {Gryaznevich}},\ and\ \bibinfo
  {author} {\bibfnamefont {S.}~\bibnamefont {McNamara}},\ }\bibfield  {title}
  {\bibinfo {title} {On the energy confinement time in spherical tokamaks:
  implications for the design of pilot plants and fusion reactors},\ }\href
  {https://doi.org/10.1088/1361-6587/aaf7e5} {\bibfield  {journal} {\bibinfo
  {journal} {Plasma Physics and Controlled Fusion}\ }\textbf {\bibinfo {volume}
  {61}},\ \bibinfo {pages} {035006} (\bibinfo {year} {2019})}\BibitemShut
  {NoStop}%
\bibitem [{\citenamefont {Wilson}\ \emph {et~al.}(2020)\citenamefont {Wilson},
  \citenamefont {Chapman}, \citenamefont {Denton}, \citenamefont {Morris},
  \citenamefont {Patel}, \citenamefont {Voss}, \citenamefont {Waldon},\ and\
  \citenamefont {{the STEP Team}}}]{Wilson2020book}%
  \BibitemOpen
  \bibfield  {author} {\bibinfo {author} {\bibfnamefont {H.}~\bibnamefont
  {Wilson}}, \bibinfo {author} {\bibfnamefont {I.}~\bibnamefont {Chapman}},
  \bibinfo {author} {\bibfnamefont {T.}~\bibnamefont {Denton}}, \bibinfo
  {author} {\bibfnamefont {W.}~\bibnamefont {Morris}}, \bibinfo {author}
  {\bibfnamefont {B.}~\bibnamefont {Patel}}, \bibinfo {author} {\bibfnamefont
  {G.}~\bibnamefont {Voss}}, \bibinfo {author} {\bibfnamefont {C.}~\bibnamefont
  {Waldon}},\ and\ \bibinfo {author} {\bibnamefont {{the STEP Team}}},\
  }\bibfield  {title} {\bibinfo {title} {{STEP}—on the pathway to fusion
  commercialization},\ }in\ \href
  {https://doi.org/10.1088/978-0-7503-2719-0ch8} {\emph {\bibinfo {booktitle}
  {Commercialising Fusion Energy}}},\ \bibinfo {series and number} {2053-2563}\
  (\bibinfo  {publisher} {IOP Publishing},\ \bibinfo {year} {2020})\ pp.\
  \bibinfo {pages} {8--1 to 8--18}\BibitemShut {NoStop}%
\bibitem [{\citenamefont {Waldon}\ \emph {et~al.}(2024)\citenamefont {Waldon},
  \citenamefont {Muldrew}, \citenamefont {Keep}, \citenamefont {Verhoeven},
  \citenamefont {Thompson},\ and\ \citenamefont
  {Kisbey-Ascott}}]{Waldon2024PTRSA}%
  \BibitemOpen
  \bibfield  {author} {\bibinfo {author} {\bibfnamefont {C.}~\bibnamefont
  {Waldon}}, \bibinfo {author} {\bibfnamefont {S.~I.}\ \bibnamefont {Muldrew}},
  \bibinfo {author} {\bibfnamefont {J.}~\bibnamefont {Keep}}, \bibinfo {author}
  {\bibfnamefont {R.}~\bibnamefont {Verhoeven}}, \bibinfo {author}
  {\bibfnamefont {T.}~\bibnamefont {Thompson}},\ and\ \bibinfo {author}
  {\bibfnamefont {M.}~\bibnamefont {Kisbey-Ascott}},\ }\bibfield  {title}
  {\bibinfo {title} {Concept design overview: a question of choices and
  compromise},\ }\href {https://doi.org/10.1098/rsta.2023.0414} {\bibfield
  {journal} {\bibinfo  {journal} {Philosophical Transactions of the Royal
  Society A: Mathematical, Physical and Engineering Sciences}\ }\textbf
  {\bibinfo {volume} {382}},\ \bibinfo {pages} {20230414} (\bibinfo {year}
  {2024})}\BibitemShut {NoStop}%
\bibitem [{\citenamefont {Meyer}(2024)}]{Meyer2024PTRA}%
  \BibitemOpen
  \bibfield  {author} {\bibinfo {author} {\bibfnamefont {H.}~\bibnamefont
  {Meyer}},\ }\bibfield  {title} {\bibinfo {title} {Plasma burn—mind the
  gap},\ }\href {https://doi.org/10.1098/rsta.2023.0406} {\bibfield  {journal}
  {\bibinfo  {journal} {Philosophical Transactions of the Royal Society A:
  Mathematical, Physical and Engineering Sciences}\ }\textbf {\bibinfo {volume}
  {382}},\ \bibinfo {pages} {20230406} (\bibinfo {year} {2024})}\BibitemShut
  {NoStop}%
\bibitem [{\citenamefont {Kingham}\ and\ \citenamefont
  {Gryaznevich}(2024)}]{Kingham2024POP}%
  \BibitemOpen
  \bibfield  {author} {\bibinfo {author} {\bibfnamefont {D.}~\bibnamefont
  {Kingham}}\ and\ \bibinfo {author} {\bibfnamefont {M.}~\bibnamefont
  {Gryaznevich}},\ }\bibfield  {title} {\bibinfo {title} {The spherical tokamak
  path to fusion power: Opportunities and challenges for development via
  public–private partnerships},\ }\href {https://doi.org/10.1063/5.0170088}
  {\bibfield  {journal} {\bibinfo  {journal} {Physics of Plasmas}\ }\textbf
  {\bibinfo {volume} {31}},\ \bibinfo {pages} {042507} (\bibinfo {year}
  {2024})}\BibitemShut {NoStop}%
\bibitem [{\citenamefont {Ono}\ and\ \citenamefont {Kaita}(2015)}]{Ono2015POP}%
  \BibitemOpen
  \bibfield  {author} {\bibinfo {author} {\bibfnamefont {M.}~\bibnamefont
  {Ono}}\ and\ \bibinfo {author} {\bibfnamefont {R.}~\bibnamefont {Kaita}},\
  }\bibfield  {title} {\bibinfo {title} {Recent progress on spherical torus
  research},\ }\href {https://doi.org/10.1063/1.4915073} {\bibfield  {journal}
  {\bibinfo  {journal} {Physics of Plasmas}\ }\textbf {\bibinfo {volume}
  {22}},\ \bibinfo {pages} {040501} (\bibinfo {year} {2015})}\BibitemShut
  {NoStop}%
\bibitem [{\citenamefont {Kaye}\ \emph {et~al.}(2007)\citenamefont {Kaye},
  \citenamefont {Levinton}, \citenamefont {Stutman}, \citenamefont {Tritz},
  \citenamefont {Yuh}, \citenamefont {Bell}, \citenamefont {Bell},
  \citenamefont {Domier}, \citenamefont {Gates}, \citenamefont {Horton},
  \citenamefont {Kim}, \citenamefont {LeBlanc}, \citenamefont {Luhmann},
  \citenamefont {Maingi}, \citenamefont {Mazzucato}, \citenamefont {Menard},
  \citenamefont {Mikkelsen}, \citenamefont {Mueller}, \citenamefont {Park},
  \citenamefont {Rewoldt}, \citenamefont {Sabbagh}, \citenamefont {Smith},\
  and\ \citenamefont {Wang}}]{Kaye2007NF}%
  \BibitemOpen
  \bibfield  {author} {\bibinfo {author} {\bibfnamefont {S.~M.}\ \bibnamefont
  {Kaye}}, \bibinfo {author} {\bibfnamefont {F.~M.}\ \bibnamefont {Levinton}},
  \bibinfo {author} {\bibfnamefont {D.}~\bibnamefont {Stutman}}, \bibinfo
  {author} {\bibfnamefont {K.}~\bibnamefont {Tritz}}, \bibinfo {author}
  {\bibfnamefont {H.}~\bibnamefont {Yuh}}, \bibinfo {author} {\bibfnamefont
  {M.~G.}\ \bibnamefont {Bell}}, \bibinfo {author} {\bibfnamefont {R.~E.}\
  \bibnamefont {Bell}}, \bibinfo {author} {\bibfnamefont {C.~W.}\ \bibnamefont
  {Domier}}, \bibinfo {author} {\bibfnamefont {D.}~\bibnamefont {Gates}},
  \bibinfo {author} {\bibfnamefont {W.}~\bibnamefont {Horton}}, \bibinfo
  {author} {\bibfnamefont {J.}~\bibnamefont {Kim}}, \bibinfo {author}
  {\bibfnamefont {B.~P.}\ \bibnamefont {LeBlanc}}, \bibinfo {author}
  {\bibfnamefont {N.~C.}\ \bibnamefont {Luhmann}}, \bibinfo {author}
  {\bibfnamefont {R.}~\bibnamefont {Maingi}}, \bibinfo {author} {\bibfnamefont
  {E.}~\bibnamefont {Mazzucato}}, \bibinfo {author} {\bibfnamefont {J.~E.}\
  \bibnamefont {Menard}}, \bibinfo {author} {\bibfnamefont {D.}~\bibnamefont
  {Mikkelsen}}, \bibinfo {author} {\bibfnamefont {D.}~\bibnamefont {Mueller}},
  \bibinfo {author} {\bibfnamefont {H.}~\bibnamefont {Park}}, \bibinfo {author}
  {\bibfnamefont {G.}~\bibnamefont {Rewoldt}}, \bibinfo {author} {\bibfnamefont
  {S.~A.}\ \bibnamefont {Sabbagh}}, \bibinfo {author} {\bibfnamefont {D.~R.}\
  \bibnamefont {Smith}},\ and\ \bibinfo {author} {\bibfnamefont
  {W.}~\bibnamefont {Wang}},\ }\bibfield  {title} {\bibinfo {title}
  {Confinement and local transport in the {National Spherical Torus Experiment
  ({NSTX})}},\ }\href {https://doi.org/10.1088/0029-5515/47/7/001} {\bibfield
  {journal} {\bibinfo  {journal} {Nuclear Fusion}\ }\textbf {\bibinfo {volume}
  {47}},\ \bibinfo {pages} {499} (\bibinfo {year} {2007})}\BibitemShut
  {NoStop}%
\bibitem [{\citenamefont {Valovi{\v{c}}}\ \emph {et~al.}(2009)\citenamefont
  {Valovi{\v{c}}}, \citenamefont {Akers}, \citenamefont {Cunningham},
  \citenamefont {Garzotti}, \citenamefont {Lloyd}, \citenamefont {Muir},
  \citenamefont {Patel}, \citenamefont {Taylor}, \citenamefont {Turnyanskiy},
  \citenamefont {Walsh},\ and\ \citenamefont {{the MAST
  team}}}]{Valovic2009NF}%
  \BibitemOpen
  \bibfield  {author} {\bibinfo {author} {\bibfnamefont {M.}~\bibnamefont
  {Valovi{\v{c}}}}, \bibinfo {author} {\bibfnamefont {R.}~\bibnamefont
  {Akers}}, \bibinfo {author} {\bibfnamefont {G.}~\bibnamefont {Cunningham}},
  \bibinfo {author} {\bibfnamefont {L.}~\bibnamefont {Garzotti}}, \bibinfo
  {author} {\bibfnamefont {B.}~\bibnamefont {Lloyd}}, \bibinfo {author}
  {\bibfnamefont {D.}~\bibnamefont {Muir}}, \bibinfo {author} {\bibfnamefont
  {A.}~\bibnamefont {Patel}}, \bibinfo {author} {\bibfnamefont
  {D.}~\bibnamefont {Taylor}}, \bibinfo {author} {\bibfnamefont
  {M.}~\bibnamefont {Turnyanskiy}}, \bibinfo {author} {\bibfnamefont
  {M.}~\bibnamefont {Walsh}},\ and\ \bibinfo {author} {\bibnamefont {{the MAST
  team}}},\ }\bibfield  {title} {\bibinfo {title} {Scaling of {H-mode} energy
  confinement with {Ip} and {BT} in the {MAST} spherical tokamak},\ }\href
  {https://doi.org/10.1088/0029-5515/49/7/075016} {\bibfield  {journal}
  {\bibinfo  {journal} {Nuclear Fusion}\ }\textbf {\bibinfo {volume} {49}},\
  \bibinfo {pages} {075016} (\bibinfo {year} {2009})}\BibitemShut {NoStop}%
\bibitem [{\citenamefont {Valovi{\v{c}}}\ \emph {et~al.}(2011)\citenamefont
  {Valovi{\v{c}}}, \citenamefont {Akers}, \citenamefont {de~Bock},
  \citenamefont {McCone}, \citenamefont {Garzotti}, \citenamefont {Michael},
  \citenamefont {Naylor}, \citenamefont {Patel}, \citenamefont {Roach},
  \citenamefont {Scannell}, \citenamefont {Turnyanskiy}, \citenamefont {Wisse},
  \citenamefont {Guttenfelder}, \citenamefont {Candy},\ and\ \citenamefont
  {{the MAST team}}}]{Valovic2011NF}%
  \BibitemOpen
  \bibfield  {author} {\bibinfo {author} {\bibfnamefont {M.}~\bibnamefont
  {Valovi{\v{c}}}}, \bibinfo {author} {\bibfnamefont {R.}~\bibnamefont
  {Akers}}, \bibinfo {author} {\bibfnamefont {M.}~\bibnamefont {de~Bock}},
  \bibinfo {author} {\bibfnamefont {J.}~\bibnamefont {McCone}}, \bibinfo
  {author} {\bibfnamefont {L.}~\bibnamefont {Garzotti}}, \bibinfo {author}
  {\bibfnamefont {C.}~\bibnamefont {Michael}}, \bibinfo {author} {\bibfnamefont
  {G.}~\bibnamefont {Naylor}}, \bibinfo {author} {\bibfnamefont
  {A.}~\bibnamefont {Patel}}, \bibinfo {author} {\bibfnamefont {C.~M.}\
  \bibnamefont {Roach}}, \bibinfo {author} {\bibfnamefont {R.}~\bibnamefont
  {Scannell}}, \bibinfo {author} {\bibfnamefont {M.}~\bibnamefont
  {Turnyanskiy}}, \bibinfo {author} {\bibfnamefont {M.}~\bibnamefont {Wisse}},
  \bibinfo {author} {\bibfnamefont {W.}~\bibnamefont {Guttenfelder}}, \bibinfo
  {author} {\bibfnamefont {J.}~\bibnamefont {Candy}},\ and\ \bibinfo {author}
  {\bibnamefont {{the MAST team}}},\ }\bibfield  {title} {\bibinfo {title}
  {Collisionality and safety factor scalings of {H-mode} energy transport in
  the {MAST} spherical tokamak},\ }\href
  {https://doi.org/10.1088/0029-5515/51/7/073045} {\bibfield  {journal}
  {\bibinfo  {journal} {Nuclear Fusion}\ }\textbf {\bibinfo {volume} {51}},\
  \bibinfo {pages} {073045} (\bibinfo {year} {2011})}\BibitemShut {NoStop}%
\bibitem [{\citenamefont {Kaye}\ \emph {et~al.}(2013)\citenamefont {Kaye},
  \citenamefont {Gerhardt}, \citenamefont {Guttenfelder}, \citenamefont
  {Maingi}, \citenamefont {Bell}, \citenamefont {Diallo}, \citenamefont
  {LeBlanc},\ and\ \citenamefont {Podesta}}]{Kaye2013NF}%
  \BibitemOpen
  \bibfield  {author} {\bibinfo {author} {\bibfnamefont {S.}~\bibnamefont
  {Kaye}}, \bibinfo {author} {\bibfnamefont {S.}~\bibnamefont {Gerhardt}},
  \bibinfo {author} {\bibfnamefont {W.}~\bibnamefont {Guttenfelder}}, \bibinfo
  {author} {\bibfnamefont {R.}~\bibnamefont {Maingi}}, \bibinfo {author}
  {\bibfnamefont {R.}~\bibnamefont {Bell}}, \bibinfo {author} {\bibfnamefont
  {A.}~\bibnamefont {Diallo}}, \bibinfo {author} {\bibfnamefont
  {B.}~\bibnamefont {LeBlanc}},\ and\ \bibinfo {author} {\bibfnamefont
  {M.}~\bibnamefont {Podesta}},\ }\bibfield  {title} {\bibinfo {title} {The
  dependence of {H-mode} energy confinement and transport on collisionality in
  {NSTX}},\ }\href {https://doi.org/10.1088/0029-5515/53/6/063005} {\bibfield
  {journal} {\bibinfo  {journal} {Nuclear Fusion}\ }\textbf {\bibinfo {volume}
  {53}},\ \bibinfo {pages} {063005} (\bibinfo {year} {2013})}\BibitemShut
  {NoStop}%
\bibitem [{\citenamefont {Kurskiev}\ \emph {et~al.}(2019)\citenamefont
  {Kurskiev}, \citenamefont {Bakharev}, \citenamefont {Bulanin}, \citenamefont
  {Chernyshev}, \citenamefont {Gusev}, \citenamefont {Khromov}, \citenamefont
  {Kiselev}, \citenamefont {Minaev}, \citenamefont {Miroshnikov}, \citenamefont
  {Mukhin}, \citenamefont {Patrov}, \citenamefont {Petrov}, \citenamefont
  {Petrov}, \citenamefont {Sakharov}, \citenamefont {Shchegolev}, \citenamefont
  {Sladkomedova}, \citenamefont {Solokha}, \citenamefont {Telnova},
  \citenamefont {Tolstyakov}, \citenamefont {Tokarev},\ and\ \citenamefont
  {Yashin}}]{Kurskiev2019NF}%
  \BibitemOpen
  \bibfield  {author} {\bibinfo {author} {\bibfnamefont {G.}~\bibnamefont
  {Kurskiev}}, \bibinfo {author} {\bibfnamefont {N.}~\bibnamefont {Bakharev}},
  \bibinfo {author} {\bibfnamefont {V.}~\bibnamefont {Bulanin}}, \bibinfo
  {author} {\bibfnamefont {F.}~\bibnamefont {Chernyshev}}, \bibinfo {author}
  {\bibfnamefont {V.}~\bibnamefont {Gusev}}, \bibinfo {author} {\bibfnamefont
  {N.}~\bibnamefont {Khromov}}, \bibinfo {author} {\bibfnamefont
  {E.}~\bibnamefont {Kiselev}}, \bibinfo {author} {\bibfnamefont
  {V.}~\bibnamefont {Minaev}}, \bibinfo {author} {\bibfnamefont
  {I.}~\bibnamefont {Miroshnikov}}, \bibinfo {author} {\bibfnamefont
  {E.}~\bibnamefont {Mukhin}}, \bibinfo {author} {\bibfnamefont
  {M.}~\bibnamefont {Patrov}}, \bibinfo {author} {\bibfnamefont
  {A.}~\bibnamefont {Petrov}}, \bibinfo {author} {\bibfnamefont
  {Y.}~\bibnamefont {Petrov}}, \bibinfo {author} {\bibfnamefont
  {N.}~\bibnamefont {Sakharov}}, \bibinfo {author} {\bibfnamefont
  {P.}~\bibnamefont {Shchegolev}}, \bibinfo {author} {\bibfnamefont
  {A.}~\bibnamefont {Sladkomedova}}, \bibinfo {author} {\bibfnamefont
  {V.}~\bibnamefont {Solokha}}, \bibinfo {author} {\bibfnamefont
  {A.}~\bibnamefont {Telnova}}, \bibinfo {author} {\bibfnamefont
  {S.}~\bibnamefont {Tolstyakov}}, \bibinfo {author} {\bibfnamefont
  {V.}~\bibnamefont {Tokarev}},\ and\ \bibinfo {author} {\bibfnamefont
  {A.}~\bibnamefont {Yashin}},\ }\bibfield  {title} {\bibinfo {title} {Thermal
  energy confinement at the {Globus-M} spherical tokamak},\ }\href
  {https://doi.org/10.1088/1741-4326/ab15c5} {\bibfield  {journal} {\bibinfo
  {journal} {Nuclear Fusion}\ }\textbf {\bibinfo {volume} {59}},\ \bibinfo
  {pages} {066032} (\bibinfo {year} {2019})}\BibitemShut {NoStop}%
\bibitem [{\citenamefont {Kurskiev}\ \emph {et~al.}(2021)\citenamefont
  {Kurskiev}, \citenamefont {Gusev}, \citenamefont {Sakharov}, \citenamefont
  {Petrov}, \citenamefont {Bakharev}, \citenamefont {Balachenkov},
  \citenamefont {Bazhenov}, \citenamefont {Chernyshev}, \citenamefont
  {Khromov}, \citenamefont {Kiselev}, \citenamefont {Krikunov}, \citenamefont
  {Minaev}, \citenamefont {Miroshnikov}, \citenamefont {Novokhatskii},
  \citenamefont {Zhiltsov}, \citenamefont {Mukhin}, \citenamefont {Patrov},
  \citenamefont {Shulyatiev}, \citenamefont {Shchegolev}, \citenamefont
  {Skrekel}, \citenamefont {Telnova}, \citenamefont {Tkachenko}, \citenamefont
  {Tukhmeneva}, \citenamefont {Tokarev}, \citenamefont {Tolstyakov},
  \citenamefont {Varfolomeev}, \citenamefont {Voronin}, \citenamefont
  {Goryainov}, \citenamefont {Bulanin}, \citenamefont {Petrov}, \citenamefont
  {Ponomarenko}, \citenamefont {Yashin}, \citenamefont {Kavin}, \citenamefont
  {Zhilin},\ and\ \citenamefont {Solovey}}]{Kurskiev2022NF}%
  \BibitemOpen
  \bibfield  {author} {\bibinfo {author} {\bibfnamefont {G.}~\bibnamefont
  {Kurskiev}}, \bibinfo {author} {\bibfnamefont {V.}~\bibnamefont {Gusev}},
  \bibinfo {author} {\bibfnamefont {N.}~\bibnamefont {Sakharov}}, \bibinfo
  {author} {\bibfnamefont {Y.}~\bibnamefont {Petrov}}, \bibinfo {author}
  {\bibfnamefont {N.}~\bibnamefont {Bakharev}}, \bibinfo {author}
  {\bibfnamefont {I.}~\bibnamefont {Balachenkov}}, \bibinfo {author}
  {\bibfnamefont {A.}~\bibnamefont {Bazhenov}}, \bibinfo {author}
  {\bibfnamefont {F.}~\bibnamefont {Chernyshev}}, \bibinfo {author}
  {\bibfnamefont {N.}~\bibnamefont {Khromov}}, \bibinfo {author} {\bibfnamefont
  {E.}~\bibnamefont {Kiselev}}, \bibinfo {author} {\bibfnamefont
  {S.}~\bibnamefont {Krikunov}}, \bibinfo {author} {\bibfnamefont
  {V.}~\bibnamefont {Minaev}}, \bibinfo {author} {\bibfnamefont
  {I.}~\bibnamefont {Miroshnikov}}, \bibinfo {author} {\bibfnamefont
  {A.}~\bibnamefont {Novokhatskii}}, \bibinfo {author} {\bibfnamefont
  {N.}~\bibnamefont {Zhiltsov}}, \bibinfo {author} {\bibfnamefont
  {E.}~\bibnamefont {Mukhin}}, \bibinfo {author} {\bibfnamefont
  {M.}~\bibnamefont {Patrov}}, \bibinfo {author} {\bibfnamefont
  {K.}~\bibnamefont {Shulyatiev}}, \bibinfo {author} {\bibfnamefont
  {P.}~\bibnamefont {Shchegolev}}, \bibinfo {author} {\bibfnamefont
  {O.}~\bibnamefont {Skrekel}}, \bibinfo {author} {\bibfnamefont
  {A.}~\bibnamefont {Telnova}}, \bibinfo {author} {\bibfnamefont
  {E.}~\bibnamefont {Tkachenko}}, \bibinfo {author} {\bibfnamefont
  {E.}~\bibnamefont {Tukhmeneva}}, \bibinfo {author} {\bibfnamefont
  {V.}~\bibnamefont {Tokarev}}, \bibinfo {author} {\bibfnamefont
  {S.}~\bibnamefont {Tolstyakov}}, \bibinfo {author} {\bibfnamefont
  {V.}~\bibnamefont {Varfolomeev}}, \bibinfo {author} {\bibfnamefont
  {A.}~\bibnamefont {Voronin}}, \bibinfo {author} {\bibfnamefont
  {V.}~\bibnamefont {Goryainov}}, \bibinfo {author} {\bibfnamefont
  {V.}~\bibnamefont {Bulanin}}, \bibinfo {author} {\bibfnamefont
  {A.}~\bibnamefont {Petrov}}, \bibinfo {author} {\bibfnamefont
  {A.}~\bibnamefont {Ponomarenko}}, \bibinfo {author} {\bibfnamefont
  {A.}~\bibnamefont {Yashin}}, \bibinfo {author} {\bibfnamefont
  {A.}~\bibnamefont {Kavin}}, \bibinfo {author} {\bibfnamefont
  {E.}~\bibnamefont {Zhilin}},\ and\ \bibinfo {author} {\bibfnamefont
  {V.}~\bibnamefont {Solovey}},\ }\bibfield  {title} {\bibinfo {title} {Energy
  confinement in the spherical tokamak {Globus-M2} with a toroidal magnetic
  field reaching {0.8 T}},\ }\href {https://doi.org/10.1088/1741-4326/ac38c9}
  {\bibfield  {journal} {\bibinfo  {journal} {Nuclear Fusion}\ }\textbf
  {\bibinfo {volume} {62}},\ \bibinfo {pages} {016011} (\bibinfo {year}
  {2021})}\BibitemShut {NoStop}%
\bibitem [{\citenamefont {Kaye}\ \emph {et~al.}(2021)\citenamefont {Kaye},
  \citenamefont {Connor},\ and\ \citenamefont {Roach}}]{Kaye2021PPCF}%
  \BibitemOpen
  \bibfield  {author} {\bibinfo {author} {\bibfnamefont {S.~M.}\ \bibnamefont
  {Kaye}}, \bibinfo {author} {\bibfnamefont {J.~W.}\ \bibnamefont {Connor}},\
  and\ \bibinfo {author} {\bibfnamefont {C.~M.}\ \bibnamefont {Roach}},\
  }\bibfield  {title} {\bibinfo {title} {Thermal confinement and transport in
  spherical tokamaks: a review},\ }\href
  {https://doi.org/10.1088/1361-6587/ac2b38} {\bibfield  {journal} {\bibinfo
  {journal} {Plasma Physics and Controlled Fusion}\ }\textbf {\bibinfo {volume}
  {63}},\ \bibinfo {pages} {123001} (\bibinfo {year} {2021})}\BibitemShut
  {NoStop}%
\bibitem [{\citenamefont {Guttenfelder}\ \emph {et~al.}(2012)\citenamefont
  {Guttenfelder}, \citenamefont {Candy}, \citenamefont {Kaye}, \citenamefont
  {Nevins}, \citenamefont {Bell}, \citenamefont {Hammett}, \citenamefont
  {LeBlanc},\ and\ \citenamefont {Yuh}}]{Guttenfelder2012POP}%
  \BibitemOpen
  \bibfield  {author} {\bibinfo {author} {\bibfnamefont {W.}~\bibnamefont
  {Guttenfelder}}, \bibinfo {author} {\bibfnamefont {J.}~\bibnamefont {Candy}},
  \bibinfo {author} {\bibfnamefont {S.~M.}\ \bibnamefont {Kaye}}, \bibinfo
  {author} {\bibfnamefont {W.~M.}\ \bibnamefont {Nevins}}, \bibinfo {author}
  {\bibfnamefont {R.~E.}\ \bibnamefont {Bell}}, \bibinfo {author}
  {\bibfnamefont {G.~W.}\ \bibnamefont {Hammett}}, \bibinfo {author}
  {\bibfnamefont {B.~P.}\ \bibnamefont {LeBlanc}},\ and\ \bibinfo {author}
  {\bibfnamefont {H.}~\bibnamefont {Yuh}},\ }\bibfield  {title} {\bibinfo
  {title} {Scaling of linear microtearing stability for a high collisionality
  {National Spherical Torus Experiment discharge}},\ }\href
  {https://doi.org/10.1063/1.3685698} {\bibfield  {journal} {\bibinfo
  {journal} {Physics of Plasmas}\ }\textbf {\bibinfo {volume} {19}},\ \bibinfo
  {pages} {022506} (\bibinfo {year} {2012})}\BibitemShut {NoStop}%
\bibitem [{\citenamefont {Guttenfelder}\ \emph {et~al.}(2013)\citenamefont
  {Guttenfelder}, \citenamefont {Peterson}, \citenamefont {Candy},
  \citenamefont {Kaye}, \citenamefont {Ren}, \citenamefont {Bell},
  \citenamefont {Hammett}, \citenamefont {LeBlanc}, \citenamefont {Mikkelsen},
  \citenamefont {Nevins},\ and\ \citenamefont {Yuh}}]{Guttenfelder2013NF}%
  \BibitemOpen
  \bibfield  {author} {\bibinfo {author} {\bibfnamefont {W.}~\bibnamefont
  {Guttenfelder}}, \bibinfo {author} {\bibfnamefont {J.~L.}\ \bibnamefont
  {Peterson}}, \bibinfo {author} {\bibfnamefont {J.}~\bibnamefont {Candy}},
  \bibinfo {author} {\bibfnamefont {S.~M.}\ \bibnamefont {Kaye}}, \bibinfo
  {author} {\bibfnamefont {Y.}~\bibnamefont {Ren}}, \bibinfo {author}
  {\bibfnamefont {R.~E.}\ \bibnamefont {Bell}}, \bibinfo {author}
  {\bibfnamefont {G.~W.}\ \bibnamefont {Hammett}}, \bibinfo {author}
  {\bibfnamefont {B.~P.}\ \bibnamefont {LeBlanc}}, \bibinfo {author}
  {\bibfnamefont {D.~R.}\ \bibnamefont {Mikkelsen}}, \bibinfo {author}
  {\bibfnamefont {W.~M.}\ \bibnamefont {Nevins}},\ and\ \bibinfo {author}
  {\bibfnamefont {H.}~\bibnamefont {Yuh}},\ }\bibfield  {title} {\bibinfo
  {title} {Progress in simulating turbulent electron thermal transport in
  {NSTX}},\ }\href {http://stacks.iop.org/0029-5515/53/i=9/a=093022} {\bibfield
   {journal} {\bibinfo  {journal} {Nuclear Fusion}\ }\textbf {\bibinfo {volume}
  {53}},\ \bibinfo {pages} {093022} (\bibinfo {year} {2013})}\BibitemShut
  {NoStop}%
\bibitem [{\citenamefont {Kaye}\ \emph {et~al.}(2014)\citenamefont {Kaye},
  \citenamefont {Guttenfelder}, \citenamefont {Bell}, \citenamefont {Gerhardt},
  \citenamefont {LeBlanc},\ and\ \citenamefont {Maingi}}]{Kaye2014POP}%
  \BibitemOpen
  \bibfield  {author} {\bibinfo {author} {\bibfnamefont {S.~M.}\ \bibnamefont
  {Kaye}}, \bibinfo {author} {\bibfnamefont {W.}~\bibnamefont {Guttenfelder}},
  \bibinfo {author} {\bibfnamefont {R.~E.}\ \bibnamefont {Bell}}, \bibinfo
  {author} {\bibfnamefont {S.~P.}\ \bibnamefont {Gerhardt}}, \bibinfo {author}
  {\bibfnamefont {B.~P.}\ \bibnamefont {LeBlanc}},\ and\ \bibinfo {author}
  {\bibfnamefont {R.}~\bibnamefont {Maingi}},\ }\bibfield  {title} {\bibinfo
  {title} {{Reduced model prediction of electron temperature profiles in
  microtearing-dominated {National Spherical Torus eXperiment} plasmas}},\
  }\href {https://doi.org/10.1063/1.4893135} {\bibfield  {journal} {\bibinfo
  {journal} {Physics of Plasmas}\ }\textbf {\bibinfo {volume} {21}},\ \bibinfo
  {pages} {082510} (\bibinfo {year} {2014})}\BibitemShut {NoStop}%
\bibitem [{\citenamefont {Clauser}\ \emph {et~al.}(2022)\citenamefont
  {Clauser}, \citenamefont {Guttenfelder}, \citenamefont {Rafiq},\ and\
  \citenamefont {Schuster}}]{Clauser2022POP}%
  \BibitemOpen
  \bibfield  {author} {\bibinfo {author} {\bibfnamefont {C.~F.}\ \bibnamefont
  {Clauser}}, \bibinfo {author} {\bibfnamefont {W.}~\bibnamefont
  {Guttenfelder}}, \bibinfo {author} {\bibfnamefont {T.}~\bibnamefont
  {Rafiq}},\ and\ \bibinfo {author} {\bibfnamefont {E.}~\bibnamefont
  {Schuster}},\ }\bibfield  {title} {\bibinfo {title} {{Linear ion-scale
  microstability analysis of high and low-collisionality {NSTX} discharges and
  {NSTX-U} projections}},\ }\href {https://doi.org/10.1063/5.0102169}
  {\bibfield  {journal} {\bibinfo  {journal} {Physics of Plasmas}\ }\textbf
  {\bibinfo {volume} {29}},\ \bibinfo {pages} {102303} (\bibinfo {year}
  {2022})}\BibitemShut {NoStop}%
\bibitem [{\citenamefont {Patel}\ \emph {et~al.}(2021)\citenamefont {Patel},
  \citenamefont {Dickinson}, \citenamefont {Roach},\ and\ \citenamefont
  {Wilson}}]{Patel2022NF}%
  \BibitemOpen
  \bibfield  {author} {\bibinfo {author} {\bibfnamefont {B.}~\bibnamefont
  {Patel}}, \bibinfo {author} {\bibfnamefont {D.}~\bibnamefont {Dickinson}},
  \bibinfo {author} {\bibfnamefont {C.}~\bibnamefont {Roach}},\ and\ \bibinfo
  {author} {\bibfnamefont {H.}~\bibnamefont {Wilson}},\ }\bibfield  {title}
  {\bibinfo {title} {Linear gyrokinetic stability of a high $\beta$
  non-inductive spherical tokamak},\ }\href
  {https://doi.org/10.1088/1741-4326/ac359c} {\bibfield  {journal} {\bibinfo
  {journal} {Nuclear Fusion}\ }\textbf {\bibinfo {volume} {62}},\ \bibinfo
  {pages} {016009} (\bibinfo {year} {2021})}\BibitemShut {NoStop}%
\bibitem [{\citenamefont {McClenaghan}\ \emph {et~al.}(2023)\citenamefont
  {McClenaghan}, \citenamefont {Slendebroek}, \citenamefont {Staebler},
  \citenamefont {Smith}, \citenamefont {Meneghini}, \citenamefont {Grierson},
  \citenamefont {Thome}, \citenamefont {Avdeeva}, \citenamefont {Lao},
  \citenamefont {Candy},\ and\ \citenamefont
  {Guttenfelder}}]{McClenaghan2023POP}%
  \BibitemOpen
  \bibfield  {author} {\bibinfo {author} {\bibfnamefont {J.}~\bibnamefont
  {McClenaghan}}, \bibinfo {author} {\bibfnamefont {T.}~\bibnamefont
  {Slendebroek}}, \bibinfo {author} {\bibfnamefont {G.~M.}\ \bibnamefont
  {Staebler}}, \bibinfo {author} {\bibfnamefont {S.~P.}\ \bibnamefont {Smith}},
  \bibinfo {author} {\bibfnamefont {O.~M.}\ \bibnamefont {Meneghini}}, \bibinfo
  {author} {\bibfnamefont {B.~A.}\ \bibnamefont {Grierson}}, \bibinfo {author}
  {\bibfnamefont {K.~E.}\ \bibnamefont {Thome}}, \bibinfo {author}
  {\bibfnamefont {G.}~\bibnamefont {Avdeeva}}, \bibinfo {author} {\bibfnamefont
  {L.~L.}\ \bibnamefont {Lao}}, \bibinfo {author} {\bibfnamefont
  {J.}~\bibnamefont {Candy}},\ and\ \bibinfo {author} {\bibfnamefont
  {W.}~\bibnamefont {Guttenfelder}},\ }\bibfield  {title} {\bibinfo {title}
  {Transition from {ITG} to {MTM} linear instabilities near pedestals of high
  density plasmas},\ }\href {https://doi.org/10.1063/5.0141179} {\bibfield
  {journal} {\bibinfo  {journal} {Physics of Plasmas}\ }\textbf {\bibinfo
  {volume} {30}},\ \bibinfo {pages} {042512} (\bibinfo {year}
  {2023})}\BibitemShut {NoStop}%
\bibitem [{\citenamefont {Kennedy}\ \emph {et~al.}(2023)\citenamefont
  {Kennedy}, \citenamefont {Giacomin}, \citenamefont {Casson}, \citenamefont
  {Dickinson}, \citenamefont {Hornsby}, \citenamefont {Patel},\ and\
  \citenamefont {Roach}}]{Kennedy2023NF}%
  \BibitemOpen
  \bibfield  {author} {\bibinfo {author} {\bibfnamefont {D.}~\bibnamefont
  {Kennedy}}, \bibinfo {author} {\bibfnamefont {M.}~\bibnamefont {Giacomin}},
  \bibinfo {author} {\bibfnamefont {F.}~\bibnamefont {Casson}}, \bibinfo
  {author} {\bibfnamefont {D.}~\bibnamefont {Dickinson}}, \bibinfo {author}
  {\bibfnamefont {W.}~\bibnamefont {Hornsby}}, \bibinfo {author} {\bibfnamefont
  {B.}~\bibnamefont {Patel}},\ and\ \bibinfo {author} {\bibfnamefont
  {C.}~\bibnamefont {Roach}},\ }\bibfield  {title} {\bibinfo {title}
  {Electromagnetic gyrokinetic instabilities in {STEP}},\ }\href
  {https://doi.org/10.1088/1741-4326/ad08e7} {\bibfield  {journal} {\bibinfo
  {journal} {Nuclear Fusion}\ }\textbf {\bibinfo {volume} {63}},\ \bibinfo
  {pages} {126061} (\bibinfo {year} {2023})}\BibitemShut {NoStop}%
\bibitem [{\citenamefont {Kennedy}\ \emph {et~al.}(2024)\citenamefont
  {Kennedy}, \citenamefont {Roach}, \citenamefont {Giacomin}, \citenamefont
  {Ivanov}, \citenamefont {Adkins}, \citenamefont {Sheffield}, \citenamefont
  {Görler}, \citenamefont {Bokshi}, \citenamefont {Dickinson}, \citenamefont
  {Dudding},\ and\ \citenamefont {Patel}}]{Kennedy2024NF}%
  \BibitemOpen
  \bibfield  {author} {\bibinfo {author} {\bibfnamefont {D.}~\bibnamefont
  {Kennedy}}, \bibinfo {author} {\bibfnamefont {C.}~\bibnamefont {Roach}},
  \bibinfo {author} {\bibfnamefont {M.}~\bibnamefont {Giacomin}}, \bibinfo
  {author} {\bibfnamefont {P.}~\bibnamefont {Ivanov}}, \bibinfo {author}
  {\bibfnamefont {T.}~\bibnamefont {Adkins}}, \bibinfo {author} {\bibfnamefont
  {F.}~\bibnamefont {Sheffield}}, \bibinfo {author} {\bibfnamefont
  {T.}~\bibnamefont {Görler}}, \bibinfo {author} {\bibfnamefont
  {A.}~\bibnamefont {Bokshi}}, \bibinfo {author} {\bibfnamefont
  {D.}~\bibnamefont {Dickinson}}, \bibinfo {author} {\bibfnamefont
  {H.}~\bibnamefont {Dudding}},\ and\ \bibinfo {author} {\bibfnamefont
  {B.}~\bibnamefont {Patel}},\ }\bibfield  {title} {\bibinfo {title} {On the
  importance of parallel magnetic-field fluctuations for electromagnetic
  instabilities in {STEP}},\ }\href {https://doi.org/10.1088/1741-4326/ad58f3}
  {\bibfield  {journal} {\bibinfo  {journal} {Nuclear Fusion}\ }\textbf
  {\bibinfo {volume} {64}},\ \bibinfo {pages} {086049} (\bibinfo {year}
  {2024})}\BibitemShut {NoStop}%
\bibitem [{\citenamefont {Giacomin}\ \emph {et~al.}(2024)\citenamefont
  {Giacomin}, \citenamefont {Kennedy}, \citenamefont {Casson}, \citenamefont
  {C~J}, \citenamefont {Dickinson}, \citenamefont {Patel},\ and\ \citenamefont
  {Roach}}]{Giacomin2024PPCF}%
  \BibitemOpen
  \bibfield  {author} {\bibinfo {author} {\bibfnamefont {M.}~\bibnamefont
  {Giacomin}}, \bibinfo {author} {\bibfnamefont {D.}~\bibnamefont {Kennedy}},
  \bibinfo {author} {\bibfnamefont {F.~J.}\ \bibnamefont {Casson}}, \bibinfo
  {author} {\bibfnamefont {A.}~\bibnamefont {C~J}}, \bibinfo {author}
  {\bibfnamefont {D.}~\bibnamefont {Dickinson}}, \bibinfo {author}
  {\bibfnamefont {B.~S.}\ \bibnamefont {Patel}},\ and\ \bibinfo {author}
  {\bibfnamefont {C.~M.}\ \bibnamefont {Roach}},\ }\bibfield  {title} {\bibinfo
  {title} {On electromagnetic turbulence and transport in {STEP}},\ }\href
  {https://doi.org/10.1088/1361-6587/ad366f} {\bibfield  {journal} {\bibinfo
  {journal} {Plasma Physics and Controlled Fusion}\ }\textbf {\bibinfo {volume}
  {66}},\ \bibinfo {pages} {055010} (\bibinfo {year} {2024})}\BibitemShut
  {NoStop}%
\bibitem [{\citenamefont {Dominski}\ \emph {et~al.}(2024)\citenamefont
  {Dominski}, \citenamefont {Guttenfelder}, \citenamefont {Hatch},
  \citenamefont {Goerler}, \citenamefont {Jenko}, \citenamefont {Munaretto},\
  and\ \citenamefont {Kaye}}]{Dominski2024POP}%
  \BibitemOpen
  \bibfield  {author} {\bibinfo {author} {\bibfnamefont {J.}~\bibnamefont
  {Dominski}}, \bibinfo {author} {\bibfnamefont {W.}~\bibnamefont
  {Guttenfelder}}, \bibinfo {author} {\bibfnamefont {D.}~\bibnamefont {Hatch}},
  \bibinfo {author} {\bibfnamefont {T.}~\bibnamefont {Goerler}}, \bibinfo
  {author} {\bibfnamefont {F.}~\bibnamefont {Jenko}}, \bibinfo {author}
  {\bibfnamefont {S.}~\bibnamefont {Munaretto}},\ and\ \bibinfo {author}
  {\bibfnamefont {S.}~\bibnamefont {Kaye}},\ }\bibfield  {title} {\bibinfo
  {title} {{Global micro-tearing modes in the wide pedestal of an {NSTX}
  plasma}},\ }\href {https://doi.org/10.1063/5.0200894} {\bibfield  {journal}
  {\bibinfo  {journal} {Physics of Plasmas}\ }\textbf {\bibinfo {volume}
  {31}},\ \bibinfo {pages} {044501} (\bibinfo {year} {2024})}\BibitemShut
  {NoStop}%
\bibitem [{\citenamefont {McClenaghan}\ \emph {et~al.}(2025)\citenamefont
  {McClenaghan}, \citenamefont {Neiser}, \citenamefont {Halpern}, \citenamefont
  {Thome}, \citenamefont {Turnbull}, \citenamefont {DeShazer}, \citenamefont
  {Meneghini}, \citenamefont {Avdeeva}, \citenamefont {Lestz},\ and\
  \citenamefont {Candy}}]{McClenaghan2025PPCF}%
  \BibitemOpen
  \bibfield  {author} {\bibinfo {author} {\bibfnamefont {J.}~\bibnamefont
  {McClenaghan}}, \bibinfo {author} {\bibfnamefont {T.~F.}\ \bibnamefont
  {Neiser}}, \bibinfo {author} {\bibfnamefont {F.~D.}\ \bibnamefont {Halpern}},
  \bibinfo {author} {\bibfnamefont {K.~E.}\ \bibnamefont {Thome}}, \bibinfo
  {author} {\bibfnamefont {A.~D.}\ \bibnamefont {Turnbull}}, \bibinfo {author}
  {\bibfnamefont {E.~W.}\ \bibnamefont {DeShazer}}, \bibinfo {author}
  {\bibfnamefont {O.~M.}\ \bibnamefont {Meneghini}}, \bibinfo {author}
  {\bibfnamefont {G.}~\bibnamefont {Avdeeva}}, \bibinfo {author} {\bibfnamefont
  {J.~B.}\ \bibnamefont {Lestz}},\ and\ \bibinfo {author} {\bibfnamefont
  {J.}~\bibnamefont {Candy}},\ }\bibfield  {title} {\bibinfo {title} {Role of
  perturbed parallel magnetic field effects in predicting turbulent transport
  in {NSTX}},\ }\href {https://doi.org/10.1088/1361-6587/adc9e2} {\bibfield
  {journal} {\bibinfo  {journal} {Plasma Physics and Controlled Fusion}\
  }\textbf {\bibinfo {volume} {67}},\ \bibinfo {pages} {055013} (\bibinfo
  {year} {2025})}\BibitemShut {NoStop}%
\bibitem [{\citenamefont {Singh}\ \emph {et~al.}(tted)\citenamefont {Singh},
  \citenamefont {Rafiq}, \citenamefont {Schuster}, \citenamefont {Lin},\ and\
  \citenamefont {Kuley}}]{Singh2025NF}%
  \BibitemOpen
  \bibfield  {author} {\bibinfo {author} {\bibfnamefont {T.}~\bibnamefont
  {Singh}}, \bibinfo {author} {\bibfnamefont {T.}~\bibnamefont {Rafiq}},
  \bibinfo {author} {\bibfnamefont {E.}~\bibnamefont {Schuster}}, \bibinfo
  {author} {\bibfnamefont {Z.}~\bibnamefont {Lin}},\ and\ \bibinfo {author}
  {\bibfnamefont {A.}~\bibnamefont {Kuley}},\ }\bibfield  {title} {\bibinfo
  {title} {Global gyrokinetic simulations of kinetic ballooning mode in
  {NSTX-U} plasmas},\ }\href@noop {} {\bibfield  {journal} {\bibinfo  {journal}
  {Nuclear Fusion}\ } (\bibinfo {year} {submitted})}\BibitemShut {NoStop}%
\bibitem [{\citenamefont {Roach}\ \emph {et~al.}(2009)\citenamefont {Roach},
  \citenamefont {Abel}, \citenamefont {Akers}, \citenamefont {Arter},
  \citenamefont {Barnes}, \citenamefont {Camenen}, \citenamefont {Casson},
  \citenamefont {Colyer}, \citenamefont {Connor}, \citenamefont {Cowley},
  \citenamefont {Dickinson}, \citenamefont {Dorland}, \citenamefont {Field},
  \citenamefont {Guttenfelder}, \citenamefont {Hammett}, \citenamefont
  {Hastie}, \citenamefont {Highcock}, \citenamefont {Loureiro}, \citenamefont
  {Peeters}, \citenamefont {Reshko}, \citenamefont {Saarelma}, \citenamefont
  {Schekochihin}, \citenamefont {Valovic},\ and\ \citenamefont
  {Wilson}}]{Roach2009PPCF}%
  \BibitemOpen
  \bibfield  {author} {\bibinfo {author} {\bibfnamefont {C.~M.}\ \bibnamefont
  {Roach}}, \bibinfo {author} {\bibfnamefont {I.~G.}\ \bibnamefont {Abel}},
  \bibinfo {author} {\bibfnamefont {R.~J.}\ \bibnamefont {Akers}}, \bibinfo
  {author} {\bibfnamefont {W.}~\bibnamefont {Arter}}, \bibinfo {author}
  {\bibfnamefont {M.}~\bibnamefont {Barnes}}, \bibinfo {author} {\bibfnamefont
  {Y.}~\bibnamefont {Camenen}}, \bibinfo {author} {\bibfnamefont {F.~J.}\
  \bibnamefont {Casson}}, \bibinfo {author} {\bibfnamefont {G.}~\bibnamefont
  {Colyer}}, \bibinfo {author} {\bibfnamefont {J.~W.}\ \bibnamefont {Connor}},
  \bibinfo {author} {\bibfnamefont {S.~C.}\ \bibnamefont {Cowley}}, \bibinfo
  {author} {\bibfnamefont {D.}~\bibnamefont {Dickinson}}, \bibinfo {author}
  {\bibfnamefont {W.}~\bibnamefont {Dorland}}, \bibinfo {author} {\bibfnamefont
  {A.~R.}\ \bibnamefont {Field}}, \bibinfo {author} {\bibfnamefont
  {W.}~\bibnamefont {Guttenfelder}}, \bibinfo {author} {\bibfnamefont {G.~W.}\
  \bibnamefont {Hammett}}, \bibinfo {author} {\bibfnamefont {R.~J.}\
  \bibnamefont {Hastie}}, \bibinfo {author} {\bibfnamefont {E.}~\bibnamefont
  {Highcock}}, \bibinfo {author} {\bibfnamefont {N.~F.}\ \bibnamefont
  {Loureiro}}, \bibinfo {author} {\bibfnamefont {A.~G.}\ \bibnamefont
  {Peeters}}, \bibinfo {author} {\bibfnamefont {M.}~\bibnamefont {Reshko}},
  \bibinfo {author} {\bibfnamefont {S.}~\bibnamefont {Saarelma}}, \bibinfo
  {author} {\bibfnamefont {A.~A.}\ \bibnamefont {Schekochihin}}, \bibinfo
  {author} {\bibfnamefont {M.}~\bibnamefont {Valovic}},\ and\ \bibinfo {author}
  {\bibfnamefont {H.~R.}\ \bibnamefont {Wilson}},\ }\bibfield  {title}
  {\bibinfo {title} {Gyrokinetic simulations of spherical tokamaks},\ }\href
  {https://doi.org/10.1088/0741-3335/51/12/124020} {\bibfield  {journal}
  {\bibinfo  {journal} {Plasma Physics and Controlled Fusion}\ }\textbf
  {\bibinfo {volume} {51}},\ \bibinfo {pages} {124020} (\bibinfo {year}
  {2009})}\BibitemShut {NoStop}%
\bibitem [{\citenamefont {Patel}\ \emph {et~al.}(2025)\citenamefont {Patel},
  \citenamefont {Hardman}, \citenamefont {Kennedy}, \citenamefont {Giacomin},
  \citenamefont {Dickinson},\ and\ \citenamefont {Roach}}]{Patel2025NF}%
  \BibitemOpen
  \bibfield  {author} {\bibinfo {author} {\bibfnamefont {B.}~\bibnamefont
  {Patel}}, \bibinfo {author} {\bibfnamefont {M.}~\bibnamefont {Hardman}},
  \bibinfo {author} {\bibfnamefont {D.}~\bibnamefont {Kennedy}}, \bibinfo
  {author} {\bibfnamefont {M.}~\bibnamefont {Giacomin}}, \bibinfo {author}
  {\bibfnamefont {D.}~\bibnamefont {Dickinson}},\ and\ \bibinfo {author}
  {\bibfnamefont {C.}~\bibnamefont {Roach}},\ }\bibfield  {title} {\bibinfo
  {title} {The impact of {E × B} shear on microtearing based transport in
  spherical tokamaks},\ }\href {https://doi.org/10.1088/1741-4326/ada627}
  {\bibfield  {journal} {\bibinfo  {journal} {Nuclear Fusion}\ }\textbf
  {\bibinfo {volume} {65}},\ \bibinfo {pages} {026063} (\bibinfo {year}
  {2025})}\BibitemShut {NoStop}%
\bibitem [{\citenamefont {Avdeeva}\ \emph {et~al.}(smas)\citenamefont
  {Avdeeva}, \citenamefont {Candy}, , \citenamefont {Thome}, \citenamefont
  {Belli}, \citenamefont {Kaye},\ and\ \citenamefont
  {Staebler}}]{Avdeeva2025pre}%
  \BibitemOpen
  \bibfield  {author} {\bibinfo {author} {\bibfnamefont {G.}~\bibnamefont
  {Avdeeva}}, \bibinfo {author} {\bibfnamefont {J.}~\bibnamefont {Candy}}, ,
  \bibinfo {author} {\bibfnamefont {K.}~\bibnamefont {Thome}}, \bibinfo
  {author} {\bibfnamefont {E.}~\bibnamefont {Belli}}, \bibinfo {author}
  {\bibfnamefont {S.}~\bibnamefont {Kaye}},\ and\ \bibinfo {author}
  {\bibfnamefont {G.}~\bibnamefont {Staebler}},\ }\bibfield  {title} {\bibinfo
  {title} {Enhanced shear-stabilization of turbulence in {NSTX}},\ }\href@noop
  {} {\  (\bibinfo {year} {submitted to Physics of Plasmas})}\BibitemShut
  {NoStop}%
\bibitem [{\citenamefont {Rafiq}\ \emph {et~al.}(2013)\citenamefont {Rafiq},
  \citenamefont {Kritz}, \citenamefont {Weiland}, \citenamefont {Pankin},\ and\
  \citenamefont {Luo}}]{Rafiq2013POP}%
  \BibitemOpen
  \bibfield  {author} {\bibinfo {author} {\bibfnamefont {T.}~\bibnamefont
  {Rafiq}}, \bibinfo {author} {\bibfnamefont {A.~H.}\ \bibnamefont {Kritz}},
  \bibinfo {author} {\bibfnamefont {J.}~\bibnamefont {Weiland}}, \bibinfo
  {author} {\bibfnamefont {A.~Y.}\ \bibnamefont {Pankin}},\ and\ \bibinfo
  {author} {\bibfnamefont {L.}~\bibnamefont {Luo}},\ }\bibfield  {title}
  {\bibinfo {title} {{Physics basis of {Multi-Mode} anomalous transport
  module}},\ }\href {https://doi.org/10.1063/1.4794288} {\bibfield  {journal}
  {\bibinfo  {journal} {Physics of Plasmas}\ }\textbf {\bibinfo {volume}
  {20}},\ \bibinfo {pages} {032506} (\bibinfo {year} {2013})}\BibitemShut
  {NoStop}%
\bibitem [{\citenamefont {Luo}\ \emph {et~al.}(2013)\citenamefont {Luo},
  \citenamefont {Rafiq},\ and\ \citenamefont {Kritz}}]{Luo2013CPC}%
  \BibitemOpen
  \bibfield  {author} {\bibinfo {author} {\bibfnamefont {L.}~\bibnamefont
  {Luo}}, \bibinfo {author} {\bibfnamefont {T.}~\bibnamefont {Rafiq}},\ and\
  \bibinfo {author} {\bibfnamefont {A.}~\bibnamefont {Kritz}},\ }\bibfield
  {title} {\bibinfo {title} {Improved {Multi-Mode} anomalous transport module
  for tokamak plasmas},\ }\href
  {https://doi.org/https://doi.org/10.1016/j.cpc.2013.05.013} {\bibfield
  {journal} {\bibinfo  {journal} {Computer Physics Communications}\ }\textbf
  {\bibinfo {volume} {184}},\ \bibinfo {pages} {2267} (\bibinfo {year}
  {2013})}\BibitemShut {NoStop}%
\bibitem [{\citenamefont {Staebler}\ \emph {et~al.}(2007)\citenamefont
  {Staebler}, \citenamefont {Kinsey},\ and\ \citenamefont
  {Waltz}}]{Staebler2007POP}%
  \BibitemOpen
  \bibfield  {author} {\bibinfo {author} {\bibfnamefont {G.~M.}\ \bibnamefont
  {Staebler}}, \bibinfo {author} {\bibfnamefont {J.~E.}\ \bibnamefont
  {Kinsey}},\ and\ \bibinfo {author} {\bibfnamefont {R.~E.}\ \bibnamefont
  {Waltz}},\ }\bibfield  {title} {\bibinfo {title} {{A theory-based transport
  model with comprehensive physics}},\ }\href
  {https://doi.org/10.1063/1.2436852} {\bibfield  {journal} {\bibinfo
  {journal} {Physics of Plasmas}\ }\textbf {\bibinfo {volume} {14}},\ \bibinfo
  {pages} {055909} (\bibinfo {year} {2007})}\BibitemShut {NoStop}%
\bibitem [{\citenamefont {Kinsey}\ \emph {et~al.}(2008)\citenamefont {Kinsey},
  \citenamefont {Staebler},\ and\ \citenamefont {Waltz}}]{Kinsey2008POP}%
  \BibitemOpen
  \bibfield  {author} {\bibinfo {author} {\bibfnamefont {J.~E.}\ \bibnamefont
  {Kinsey}}, \bibinfo {author} {\bibfnamefont {G.~M.}\ \bibnamefont
  {Staebler}},\ and\ \bibinfo {author} {\bibfnamefont {R.~E.}\ \bibnamefont
  {Waltz}},\ }\bibfield  {title} {\bibinfo {title} {{The first transport code
  simulations using the {trapped gyro-Landau-fluid model}}},\ }\href
  {https://doi.org/10.1063/1.2889008} {\bibfield  {journal} {\bibinfo
  {journal} {Physics of Plasmas}\ }\textbf {\bibinfo {volume} {15}},\ \bibinfo
  {pages} {055908} (\bibinfo {year} {2008})}\BibitemShut {NoStop}%
\bibitem [{\citenamefont {Rafiq}\ \emph {et~al.}(2022)\citenamefont {Rafiq},
  \citenamefont {Wilson}, \citenamefont {Luo}, \citenamefont {Weiland},
  \citenamefont {Schuster}, \citenamefont {Pankin}, \citenamefont
  {Guttenfelder},\ and\ \citenamefont {Kaye}}]{Rafiq2022POP}%
  \BibitemOpen
  \bibfield  {author} {\bibinfo {author} {\bibfnamefont {T.}~\bibnamefont
  {Rafiq}}, \bibinfo {author} {\bibfnamefont {C.}~\bibnamefont {Wilson}},
  \bibinfo {author} {\bibfnamefont {L.}~\bibnamefont {Luo}}, \bibinfo {author}
  {\bibfnamefont {J.}~\bibnamefont {Weiland}}, \bibinfo {author} {\bibfnamefont
  {E.}~\bibnamefont {Schuster}}, \bibinfo {author} {\bibfnamefont {A.~Y.}\
  \bibnamefont {Pankin}}, \bibinfo {author} {\bibfnamefont {W.}~\bibnamefont
  {Guttenfelder}},\ and\ \bibinfo {author} {\bibfnamefont {S.}~\bibnamefont
  {Kaye}},\ }\bibfield  {title} {\bibinfo {title} {{Electron temperature
  gradient driven transport model for tokamak plasmas}},\ }\href
  {https://doi.org/10.1063/5.0104672} {\bibfield  {journal} {\bibinfo
  {journal} {Physics of Plasmas}\ }\textbf {\bibinfo {volume} {29}},\ \bibinfo
  {pages} {092503} (\bibinfo {year} {2022})}\BibitemShut {NoStop}%
\bibitem [{\citenamefont {Rafiq}\ \emph {et~al.}(2016)\citenamefont {Rafiq},
  \citenamefont {Weiland}, \citenamefont {Kritz}, \citenamefont {Luo},\ and\
  \citenamefont {Pankin}}]{Rafiq2016POP}%
  \BibitemOpen
  \bibfield  {author} {\bibinfo {author} {\bibfnamefont {T.}~\bibnamefont
  {Rafiq}}, \bibinfo {author} {\bibfnamefont {J.}~\bibnamefont {Weiland}},
  \bibinfo {author} {\bibfnamefont {A.~H.}\ \bibnamefont {Kritz}}, \bibinfo
  {author} {\bibfnamefont {L.}~\bibnamefont {Luo}},\ and\ \bibinfo {author}
  {\bibfnamefont {A.~Y.}\ \bibnamefont {Pankin}},\ }\bibfield  {title}
  {\bibinfo {title} {{Microtearing modes in tokamak discharges}},\ }\href
  {https://doi.org/10.1063/1.4953609} {\bibfield  {journal} {\bibinfo
  {journal} {Physics of Plasmas}\ }\textbf {\bibinfo {volume} {23}},\ \bibinfo
  {pages} {062507} (\bibinfo {year} {2016})}\BibitemShut {NoStop}%
\bibitem [{Wei(2012)}]{Weiland2012text}%
  \BibitemOpen
  \href@noop {} {\emph {\bibinfo {title} {Stability and Transport in Magnetic
  Confinement Systems}}}\ (\bibinfo  {publisher} {Springer Science \& Business
  Media},\ \bibinfo {year} {2012})\BibitemShut {NoStop}%
\bibitem [{\citenamefont {Rafiq}\ \emph {et~al.}(2010)\citenamefont {Rafiq},
  \citenamefont {Bateman}, \citenamefont {Kritz},\ and\ \citenamefont
  {Pankin}}]{Rafiq2010POP}%
  \BibitemOpen
  \bibfield  {author} {\bibinfo {author} {\bibfnamefont {T.}~\bibnamefont
  {Rafiq}}, \bibinfo {author} {\bibfnamefont {G.}~\bibnamefont {Bateman}},
  \bibinfo {author} {\bibfnamefont {A.~H.}\ \bibnamefont {Kritz}},\ and\
  \bibinfo {author} {\bibfnamefont {A.~Y.}\ \bibnamefont {Pankin}},\ }\bibfield
   {title} {\bibinfo {title} {{Development of drift-resistive-inertial
  ballooning transport model for tokamak edge plasmas}},\ }\href
  {https://doi.org/10.1063/1.3478979} {\bibfield  {journal} {\bibinfo
  {journal} {Physics of Plasmas}\ }\textbf {\bibinfo {volume} {17}},\ \bibinfo
  {pages} {082511} (\bibinfo {year} {2010})}\BibitemShut {NoStop}%
\bibitem [{\citenamefont {Rafiq}\ \emph {et~al.}(2024)\citenamefont {Rafiq},
  \citenamefont {Wilson}, \citenamefont {Clauser}, \citenamefont {Schuster},
  \citenamefont {Weiland}, \citenamefont {Anderson}, \citenamefont {Kaye},
  \citenamefont {Pankin}, \citenamefont {LeBlanc},\ and\ \citenamefont
  {Bell}}]{Rafiq2024NF}%
  \BibitemOpen
  \bibfield  {author} {\bibinfo {author} {\bibfnamefont {T.}~\bibnamefont
  {Rafiq}}, \bibinfo {author} {\bibfnamefont {C.}~\bibnamefont {Wilson}},
  \bibinfo {author} {\bibfnamefont {C.}~\bibnamefont {Clauser}}, \bibinfo
  {author} {\bibfnamefont {E.}~\bibnamefont {Schuster}}, \bibinfo {author}
  {\bibfnamefont {J.}~\bibnamefont {Weiland}}, \bibinfo {author} {\bibfnamefont
  {J.}~\bibnamefont {Anderson}}, \bibinfo {author} {\bibfnamefont
  {S.}~\bibnamefont {Kaye}}, \bibinfo {author} {\bibfnamefont {A.}~\bibnamefont
  {Pankin}}, \bibinfo {author} {\bibfnamefont {B.}~\bibnamefont {LeBlanc}},\
  and\ \bibinfo {author} {\bibfnamefont {R.}~\bibnamefont {Bell}},\ }\bibfield
  {title} {\bibinfo {title} {Predictive modeling of {NSTX} discharges with the
  updated multi-mode anomalous transport module},\ }\href
  {https://doi.org/10.1088/1741-4326/ad4d01} {\bibfield  {journal} {\bibinfo
  {journal} {Nuclear Fusion}\ }\textbf {\bibinfo {volume} {64}},\ \bibinfo
  {pages} {076024} (\bibinfo {year} {2024})}\BibitemShut {NoStop}%
\bibitem [{\citenamefont {Weiland}\ \emph {et~al.}(2023)\citenamefont
  {Weiland}, \citenamefont {Rafiq},\ and\ \citenamefont
  {Schuster}}]{Weiland2023POP}%
  \BibitemOpen
  \bibfield  {author} {\bibinfo {author} {\bibfnamefont {J.}~\bibnamefont
  {Weiland}}, \bibinfo {author} {\bibfnamefont {T.}~\bibnamefont {Rafiq}},\
  and\ \bibinfo {author} {\bibfnamefont {E.}~\bibnamefont {Schuster}},\
  }\bibfield  {title} {\bibinfo {title} {Fast particles in drift wave
  turbulence},\ }\href {https://doi.org/10.1063/5.0147320} {\bibfield
  {journal} {\bibinfo  {journal} {Physics of Plasmas}\ }\textbf {\bibinfo
  {volume} {30}},\ \bibinfo {pages} {042517} (\bibinfo {year}
  {2023})}\BibitemShut {NoStop}%
\bibitem [{\citenamefont {Rafiq}\ \emph
  {et~al.}(2023{\natexlab{a}})\citenamefont {Rafiq}, \citenamefont {Wilson},
  \citenamefont {Weiland}, \citenamefont {Whitall},\ and\ \citenamefont
  {Schuster}}]{Rafiq2023APS}%
  \BibitemOpen
  \bibfield  {author} {\bibinfo {author} {\bibfnamefont {T.}~\bibnamefont
  {Rafiq}}, \bibinfo {author} {\bibfnamefont {C.}~\bibnamefont {Wilson}},
  \bibinfo {author} {\bibfnamefont {J.}~\bibnamefont {Weiland}}, \bibinfo
  {author} {\bibfnamefont {L.}~\bibnamefont {Whitall}},\ and\ \bibinfo {author}
  {\bibfnamefont {E.}~\bibnamefont {Schuster}},\ }\bibfield  {title} {\bibinfo
  {title} {Towards efficient and practical models for energetic particle
  transport in tokamaks},\ }in\ \href
  {https://meetings.aps.org/Meeting/DPP23/Session/JP11.102} {\emph {\bibinfo
  {booktitle} {63$^{rd}$ APS DPP Meeting}}}\ (\bibinfo {address} {Denver, CO},\
  \bibinfo {year} {2023})\BibitemShut {NoStop}%
\bibitem [{\citenamefont {Rafiq}\ \emph {et~al.}(2021)\citenamefont {Rafiq},
  \citenamefont {Kaye}, \citenamefont {Guttenfelder}, \citenamefont {Weiland},
  \citenamefont {Schuster}, \citenamefont {Anderson},\ and\ \citenamefont
  {Luo}}]{Rafiq2021POP}%
  \BibitemOpen
  \bibfield  {author} {\bibinfo {author} {\bibfnamefont {T.}~\bibnamefont
  {Rafiq}}, \bibinfo {author} {\bibfnamefont {S.}~\bibnamefont {Kaye}},
  \bibinfo {author} {\bibfnamefont {W.}~\bibnamefont {Guttenfelder}}, \bibinfo
  {author} {\bibfnamefont {J.}~\bibnamefont {Weiland}}, \bibinfo {author}
  {\bibfnamefont {E.}~\bibnamefont {Schuster}}, \bibinfo {author}
  {\bibfnamefont {J.}~\bibnamefont {Anderson}},\ and\ \bibinfo {author}
  {\bibfnamefont {L.}~\bibnamefont {Luo}},\ }\bibfield  {title} {\bibinfo
  {title} {{Microtearing instabilities and electron thermal transport in low
  and high collisionality {NSTX} discharges}},\ }\href
  {https://doi.org/10.1063/5.0029120} {\bibfield  {journal} {\bibinfo
  {journal} {Physics of Plasmas}\ }\textbf {\bibinfo {volume} {28}},\ \bibinfo
  {pages} {022504} (\bibinfo {year} {2021})}\BibitemShut {NoStop}%
\bibitem [{\citenamefont {Hawryluk}(1980)}]{Hawryluk1980transp}%
  \BibitemOpen
  \bibfield  {author} {\bibinfo {author} {\bibfnamefont {R.~J.}\ \bibnamefont
  {Hawryluk}},\ }\bibfield  {title} {\bibinfo {title} {An empirical approach to
  tokamak transport},\ }in\ \href {https://transp.pppl.gov/files/Hawryluk.pdf}
  {\emph {\bibinfo {booktitle} {Physics of Plasmas Close to Thermonuclear
  Conditions}}},\ Vol.~\bibinfo {volume} {1},\ \bibinfo {editor} {edited by\
  \bibinfo {editor} {\bibfnamefont {B.}~\bibnamefont {Coppi}}, \bibinfo
  {editor} {\bibfnamefont {G.~G.}\ \bibnamefont {Leotta}}, \bibinfo {editor}
  {\bibfnamefont {D.}~\bibnamefont {Pfirsch}}, \bibinfo {editor} {\bibfnamefont
  {R.}~\bibnamefont {Pozzoli}},\ and\ \bibinfo {editor} {\bibfnamefont
  {E.}~\bibnamefont {Sindoni}}}\ (\bibinfo  {publisher} {CEC},\ \bibinfo
  {address} {Brussels},\ \bibinfo {year} {1980})\ pp.\ \bibinfo {pages}
  {19--46}\BibitemShut {NoStop}%
\bibitem [{\citenamefont {Goldston}\ \emph {et~al.}(1981)\citenamefont
  {Goldston}, \citenamefont {McCune}, \citenamefont {Towner}, \citenamefont
  {Davis}, \citenamefont {Hawryluk},\ and\ \citenamefont
  {Schmidt}}]{Goldston1981JCP}%
  \BibitemOpen
  \bibfield  {author} {\bibinfo {author} {\bibfnamefont {R.~J.}\ \bibnamefont
  {Goldston}}, \bibinfo {author} {\bibfnamefont {D.~C.}\ \bibnamefont
  {McCune}}, \bibinfo {author} {\bibfnamefont {H.~H.}\ \bibnamefont {Towner}},
  \bibinfo {author} {\bibfnamefont {S.~L.}\ \bibnamefont {Davis}}, \bibinfo
  {author} {\bibfnamefont {R.~J.}\ \bibnamefont {Hawryluk}},\ and\ \bibinfo
  {author} {\bibfnamefont {G.~L.}\ \bibnamefont {Schmidt}},\ }\bibfield
  {title} {\bibinfo {title} {New techniques for calculating heat and particle
  source rates due to neutral beam injection in axisymmetric tokamaks},\ }\href
  {https://doi.org/http://dx.doi.org/10.1016/0021-9991(81)90111-X} {\bibfield
  {journal} {\bibinfo  {journal} {Journal of Computational Physics}\ }\textbf
  {\bibinfo {volume} {43}},\ \bibinfo {pages} {61 } (\bibinfo {year}
  {1981})}\BibitemShut {NoStop}%
\bibitem [{\citenamefont {Poli}\ \emph {et~al.}(2018)\citenamefont {Poli},
  \citenamefont {Sachdev}, \citenamefont {Breslau}, \citenamefont
  {Gorelenkova},\ and\ \citenamefont {Yuan}}]{transp2018}%
  \BibitemOpen
  \bibfield  {author} {\bibinfo {author} {\bibfnamefont {F.}~\bibnamefont
  {Poli}}, \bibinfo {author} {\bibfnamefont {J.}~\bibnamefont {Sachdev}},
  \bibinfo {author} {\bibfnamefont {J.}~\bibnamefont {Breslau}}, \bibinfo
  {author} {\bibfnamefont {M.}~\bibnamefont {Gorelenkova}},\ and\ \bibinfo
  {author} {\bibfnamefont {X.}~\bibnamefont {Yuan}},\ }\href
  {https://dx.doi.org/10.11578/dc.20180627.4} {\bibinfo {title} {{TRANSP}
  v18.2}},\ \bibinfo {howpublished} {[Computer Software]
  \url{https://dx.doi.org/10.11578/dc.20180627.4}} (\bibinfo {year}
  {2018})\BibitemShut {NoStop}%
\bibitem [{\citenamefont {Grierson}\ \emph {et~al.}(2018)\citenamefont
  {Grierson}, \citenamefont {Yuan}, \citenamefont {Gorelenkova}, \citenamefont
  {Kaye}, \citenamefont {Logan}, \citenamefont {Meneghini}, \citenamefont
  {Haskey}, \citenamefont {Buchanan}, \citenamefont {Fitzgerald}, \citenamefont
  {Smith}, \citenamefont {Cui}, \citenamefont {Budny},\ and\ \citenamefont
  {Poli}}]{Grierson2018FST}%
  \BibitemOpen
  \bibfield  {author} {\bibinfo {author} {\bibfnamefont {B.}~\bibnamefont
  {Grierson}}, \bibinfo {author} {\bibfnamefont {X.}~\bibnamefont {Yuan}},
  \bibinfo {author} {\bibfnamefont {M.}~\bibnamefont {Gorelenkova}}, \bibinfo
  {author} {\bibfnamefont {S.}~\bibnamefont {Kaye}}, \bibinfo {author}
  {\bibfnamefont {N.}~\bibnamefont {Logan}}, \bibinfo {author} {\bibfnamefont
  {O.}~\bibnamefont {Meneghini}}, \bibinfo {author} {\bibfnamefont
  {S.}~\bibnamefont {Haskey}}, \bibinfo {author} {\bibfnamefont
  {J.}~\bibnamefont {Buchanan}}, \bibinfo {author} {\bibfnamefont
  {M.}~\bibnamefont {Fitzgerald}}, \bibinfo {author} {\bibfnamefont
  {S.}~\bibnamefont {Smith}}, \bibinfo {author} {\bibfnamefont
  {L.}~\bibnamefont {Cui}}, \bibinfo {author} {\bibfnamefont {R.}~\bibnamefont
  {Budny}},\ and\ \bibinfo {author} {\bibfnamefont {F.}~\bibnamefont {Poli}},\
  }\bibfield  {title} {\bibinfo {title} {Orchestrating {TRANSP} simulations for
  interpretative and predictive tokamak modeling with {OMFIT}},\ }\href
  {https://doi.org/10.1080/15361055.2017.1398585} {\bibfield  {journal}
  {\bibinfo  {journal} {Fusion Science and Technology}\ }\textbf {\bibinfo
  {volume} {74}},\ \bibinfo {pages} {101} (\bibinfo {year} {2018})}\BibitemShut
  {NoStop}%
\bibitem [{\citenamefont {Pankin}\ \emph {et~al.}(2025)\citenamefont {Pankin},
  \citenamefont {Breslau}, \citenamefont {Gorelenkova}, \citenamefont {Andre},
  \citenamefont {Grierson}, \citenamefont {Sachdev}, \citenamefont {Goliyad},\
  and\ \citenamefont {Perumpilly}}]{Pankin2025CPC}%
  \BibitemOpen
  \bibfield  {author} {\bibinfo {author} {\bibfnamefont {A.}~\bibnamefont
  {Pankin}}, \bibinfo {author} {\bibfnamefont {J.}~\bibnamefont {Breslau}},
  \bibinfo {author} {\bibfnamefont {M.}~\bibnamefont {Gorelenkova}}, \bibinfo
  {author} {\bibfnamefont {R.}~\bibnamefont {Andre}}, \bibinfo {author}
  {\bibfnamefont {B.}~\bibnamefont {Grierson}}, \bibinfo {author}
  {\bibfnamefont {J.}~\bibnamefont {Sachdev}}, \bibinfo {author} {\bibfnamefont
  {M.}~\bibnamefont {Goliyad}},\ and\ \bibinfo {author} {\bibfnamefont
  {G.}~\bibnamefont {Perumpilly}},\ }\bibfield  {title} {\bibinfo {title}
  {{TRANSP} integrated modeling code for interpretive and predictive analysis
  of tokamak plasmas},\ }\href
  {https://doi.org/https://doi.org/10.1016/j.cpc.2025.109611} {\bibfield
  {journal} {\bibinfo  {journal} {Computer Physics Communications}\ }\textbf
  {\bibinfo {volume} {312}},\ \bibinfo {pages} {109611} (\bibinfo {year}
  {2025})}\BibitemShut {NoStop}%
\bibitem [{\citenamefont {Jardin}\ \emph {et~al.}(2008)\citenamefont {Jardin},
  \citenamefont {Bateman}, \citenamefont {Hammett},\ and\ \citenamefont
  {Ku}}]{Jardin2008JCP}%
  \BibitemOpen
  \bibfield  {author} {\bibinfo {author} {\bibfnamefont {S.~C.}\ \bibnamefont
  {Jardin}}, \bibinfo {author} {\bibfnamefont {G.}~\bibnamefont {Bateman}},
  \bibinfo {author} {\bibfnamefont {G.~W.}\ \bibnamefont {Hammett}},\ and\
  \bibinfo {author} {\bibfnamefont {L.~P.}\ \bibnamefont {Ku}},\ }\bibfield
  {title} {\bibinfo {title} {Short note: On {1D} diffusion problems with a
  gradient-dependent diffusion coefficient},\ }\href
  {https://doi.org/10.1016/j.jcp.2008.06.032} {\bibfield  {journal} {\bibinfo
  {journal} {J. Comput. Phys.}\ }\textbf {\bibinfo {volume} {227}},\ \bibinfo
  {pages} {8769–8775} (\bibinfo {year} {2008})}\BibitemShut {NoStop}%
\bibitem [{\citenamefont {Yuan}\ \emph {et~al.}(2011)\citenamefont {Yuan},
  \citenamefont {McCune}, \citenamefont {Jardin}, \citenamefont {Budny},\ and\
  \citenamefont {Hammett}}]{Yuan2011APS}%
  \BibitemOpen
  \bibfield  {author} {\bibinfo {author} {\bibfnamefont {X.}~\bibnamefont
  {Yuan}}, \bibinfo {author} {\bibfnamefont {D.}~\bibnamefont {McCune}},
  \bibinfo {author} {\bibfnamefont {S.}~\bibnamefont {Jardin}}, \bibinfo
  {author} {\bibfnamefont {R.}~\bibnamefont {Budny}},\ and\ \bibinfo {author}
  {\bibfnamefont {G.}~\bibnamefont {Hammett}},\ }\bibfield  {title} {\bibinfo
  {title} {A modular, parallel, multi-region, predictive transport equation
  solver, installed and available in {PTRANSP}},\ }in\ \href
  {https://meetings.aps.org/Meeting/DPP11/Session/JP9.139} {\emph {\bibinfo
  {booktitle} {53$^\text{rd}$ APS DPP Meeting}}}\ (\bibinfo {address} {Salt
  Lake City, UT},\ \bibinfo {year} {2011})\BibitemShut {NoStop}%
\bibitem [{\citenamefont {Yuan}\ \emph {et~al.}(2012)\citenamefont {Yuan},
  \citenamefont {Jardin}, \citenamefont {Budny}, \citenamefont {Staebler},\
  and\ \citenamefont {Hammett}}]{Yuan2012APS}%
  \BibitemOpen
  \bibfield  {author} {\bibinfo {author} {\bibfnamefont {X.}~\bibnamefont
  {Yuan}}, \bibinfo {author} {\bibfnamefont {S.}~\bibnamefont {Jardin}},
  \bibinfo {author} {\bibfnamefont {R.}~\bibnamefont {Budny}}, \bibinfo
  {author} {\bibfnamefont {G.}~\bibnamefont {Staebler}},\ and\ \bibinfo
  {author} {\bibfnamefont {G.}~\bibnamefont {Hammett}},\ }\bibfield  {title}
  {\bibinfo {title} {New predictive capabilities in {PTRANSP} with
  {PTSOLVER}},\ }in\ \href
  {https://meetings.aps.org/Meeting/DPP12/Session/BP8.131} {\emph {\bibinfo
  {booktitle} {54$^\text{th}$ APS DPP Meeting}}}\ (\bibinfo {address}
  {Providence, RI},\ \bibinfo {year} {2012})\BibitemShut {NoStop}%
\bibitem [{\citenamefont {Yuan}\ \emph {et~al.}(2013)\citenamefont {Yuan},
  \citenamefont {Jardin}, \citenamefont {Hammett}, \citenamefont {Budny},\ and\
  \citenamefont {Staebler}}]{Yuan2013APS}%
  \BibitemOpen
  \bibfield  {author} {\bibinfo {author} {\bibfnamefont {X.}~\bibnamefont
  {Yuan}}, \bibinfo {author} {\bibfnamefont {S.}~\bibnamefont {Jardin}},
  \bibinfo {author} {\bibfnamefont {G.}~\bibnamefont {Hammett}}, \bibinfo
  {author} {\bibfnamefont {R.}~\bibnamefont {Budny}},\ and\ \bibinfo {author}
  {\bibfnamefont {G.}~\bibnamefont {Staebler}},\ }\bibfield  {title} {\bibinfo
  {title} {Parallel computing aspect in {TRANSP} with {PT-SOLVER}},\ }in\ \href
  {https://meetings.aps.org/Meeting/DPP13/Session/JP8.119} {\emph {\bibinfo
  {booktitle} {55$^\text{th}$ APS DPP Meeting}}}\ (\bibinfo {address} {Denver,
  CO},\ \bibinfo {year} {2013})\BibitemShut {NoStop}%
\bibitem [{\citenamefont {Poli}\ \emph {et~al.}(2015)\citenamefont {Poli},
  \citenamefont {Andre}, \citenamefont {Bertelli}, \citenamefont {Gerhardt},
  \citenamefont {Mueller},\ and\ \citenamefont {Taylor}}]{Poli2015NF}%
  \BibitemOpen
  \bibfield  {author} {\bibinfo {author} {\bibfnamefont {F.}~\bibnamefont
  {Poli}}, \bibinfo {author} {\bibfnamefont {R.}~\bibnamefont {Andre}},
  \bibinfo {author} {\bibfnamefont {N.}~\bibnamefont {Bertelli}}, \bibinfo
  {author} {\bibfnamefont {S.}~\bibnamefont {Gerhardt}}, \bibinfo {author}
  {\bibfnamefont {D.}~\bibnamefont {Mueller}},\ and\ \bibinfo {author}
  {\bibfnamefont {G.}~\bibnamefont {Taylor}},\ }\bibfield  {title} {\bibinfo
  {title} {Simulations towards the achievement of non-inductive current ramp-up
  and sustainment in the {National Spherical Torus Experiment Upgrade}},\
  }\href {https://doi.org/10.1088/0029-5515/55/12/123011} {\bibfield  {journal}
  {\bibinfo  {journal} {Nuclear Fusion}\ }\textbf {\bibinfo {volume} {55}},\
  \bibinfo {pages} {123011} (\bibinfo {year} {2015})}\BibitemShut {NoStop}%
\bibitem [{\citenamefont {Lopez}\ and\ \citenamefont
  {Poli}(2018)}]{Lopez2018PPCF}%
  \BibitemOpen
  \bibfield  {author} {\bibinfo {author} {\bibfnamefont {N.~A.}\ \bibnamefont
  {Lopez}}\ and\ \bibinfo {author} {\bibfnamefont {F.~M.}\ \bibnamefont
  {Poli}},\ }\bibfield  {title} {\bibinfo {title} {Regarding the optimization
  of {O1-mode ECRH} and the feasibility of {EBW} startup on {NSTX-U}},\ }\href
  {https://doi.org/10.1088/1361-6587/aabaa8} {\bibfield  {journal} {\bibinfo
  {journal} {Plasma Physics and Controlled Fusion}\ }\textbf {\bibinfo {volume}
  {60}},\ \bibinfo {pages} {065007} (\bibinfo {year} {2018})}\BibitemShut
  {NoStop}%
\bibitem [{\citenamefont {Podestà}\ \emph {et~al.}(2024)\citenamefont
  {Podestà}, \citenamefont {Cruz-Zabala}, \citenamefont {Poli}, \citenamefont
  {Dominguez-Palacios}, \citenamefont {Berkery}, \citenamefont {Garcia-Muñoz},
  \citenamefont {Viezzer}, \citenamefont {Mancini}, \citenamefont {Segado},
  \citenamefont {Velarde},\ and\ \citenamefont {Kaye}}]{Podesta2024PPCF}%
  \BibitemOpen
  \bibfield  {author} {\bibinfo {author} {\bibfnamefont {M.}~\bibnamefont
  {Podestà}}, \bibinfo {author} {\bibfnamefont {D.~J.}\ \bibnamefont
  {Cruz-Zabala}}, \bibinfo {author} {\bibfnamefont {F.~M.}\ \bibnamefont
  {Poli}}, \bibinfo {author} {\bibfnamefont {J.}~\bibnamefont
  {Dominguez-Palacios}}, \bibinfo {author} {\bibfnamefont {J.~W.}\ \bibnamefont
  {Berkery}}, \bibinfo {author} {\bibfnamefont {M.}~\bibnamefont
  {Garcia-Muñoz}}, \bibinfo {author} {\bibfnamefont {E.}~\bibnamefont
  {Viezzer}}, \bibinfo {author} {\bibfnamefont {A.}~\bibnamefont {Mancini}},
  \bibinfo {author} {\bibfnamefont {J.}~\bibnamefont {Segado}}, \bibinfo
  {author} {\bibfnamefont {L.}~\bibnamefont {Velarde}},\ and\ \bibinfo {author}
  {\bibfnamefont {S.~M.}\ \bibnamefont {Kaye}},\ }\bibfield  {title} {\bibinfo
  {title} {{NBI} optimization on {SMART} and implications for scenario
  development},\ }\href {https://doi.org/10.1088/1361-6587/ad2edc} {\bibfield
  {journal} {\bibinfo  {journal} {Plasma Physics and Controlled Fusion}\
  }\textbf {\bibinfo {volume} {66}},\ \bibinfo {pages} {045021} (\bibinfo
  {year} {2024})}\BibitemShut {NoStop}%
\bibitem [{\citenamefont {Cruz-Zabala}\ \emph {et~al.}(2024)\citenamefont
  {Cruz-Zabala}, \citenamefont {Podesta}, \citenamefont {Poli}, \citenamefont
  {Kaye}, \citenamefont {Garcia-Munoz}, \citenamefont {Viezzer},\ and\
  \citenamefont {Berkery}}]{CruzZabala2024NF}%
  \BibitemOpen
  \bibfield  {author} {\bibinfo {author} {\bibfnamefont {D.~J.}\ \bibnamefont
  {Cruz-Zabala}}, \bibinfo {author} {\bibfnamefont {M.}~\bibnamefont
  {Podesta}}, \bibinfo {author} {\bibfnamefont {F.~M.}\ \bibnamefont {Poli}},
  \bibinfo {author} {\bibfnamefont {S.~M.}\ \bibnamefont {Kaye}}, \bibinfo
  {author} {\bibfnamefont {M.}~\bibnamefont {Garcia-Munoz}}, \bibinfo {author}
  {\bibfnamefont {E.}~\bibnamefont {Viezzer}},\ and\ \bibinfo {author}
  {\bibfnamefont {J.~W.}\ \bibnamefont {Berkery}},\ }\bibfield  {title}
  {\bibinfo {title} {Performance prediction applying different reduced
  turbulence models to the {SMART} tokamak},\ }\href
  {http://iopscience.iop.org/article/10.1088/1741-4326/ad8a70} {\bibfield
  {journal} {\bibinfo  {journal} {Nuclear Fusion}\ } (\bibinfo {year}
  {2024})}\BibitemShut {NoStop}%
\bibitem [{\citenamefont {Rafiq}\ \emph
  {et~al.}(2023{\natexlab{b}})\citenamefont {Rafiq}, \citenamefont {Wang},
  \citenamefont {Morosohk}, \citenamefont {Schuster}, \citenamefont {Weiland},
  \citenamefont {Choi},\ and\ \citenamefont {Kim}}]{Rafiq2023plasma}%
  \BibitemOpen
  \bibfield  {author} {\bibinfo {author} {\bibfnamefont {T.}~\bibnamefont
  {Rafiq}}, \bibinfo {author} {\bibfnamefont {Z.}~\bibnamefont {Wang}},
  \bibinfo {author} {\bibfnamefont {S.}~\bibnamefont {Morosohk}}, \bibinfo
  {author} {\bibfnamefont {E.}~\bibnamefont {Schuster}}, \bibinfo {author}
  {\bibfnamefont {J.}~\bibnamefont {Weiland}}, \bibinfo {author} {\bibfnamefont
  {W.}~\bibnamefont {Choi}},\ and\ \bibinfo {author} {\bibfnamefont {H.-T.}\
  \bibnamefont {Kim}},\ }\bibfield  {title} {\bibinfo {title} {Validating the
  {Multi-Mode Model’s} ability to reproduce diverse tokamak scenarios},\
  }\href {https://doi.org/10.3390/plasma6030030} {\bibfield  {journal}
  {\bibinfo  {journal} {Plasma}\ }\textbf {\bibinfo {volume} {6}},\ \bibinfo
  {pages} {435} (\bibinfo {year} {2023}{\natexlab{b}})}\BibitemShut {NoStop}%
\bibitem [{\citenamefont {Pankin}\ \emph {et~al.}(2018)\citenamefont {Pankin},
  \citenamefont {Kritz}, \citenamefont {Rafiq}, \citenamefont {Garofalo},
  \citenamefont {Holod},\ and\ \citenamefont {Weiland}}]{Pankin2018POP}%
  \BibitemOpen
  \bibfield  {author} {\bibinfo {author} {\bibfnamefont {A.~Y.}\ \bibnamefont
  {Pankin}}, \bibinfo {author} {\bibfnamefont {A.~H.}\ \bibnamefont {Kritz}},
  \bibinfo {author} {\bibfnamefont {T.}~\bibnamefont {Rafiq}}, \bibinfo
  {author} {\bibfnamefont {A.~M.}\ \bibnamefont {Garofalo}}, \bibinfo {author}
  {\bibfnamefont {I.}~\bibnamefont {Holod}},\ and\ \bibinfo {author}
  {\bibfnamefont {J.}~\bibnamefont {Weiland}},\ }\bibfield  {title} {\bibinfo
  {title} {{Extending the validation of multi-mode model for anomalous
  transport to high beta poloidal tokamak scenario in {DIII-D}}},\ }\href
  {https://doi.org/10.1063/1.5010339} {\bibfield  {journal} {\bibinfo
  {journal} {Physics of Plasmas}\ }\textbf {\bibinfo {volume} {25}},\ \bibinfo
  {pages} {052505} (\bibinfo {year} {2018})}\BibitemShut {NoStop}%
\bibitem [{\citenamefont {Houlberg}\ \emph {et~al.}(1997)\citenamefont
  {Houlberg}, \citenamefont {Shaing}, \citenamefont {Hirshman},\ and\
  \citenamefont {Zarnstorff}}]{Houlberg1997POP}%
  \BibitemOpen
  \bibfield  {author} {\bibinfo {author} {\bibfnamefont {W.~A.}\ \bibnamefont
  {Houlberg}}, \bibinfo {author} {\bibfnamefont {K.~C.}\ \bibnamefont
  {Shaing}}, \bibinfo {author} {\bibfnamefont {S.~P.}\ \bibnamefont
  {Hirshman}},\ and\ \bibinfo {author} {\bibfnamefont {M.~C.}\ \bibnamefont
  {Zarnstorff}},\ }\bibfield  {title} {\bibinfo {title} {{Bootstrap current and
  neoclassical transport in tokamaks of arbitrary collisionality and aspect
  ratio}},\ }\href {https://doi.org/10.1063/1.872465} {\bibfield  {journal}
  {\bibinfo  {journal} {Physics of Plasmas}\ }\textbf {\bibinfo {volume} {4}},\
  \bibinfo {pages} {3230} (\bibinfo {year} {1997})}\BibitemShut {NoStop}%
\bibitem [{\citenamefont {Avdeeva}\ \emph {et~al.}(2023)\citenamefont
  {Avdeeva}, \citenamefont {Thome}, \citenamefont {Smith}, \citenamefont
  {Battaglia}, \citenamefont {Clauser}, \citenamefont {Guttenfelder},
  \citenamefont {Kaye}, \citenamefont {McClenaghan}, \citenamefont {Meneghini},
  \citenamefont {Odstrcil},\ and\ \citenamefont {Staebler}}]{Avdeeva2023NF}%
  \BibitemOpen
  \bibfield  {author} {\bibinfo {author} {\bibfnamefont {G.}~\bibnamefont
  {Avdeeva}}, \bibinfo {author} {\bibfnamefont {K.}~\bibnamefont {Thome}},
  \bibinfo {author} {\bibfnamefont {S.}~\bibnamefont {Smith}}, \bibinfo
  {author} {\bibfnamefont {D.}~\bibnamefont {Battaglia}}, \bibinfo {author}
  {\bibfnamefont {C.}~\bibnamefont {Clauser}}, \bibinfo {author} {\bibfnamefont
  {W.}~\bibnamefont {Guttenfelder}}, \bibinfo {author} {\bibfnamefont
  {S.}~\bibnamefont {Kaye}}, \bibinfo {author} {\bibfnamefont {J.}~\bibnamefont
  {McClenaghan}}, \bibinfo {author} {\bibfnamefont {O.}~\bibnamefont
  {Meneghini}}, \bibinfo {author} {\bibfnamefont {T.}~\bibnamefont
  {Odstrcil}},\ and\ \bibinfo {author} {\bibfnamefont {G.}~\bibnamefont
  {Staebler}},\ }\bibfield  {title} {\bibinfo {title} {Energy transport
  analysis of {NSTX} plasmas with the {TGLF} turbulent and {NEO} neoclassical
  transport models},\ }\href {https://doi.org/10.1088/1741-4326/acfc56}
  {\bibfield  {journal} {\bibinfo  {journal} {Nuclear Fusion}\ }\textbf
  {\bibinfo {volume} {63}},\ \bibinfo {pages} {126020} (\bibinfo {year}
  {2023})}\BibitemShut {NoStop}%
\bibitem [{\citenamefont {Avdeeva}\ \emph {et~al.}(2024)\citenamefont
  {Avdeeva}, \citenamefont {Thome}, \citenamefont {Berkery}, \citenamefont
  {Kaye}, \citenamefont {McClenaghan}, \citenamefont {Meneghini}, \citenamefont
  {Odstrcil}, \citenamefont {Sabbagh}, \citenamefont {Smith},\ and\
  \citenamefont {Turnbull}}]{Avdeeva2024PPCF}%
  \BibitemOpen
  \bibfield  {author} {\bibinfo {author} {\bibfnamefont {G.}~\bibnamefont
  {Avdeeva}}, \bibinfo {author} {\bibfnamefont {K.~E.}\ \bibnamefont {Thome}},
  \bibinfo {author} {\bibfnamefont {J.~W.}\ \bibnamefont {Berkery}}, \bibinfo
  {author} {\bibfnamefont {S.~M.}\ \bibnamefont {Kaye}}, \bibinfo {author}
  {\bibfnamefont {J.}~\bibnamefont {McClenaghan}}, \bibinfo {author}
  {\bibfnamefont {O.}~\bibnamefont {Meneghini}}, \bibinfo {author}
  {\bibfnamefont {T.}~\bibnamefont {Odstrcil}}, \bibinfo {author}
  {\bibfnamefont {S.~A.}\ \bibnamefont {Sabbagh}}, \bibinfo {author}
  {\bibfnamefont {S.~P.}\ \bibnamefont {Smith}},\ and\ \bibinfo {author}
  {\bibfnamefont {A.~D.}\ \bibnamefont {Turnbull}},\ }\bibfield  {title}
  {\bibinfo {title} {Accuracy of kinetic equilibrium reconstruction of {NSTX}
  and {NSTX-U} plasmas and its impact on the transport and stability
  analysis},\ }\href {https://doi.org/10.1088/1361-6587/ad788a} {\bibfield
  {journal} {\bibinfo  {journal} {Plasma Physics and Controlled Fusion}\
  }\textbf {\bibinfo {volume} {66}},\ \bibinfo {pages} {115003} (\bibinfo
  {year} {2024})}\BibitemShut {NoStop}%
\bibitem [{\citenamefont {Gerhardt}\ \emph
  {et~al.}(2011{\natexlab{a}})\citenamefont {Gerhardt}, \citenamefont {Gates},
  \citenamefont {Kaye}, \citenamefont {Maingi}, \citenamefont {Menard},
  \citenamefont {Sabbagh}, \citenamefont {Soukhanovskii}, \citenamefont {Bell},
  \citenamefont {Bell}, \citenamefont {Canik}, \citenamefont {Fredrickson},
  \citenamefont {Kaita}, \citenamefont {Kolemen}, \citenamefont {Kugel},
  \citenamefont {Blanc}, \citenamefont {Mastrovito}, \citenamefont {Mueller},\
  and\ \citenamefont {Yuh}}]{Gerhardt2011NFat}%
  \BibitemOpen
  \bibfield  {author} {\bibinfo {author} {\bibfnamefont {S.}~\bibnamefont
  {Gerhardt}}, \bibinfo {author} {\bibfnamefont {D.}~\bibnamefont {Gates}},
  \bibinfo {author} {\bibfnamefont {S.}~\bibnamefont {Kaye}}, \bibinfo {author}
  {\bibfnamefont {R.}~\bibnamefont {Maingi}}, \bibinfo {author} {\bibfnamefont
  {J.}~\bibnamefont {Menard}}, \bibinfo {author} {\bibfnamefont
  {S.}~\bibnamefont {Sabbagh}}, \bibinfo {author} {\bibfnamefont
  {V.}~\bibnamefont {Soukhanovskii}}, \bibinfo {author} {\bibfnamefont
  {M.}~\bibnamefont {Bell}}, \bibinfo {author} {\bibfnamefont {R.}~\bibnamefont
  {Bell}}, \bibinfo {author} {\bibfnamefont {J.}~\bibnamefont {Canik}},
  \bibinfo {author} {\bibfnamefont {E.}~\bibnamefont {Fredrickson}}, \bibinfo
  {author} {\bibfnamefont {R.}~\bibnamefont {Kaita}}, \bibinfo {author}
  {\bibfnamefont {E.}~\bibnamefont {Kolemen}}, \bibinfo {author} {\bibfnamefont
  {H.}~\bibnamefont {Kugel}}, \bibinfo {author} {\bibfnamefont {B.~L.}\
  \bibnamefont {Blanc}}, \bibinfo {author} {\bibfnamefont {D.}~\bibnamefont
  {Mastrovito}}, \bibinfo {author} {\bibfnamefont {D.}~\bibnamefont
  {Mueller}},\ and\ \bibinfo {author} {\bibfnamefont {H.}~\bibnamefont {Yuh}},\
  }\bibfield  {title} {\bibinfo {title} {Recent progress towards an advanced
  spherical torus operating point in {NSTX}},\ }\href
  {https://doi.org/10.1088/0029-5515/51/7/073031} {\bibfield  {journal}
  {\bibinfo  {journal} {Nuclear Fusion}\ }\textbf {\bibinfo {volume} {51}},\
  \bibinfo {pages} {073031} (\bibinfo {year} {2011}{\natexlab{a}})}\BibitemShut
  {NoStop}%
\bibitem [{\citenamefont {Gerhardt}\ \emph
  {et~al.}(2011{\natexlab{b}})\citenamefont {Gerhardt}, \citenamefont
  {Fredrickson}, \citenamefont {Gates}, \citenamefont {Kaye}, \citenamefont
  {Menard}, \citenamefont {Bell}, \citenamefont {Bell}, \citenamefont {Blanc},
  \citenamefont {Kugel}, \citenamefont {Sabbagh},\ and\ \citenamefont
  {Yuh}}]{Gerhardt2011NFcur}%
  \BibitemOpen
  \bibfield  {author} {\bibinfo {author} {\bibfnamefont {S.}~\bibnamefont
  {Gerhardt}}, \bibinfo {author} {\bibfnamefont {E.}~\bibnamefont
  {Fredrickson}}, \bibinfo {author} {\bibfnamefont {D.}~\bibnamefont {Gates}},
  \bibinfo {author} {\bibfnamefont {S.}~\bibnamefont {Kaye}}, \bibinfo {author}
  {\bibfnamefont {J.}~\bibnamefont {Menard}}, \bibinfo {author} {\bibfnamefont
  {M.}~\bibnamefont {Bell}}, \bibinfo {author} {\bibfnamefont {R.}~\bibnamefont
  {Bell}}, \bibinfo {author} {\bibfnamefont {B.~L.}\ \bibnamefont {Blanc}},
  \bibinfo {author} {\bibfnamefont {H.}~\bibnamefont {Kugel}}, \bibinfo
  {author} {\bibfnamefont {S.}~\bibnamefont {Sabbagh}},\ and\ \bibinfo {author}
  {\bibfnamefont {H.}~\bibnamefont {Yuh}},\ }\bibfield  {title} {\bibinfo
  {title} {Calculation of the non-inductive current profile in high-performance
  {NSTX} plasmas},\ }\href {https://doi.org/10.1088/0029-5515/51/3/033004}
  {\bibfield  {journal} {\bibinfo  {journal} {Nuclear Fusion}\ }\textbf
  {\bibinfo {volume} {51}},\ \bibinfo {pages} {033004} (\bibinfo {year}
  {2011}{\natexlab{b}})}\BibitemShut {NoStop}%
\bibitem [{\citenamefont {Ren}\ \emph {et~al.}(2012)\citenamefont {Ren},
  \citenamefont {Guttenfelder}, \citenamefont {Kaye}, \citenamefont
  {Mazzucato}, \citenamefont {Bell}, \citenamefont {Diallo}, \citenamefont
  {Domier}, \citenamefont {LeBlanc}, \citenamefont {Lee}, \citenamefont
  {Smith},\ and\ \citenamefont {Yuh}}]{Ren2012POP}%
  \BibitemOpen
  \bibfield  {author} {\bibinfo {author} {\bibfnamefont {Y.}~\bibnamefont
  {Ren}}, \bibinfo {author} {\bibfnamefont {W.}~\bibnamefont {Guttenfelder}},
  \bibinfo {author} {\bibfnamefont {S.~M.}\ \bibnamefont {Kaye}}, \bibinfo
  {author} {\bibfnamefont {E.}~\bibnamefont {Mazzucato}}, \bibinfo {author}
  {\bibfnamefont {R.~E.}\ \bibnamefont {Bell}}, \bibinfo {author}
  {\bibfnamefont {A.}~\bibnamefont {Diallo}}, \bibinfo {author} {\bibfnamefont
  {C.~W.}\ \bibnamefont {Domier}}, \bibinfo {author} {\bibfnamefont {B.~P.}\
  \bibnamefont {LeBlanc}}, \bibinfo {author} {\bibfnamefont {K.~C.}\
  \bibnamefont {Lee}}, \bibinfo {author} {\bibfnamefont {D.~R.}\ \bibnamefont
  {Smith}},\ and\ \bibinfo {author} {\bibfnamefont {H.}~\bibnamefont {Yuh}},\
  }\bibfield  {title} {\bibinfo {title} {{Experimental study of parametric
  dependence of electron-scale turbulence in a spherical tokamak}},\ }\href
  {https://doi.org/10.1063/1.4719689} {\bibfield  {journal} {\bibinfo
  {journal} {Physics of Plasmas}\ }\textbf {\bibinfo {volume} {19}},\ \bibinfo
  {pages} {056125} (\bibinfo {year} {2012})}\BibitemShut {NoStop}%
\bibitem [{\citenamefont {Ren}\ \emph {et~al.}(2013)\citenamefont {Ren},
  \citenamefont {Guttenfelder}, \citenamefont {Kaye}, \citenamefont
  {Mazzucato}, \citenamefont {Bell}, \citenamefont {Diallo}, \citenamefont
  {Domier}, \citenamefont {LeBlanc}, \citenamefont {Lee}, \citenamefont
  {Podesta}, \citenamefont {Smith},\ and\ \citenamefont {Yuh}}]{Ren2013NF}%
  \BibitemOpen
  \bibfield  {author} {\bibinfo {author} {\bibfnamefont {Y.}~\bibnamefont
  {Ren}}, \bibinfo {author} {\bibfnamefont {W.}~\bibnamefont {Guttenfelder}},
  \bibinfo {author} {\bibfnamefont {S.}~\bibnamefont {Kaye}}, \bibinfo {author}
  {\bibfnamefont {E.}~\bibnamefont {Mazzucato}}, \bibinfo {author}
  {\bibfnamefont {R.}~\bibnamefont {Bell}}, \bibinfo {author} {\bibfnamefont
  {A.}~\bibnamefont {Diallo}}, \bibinfo {author} {\bibfnamefont
  {C.}~\bibnamefont {Domier}}, \bibinfo {author} {\bibfnamefont
  {B.}~\bibnamefont {LeBlanc}}, \bibinfo {author} {\bibfnamefont
  {K.}~\bibnamefont {Lee}}, \bibinfo {author} {\bibfnamefont {M.}~\bibnamefont
  {Podesta}}, \bibinfo {author} {\bibfnamefont {D.}~\bibnamefont {Smith}},\
  and\ \bibinfo {author} {\bibfnamefont {H.}~\bibnamefont {Yuh}},\ }\bibfield
  {title} {\bibinfo {title} {Electron-scale turbulence spectra and plasma
  thermal transport responding to continuous {E × B} shear ramp-up in a
  spherical tokamak},\ }\href {https://doi.org/10.1088/0029-5515/53/8/083007}
  {\bibfield  {journal} {\bibinfo  {journal} {Nuclear Fusion}\ }\textbf
  {\bibinfo {volume} {53}},\ \bibinfo {pages} {083007} (\bibinfo {year}
  {2013})}\BibitemShut {NoStop}%
\bibitem [{\citenamefont {Gerhardt}\ \emph {et~al.}(2014)\citenamefont
  {Gerhardt}, \citenamefont {Canik}, \citenamefont {Maingi}, \citenamefont
  {Battaglia}, \citenamefont {Bell}, \citenamefont {Guttenfelder},
  \citenamefont {LeBlanc}, \citenamefont {Smith}, \citenamefont {Yuh},\ and\
  \citenamefont {Sabbagh}}]{Gerhardt2014NF}%
  \BibitemOpen
  \bibfield  {author} {\bibinfo {author} {\bibfnamefont {S.}~\bibnamefont
  {Gerhardt}}, \bibinfo {author} {\bibfnamefont {J.}~\bibnamefont {Canik}},
  \bibinfo {author} {\bibfnamefont {R.}~\bibnamefont {Maingi}}, \bibinfo
  {author} {\bibfnamefont {D.}~\bibnamefont {Battaglia}}, \bibinfo {author}
  {\bibfnamefont {R.}~\bibnamefont {Bell}}, \bibinfo {author} {\bibfnamefont
  {W.}~\bibnamefont {Guttenfelder}}, \bibinfo {author} {\bibfnamefont
  {B.}~\bibnamefont {LeBlanc}}, \bibinfo {author} {\bibfnamefont
  {D.}~\bibnamefont {Smith}}, \bibinfo {author} {\bibfnamefont
  {H.}~\bibnamefont {Yuh}},\ and\ \bibinfo {author} {\bibfnamefont
  {S.}~\bibnamefont {Sabbagh}},\ }\bibfield  {title} {\bibinfo {title}
  {Progress in understanding the enhanced pedestal {H-mode} in {NSTX}},\ }\href
  {https://doi.org/10.1088/0029-5515/54/8/083021} {\bibfield  {journal}
  {\bibinfo  {journal} {Nuclear Fusion}\ }\textbf {\bibinfo {volume} {54}},\
  \bibinfo {pages} {083021} (\bibinfo {year} {2014})}\BibitemShut {NoStop}%
\bibitem [{\citenamefont {{Ruiz Ruiz}}\ \emph {et~al.}(2019)\citenamefont
  {{Ruiz Ruiz}}, \citenamefont {Guttenfelder}, \citenamefont {White},
  \citenamefont {Howard}, \citenamefont {Candy}, \citenamefont {Ren},
  \citenamefont {Smith}, \citenamefont {Loureiro}, \citenamefont {Holland},\
  and\ \citenamefont {Domier}}]{RuizRuiz2019PPCF}%
  \BibitemOpen
  \bibfield  {author} {\bibinfo {author} {\bibfnamefont {J.}~\bibnamefont
  {{Ruiz Ruiz}}}, \bibinfo {author} {\bibfnamefont {W.}~\bibnamefont
  {Guttenfelder}}, \bibinfo {author} {\bibfnamefont {A.~E.}\ \bibnamefont
  {White}}, \bibinfo {author} {\bibfnamefont {N.~T.}\ \bibnamefont {Howard}},
  \bibinfo {author} {\bibfnamefont {J.}~\bibnamefont {Candy}}, \bibinfo
  {author} {\bibfnamefont {Y.}~\bibnamefont {Ren}}, \bibinfo {author}
  {\bibfnamefont {D.~R.}\ \bibnamefont {Smith}}, \bibinfo {author}
  {\bibfnamefont {N.~F.}\ \bibnamefont {Loureiro}}, \bibinfo {author}
  {\bibfnamefont {C.}~\bibnamefont {Holland}},\ and\ \bibinfo {author}
  {\bibfnamefont {C.~W.}\ \bibnamefont {Domier}},\ }\bibfield  {title}
  {\bibinfo {title} {Validation of gyrokinetic simulations of a {National
  Spherical Torus eXperiment} {H-mode} plasma and comparisons with a high-k
  scattering synthetic diagnostic},\ }\href
  {https://doi.org/10.1088/1361-6587/ab4742} {\bibfield  {journal} {\bibinfo
  {journal} {Plasma Physics and Controlled Fusion}\ }\textbf {\bibinfo {volume}
  {61}},\ \bibinfo {pages} {115015} (\bibinfo {year} {2019})}\BibitemShut
  {NoStop}%
\bibitem [{\citenamefont {Ren}\ \emph {et~al.}(2019)\citenamefont {Ren},
  \citenamefont {Wang}, \citenamefont {Guttenfelder}, \citenamefont {Kaye},
  \citenamefont {Ruiz-Ruiz}, \citenamefont {Ethier}, \citenamefont {Bell},
  \citenamefont {LeBlanc}, \citenamefont {Mazzucato}, \citenamefont {Smith},
  \citenamefont {Domier},\ and\ \citenamefont {Yuh}}]{Ren2020NF}%
  \BibitemOpen
  \bibfield  {author} {\bibinfo {author} {\bibfnamefont {Y.}~\bibnamefont
  {Ren}}, \bibinfo {author} {\bibfnamefont {W.}~\bibnamefont {Wang}}, \bibinfo
  {author} {\bibfnamefont {W.}~\bibnamefont {Guttenfelder}}, \bibinfo {author}
  {\bibfnamefont {S.}~\bibnamefont {Kaye}}, \bibinfo {author} {\bibfnamefont
  {J.}~\bibnamefont {Ruiz-Ruiz}}, \bibinfo {author} {\bibfnamefont
  {S.}~\bibnamefont {Ethier}}, \bibinfo {author} {\bibfnamefont
  {R.}~\bibnamefont {Bell}}, \bibinfo {author} {\bibfnamefont {B.}~\bibnamefont
  {LeBlanc}}, \bibinfo {author} {\bibfnamefont {E.}~\bibnamefont {Mazzucato}},
  \bibinfo {author} {\bibfnamefont {D.}~\bibnamefont {Smith}}, \bibinfo
  {author} {\bibfnamefont {C.}~\bibnamefont {Domier}},\ and\ \bibinfo {author}
  {\bibfnamefont {H.}~\bibnamefont {Yuh}},\ }\bibfield  {title} {\bibinfo
  {title} {Exploring the regime of validity of global gyrokinetic simulations
  with spherical tokamak plasmas},\ }\href
  {https://doi.org/10.1088/1741-4326/ab5bf5} {\bibfield  {journal} {\bibinfo
  {journal} {Nuclear Fusion}\ }\textbf {\bibinfo {volume} {60}},\ \bibinfo
  {pages} {026005} (\bibinfo {year} {2019})}\BibitemShut {NoStop}%
\bibitem [{\citenamefont {Battaglia}\ \emph {et~al.}(2020)\citenamefont
  {Battaglia}, \citenamefont {Guttenfelder}, \citenamefont {Bell},
  \citenamefont {Diallo}, \citenamefont {Ferraro}, \citenamefont {Fredrickson},
  \citenamefont {Gerhardt}, \citenamefont {Kaye}, \citenamefont {Maingi},\ and\
  \citenamefont {Smith}}]{Battaglia2020POP}%
  \BibitemOpen
  \bibfield  {author} {\bibinfo {author} {\bibfnamefont {D.~J.}\ \bibnamefont
  {Battaglia}}, \bibinfo {author} {\bibfnamefont {W.}~\bibnamefont
  {Guttenfelder}}, \bibinfo {author} {\bibfnamefont {R.~E.}\ \bibnamefont
  {Bell}}, \bibinfo {author} {\bibfnamefont {A.}~\bibnamefont {Diallo}},
  \bibinfo {author} {\bibfnamefont {N.}~\bibnamefont {Ferraro}}, \bibinfo
  {author} {\bibfnamefont {E.}~\bibnamefont {Fredrickson}}, \bibinfo {author}
  {\bibfnamefont {S.~P.}\ \bibnamefont {Gerhardt}}, \bibinfo {author}
  {\bibfnamefont {S.~M.}\ \bibnamefont {Kaye}}, \bibinfo {author}
  {\bibfnamefont {R.}~\bibnamefont {Maingi}},\ and\ \bibinfo {author}
  {\bibfnamefont {D.~R.}\ \bibnamefont {Smith}},\ }\bibfield  {title} {\bibinfo
  {title} {{Enhanced pedestal {H-mode} at low edge ion collisionality on
  {NSTX}}},\ }\href {https://doi.org/10.1063/5.0011614} {\bibfield  {journal}
  {\bibinfo  {journal} {Physics of Plasmas}\ }\textbf {\bibinfo {volume}
  {27}},\ \bibinfo {pages} {072511} (\bibinfo {year} {2020})}\BibitemShut
  {NoStop}%
\bibitem [{\citenamefont {Clauser}\ \emph {et~al.}(2025)\citenamefont
  {Clauser}, \citenamefont {Rafiq}, \citenamefont {Parisi}, \citenamefont
  {Avdeeva}, \citenamefont {Guttenfelder}, \citenamefont {Schuster},\ and\
  \citenamefont {Wilson}}]{Clauser2025POP}%
  \BibitemOpen
  \bibfield  {author} {\bibinfo {author} {\bibfnamefont {C.~F.}\ \bibnamefont
  {Clauser}}, \bibinfo {author} {\bibfnamefont {T.}~\bibnamefont {Rafiq}},
  \bibinfo {author} {\bibfnamefont {J.}~\bibnamefont {Parisi}}, \bibinfo
  {author} {\bibfnamefont {G.}~\bibnamefont {Avdeeva}}, \bibinfo {author}
  {\bibfnamefont {W.}~\bibnamefont {Guttenfelder}}, \bibinfo {author}
  {\bibfnamefont {E.}~\bibnamefont {Schuster}},\ and\ \bibinfo {author}
  {\bibfnamefont {C.}~\bibnamefont {Wilson}},\ }\bibfield  {title} {\bibinfo
  {title} {Electron temperature gradient instability and transport analysis in
  {NSTX} and {NSTX-U} plasmas},\ }\href {https://doi.org/10.1063/5.0232697}
  {\bibfield  {journal} {\bibinfo  {journal} {Physics of Plasmas}\ }\textbf
  {\bibinfo {volume} {32}},\ \bibinfo {pages} {022305} (\bibinfo {year}
  {2025})}\BibitemShut {NoStop}%
\bibitem [{\citenamefont {{ITER Physics Expert Group on Confinement and
  Transport}}\ \emph {et~al.}(1999)\citenamefont {{ITER Physics Expert Group on
  Confinement and Transport}}, \citenamefont {{ITER Physics Expert Group on
  Confinement Modelling and Database}},\ and\ \citenamefont {{ITER Physics
  Basis Editors}}}]{ITER1999NF}%
  \BibitemOpen
  \bibfield  {author} {\bibinfo {author} {\bibnamefont {{ITER Physics Expert
  Group on Confinement and Transport}}}, \bibinfo {author} {\bibnamefont {{ITER
  Physics Expert Group on Confinement Modelling and Database}}},\ and\ \bibinfo
  {author} {\bibnamefont {{ITER Physics Basis Editors}}},\ }\bibfield  {title}
  {\bibinfo {title} {Chapter 2: Plasma confinement and transport},\ }\href
  {https://doi.org/10.1088/0029-5515/39/12/302} {\bibfield  {journal} {\bibinfo
   {journal} {Nuclear Fusion}\ }\textbf {\bibinfo {volume} {39}},\ \bibinfo
  {pages} {2175} (\bibinfo {year} {1999})}\BibitemShut {NoStop}%
\bibitem [{\citenamefont {Abbate}\ \emph {et~al.}(2024)\citenamefont {Abbate},
  \citenamefont {Fable}, \citenamefont {Grierson}, \citenamefont {Pankin},
  \citenamefont {Tardini},\ and\ \citenamefont {Kolemen}}]{Abbate2024POP}%
  \BibitemOpen
  \bibfield  {author} {\bibinfo {author} {\bibfnamefont {J.}~\bibnamefont
  {Abbate}}, \bibinfo {author} {\bibfnamefont {E.}~\bibnamefont {Fable}},
  \bibinfo {author} {\bibfnamefont {B.}~\bibnamefont {Grierson}}, \bibinfo
  {author} {\bibfnamefont {A.}~\bibnamefont {Pankin}}, \bibinfo {author}
  {\bibfnamefont {G.}~\bibnamefont {Tardini}},\ and\ \bibinfo {author}
  {\bibfnamefont {E.}~\bibnamefont {Kolemen}},\ }\bibfield  {title} {\bibinfo
  {title} {{Large-database cross-verification and validation of tokamak
  transport models using baselines for comparison}},\ }\href
  {https://doi.org/10.1063/5.0190908} {\bibfield  {journal} {\bibinfo
  {journal} {Physics of Plasmas}\ }\textbf {\bibinfo {volume} {31}},\ \bibinfo
  {pages} {042506} (\bibinfo {year} {2024})}\BibitemShut {NoStop}%
\bibitem [{\citenamefont {Spearman}(1904)}]{Spearman1904AJP}%
  \BibitemOpen
  \bibfield  {author} {\bibinfo {author} {\bibfnamefont {C.}~\bibnamefont
  {Spearman}},\ }\bibfield  {title} {\bibinfo {title} {The proof and
  measurement of association between two things},\ }\href
  {http://www.jstor.org/stable/1412159} {\bibfield  {journal} {\bibinfo
  {journal} {The American Journal of Psychology}\ }\textbf {\bibinfo {volume}
  {15}},\ \bibinfo {pages} {72} (\bibinfo {year} {1904})}\BibitemShut {NoStop}%
\bibitem [{\citenamefont {Kinsey}\ \emph {et~al.}(2007)\citenamefont {Kinsey},
  \citenamefont {Waltz},\ and\ \citenamefont {Candy}}]{Kinsey2007POP}%
  \BibitemOpen
  \bibfield  {author} {\bibinfo {author} {\bibfnamefont {J.~E.}\ \bibnamefont
  {Kinsey}}, \bibinfo {author} {\bibfnamefont {R.~E.}\ \bibnamefont {Waltz}},\
  and\ \bibinfo {author} {\bibfnamefont {J.}~\bibnamefont {Candy}},\ }\bibfield
   {title} {\bibinfo {title} {The effect of plasma shaping on turbulent
  transport and {E×B} shear quenching in nonlinear gyrokinetic simulations},\
  }\href {https://doi.org/10.1063/1.2786857} {\bibfield  {journal} {\bibinfo
  {journal} {Physics of Plasmas}\ }\textbf {\bibinfo {volume} {14}},\ \bibinfo
  {pages} {102306} (\bibinfo {year} {2007})}\BibitemShut {NoStop}%
\bibitem [{\citenamefont {Candy}\ \emph {et~al.}(2016)\citenamefont {Candy},
  \citenamefont {Belli},\ and\ \citenamefont {Bravenec}}]{Candy2016JCP}%
  \BibitemOpen
  \bibfield  {author} {\bibinfo {author} {\bibfnamefont {J.}~\bibnamefont
  {Candy}}, \bibinfo {author} {\bibfnamefont {E.}~\bibnamefont {Belli}},\ and\
  \bibinfo {author} {\bibfnamefont {R.}~\bibnamefont {Bravenec}},\ }\bibfield
  {title} {\bibinfo {title} {A high-accuracy eulerian gyrokinetic solver for
  collisional plasmas},\ }\href
  {https://doi.org/https://doi.org/10.1016/j.jcp.2016.07.039} {\bibfield
  {journal} {\bibinfo  {journal} {Journal of Computational Physics}\ }\textbf
  {\bibinfo {volume} {324}},\ \bibinfo {pages} {73} (\bibinfo {year}
  {2016})}\BibitemShut {NoStop}%
\bibitem [{\citenamefont {Meneghini}\ \emph {et~al.}(2024)\citenamefont
  {Meneghini}, \citenamefont {Slendebroek}, \citenamefont {Lyons},
  \citenamefont {McLaughlin}, \citenamefont {McClenaghan}, \citenamefont
  {Stagner}, \citenamefont {Harvey}, \citenamefont {Neiser}, \citenamefont
  {Ghiozzi}, \citenamefont {Dose}, \citenamefont {Guterl}, \citenamefont
  {Zalzali}, \citenamefont {Cote}, \citenamefont {Shi}, \citenamefont
  {Weisberg}, \citenamefont {Smith}, \citenamefont {Grierson},\ and\
  \citenamefont {Candy}}]{Meneghini2024arxiv}%
  \BibitemOpen
  \bibfield  {author} {\bibinfo {author} {\bibfnamefont {O.}~\bibnamefont
  {Meneghini}}, \bibinfo {author} {\bibfnamefont {T.}~\bibnamefont
  {Slendebroek}}, \bibinfo {author} {\bibfnamefont {B.~C.}\ \bibnamefont
  {Lyons}}, \bibinfo {author} {\bibfnamefont {K.}~\bibnamefont {McLaughlin}},
  \bibinfo {author} {\bibfnamefont {J.}~\bibnamefont {McClenaghan}}, \bibinfo
  {author} {\bibfnamefont {L.}~\bibnamefont {Stagner}}, \bibinfo {author}
  {\bibfnamefont {J.}~\bibnamefont {Harvey}}, \bibinfo {author} {\bibfnamefont
  {T.~F.}\ \bibnamefont {Neiser}}, \bibinfo {author} {\bibfnamefont
  {A.}~\bibnamefont {Ghiozzi}}, \bibinfo {author} {\bibfnamefont
  {G.}~\bibnamefont {Dose}}, \bibinfo {author} {\bibfnamefont {J.}~\bibnamefont
  {Guterl}}, \bibinfo {author} {\bibfnamefont {A.}~\bibnamefont {Zalzali}},
  \bibinfo {author} {\bibfnamefont {T.}~\bibnamefont {Cote}}, \bibinfo {author}
  {\bibfnamefont {N.}~\bibnamefont {Shi}}, \bibinfo {author} {\bibfnamefont
  {D.}~\bibnamefont {Weisberg}}, \bibinfo {author} {\bibfnamefont {S.~P.}\
  \bibnamefont {Smith}}, \bibinfo {author} {\bibfnamefont {B.~A.}\ \bibnamefont
  {Grierson}},\ and\ \bibinfo {author} {\bibfnamefont {J.}~\bibnamefont
  {Candy}},\ }\href {https://arxiv.org/abs/2409.05894} {\bibinfo {title}
  {{FUSE} (fusion synthesis engine): A next generation framework for integrated
  design of fusion pilot plants}} (\bibinfo {year} {2024}),\ \Eprint
  {https://arxiv.org/abs/2409.05894} {arXiv:2409.05894 [physics.plasm-ph]}
  \BibitemShut {NoStop}%
\bibitem [{\citenamefont {Neiser}\ \emph {et~al.}(2024)\citenamefont {Neiser},
  \citenamefont {Meneghini}, \citenamefont {Slendebroek}, \citenamefont
  {Smith}, \citenamefont {McClenaghan}, \citenamefont {Ghiozzi}, \citenamefont
  {Patel}, \citenamefont {Nelson}, \citenamefont {Avdeeva}, \citenamefont
  {Roach}, \citenamefont {Casson}, \citenamefont {Dudding}, \citenamefont
  {Staebler}, \citenamefont {Hall}, \citenamefont {Lyons}, \citenamefont
  {Belli}, \citenamefont {Candy}, \citenamefont {Yu}, \citenamefont {Sammuli},
  \citenamefont {Nazikian},\ and\ \citenamefont {Osborne}}]{Neiser2024APS}%
  \BibitemOpen
  \bibfield  {author} {\bibinfo {author} {\bibfnamefont {T.}~\bibnamefont
  {Neiser}}, \bibinfo {author} {\bibfnamefont {O.}~\bibnamefont {Meneghini}},
  \bibinfo {author} {\bibfnamefont {T.}~\bibnamefont {Slendebroek}}, \bibinfo
  {author} {\bibfnamefont {S.}~\bibnamefont {Smith}}, \bibinfo {author}
  {\bibfnamefont {J.}~\bibnamefont {McClenaghan}}, \bibinfo {author}
  {\bibfnamefont {A.}~\bibnamefont {Ghiozzi}}, \bibinfo {author} {\bibfnamefont
  {B.}~\bibnamefont {Patel}}, \bibinfo {author} {\bibfnamefont
  {A.}~\bibnamefont {Nelson}}, \bibinfo {author} {\bibfnamefont
  {G.}~\bibnamefont {Avdeeva}}, \bibinfo {author} {\bibfnamefont
  {C.}~\bibnamefont {Roach}}, \bibinfo {author} {\bibfnamefont
  {F.}~\bibnamefont {Casson}}, \bibinfo {author} {\bibfnamefont
  {H.}~\bibnamefont {Dudding}}, \bibinfo {author} {\bibfnamefont
  {G.}~\bibnamefont {Staebler}}, \bibinfo {author} {\bibfnamefont
  {J.}~\bibnamefont {Hall}}, \bibinfo {author} {\bibfnamefont {B.}~\bibnamefont
  {Lyons}}, \bibinfo {author} {\bibfnamefont {E.}~\bibnamefont {Belli}},
  \bibinfo {author} {\bibfnamefont {J.}~\bibnamefont {Candy}}, \bibinfo
  {author} {\bibfnamefont {R.}~\bibnamefont {Yu}}, \bibinfo {author}
  {\bibfnamefont {B.}~\bibnamefont {Sammuli}}, \bibinfo {author} {\bibfnamefont
  {R.}~\bibnamefont {Nazikian}},\ and\ \bibinfo {author} {\bibfnamefont
  {T.}~\bibnamefont {Osborne}},\ }\bibfield  {title} {\bibinfo {title} {Large
  database validation of {TGLF} on {DIII-D} and {MAST-U} plasmas},\ }in\ \href
  {https://meetings.aps.org/Meeting/DPP24/Session/PP12.97} {\emph {\bibinfo
  {booktitle} {66$^{th}$ APS DPP Meeting}}}\ (\bibinfo {address} {Atlanta,
  GA},\ \bibinfo {year} {2024})\BibitemShut {NoStop}%
\bibitem [{\citenamefont {Belli}\ and\ \citenamefont
  {Candy}(2008)}]{Belli2008PPCF}%
  \BibitemOpen
  \bibfield  {author} {\bibinfo {author} {\bibfnamefont {E.~A.}\ \bibnamefont
  {Belli}}\ and\ \bibinfo {author} {\bibfnamefont {J.}~\bibnamefont {Candy}},\
  }\bibfield  {title} {\bibinfo {title} {Kinetic calculation of neoclassical
  transport including self-consistent electron and impurity dynamics},\ }\href
  {https://doi.org/10.1088/0741-3335/50/9/095010} {\bibfield  {journal}
  {\bibinfo  {journal} {Plasma Physics and Controlled Fusion}\ }\textbf
  {\bibinfo {volume} {50}},\ \bibinfo {pages} {095010} (\bibinfo {year}
  {2008})}\BibitemShut {NoStop}%
\bibitem [{\citenamefont {Belli}\ and\ \citenamefont
  {Candy}(2011)}]{Belli2011PPCF}%
  \BibitemOpen
  \bibfield  {author} {\bibinfo {author} {\bibfnamefont {E.~A.}\ \bibnamefont
  {Belli}}\ and\ \bibinfo {author} {\bibfnamefont {J.}~\bibnamefont {Candy}},\
  }\bibfield  {title} {\bibinfo {title} {Full linearized fokker–planck
  collisions in neoclassical transport simulations},\ }\href
  {https://doi.org/10.1088/0741-3335/54/1/015015} {\bibfield  {journal}
  {\bibinfo  {journal} {Plasma Physics and Controlled Fusion}\ }\textbf
  {\bibinfo {volume} {54}},\ \bibinfo {pages} {015015} (\bibinfo {year}
  {2011})}\BibitemShut {NoStop}%
\bibitem [{\citenamefont {Pankin}\ \emph {et~al.}(2004)\citenamefont {Pankin},
  \citenamefont {McCune}, \citenamefont {Andre}, \citenamefont {Bateman},\ and\
  \citenamefont {Kritz}}]{Pankin2004CPC}%
  \BibitemOpen
  \bibfield  {author} {\bibinfo {author} {\bibfnamefont {A.}~\bibnamefont
  {Pankin}}, \bibinfo {author} {\bibfnamefont {D.}~\bibnamefont {McCune}},
  \bibinfo {author} {\bibfnamefont {R.}~\bibnamefont {Andre}}, \bibinfo
  {author} {\bibfnamefont {G.}~\bibnamefont {Bateman}},\ and\ \bibinfo {author}
  {\bibfnamefont {A.}~\bibnamefont {Kritz}},\ }\bibfield  {title} {\bibinfo
  {title} {The tokamak {Monte Carlo} fast ion module nubeam in the {National
  Transport Code Collaboration} library},\ }\href
  {https://doi.org/http://doi.org/10.1016/j.cpc.2003.11.002} {\bibfield
  {journal} {\bibinfo  {journal} {Computer Physics Communications}\ }\textbf
  {\bibinfo {volume} {159}},\ \bibinfo {pages} {157 } (\bibinfo {year}
  {2004})}\BibitemShut {NoStop}%
\bibitem [{\citenamefont {Coppi}(1988)}]{Coppi1988PLA}%
  \BibitemOpen
  \bibfield  {author} {\bibinfo {author} {\bibfnamefont {B.}~\bibnamefont
  {Coppi}},\ }\bibfield  {title} {\bibinfo {title} {Profile consistency: Global
  and nonlinear transport},\ }\href
  {https://doi.org/https://doi.org/10.1016/0375-9601(88)90908-5} {\bibfield
  {journal} {\bibinfo  {journal} {Physics Letters A}\ }\textbf {\bibinfo
  {volume} {128}},\ \bibinfo {pages} {193} (\bibinfo {year}
  {1988})}\BibitemShut {NoStop}%
\bibitem [{\citenamefont {Stutman}\ \emph {et~al.}(2009)\citenamefont
  {Stutman}, \citenamefont {Delgado-Aparicio}, \citenamefont {Gorelenkov},
  \citenamefont {Finkenthal}, \citenamefont {Fredrickson}, \citenamefont
  {Kaye}, \citenamefont {Mazzucato},\ and\ \citenamefont
  {Tritz}}]{Stutman2009PRL}%
  \BibitemOpen
  \bibfield  {author} {\bibinfo {author} {\bibfnamefont {D.}~\bibnamefont
  {Stutman}}, \bibinfo {author} {\bibfnamefont {L.}~\bibnamefont
  {Delgado-Aparicio}}, \bibinfo {author} {\bibfnamefont {N.}~\bibnamefont
  {Gorelenkov}}, \bibinfo {author} {\bibfnamefont {M.}~\bibnamefont
  {Finkenthal}}, \bibinfo {author} {\bibfnamefont {E.}~\bibnamefont
  {Fredrickson}}, \bibinfo {author} {\bibfnamefont {S.}~\bibnamefont {Kaye}},
  \bibinfo {author} {\bibfnamefont {E.}~\bibnamefont {Mazzucato}},\ and\
  \bibinfo {author} {\bibfnamefont {K.}~\bibnamefont {Tritz}},\ }\bibfield
  {title} {\bibinfo {title} {Correlation between electron transport and shear
  {\Alfven} activity in the {National Spherical Torus Experiment}},\ }\href
  {https://doi.org/10.1103/PhysRevLett.102.115002} {\bibfield  {journal}
  {\bibinfo  {journal} {Phys. Rev. Lett.}\ }\textbf {\bibinfo {volume} {102}},\
  \bibinfo {pages} {115002} (\bibinfo {year} {2009})}\BibitemShut {NoStop}%
\bibitem [{\citenamefont {Ren}\ \emph {et~al.}(2017)\citenamefont {Ren},
  \citenamefont {Belova}, \citenamefont {Gorelenkov}, \citenamefont
  {Guttenfelder}, \citenamefont {Kaye}, \citenamefont {Mazzucato},
  \citenamefont {Peterson}, \citenamefont {Smith}, \citenamefont {Stutman},
  \citenamefont {Tritz}, \citenamefont {Wang}, \citenamefont {Yuh},
  \citenamefont {Bell}, \citenamefont {Domier},\ and\ \citenamefont
  {LeBlanc}}]{Ren2017NF}%
  \BibitemOpen
  \bibfield  {author} {\bibinfo {author} {\bibfnamefont {Y.}~\bibnamefont
  {Ren}}, \bibinfo {author} {\bibfnamefont {E.}~\bibnamefont {Belova}},
  \bibinfo {author} {\bibfnamefont {N.}~\bibnamefont {Gorelenkov}}, \bibinfo
  {author} {\bibfnamefont {W.}~\bibnamefont {Guttenfelder}}, \bibinfo {author}
  {\bibfnamefont {S.~M.}\ \bibnamefont {Kaye}}, \bibinfo {author}
  {\bibfnamefont {E.}~\bibnamefont {Mazzucato}}, \bibinfo {author}
  {\bibfnamefont {J.~L.}\ \bibnamefont {Peterson}}, \bibinfo {author}
  {\bibfnamefont {D.~R.}\ \bibnamefont {Smith}}, \bibinfo {author}
  {\bibfnamefont {D.}~\bibnamefont {Stutman}}, \bibinfo {author} {\bibfnamefont
  {K.}~\bibnamefont {Tritz}}, \bibinfo {author} {\bibfnamefont {W.~X.}\
  \bibnamefont {Wang}}, \bibinfo {author} {\bibfnamefont {H.}~\bibnamefont
  {Yuh}}, \bibinfo {author} {\bibfnamefont {R.~E.}\ \bibnamefont {Bell}},
  \bibinfo {author} {\bibfnamefont {C.~W.}\ \bibnamefont {Domier}},\ and\
  \bibinfo {author} {\bibfnamefont {B.~P.}\ \bibnamefont {LeBlanc}},\
  }\bibfield  {title} {\bibinfo {title} {Recent progress in understanding
  electron thermal transport in {NSTX}},\ }\href
  {http://stacks.iop.org/0029-5515/57/i=7/a=072002} {\bibfield  {journal}
  {\bibinfo  {journal} {Nuclear Fusion}\ }\textbf {\bibinfo {volume} {57}},\
  \bibinfo {pages} {072002} (\bibinfo {year} {2017})}\BibitemShut {NoStop}%
\bibitem [{\citenamefont {Gorelenkov}\ \emph {et~al.}(2010)\citenamefont
  {Gorelenkov}, \citenamefont {Stutman}, \citenamefont {Tritz}, \citenamefont
  {Boozer}, \citenamefont {Delgado-Aparicio}, \citenamefont {Fredrickson},
  \citenamefont {Kaye},\ and\ \citenamefont {White}}]{Gorelenkov2010NF}%
  \BibitemOpen
  \bibfield  {author} {\bibinfo {author} {\bibfnamefont {N.~N.}\ \bibnamefont
  {Gorelenkov}}, \bibinfo {author} {\bibfnamefont {D.}~\bibnamefont {Stutman}},
  \bibinfo {author} {\bibfnamefont {K.}~\bibnamefont {Tritz}}, \bibinfo
  {author} {\bibfnamefont {A.}~\bibnamefont {Boozer}}, \bibinfo {author}
  {\bibfnamefont {L.}~\bibnamefont {Delgado-Aparicio}}, \bibinfo {author}
  {\bibfnamefont {E.}~\bibnamefont {Fredrickson}}, \bibinfo {author}
  {\bibfnamefont {S.}~\bibnamefont {Kaye}},\ and\ \bibinfo {author}
  {\bibfnamefont {R.}~\bibnamefont {White}},\ }\bibfield  {title} {\bibinfo
  {title} {Anomalous electron transport due to multiple high frequency beam ion
  driven {\Alfven} eigenmodes},\ }\href
  {http://stacks.iop.org/0029-5515/50/i=8/a=084012} {\bibfield  {journal}
  {\bibinfo  {journal} {Nuclear Fusion}\ }\textbf {\bibinfo {volume} {50}},\
  \bibinfo {pages} {084012} (\bibinfo {year} {2010})}\BibitemShut {NoStop}%
\bibitem [{\citenamefont {Kolesnichenko}\ \emph
  {et~al.}(2010{\natexlab{a}})\citenamefont {Kolesnichenko}, \citenamefont
  {Yakovenko},\ and\ \citenamefont {Lutsenko}}]{Kolesnichenko2010PRL}%
  \BibitemOpen
  \bibfield  {author} {\bibinfo {author} {\bibfnamefont {Y.~I.}\ \bibnamefont
  {Kolesnichenko}}, \bibinfo {author} {\bibfnamefont {Y.~V.}\ \bibnamefont
  {Yakovenko}},\ and\ \bibinfo {author} {\bibfnamefont {V.~V.}\ \bibnamefont
  {Lutsenko}},\ }\bibfield  {title} {\bibinfo {title} {Channeling of the energy
  and momentum during energetic-ion-driven instabilities in fusion plasmas},\
  }\href {https://doi.org/10.1103/PhysRevLett.104.075001} {\bibfield  {journal}
  {\bibinfo  {journal} {Phys. Rev. Lett.}\ }\textbf {\bibinfo {volume} {104}},\
  \bibinfo {pages} {075001} (\bibinfo {year} {2010}{\natexlab{a}})}\BibitemShut
  {NoStop}%
\bibitem [{\citenamefont {Kolesnichenko}\ \emph
  {et~al.}(2010{\natexlab{b}})\citenamefont {Kolesnichenko}, \citenamefont
  {Yakovenko}, \citenamefont {Lutsenko}, \citenamefont {White},\ and\
  \citenamefont {Weller}}]{Kolesnichenko2010NF}%
  \BibitemOpen
  \bibfield  {author} {\bibinfo {author} {\bibfnamefont {Y.}~\bibnamefont
  {Kolesnichenko}}, \bibinfo {author} {\bibfnamefont {Y.}~\bibnamefont
  {Yakovenko}}, \bibinfo {author} {\bibfnamefont {V.~V.}\ \bibnamefont
  {Lutsenko}}, \bibinfo {author} {\bibfnamefont {R.~B.}\ \bibnamefont
  {White}},\ and\ \bibinfo {author} {\bibfnamefont {A.}~\bibnamefont
  {Weller}},\ }\bibfield  {title} {\bibinfo {title} {Effects of
  energetic-ion-driven instabilities on plasma heating, transport and rotation
  in toroidal systems},\ }\href
  {http://stacks.iop.org/0029-5515/50/i=8/a=084011} {\bibfield  {journal}
  {\bibinfo  {journal} {Nuclear Fusion}\ }\textbf {\bibinfo {volume} {50}},\
  \bibinfo {pages} {084011} (\bibinfo {year} {2010}{\natexlab{b}})}\BibitemShut
  {NoStop}%
\bibitem [{\citenamefont {Belova}\ \emph {et~al.}(2015)\citenamefont {Belova},
  \citenamefont {Gorelenkov}, \citenamefont {Fredrickson}, \citenamefont
  {Tritz},\ and\ \citenamefont {Crocker}}]{Belova2015PRL}%
  \BibitemOpen
  \bibfield  {author} {\bibinfo {author} {\bibfnamefont {E.~V.}\ \bibnamefont
  {Belova}}, \bibinfo {author} {\bibfnamefont {N.~N.}\ \bibnamefont
  {Gorelenkov}}, \bibinfo {author} {\bibfnamefont {E.~D.}\ \bibnamefont
  {Fredrickson}}, \bibinfo {author} {\bibfnamefont {K.}~\bibnamefont {Tritz}},\
  and\ \bibinfo {author} {\bibfnamefont {N.~A.}\ \bibnamefont {Crocker}},\
  }\bibfield  {title} {\bibinfo {title} {Coupling of neutral-beam-driven
  compressional {\Alfven} eigenmodes to kinetic {\Alfven} waves in {NSTX}
  tokamak and energy channeling},\ }\href
  {https://doi.org/10.1103/PhysRevLett.115.015001} {\bibfield  {journal}
  {\bibinfo  {journal} {Phys. Rev. Lett.}\ }\textbf {\bibinfo {volume} {115}},\
  \bibinfo {pages} {015001} (\bibinfo {year} {2015})}\BibitemShut {NoStop}%
\bibitem [{\citenamefont {Belova}\ \emph {et~al.}(2017)\citenamefont {Belova},
  \citenamefont {Gorelenkov}, \citenamefont {Crocker}, \citenamefont {Lestz},
  \citenamefont {Fredrickson}, \citenamefont {Tang},\ and\ \citenamefont
  {Tritz}}]{Belova2017POP}%
  \BibitemOpen
  \bibfield  {author} {\bibinfo {author} {\bibfnamefont {E.~V.}\ \bibnamefont
  {Belova}}, \bibinfo {author} {\bibfnamefont {N.~N.}\ \bibnamefont
  {Gorelenkov}}, \bibinfo {author} {\bibfnamefont {N.~A.}\ \bibnamefont
  {Crocker}}, \bibinfo {author} {\bibfnamefont {J.~B.}\ \bibnamefont {Lestz}},
  \bibinfo {author} {\bibfnamefont {E.~D.}\ \bibnamefont {Fredrickson}},
  \bibinfo {author} {\bibfnamefont {S.}~\bibnamefont {Tang}},\ and\ \bibinfo
  {author} {\bibfnamefont {K.}~\bibnamefont {Tritz}},\ }\bibfield  {title}
  {\bibinfo {title} {Nonlinear simulations of beam-driven compressional
  {\Alfven} eigenmodes in {NSTX}},\ }\href {https://doi.org/10.1063/1.4979278}
  {\bibfield  {journal} {\bibinfo  {journal} {Physics of Plasmas}\ }\textbf
  {\bibinfo {volume} {24}},\ \bibinfo {pages} {042505} (\bibinfo {year}
  {2017})}\BibitemShut {NoStop}%
\bibitem [{\citenamefont {Lestz}\ \emph {et~al.}(2021)\citenamefont {Lestz},
  \citenamefont {Belova},\ and\ \citenamefont {Gorelenkov}}]{Lestz2021NF}%
  \BibitemOpen
  \bibfield  {author} {\bibinfo {author} {\bibfnamefont {J.~B.}\ \bibnamefont
  {Lestz}}, \bibinfo {author} {\bibfnamefont {E.~V.}\ \bibnamefont {Belova}},\
  and\ \bibinfo {author} {\bibfnamefont {N.~N.}\ \bibnamefont {Gorelenkov}},\
  }\bibfield  {title} {\bibinfo {title} {Hybrid simulations of sub-cyclotron
  compressional and global {\Alfven} eigenmode stability in spherical
  tokamaks},\ }\href {https://doi.org/10.1088/1741-4326/abf028} {\bibfield
  {journal} {\bibinfo  {journal} {Nuclear Fusion}\ }\textbf {\bibinfo {volume}
  {61}},\ \bibinfo {pages} {086016} (\bibinfo {year} {2021})}\BibitemShut
  {NoStop}%
\bibitem [{\citenamefont {Jardin}\ \emph {et~al.}(2022)\citenamefont {Jardin},
  \citenamefont {Ferraro}, \citenamefont {Guttenfelder}, \citenamefont {Kaye},\
  and\ \citenamefont {Munaretto}}]{Jardin2022PRL}%
  \BibitemOpen
  \bibfield  {author} {\bibinfo {author} {\bibfnamefont {S.~C.}\ \bibnamefont
  {Jardin}}, \bibinfo {author} {\bibfnamefont {N.~M.}\ \bibnamefont {Ferraro}},
  \bibinfo {author} {\bibfnamefont {W.}~\bibnamefont {Guttenfelder}}, \bibinfo
  {author} {\bibfnamefont {S.~M.}\ \bibnamefont {Kaye}},\ and\ \bibinfo
  {author} {\bibfnamefont {S.}~\bibnamefont {Munaretto}},\ }\bibfield  {title}
  {\bibinfo {title} {Ideal {MHD} limited electron temperature in spherical
  tokamaks},\ }\href {https://doi.org/10.1103/PhysRevLett.128.245001}
  {\bibfield  {journal} {\bibinfo  {journal} {Phys. Rev. Lett.}\ }\textbf
  {\bibinfo {volume} {128}},\ \bibinfo {pages} {245001} (\bibinfo {year}
  {2022})}\BibitemShut {NoStop}%
\bibitem [{\citenamefont {Jardin}\ \emph {et~al.}(2023)\citenamefont {Jardin},
  \citenamefont {Ferraro}, \citenamefont {Guttenfelder}, \citenamefont {Kaye},\
  and\ \citenamefont {Munaretto}}]{Jardin2023POP}%
  \BibitemOpen
  \bibfield  {author} {\bibinfo {author} {\bibfnamefont {S.~C.}\ \bibnamefont
  {Jardin}}, \bibinfo {author} {\bibfnamefont {N.~M.}\ \bibnamefont {Ferraro}},
  \bibinfo {author} {\bibfnamefont {W.}~\bibnamefont {Guttenfelder}}, \bibinfo
  {author} {\bibfnamefont {S.~M.}\ \bibnamefont {Kaye}},\ and\ \bibinfo
  {author} {\bibfnamefont {S.}~\bibnamefont {Munaretto}},\ }\bibfield  {title}
  {\bibinfo {title} {{Ideal MHD induced temperature flattening in spherical
  tokamaks}},\ }\href {https://doi.org/10.1063/5.0141858} {\bibfield  {journal}
  {\bibinfo  {journal} {Physics of Plasmas}\ }\textbf {\bibinfo {volume}
  {30}},\ \bibinfo {pages} {042507} (\bibinfo {year} {2023})}\BibitemShut
  {NoStop}%
\bibitem [{\citenamefont {Meneghini}\ \emph {et~al.}(2015)\citenamefont
  {Meneghini}, \citenamefont {Smith}, \citenamefont {Lao}, \citenamefont
  {Izacard}, \citenamefont {Ren}, \citenamefont {Park}, \citenamefont {Candy},
  \citenamefont {Wang}, \citenamefont {Luna}, \citenamefont {Izzo},
  \citenamefont {Grierson}, \citenamefont {Snyder}, \citenamefont {Holland},
  \citenamefont {Penna}, \citenamefont {Lu}, \citenamefont {Raum},
  \citenamefont {McCubbin}, \citenamefont {Orlov}, \citenamefont {Belli},
  \citenamefont {Ferraro}, \citenamefont {Prater}, \citenamefont {Osborne},
  \citenamefont {Turnbull},\ and\ \citenamefont {Staebler}}]{Meneghini2015NF}%
  \BibitemOpen
  \bibfield  {author} {\bibinfo {author} {\bibfnamefont {O.}~\bibnamefont
  {Meneghini}}, \bibinfo {author} {\bibfnamefont {S.}~\bibnamefont {Smith}},
  \bibinfo {author} {\bibfnamefont {L.}~\bibnamefont {Lao}}, \bibinfo {author}
  {\bibfnamefont {O.}~\bibnamefont {Izacard}}, \bibinfo {author} {\bibfnamefont
  {Q.}~\bibnamefont {Ren}}, \bibinfo {author} {\bibfnamefont {J.}~\bibnamefont
  {Park}}, \bibinfo {author} {\bibfnamefont {J.}~\bibnamefont {Candy}},
  \bibinfo {author} {\bibfnamefont {Z.}~\bibnamefont {Wang}}, \bibinfo {author}
  {\bibfnamefont {C.}~\bibnamefont {Luna}}, \bibinfo {author} {\bibfnamefont
  {V.}~\bibnamefont {Izzo}}, \bibinfo {author} {\bibfnamefont {B.}~\bibnamefont
  {Grierson}}, \bibinfo {author} {\bibfnamefont {P.}~\bibnamefont {Snyder}},
  \bibinfo {author} {\bibfnamefont {C.}~\bibnamefont {Holland}}, \bibinfo
  {author} {\bibfnamefont {J.}~\bibnamefont {Penna}}, \bibinfo {author}
  {\bibfnamefont {G.}~\bibnamefont {Lu}}, \bibinfo {author} {\bibfnamefont
  {P.}~\bibnamefont {Raum}}, \bibinfo {author} {\bibfnamefont {A.}~\bibnamefont
  {McCubbin}}, \bibinfo {author} {\bibfnamefont {D.}~\bibnamefont {Orlov}},
  \bibinfo {author} {\bibfnamefont {E.}~\bibnamefont {Belli}}, \bibinfo
  {author} {\bibfnamefont {N.}~\bibnamefont {Ferraro}}, \bibinfo {author}
  {\bibfnamefont {R.}~\bibnamefont {Prater}}, \bibinfo {author} {\bibfnamefont
  {T.}~\bibnamefont {Osborne}}, \bibinfo {author} {\bibfnamefont
  {A.}~\bibnamefont {Turnbull}},\ and\ \bibinfo {author} {\bibfnamefont
  {G.}~\bibnamefont {Staebler}},\ }\bibfield  {title} {\bibinfo {title}
  {Integrated modeling applications for tokamak experiments with {OMFIT}},\
  }\href {http://iopscience.iop.org/article/10.1088/0029-5515/55/8/083008/meta}
  {\bibfield  {journal} {\bibinfo  {journal} {Nuclear Fusion}\ }\textbf
  {\bibinfo {volume} {55}},\ \bibinfo {pages} {083008} (\bibinfo {year}
  {2015})}\BibitemShut {NoStop}%
\end{thebibliography}%

\end{document}